\documentclass[%
 aip,
 amsmath,amssymb,
 reprint,%
]{revtex4-1}
\usepackage{amsthm}
\usepackage{graphicx}
\usepackage{dcolumn}
\usepackage{bm}

\usepackage[utf8]{inputenc}
\usepackage[T1]{fontenc}
\usepackage{mathptmx}
\usepackage{etoolbox}
\usepackage{subcaption}
\usepackage[utf8]{inputenc}  
\usepackage[T1]{fontenc}     
\usepackage{textcomp}        
\usepackage{mathrsfs}
\newtheorem{theorem}{Theorem}

\newtheorem{remark}{Remark}
\newtheorem{assumption}{Assumption}

\makeatletter
\def\@email#1#2{%
 \endgroup
 \patchcmd{\titleblock@produce}
  {\frontmatter@RRAPformat}
  {\frontmatter@RRAPformat{\produce@RRAP{*#1\href{mailto:#2}{#2}}}\frontmatter@RRAPformat}
  {}{}
}%
\makeatother
\begin{document}

\preprint{AIP/123-QED}

\title[]{Hybrid Compact Least-Squares and Central Weighted Essentially Non-Oscillatory Schemes for Hyperbolic Conservation Laws on Structured Curvilinear Grids}
\author{Jianhua Pan}
 \altaffiliation[Also at ]{Key Laboratory of Impact and Safety Engineering, Ministry of Education, Ningbo University, Ningbo 315211, China}
 \email{panjianhua@nbu.edu.cn}
\author{Luxin Li}%
 \altaffiliation[Also at ]{Key Laboratory of Impact and Safety Engineering, Ministry of Education, Ningbo University, Ningbo 315211, China}
\affiliation{ 
Zhejiang Provincial Engineering Research Center for the Safety of Pressure Vessel and Pipeline, Ningbo University, Ningbo, 315211, China
}%
\author{Ji Yin}%
 \altaffiliation[Also at ]{Key Laboratory of Impact and Safety Engineering, Ministry of Education, Ningbo University, Ningbo 315211, China}
\affiliation{ 
Zhejiang Provincial Engineering Research Center for the Safety of Pressure Vessel and Pipeline, Ningbo University, Ningbo, 315211, China
}%

\author{Wei-Gang Zeng}
\affiliation{%
Academy for Advanced Interdisciplinary Studies, Northeast Normal University, Changchun, 130024, China
}%


\begin{abstract}
  A series of third- and fifth-order hybrid compact least-squares central weighted essentially non-oscillatory schemes are proposed and applied to curvilinear structured grids for the finite volume method. In smooth regions, compact least-squares schemes based on interfacial differences of derivatives are utilized to keep the high resolution of multiscale structures, whereas in discontinuous regions, central weighted essentially non-oscillatory schemes are included to enable the oscillation free shock capturing capability of the hybrid schemes. Free parameters in the proposed compact least-squares schemes are firstly optimized to reach a broad range of resolved bandwidths with different levels of dissipation. In addition, a shock detector proposed in our previous work is introduced and validated to be not only able to detect smooth first-order extrema but also robust in detecting discontinuities. Through the solution of block tridiagonal reconstruction linear systems, the resulting schemes can give an explicit polynomial for each control volume and are efficient with an acceptable computational overhead when compared to the pure central weighted essentially non-oscillatory schemes. Eventually, benchmarks including one-dimensional and two-dimensional, linear and nonlinear, inviscid and viscous problems on uniform and nonuniform curvilinear grids demonstrate the proposed schemes' promising applicability in compressible flows which have both multiscale structures and discontinuities. 
\end{abstract}

\maketitle

\section{Introduction}
Direct numerical simulations of compressible flows pose two challenges for the underlying numerical schemes. The first one is that the numerical schemes must resolve a broad range of bandwidths in Fourier space; the second is that they should robustly capture the discontinuities and suppress the unnecessary numerical oscillations stemming from shocks or contact discontinuities. 

The work aimed at broadening the bandwidth of resolved wavenumber in Fourier space, i.e., improving the resolution of the numerical scheme in resolving multiscale flow structures, originated from the seminal 1993 paper by Tam and Webb\cite{tam1993dispersion}. In Tam and Webb's paper \cite{tam1993dispersion}, the dispersion and dissipation properties of a seven-point finite difference scheme were optimized by minimizing the integrated error of dispersion in Fourier space. Tam and Webb's work established a general framework for subsequent high-resolution schemes \cite{cheong2001grid,sun2011class,sun2014sixth,zeng2025high,li2023scale,sengupta2009new,popescu2005finite,wang2013low}. On the other hand, Lele \cite{lele1992compact} proposed a series of compact finite difference schemes with superior spectral properties compared to non-compact methods. The idea of compact finite difference schemes is to relate the variable derivatives to variable values through a sparse matrix where the non-diagonal elements are potentially non-zero. Lele's coefficients were derived for uniform grids. When applied to non-uniform curvilinear grids, two approaches exist. The first approach\cite{zhong2003high,spotz1998formulation,ge2001high,wang2023efficient} maps differential equations from the physical domain to a uniform computational domain, where compact schemes for uniform grids are applied. The second approach bypasses coordinate transformation, instead deriving sufficient and necessary conditions for compact schemes to achieve a certain prior order of accuracy via Taylor expansion. Coefficients are determined using the method of undetermined coefficients \cite{gamet1999compact}. Further improvements include the utilization of the interpolation function which was more robust and stable than the method of undermined coefficients \cite{shukla2007very,shukla2005derivation}. The compact finite difference schemes based on coordinate transformation require that the mesh should be smooth enough, otherwise, the numerical accuracy would be significantly reduced \cite{gamet1999compact}. Additionally, some compact finite difference schemes on non-uniform grids \cite{gamet1999compact,shukla2007very,shukla2005derivation} lack the property of conservation, risking non-physical solutions for shock-related problems. 

Meanwhile, the finite volume method on curvilinear grids is more robust and genuinely conservative for hyperbolic conservation laws. Gaitonde and Shang \cite{gaitonde1997optimized} and Kobayashi \cite{kobayashi1999class} were pioneers who extended compact schemes from finite difference methods to the finite volume framework using primitive functions. Pereira et al. \cite{pereira2001fourth} applied the compact finite volume method to incompressible Navier-Stokes equations. When dealing with non-uniform curvilinear grids, either coordinate transformation like the finite difference schemes \cite{piller2008compact} or the accuracy conditions in physical domain \cite{fosso2010curvilinear,lacor2004finite} can be incorporated.

In all the aforementioned compact finite difference or finite volume schemes, variable derivatives of different orders are solved through distinct sparse linear systems. Beyond these, alternative compact finite volume schemes were proposed by Wang and Ren \cite{wang2015accurate}, Huang et al. \cite{huang2018high,huang2022adaptive} and Wang et al. \cite{wang2016compact1, wang2016compact2}. In these alternative compact schemes,  explicit polynomials for each control volume are first assumed and then solved by either spline reconstruction or compact least-squares (CLS) reconstruction. The reconstruction linear system is block tridiagonal, allowing simultaneous computation of all derivatives. In the spline reconstruction \cite{wang2015accurate,huang2018high,huang2022adaptive}, polynomials are continuous across cell interfaces in smooth regions and artificial viscosity is required to stable the schemes. In contrast, CLS-reconstructed polynomials are discontinuous across cell interfaces, which is the main difference between spline reconstruction and CLS reconstruction. In the work of Wang et al. \cite{wang2016compact1}, the CLS reconstruction was applied to one-dimensional (1D) cases and extended later to unstructured grids \cite{wang2016compact2}. 

Although the compact schemes can capture a broad range of bandwidths for problems with multiscale structures, appropriate treatments near discontinuities must be introduced for robust simulations of compressible flows. Three different approaches exist. In the first approach, for example in the work \cite{cockburn1994nonlinearly,adams1996high,pirozzoli2002conservative,huang2018high,huang2022adaptive}, different schemes like TVD scheme, ENO scheme or WENO scheme are applied in discontinuous regions with a switch function. In the second approach, lower-order coefficients which automatically select the least oscillating stencils are utilized in the discontinuous region \cite{deng1997compact,guo2016fifth} for the compact schemes. This approach can be treated as an extension of WENO concept to compact schemes. In the third approach, i.e., the approach in CLS schemes \cite{wang2016compact1, wang2016compact2}, a separate gradient limiter is applied subsequently after the reconstruction step which does not incorporate nonlinear mechanism near discontinuities. 

This paper focuses on the CLS schemes and proposes three key improvements. Firstly, the free parameters in the CLS reconstruction are optimized such that the resulting schemes tend to resolve the wavenumber as broad as possible while maintaining different levels of dissipation. Specially, it is the first time to the authors' knowledge that the fifth-order CLS schemes are optimzied and applied to curvilinear grids. Secondly, to improve the resolution of overall method, instead of using a gradient limiter \cite{wang2016compact1, wang2016compact2}, CLS schemes are hybridized with central weighted essentially non-oscillatory (CWENO) schemes. Thirdly, a novel switch function which is proposed in our previous work \cite{PANWCLS} is introduced and validated here. The switch function denoted as shock detector not only can capture the first-order extrema in smooth regions but also is robust in discontinuous regions.

The remainder of the paper is organized as follows: In Section \ref{sec:cls}, the fundamental principles of compact least squares reconstruction is reviewed; In Section \ref{sec:optimization}, the free parameters in CLS schemes are optimized to maximize the wavenumber resolution. In Section \ref{sec:hybrid}, the shock detector proposed in our previous work \cite{PANWCLS} is introduced and the CLS schemes are hybridized with CWENO schemes. In Section \ref{sec:examples}, a series of benchmarks including one-dimensional (1D) and two-dimensional (2D), linear and nonlinear, inviscid and viscous problems on uniform and nonuniform curvilinear grids are carried to validate the high resolution and high robustness of our proposed schemes.

\section{Compact least-squares reconstruction \label{sec:cls}}
The fundamental principle of CLS reconstruction \cite{wang2016compact1} is briefly reviewed in this section. Consider the following 1D linear convection equation,
\begin{equation}
  \frac{\partial u}{\partial t} + \frac{\partial f(u)}{\partial x} = 0, \label{eq:1dlinear}
\end{equation}
where $f(u) = a u$ with constant wavespeed $a$. The computational domain $\Omega = [x_l, x_r]$ is discretized into $N$ non-overlapping control volumes $\Omega_i,\,\, i=1,2,\cdots, N$. $\Omega_i = [x_{i-1/2}, x_{i+1/2}]$. The center and the width of $\Omega_i$ are given by $x_i = (x_{i-1/2} + x_{i+1/2})/2$ and $\Delta x_i = x_{i+1/2}-x_{i-1/2}$, respectively.
  In finite volume method, the integration of Eq. (\ref{eq:1dlinear}) over $\Omega_i$ results in
  \begin{equation}
    \frac{\partial \overline{u}_i}{\partial t} + \frac{1}{\Delta x_i}\left(f_{i+1/2}-f_{i-1/2}\right) = 0,
    \label{eq:1dlinearsemidiscrete}
  \end{equation}
where $\overline{u}_i = \frac{1}{\Delta x_i}\int_{x_{i-1/2}}^{x_{i+1/2}}{u\,\mathrm{d}x}$.

The key step in finite volume method is the reconstruction from cell averaged values $\overline{u}_i$ to point values at each interface, i.e., $u_{i+1/2}$ at $x_{i+1/2},\,\,i=0, 1,2,\cdots, N$. In CLS method, a $k$-th order polynomial is assumed for control volume $\Omega_i$ as
\begin{equation}
P_i(x) = \overline{u}_i + \sum_{j=1}^{k}{a_i^{(j)} \phi_i^{(j)}(x)},
\end{equation}
where $\phi_i^{(j)}$ is the $j$-th zero-mean basis function defined in $\Omega_i$ as
\begin{equation}
\phi_i^{(j)}(x)  =   \left(\frac{x-x_i}{\Delta x_i}\right)^j - \frac{1}{\Delta x_i}\int_{x_{i-1/2}}^{x_{i+1/2}}{\left(\frac{x-x_i}{\Delta x_i}\right)^j\,\mathrm{d}x}, 
\end{equation}
The unknowns $a_i^{(j)},\,\,i=1,2,\cdots, N, \,j =1,2,\cdots, k$ are determined through a series of local least-squares problem. The least-squares problem defined on cell $\Omega_i$ constitutes the squared differences of all the derivatives between $P_i(x)$ and the polynomials on neighboring cells $\Omega_{i\pm 1}$ at interfaces $x_{i\pm 1/2}$, which is different from the least-squares problem proposed in the work of Wang et al. \cite{wang2016compact1} but very similar as the work in Wang et al.\cite{wang_compact_2017} In detail, the least-squares problem defined on control volume $\Omega_i$ in our work is,
\begin{equation}
  \begin{aligned}
  I_i & = \sum_{j=0}^{k}{W_{j}l_{i,-1/2}^{2j}\left(\left.\frac{\mathrm{d}^j P_i}{\mathrm{d}x^j}\right|_{x_{i-1/2}} -\left.\frac{\mathrm{d}^j P_{i-1}}{\mathrm{d}x^j} \right|_{x_{i-1/2}}\right)^2}\\
  & + \sum_{j=0}^{k}{W_{j} l_{i,1/2}^{2j}\left(\left.\frac{\mathrm{d}^j P_i}{\mathrm{d}x^j}\right|_{x_{i+1/2}} -\left.\frac{\mathrm{d}^j P_{i+1}}{\mathrm{d}x^j} \right|_{x_{i+1/2}}\right)^2},
  \end{aligned}
  \label{eq:clsmywork}
\end{equation}
where $W_j,\,\,j=0,1,\cdots,k$ are free parameters that can be optimized; $l_{i,-1/2}$ and $l_{i, 1/2}$ are the characteristic lengths which make the terms in $I_i$ dimensionless.  In this work, we choose $l_{i, -1/2} = l_{i, 1/2} = \Delta x_i$.
On the other hand, the least-squares problem defined in the work of Wang \cite{wang2016compact1} is as follows,
\begin{equation}
  \begin{aligned}
  I_i & = \sum_{\substack{m \in\\  \left\{i-1, i+1\right\}} } \sum_{j=0}^{k}{W_{j}\Delta x_i^{2j}\cdot} \\
  & \quad \quad \quad \left( \frac{1}{\Delta x_{m}}\int_{\Omega_{m}}{\frac{\mathrm{d}^j P_i}{\mathrm{d}x^j} \mathrm{d}x}  -  \frac{1}{\Delta x_{m}}\int_{\Omega_{m}}{\frac{\mathrm{d}^j P_{m}}{\mathrm{d}x^j} \mathrm{d}x}  \right)^2.
  \end{aligned}\label{eq:clswang}
\end{equation}
Without loss of generality, $W_0$ in Eqs. (\ref{eq:clsmywork}) and (\ref{eq:clswang}) is assigned to be 1. For ease of presentation, the CLS scheme based on Eq. (\ref{eq:clsmywork}) is denoted as CLS-ID and the CLS scheme based on Eq. (\ref{eq:clswang}) is denoted as CLS-CD.

Section \ref{sec:optimization} demonstrates that the dispersion error in CLS-ID schemes using interfacial differences as in Eq. (\ref{eq:clsmywork}) is significantly smaller than in CLS-CD schemes employing averaged derivative differences across neighboring control volumes  as in Eq. (\ref{eq:clswang}).

Taking derivatives of Eq. (\ref{eq:clsmywork}) to $a_i^{(j)}, \,\,j=1,2,\cdots, k$ results in a reconstruction linear system on control volume $\Omega_i$ in the form as
\begin{equation}
\mathbf{M}_i^{(-1)} \overrightarrow{a}_{i-1}
+\mathbf{M}_i^{(0)} \overrightarrow{a}_{i}
+\mathbf{M}_i^{(1)} \overrightarrow{a}_{i+1} = \overrightarrow{b}_i, \label{eq:localcls}
\end{equation}
where $\overrightarrow{a}_i = \left[a_i^{(1)},a_i^{(2)},\cdots, a_i^{(k)}\right]^T$ is a column vector constituting the unknowns of $\Omega_i$.

 In this work, the third- and fifth-order CLS schemes are focused. For the third-order schemes,
 \begin{equation}
 \mathbf{M}_i^{(-1)} =
\begin{bmatrix}
 -\frac{4 W_1-h_{i-1} }{4 h_{i-1} } & -\frac{12 W_1-h_{i-1} }{12 h_{i-1} } \\
 \frac{12 W_1-h_{i-1} }{12 h_{i-1} } & -\frac{-36 h_{i-1} W_1+144  W_2+h_{i-1}^2 }{36 h_{i-1}^2 } \\
\end{bmatrix}
 ,\label{eq:m-13rd}
 \end{equation}

 \begin{equation}
 \mathbf{M}_i^{(0)} =
 \begin{bmatrix}
  \frac{1}{2} (1+4 W_1) & 0\\
  0& \frac{1}{18} (1+36 W_1+144 W_2)
 \end{bmatrix}.
 \end{equation}

 \begin{equation}
 \mathbf{M}_i^{(1)} =
\begin{bmatrix}
 -\frac{4  W_1- h_{i+1}}{4 h_{i+1}} & \frac{12  W_1- h_{i+1}}{12 h_{i+1}} \\
 -\frac{12  W_1- h_{i+1}}{12  h_{i+1}} & -\frac{144  W_2-36 h_{i+1}  W_1+h_{i+1}^2 }{36 h_{i+1}^2} \\
\end{bmatrix}
 ,
 \end{equation}

 \begin{equation}
  \overrightarrow{b}_i = 
  \left[
\begin{array}{c}
 \frac{1}{2} \left(\overline{u}_{i+1}-\overline{u}_{i-1}\right) \\
 \frac{1}{6} \left(\overline{u}_{i-1}-2 u_{i}+\overline{u}_{i+1}\right) \\
\end{array}
\right].
 \end{equation}
And for the fifth-order schemes,
\begin{equation}
  \begin{aligned}
 \mathbf{M}_i^{(-1)} & =
\left[
\begin{array}{cc}
 \frac{1}{4}-\frac{W_1}{h_{i-1}} & \frac{1}{12}-\frac{W_1}{h_{i-1}} \\
 \frac{W_1}{h_{i-1}}-\frac{1}{12} & \frac{W_1}{h_{i-1}}-\frac{4 W_2}{h_{i-1}^2}-\frac{1}{36} \\
 \frac{1}{16}-\frac{3 W_1}{4 h_{i-1}} & -\frac{3 W_1}{4 h_{i-1}}+\frac{6 W_2}{h_{i-1}^2}+\frac{1}{48} \\
 \frac{W_1}{2 h_{i-1}}-\frac{1}{40} & \frac{W_1}{2 h_{i-1}}-\frac{6 W_2}{h_{i-1}^2}-\frac{1}{120} 
\end{array}
\right. \\
&
\begin{array}{c}
  \frac{1}{16}-\frac{3 W_1}{4 h_{i-1}}  \\
  \frac{3 W_1}{4 h_{i-1}}-\frac{6 W_2}{h_{i-1}^2}-\frac{1}{48}  \\
  -\frac{9 W_1}{16 h_{i-1}}+\frac{9 W_2}{h_{i-1}^2}-\frac{36 W_3}{h_{i-1}^3}+\frac{1}{64}  \\
  \frac{3 W_1}{8 h_{i-1}}+\frac{72 W_3}{h_{i-1}^3}-\frac{9 W_2}{h_{i-1}^2}-\frac{1}{160}  \\
\end{array}\\
&
\left.
\begin{array}{c}
  \frac{1}{40}-\frac{W_1}{2 h_{i-1}} \\
  \frac{W_1}{2 h_{i-1}}-\frac{6 W_2}{h_{i-1}^2}-\frac{1}{120} \\
  -\frac{3 W_1}{8 h_{i-1}}+\frac{9 W_2}{h_{i-1}^2}-\frac{72 W_3}{h_{i-1}^3}+\frac{1}{160} \\
  \frac{W_1}{4 h_{i-1}}+\frac{144 W_3}{h_{i-1}^3}-\frac{9 W_2}{h_{i-1}^2}-\frac{576 W_4}{h_{i-1}^4}-\frac{1}{400} \\
\end{array}
\right],
\end{aligned}
\end{equation}

\begin{equation}
  \begin{aligned}
 \mathbf{M}_i^{(0)}  & =
 \left[
\begin{array}{cc}
 2 W_1+\frac{1}{2} & 0  \\
 0 & 2 W_1+8 W_2+\frac{1}{18}   \\
 \frac{3 W_1}{2}+\frac{1}{8}   \\
 0 & W_1+12 W_2+\frac{1}{60} 
\end{array} 
\right. \\
\vspace{10pt}
& \left.
\begin{array}{c}
  \frac{3 W_1}{2}+\frac{1}{8}  \\
  0 \\
  \frac{9 W_1}{8}+18 W_2+72 W_3+\frac{1}{32}  \\
  0 \\
\end{array} \right.
 \\
\vspace{10pt}
& \left.
\begin{array}{cc}
   0 \\
   W_1+12 W_2+\frac{1}{60} \\
   0 \\
   \frac{W_1}{2}+18 W_2+288 W_3+1152 W_4+\frac{1}{200} \\
\end{array} \right],
  \end{aligned}
\end{equation}

\begin{equation}
  \begin{aligned}
 \mathbf{M}_i^{(1)}  & =
\left[
\begin{array}{cc}
 \frac{1}{4}-\frac{W_1}{h_{i+1}} & \frac{W_1}{h_{i+1}}-\frac{1}{12}  \\
 \frac{1}{12}-\frac{W_1}{h_{i+1}} & \frac{W_1}{h_{i+1}}-\frac{4 W_2}{h_{i+1}^2}-\frac{1}{36}  \\
 \frac{1}{16}-\frac{3 W_1}{4 h_{i+1}} & \frac{3 W_1}{4 h_{i+1}}-\frac{6 W_2}{h_{i+1}^2}-\frac{1}{48}  \\
 \frac{1}{40}-\frac{W_1}{2 h_{i+1}} & \frac{W_1}{2 h_{i+1}}-\frac{6 W_2}{h_{i+1}^2}-\frac{1}{120}  \\
\end{array}
\right. \\
  & 
\left.
\begin{array}{c}
  \frac{1}{16}-\frac{3 W_1}{4 h_{i+1}}  \\
  -\frac{3 W_1}{4 h_{i+1}}+\frac{6 W_2}{h_{i+1}^2}+\frac{1}{48}  \\
  -\frac{9 W_1}{16 h_{i+1}}+\frac{9 W_2}{h_{i+1}^2}-\frac{36 W_3}{h_{i+1}^3}+\frac{1}{64}  \\
  -\frac{3 W_1}{8 h_{i+1}}+\frac{9 W_2}{h_{i+1}^2}-\frac{72 W_3}{h_{i+1}^3}+\frac{1}{160}  \\
\end{array}
\right. \\
& \left.
\begin{array}{c}
   \frac{W_1}{2 h_{i+1}}-\frac{1}{40} \\
   \frac{W_1}{2 h_{i+1}}-\frac{6 W_2}{h_{i+1}^2}-\frac{1}{120} \\
   \frac{3 W_1}{8 h_{i+1}}+\frac{72 W_3}{h_{i+1}^3}-\frac{9 W_2}{h_{i+1}^2}-\frac{1}{160} \\
   \frac{W_1}{4 h_{i+1}}+\frac{144 W_3}{h_{i+1}^3}-\frac{9 W_2}{h_{i+1}^2}-\frac{576 W_4}{h_{i+1}^4}-\frac{1}{400} \\
\end{array}
\right],\\
  \end{aligned}
\end{equation}
\begin{equation}
  \overrightarrow{b}_i = 
\left[
\begin{array}{c}
 \frac{1}{2} \left(\overline{u}_2-\overline{u}_0\right) \\
 \frac{1}{6} \left(\overline{u}_0-2 \overline{u}_1+\overline{u}_2\right) \\
 \frac{1}{8} \left(\overline{u}_2-\overline{u}_0\right) \\
 \frac{1}{20} \left(\overline{u}_0-2 \overline{u}_1+\overline{u}_2\right) \\
\end{array}
\right].
 \label{eq:b5th}
\end{equation}

In Eqs. (\ref{eq:m-13rd}-\ref{eq:b5th}), 
\begin{equation}
  h_{i-1} = \frac{\Delta x_{i-1}}{\Delta x_i}, \,\,
  h_{i+1} = \frac{\Delta x_{i+1}}{\Delta x_i}. 
\end{equation}

Coupling the reconstruction linear systems for all control volumes, a block tridiagonal linear system is obtained for the computational domain as
\begin{equation}
\mathbb{M} \overrightarrow{\alpha} =\overrightarrow{\chi}, \label{eq:clsglobal}
\end{equation}
where 
$\overrightarrow{\alpha} = \left[\overrightarrow{a}_1^T,\overrightarrow{a}_2^T,\cdots,\overrightarrow{a}_N^T\right]^T$ and $\overrightarrow{\chi} = \left[\overrightarrow{b}_1^T,\overrightarrow{b}_2^T,\cdots,\overrightarrow{b}_N^T\right]^T$.

It should be noted that 
if $l_{i, -1/2} = \frac{\Delta x_i + \Delta x_{i-1}}{2}$ and $l_{i, 1/2} = \frac{\Delta x_i + \Delta x_{i+1}}{2}$ in Eq. (\ref{eq:clsmywork}), the $\mathbb{M}$ matrix in Eq. (\ref{eq:clsglobal}) becomes symmetric, making the CLS-ID scheme equivalent to the variational finite volume (VFV) method proposed by Wang et al.\cite{wang_compact_2017} In other words, the CLS-FV method redues to the VFV method if the terms regarding two neighboring control volumes are the same in the neighboring least-squares problems. For consistency, we collectively refer to both CLS-FV and VFV schemes as CLS-FV schemes in this work.
If a periodic boundary condition is applied, $\mathbb{M}$ is a cyclic block tridiagonal matrix as
\begin{equation}
  \begin{bmatrix}
 \mathbf{M}_1^{(0)} &  \mathbf{M}_1^{(1)}& \mathbf{0} & \cdots &\cdots& \mathbf{0}& \mathbf{M}_1^{(-1)}  \\
 \mathbf{M}_2^{(-1)} & \mathbf{M}_2^{(0)}& \mathbf{M}_2^{(1)}& \mathbf{0} &\cdots& \cdots& \mathbf{0} \\
 \mathbf{0} & \mathbf{M}_3^{(-1)} &  \mathbf{M}_3^{(0)}& \mathbf{M}_3^{(1)} & \mathbf{0} & \cdots&  \mathbf{0} \\
 \vdots & \vdots & \vdots & \vdots &\vdots & \ddots & \vdots \\
 \mathbf{M}_{N}^{(1)} & 0 &\cdots&\cdots&0&\mathbf{M}_{N}^{(-1)} & \mathbf{M}_{N}^{(0)}
  \end{bmatrix}.
\end{equation}

If a Dirichlet boundary condition is applied, for instance, at the left side of the computational domain, the least-squares problem for $\Omega_1$ is adjusted as 
\begin{equation}
  \begin{aligned}
  I_1 & = {\lambda \left(\left.P_1\right|_{x_{1/2}} -\left. u \right|_{x_{1/2}}\right)^2}\\
  & + \sum_{j=0}^{k}{W_{j}\Delta x_i^{2j}\left(\left.\frac{\mathrm{d}^j P_i}{\mathrm{d}x^j}\right|_{x_{i+1/2}} -\left.\frac{\mathrm{d}^j P_{i+1}}{\mathrm{d}x^j} \right|_{x_{i+1/2}}\right)^2},
  \end{aligned}
  \label{eq:clsdirechlet}
\end{equation}
where $\left.u\right|_{x_{1/2}}$ is the desired boundary value of $u$ and $\lambda$ is the weight of the boundary condition. $\lambda \rightarrow \infty$ corresponds to a constrained least-squares problem. 

Similarly, if a Neumann boundary condition is applied at the left side of the computational domain, the least-squares problem for $\Omega_1$ can be modified as
\begin{equation}
  \begin{aligned}
  I_1 & = {\lambda \left(\left.\frac{\mathrm{d}P_1}{\mathrm{d}x}\right|_{x_{1/2}} -\left. \frac{\mathrm{d}u}{\mathrm{d}x} \right|_{x_{1/2}}\right)^2}\\
  & + \sum_{j=0}^{k}{W_{j}\Delta x_i^{2j}\left(\left.\frac{\mathrm{d}^j P_i}{\mathrm{d}x^j}\right|_{x_{i+1/2}} -\left.\frac{\mathrm{d}^j P_{i+1}}{\mathrm{d}x^j} \right|_{x_{i+1/2}}\right)^2},
  \end{aligned}
  \label{eq:clsneumann}
\end{equation}
where $\left. \frac{\mathrm{d}u}{\mathrm{d}x} \right|_{x_{1/2}}$ is the desired first-order derivative of $u$ at the boundary.

Other complex boundary conditions can be applied by adding penalty terms like in Eqs. (\ref{eq:clsdirechlet}) and (\ref{eq:clsneumann}). 

For the finite volume method, the boundary condition is strictly applied when calculating the flux terms at boundary faces. Thus, a weak constrained boundary condition during reconstruction is feasible since the exact boundary conditions are enforced later when calculating the flux terms. In this work, $\lambda$ is chosen to be 1. For either Dirichlet boundary or Neumann boundary, $\mathbb{M}$ is a block tridiagonal matrix which can be solved efficiently.

After the solution of Eq. (\ref{eq:clsglobal}), cell-wise smooth polynomials for all control volumes are obtained, and the flux $f_{i+1/2}$ in Eq. (\ref{eq:1dlinearsemidiscrete}) is calculated by
\begin{equation}
  f_{i+1/2} = \frac{a}{2}\left(u_{i+1/2}^L + u_{i+1/2}^R\right) - \frac{|a|}{2}\left(u_{i+1/2}^R - u_{i+1/2}^L\right),\label{eq:fluxscalar}
\end{equation}
where $u_{i+1/2}^L = P_{i}(x_{i+1/2})$ and $u_{i+1/2}^R = P_{i+1}(x_{i+1/2})$.

Eventually, a multi-stage Runge-Kutta method, e.g., the third-order strong stability preserving Runge-Kutta method (SSP-RK3) \cite{gottlieb2011strong,PAN202324} is utilized to advance Eq. (\ref{eq:1dlinearsemidiscrete}) from time $t^n$ to $t^{n+1}$.
 
\section{Optimization of dispersion and dissipation properties\label{sec:optimization}}
To improve the resolution of the CLS schemes, the free parameters in Eq. (\ref{eq:clsmywork}) are optimized in this section.

Firstly, let us derive the modified non-dimensional wavenumber for the semi-discretized CLS  finite volume method of Eq. (\ref{eq:1dlinearsemidiscrete}). Assuming uniform grids over the computational domain $\Omega$ and a harmonic distribution of $u(x,t)=  A e^{Ikx}$, where $I=\sqrt{-1}$, the following equations can be obtained,
  \begin{align}
\overline{u}_{i-1} & = \frac{1}{\Delta x}\int_{x_i-3\Delta x/2}^{x_i-\Delta x/2}{u(x) \mathrm{d}x} = 2 \frac{A}{\kappa} e^{-I \kappa}\sin \left(\frac{\kappa }{2}\right) e^{Ikx_i},\label{eq:averagedu-1}\\
\overline{u}_{i} & = \frac{1}{\Delta x}\int_{x_i-\Delta x/2}^{x_i+\Delta x/2}{u(x) \mathrm{d}x} = 2 \frac{A}{\kappa} \sin \left(\frac{\kappa }{2}\right) e^{Ikx_i},\label{eq:averagedu}\\
\overline{u}_{i+1} & = \frac{1}{\Delta x}\int_{x_i+\Delta x/2}^{x_i+3\Delta x/2}{u(x) \mathrm{d}x} = 2 \frac{A}{\kappa} e^{+I \kappa}\sin \left(\frac{\kappa }{2}\right) e^{Ikx_i},\label{eq:averagedu+1}
  \end{align}
  where $\kappa = k\Delta x$ is the non-dimensional wavenumber.

Meanwhile, assuming a periodic boundary condition, the unknowns on $\Omega_{i-1}, \Omega_{i}$ and $\Omega_{i+1}$ satisfy
\begin{equation}
  \begin{aligned}
  \overrightarrow{a}_{i-1} & = \overrightarrow{a}_{i} e^{-I\kappa}, \\
  \overrightarrow{a}_{i+1} & = \overrightarrow{a}_{i} e^{I\kappa},
  \end{aligned}
  \label{eq:unknowns}
\end{equation}
Substituting Eqs. (\ref{eq:averagedu-1}-\ref{eq:unknowns}) into Eq. (\ref{eq:localcls}), the unknowns $\overrightarrow{a}_i$ can be solved. In turn $\overrightarrow{a}_{i-1}$ and $\overrightarrow{a}_{i+1}$ can be obtained by Eq. (\ref{eq:unknowns}). 
For the third-order CLS-ID scheme,
\begin{equation}
  \overrightarrow{a}_i = \left[\frac{X_1}{X_2}, \frac{X_3}{X_4}\right]^T,
\end{equation}
where
\begin{equation}
\begin{aligned}
 X_1  =&\,\, 6 A \left(-1+e^{2 i \kappa }\right) \sin \left(\frac{\kappa }{2}\right) \left(\left(4 e^{i \kappa }+e^{2 i \kappa }+1\right) W_1\right.\\&\left.-6 \left(-1+e^{i \kappa }\right)^2 W_2\right) e^{Ikx_i},\\
 X_2  =&\,\, \kappa  \left(W_1 \left(\left(4 e^{i \kappa }+e^{2 i \kappa }+1\right)^2+36 \left(-1+e^{i \kappa }\right)^4 W_2\right)\right.\\&\left.-9 \left(-1+e^{2 i \kappa }\right)^2 W_2\right), \\
X_3  =&\,\, 6 A \left(-1+e^{i \kappa }\right)^2 \left(4 e^{i \kappa }+e^{2 i \kappa }+1\right) W_1 \sin \left(\frac{\kappa }{2}\right)e^{Ikx_i},\\
X_4 = &\,\,X_2.
\\
\end{aligned}
\end{equation}
For the fifth-order schemes, the detailed expression of $\overrightarrow{a}_i$ can be derived through Mathematica, and is not presented here for brevity.

Substituting the interfacial values $u_{i-1/2}^L$, $u_{i-1/2}^R$, $u_{i+1/2}^L$ and $u_{i+1/2}^R$ into the flux of Eq. (\ref{eq:fluxscalar}), which is in turn substituted into Eq. (\ref{eq:1dlinearsemidiscrete}),  results in
\begin{equation}
  \begin{aligned}
  \frac{\partial \overline{u}_i}{\partial t} & = \frac{\partial A}{\partial t} \frac{2}{\kappa} \sin(\frac{\kappa}{2}) e^{Ikx_i} \\
  & = a\frac{(P_{i-1}(x_{i-1/2}) - P_{i}(x_{i+1/2}))}{\Delta x}.
  \end{aligned}
\end{equation}

Furthermore, we have
\begin{equation}
  \frac{\partial A}{\partial t} =  a\frac{(P_{i-1}(x_{i-1/2}) - P_{i}(x_{i+1/2}))\kappa}{2 \sin(\kappa/2) e^{Ikx_i}\Delta x}.
  \label{eq:modifiedwavenumber}
\end{equation}

On the other hand, for the exact solution,
\begin{equation}
  \frac{\partial A }{\partial t} = -I k a A.
  \label{eq:wavenumber}
\end{equation}

Comparing Eqs. (\ref{eq:modifiedwavenumber})  and (\ref{eq:wavenumber}), the modified non-dimensional wavenumber $\kappa '$ is derived as
\begin{equation}
   \kappa' = -\frac{(P_{i-1}(x_{i-1/2}) - P_{i}(x_{i+1/2}))\kappa}{2 \sin(\kappa/2) e^{Ikx_i}I A}.
\end{equation}

For the third-order CLS-ID scheme based on interfacial differences as in Eq. (\ref{eq:clsmywork}), 
\begin{equation}
  \begin{aligned}
Re(\kappa ') & = \frac{-6 \sin(\kappa)\left(
  n_1 + n_2 \cos(\kappa) + n_3 \cos(2 \kappa)
\right)}{
  d_1 + d_2 \cos(\kappa) + d_3 \cos(2\kappa)
},\\
Im(\kappa ') & = \frac{576 W_1 W_2 \left(\sin^6(\kappa /2)\right)}{ d_1 + d_2 \cos(\kappa) + d_3 \cos(2\kappa)},
  \end{aligned}
  \label{eq:dispdissmywork3}
\end{equation}
where
\begin{align}
n_1 & = 2 \left(W_1 + 3W_2 + 9 W_1 W_2\right),\\
n_2 & = W_1 -6\left(1+4W_1\right) W_2,\\
n_3 & = 6 W_1 W_2,\\
d_1 & = -9\left(W_1 + W_2 + 12 W_1 W_2\right),\\
d_2 &= 8 W_1 \left(-1+18 W_2\right),\\
d_3 &=-\left(W_1 + 9\left(-1+4W_1\right)W_2\right).
\end{align}

In comparison, for the third-order CLS-CD scheme based on differences of averaged derivatives on neighboring control volumes of Eq. (\ref{eq:clswang}),
\begin{equation}
  \begin{aligned}
  Re(\kappa') & = \frac{n_4 \sin(\kappa) + n_5 \sin(2\kappa)+ n_6 \sin(3\kappa)}{d_4 + d_5 \cos(\kappa) + d_6 \cos(2\kappa)},\\
  Im(\kappa') & = \frac{n_7(n_8 + n_9 \cos(\kappa))}{d_4 + d_5 \cos(\kappa) + d_6 \cos(2\kappa)},
  \end{aligned}
  \label{eq:dispdisswang3}
\end{equation}
where
\begin{align}
  n_4 &= -4 W_1^2+\left(-5 W_2-\frac{157}{24}\right) W_1+\frac{1}{6} \left(-39 W_2-8\right),\\
  n_5 & =2 W_1^2+\left(4 W_2+\frac{4}{3}\right) W_1+4 W_2+\frac{1}{6},\\
  n_6 & = W_1 \left(-W_2-\frac{1}{24}\right)-\frac{W_2}{2},\\
  n_7&=\frac{2}{3} \sin ^4\left(\frac{\kappa }{2}\right),\\
  n_8 & =24 W_1^2+3 \left(8 W_2+5\right) W_1+12 W_2+2 ,\\
  n_9 & =-12 W_2-W_1 \left(24 W_2+1\right) ,\\
  d_4 & = -4 W_1^2-\left(6 W_2+5\right) W_1-4 W_2-1,\\
  d_5 & =4 W_1^2+\left(8 W_2+1\right) W_1+4 W_2,\\
  d_6 &= -2 W_1 W_2.
\end{align}

For the fifth-order CLS-ID scheme based on the interfacial differences of derivatives,
\begin{equation}
\begin{aligned}
Re(\kappa') &= \frac{ \left(
  \begin{array}{c}
  n_{10} \sin(\kappa) +n_{11} \sin(2 \kappa) +n_{12} \sin(3 \kappa) \\ 
  \quad\quad\quad +n_{13} \sin(4 \kappa) +n_{14} \sin(5 \kappa) 
  \end{array}
  \right)
}{
  \left(
    \begin{array}{c}
  d_7 + d_8 \cos (\kappa ) + d_{9} \cos (2 \kappa ) \\
  \quad\quad\quad + d_{10} \cos (3 \kappa )+ d_{11} \cos (4 \kappa )
    \end{array}
  \right)
  },\\
Im(\kappa') & = \frac{-3686400 W_1 W_2 W_3 W_4 \sin ^{10}\left(\frac{\kappa }{2}\right)}{ 
  \left(
    \begin{array}{c}
  d_7 + d_8 \cos (\kappa ) + d_{9} \cos (2 \kappa ) \\
  \quad\quad\quad + d_{10} \cos (3 \kappa )+ d_{11} \cos (4 \kappa )
    \end{array}
  \right)
  },
\end{aligned}
  \label{eq:dispdissmywork5}
\end{equation}
where
  \begin{align}
    n_{10}&=20 \left(5040 W_2 W_3 W_4+W_1 \left(960 W_3 W_4 \right.\right.\nonumber\\
    &\quad\quad\left.\left.+W_2 \left(340 W_4+W_3 \left(15120 W_4+169\right)\right)\right)\right),\\
    n_{11}&=-10 \left(10080 W_2 W_3 W_4+W_1 \left(1200 W_3 W_4\right.\right.\nonumber\\
    &\quad\quad\left.\left.+W_2 \left(140 W_4+W_3 \left(34560 W_4-163\right)\right)\right)\right),\\
    n_{12}&=60 W_2 \left(720 W_3 W_4+W_1 \left(W_3 \left(3240 W_4+3\right)\right.\right.\nonumber\\
    &\quad\quad\left.\left.-20 W_4\right)\right),\\
    n_{13}&=W_1 \left(240 W_3 W_4+W_2 \left(W_3 \left(1-11520 W_4\right)\right.\right.\nonumber\\
    &\quad\quad\left.\left.-20 W_4\right)\right)-7200 W_2 W_3 W_4,\\
    n_{14}&=7200 W_1 W_2 W_3 W_4,
  \end{align}
  \begin{align}
d_7 & = 5 \left(3600 W_2 W_3 W_4+W_1 \left(1840 W_3 W_4 \right.\right.\nonumber\\
 & \quad \quad \left.\left. +W_2 \left(505 W_4+W_3 \left(100800 W_4+571\right)\right)\right)\right), \\
d_8 &= -4 \left(3600 W_2 W_3 W_4+W_1 \left(25 \left(80 W_3 \right.\right.\right.\nonumber\\
&\quad \quad \left.\left.\left.+W_2 \left(8064 W_3-5\right)\right) W_4-871 W_2 W_3\right)\right),\\
d_9 &=4 \left(W_1 \left(202 W_2 W_3+25 \left(W_2 \left(4032 W_3-25\right)\right.\right.\right.\nonumber\\
&\quad \quad \left.\left.\left.-32 W_3\right) W_4\right)-3600 W_2 W_3 W_4\right), \\
d_{10}&=4 \left(3600 W_2 W_3 W_4+W_1 \left(13 W_2 W_3-25 \left(5 W_2\right.\right.\right.\nonumber\\
&\quad \quad \left.\left.\left.+16 \left(72 W_2-1\right) W_3\right) W_4\right)\right), \\
d_{11} &=W_1 \left(400 W_3 W_4+W_2 \left(W_3+25 \left(576 W_3-1\right) W_4\right)\right)\nonumber\\
&\quad \quad -3600 W_2 W_3 W_4.
  \end{align}

And for the fifth-order CLS-CD scheme based on the differences of averaged derivatives on neighboring cells,
\begin{equation}
\begin{aligned}
Re(\kappa') &= \frac{ \left(
  \begin{array}{c}
  n_{15} \sin(\kappa) +n_{16} \sin(2 \kappa) +n_{17} \sin(3 \kappa) \\ 
  \quad\quad\quad +n_{18} \sin(4 \kappa) +n_{19} \sin(5 \kappa) 
  \end{array}
  \right)
}{
  \left(
    \begin{array}{c}
  d_{12} + d_{13} \cos (\kappa ) + d_{14} \cos (2 \kappa ) \\
  \quad\quad\quad + d_{15} \cos (3 \kappa )+ d_{16} \cos (4 \kappa )
    \end{array}
  \right)
  },\\
Im(\kappa') & = \frac{
  \left(
    \begin{array}{c}
  n_{20} + n_{21} \cos (\kappa ) + n_{22} \cos (2 \kappa ) \\
  + n_{23} \cos (3 \kappa )+ n_{24} \cos (4 \kappa ) + n_{25} \cos (5 \kappa )
    \end{array}
  \right)
  }{ 
  \left(
    \begin{array}{c}
  d_{12} + d_{13} \cos (\kappa ) + d_{14} \cos (2 \kappa ) \\
  \quad\quad\quad + d_{15} \cos (3 \kappa )+ d_{16} \cos (4 \kappa )
    \end{array}
  \right)
  }.
\end{aligned}
  \label{eq:dispdisswang5}
\end{equation}
The detailed expressions of $n_{15}, n_{16}, \cdots, n_{25}$ and $d_{12}, d_{13}, \cdots, d_{16}$ are listed in the appendix.

It can be verified that when $W_2 = 0$, the real and imaginary parts of the modified non-dimensional wavenumber for third-order CLS-ID scheme of Eq. (\ref{eq:clsmywork}) are
\begin{equation}
  \begin{aligned}
  Re(\kappa') & = \frac{3 \sin (\kappa )}{\cos (\kappa )+2},\\
  Im(\kappa') & = 0.
  \end{aligned}
  \label{eq:dispdissspline3}
\end{equation}
And when $W_3 = 0, W_4 = 0$, the real and imaginary parts of the modified non-dimensional wavenumber for the fifth-order CLS-ID scheme of Eq. (\ref{eq:clsmywork}) are
\begin{equation}
  \begin{aligned}
  Re(\kappa') & = \frac{5 (10 \sin (\kappa )+\sin (2 \kappa ))}{26 \cos (\kappa )+\cos (2 \kappa )+33},\\
  Im(\kappa') & = 0.
  \end{aligned}
  \label{eq:dispdissspline5}
\end{equation}
In both cases, the dissipation error of the CLS-ID schemes becomes zero and the CLS-ID schemes are equivalent to the spline reconstruction proposed in the work of Wang and Ren\cite{wang2015accurate}, Huang et al. \cite{huang2018high}
Equations (\ref{eq:dispdissspline3}) and (\ref{eq:dispdissspline5}) are consistent with the spectral property of the third- and fifth-order spline reconstruction-based finite volume methods \cite{wang2015accurate,huang2018high}, respectively.
Besides, it can be concluded from Eqs. (\ref{eq:dispdissmywork3}) and (\ref{eq:dispdissmywork5}) that for the CLS-ID schemes, the dissipation error approaches zero when $W_k \rightarrow 0, k = 1,2,\cdots$.
In this work, the values of $W_k, k=1,2,\cdots$ are constrained to be greater than zero such that the resulting CLS schemes have enough numerical dissipation. 

Defining the maximum resolved non-dimensional wavenumber as $\kappa_m$ such that
\begin{equation}
  \begin{aligned}
  \forall \kappa \leq \kappa_m,&  \left|\frac{Re(\kappa')}{\kappa}-1\right| \leq 0.5\%, \,\,\mathrm{and}\,\, -Im(\kappa') \leq 0.5\%,\\
  \exists \kappa > \kappa_m, &  \left|\frac{Re(\kappa')}{\kappa}-1\right| > 0.5\%, \,\,\mathrm{or}\,\, -Im(\kappa') > 0.5\%.
  \end{aligned}
\end{equation}

For the third-order schemes, given a critical non-dimensional wavenumber $\kappa_c$, the optimized $W_1$ and $W_2$ are obtained such that $W_2$ has the maximum value in the set defined by 
\begin{equation}
\left\{\left.(W_1, W_2)\right| \kappa_m \geq \kappa_c, W_1 \geq W_2, \frac{Re(\kappa')}{\kappa}-1 \leq 0.5\% \right\}.
\label{eq:constrained3rd}
\end{equation}
The constraint $\frac{Re(\kappa')}{\kappa}-1 \leq 0.5\%$ in Eq. (\ref{eq:constrained3rd}) is introduced to eliminate the overshoot phenomenon in the dispersion curve.
\begin{table}
\caption{Optimized $W_1$ and $W_2$ for the third-order CLS-ID scheme of Eq. (\ref{eq:clsmywork}) under various $\kappa_c$.}\label{tab:w1w2}
\begin{ruledtabular}
\begin{tabular}{lcr}
$\kappa_c$&$W_1$&$W_2$\\
\hline
1.0 & $8.2382\times 10^{-2}$ & $1.0720\times 10^{-2}$ \\
1.2 & $3.4334\times 10^{-2}$ & $3.4712\times 10^{-3}$ \\
2.0 & $1.8619\times 10^{-3}$ & $1.5469\times 10^{-4}$ \\
\end{tabular}
\end{ruledtabular}
\end{table}

\begin{table}
\caption{Optimized $W_1$ and $W_2$ for the third-order CLS-ID scheme in the literature.}\label{tab:w1w2huang}
\begin{ruledtabular}
\begin{tabular}{lcr}
  &$W_1$&$W_2$\\
\hline
Huang \cite{huang_high-order_2022} & $2.1557\times 10^{-1}$ & $2.4305 \times 10^{-2}$ \\
\end{tabular}
\end{ruledtabular}
\end{table}

When $\kappa_c = 1.0, 1.2, 2.0$, the obtained $W_1$ and $W_2$ for the third-order CLS-ID scheme are listed in Tab. \ref{tab:w1w2}. It can be concluded that if the desired maximum resolved non-dimensional wavenumber increases from $1.0$ to $2.0$, the maximum allowable values for the weights $W_1$ and $W_2$ decrease gradually. 
The values of $W_1$ and $W_2$ in Tab. \ref{tab:w1w2} are obtained through a straightforward grid searching algorithm. Table \ref{tab:w1w2huang} also lists the weights optimized by Huang et al. \cite{huang_high-order_2022}

On the other hand, for the third-order CLS-CD scheme, the set defined by Eq. (\ref{eq:constrained3rd}) is empty with $\kappa_c = 1.0, 1.2$ or $2.0$.

\begin{figure}[htbp]
  \centering
    \begin{subfigure}[b]{\columnwidth}
    \includegraphics[width=0.6\columnwidth]{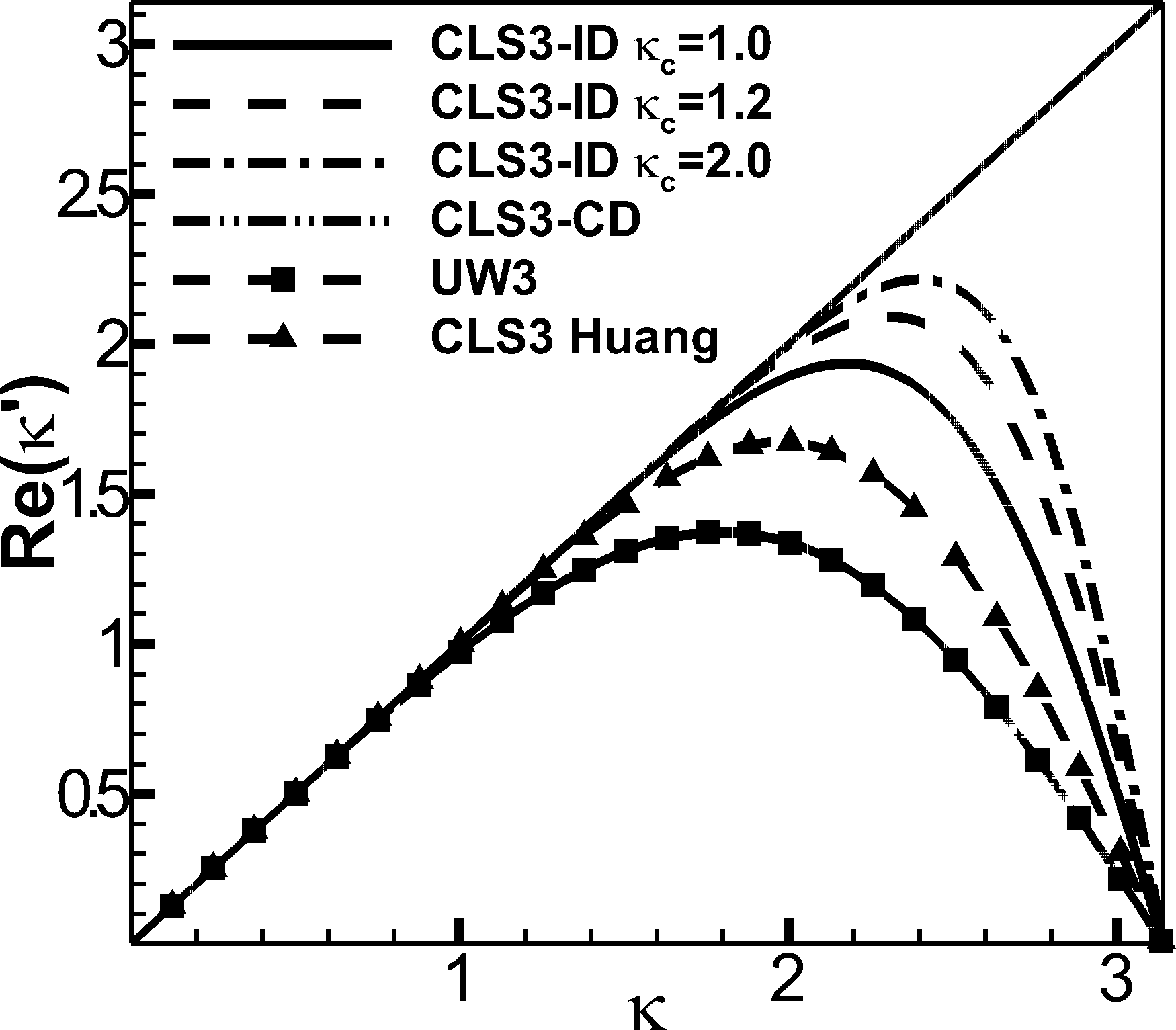}
    \caption{Dispersion error for the third-order CLS schemes.}
    \label{fig:dispcls3}
    \end{subfigure}
    \hfill 
    \begin{subfigure}[b]{\columnwidth}
      \includegraphics[width=0.6\textwidth]{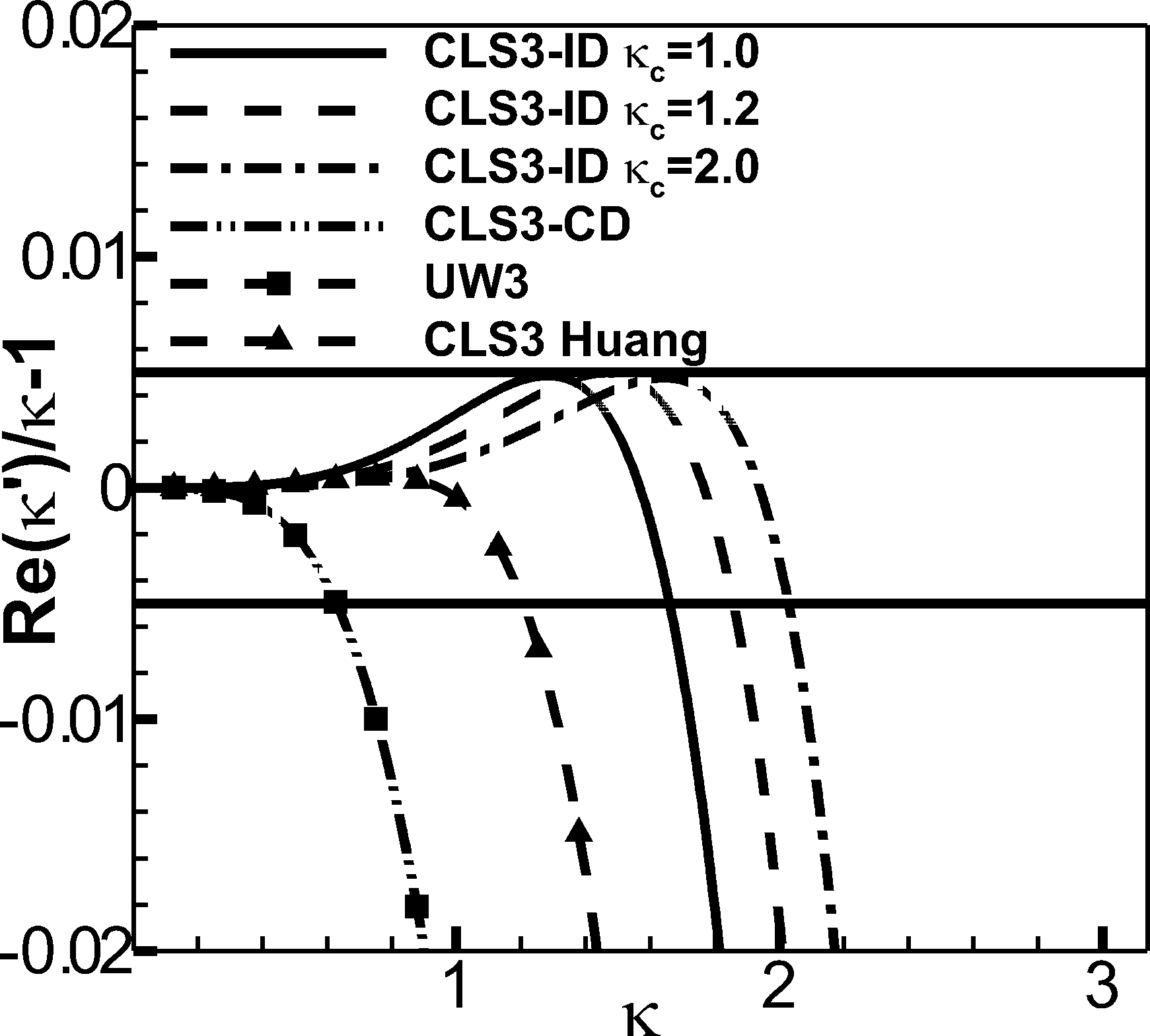}
      \caption{Relative dispersion error for the third-order CLS schemes.}
      \label{fig:redispcls3}
    \end{subfigure}
    \hfill
    \begin{subfigure}[b]{\columnwidth}
      \includegraphics[width=0.6\textwidth]{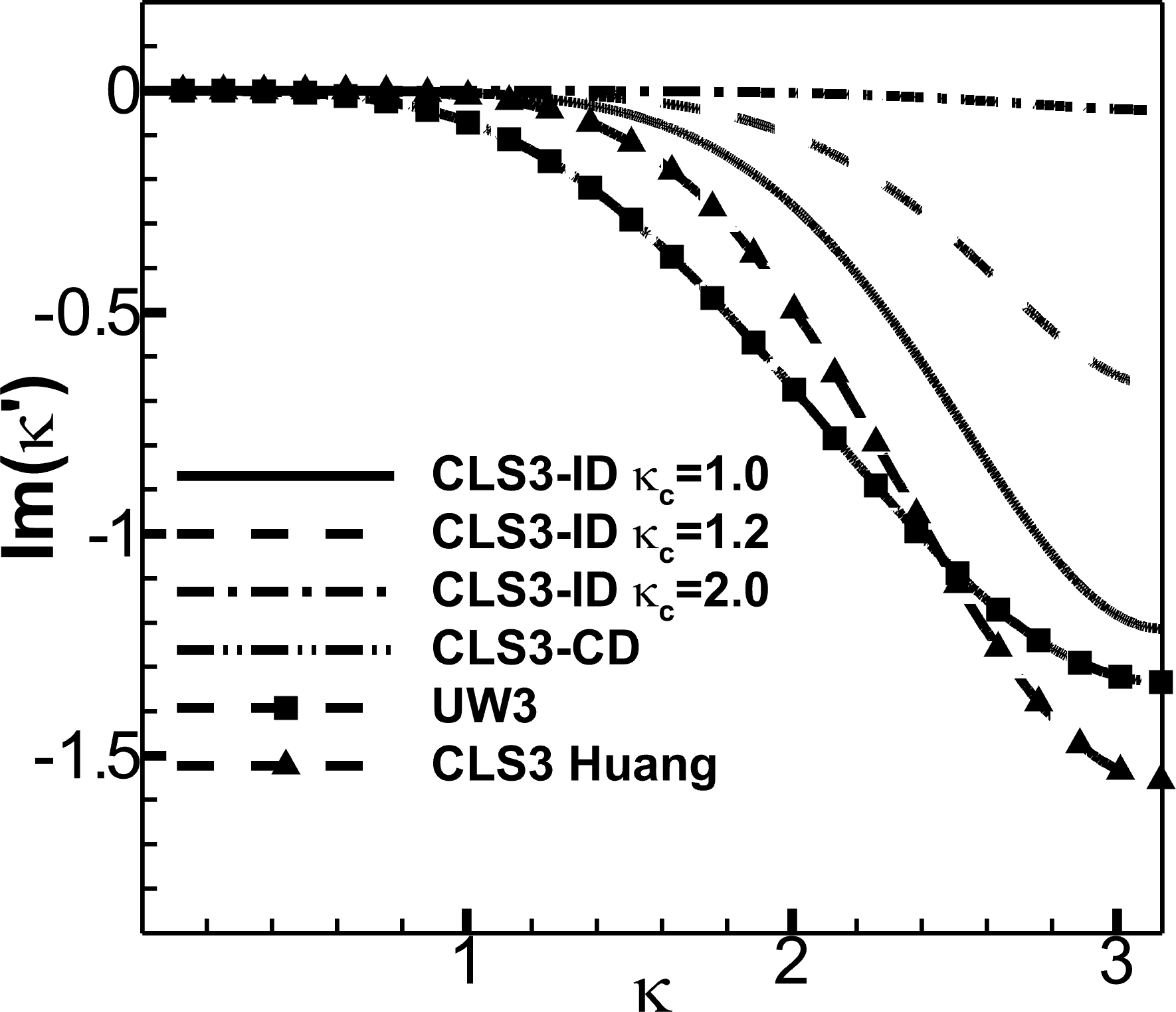}
      \caption{Dissipation error for the third-order CLS schemes.}
      \label{fig:disscls3}
    \end{subfigure}
    \caption{\label{fig:dispdisscls3} Spectral properties for the third-order CLS schemes. The curves of the third-order CLS-CD scheme coincide with the curves of third-order upwind scheme.}
\end{figure}

Figure \ref{fig:dispcls3} shows the dispersion and dissipation error of the third-order CLS-ID scheme. As a comparison, the curves for the optimized third-order CLS-CD scheme in the work of Wang et al.\cite{wang2016compact1} with $W_1 = W_2 = 0$, and the curves of third- and fifth-order upwind schemes are also presented. It should be noted that the third-order CLS-CD scheme with $W_1= W_2 = 0$ is equivalent to the third-order upwind scheme and their curves are the same. The dispersion error of the third-order CLS-ID scheme proposed in this paper is much smaller than the third-order CLS-CD scheme based on the differences of averaged derivatives on neighboring cells and even better than the fifth-order upwind scheme.

\begin{figure}[htbp]
  \centering
  \includegraphics[width=0.8\columnwidth]{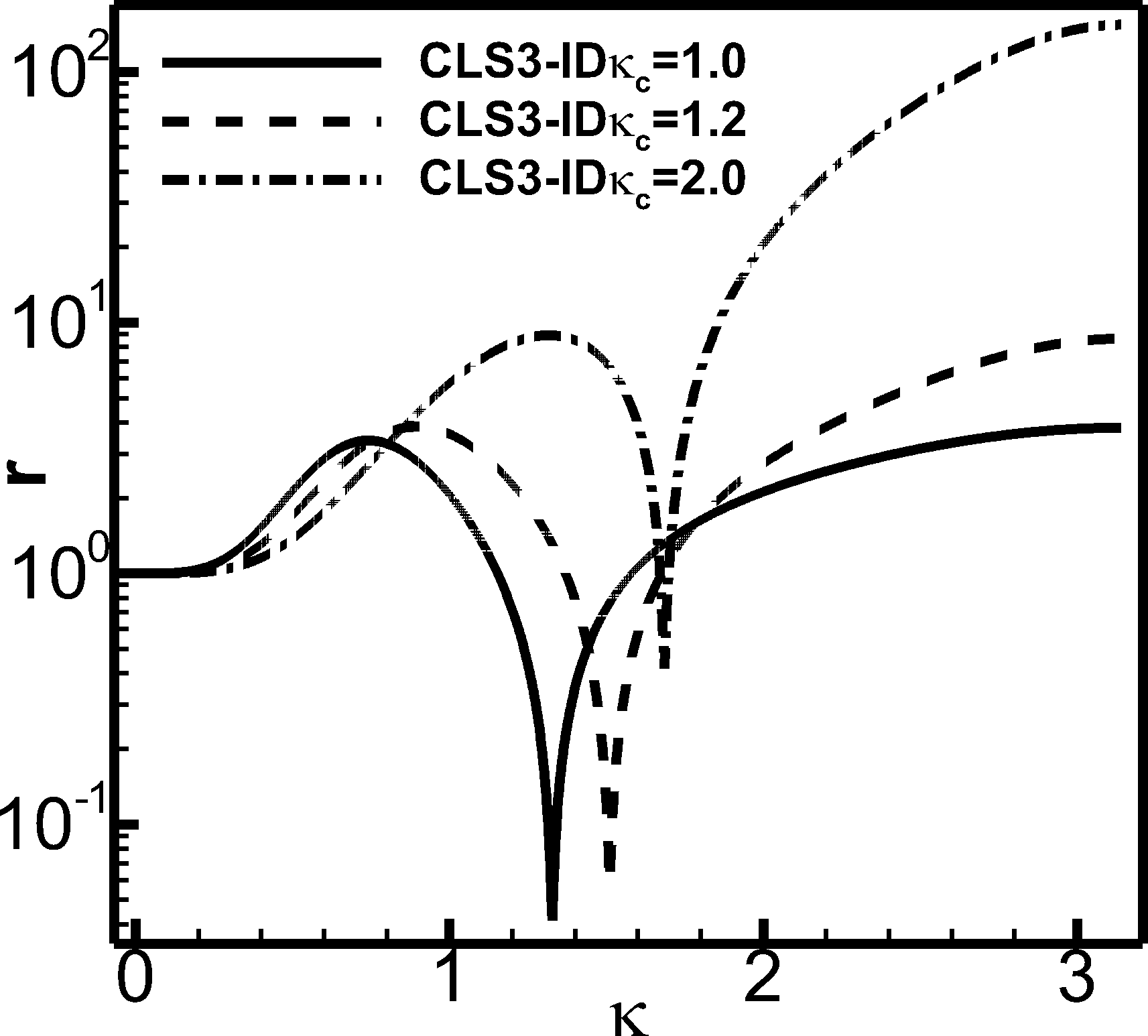}
  \caption{\label{fig:rindexcls3} The $r$-index for the third-order CLS-ID scheme.}
\end{figure}

Additionally, it is demonstrated by Fig. (\ref{fig:disscls3}) that the dissipation error of third-order CLS-ID scheme becomes smaller with the increasing of $\kappa_c$. 
As pointed out by Hu et al. \cite{hu2012dispersion}, enough dissipation should be applied to stabilize the numerical solution, and Eq. (\ref{eq:rindex}) denoted as the $r$-index should be smaller than O(10), 
\begin{equation}
  r = \frac{\left|\frac{\partial Re(\kappa')}{\partial \kappa} -1\right|+0.001}{\left|-Im(\kappa')\right|+0.001}.
  \label{eq:rindex}
\end{equation}
Figure \ref{fig:rindexcls3} shows the $r$-index for the third-order CLS-ID scheme. The third-order CLS-ID scheme with weights corresponding to $\kappa_c = 1.0$ and $1.2$ does have enough numerical dissipation while the scheme with weights corresponding to $\kappa_c = 2.0$ lacks enough dissipation in high-wave-number region.

For the fifth-order CLS-ID scheme, it is time-consuming to optimize the four free parameters through the grid searching algorithm. Thus, a genetic algorithm is utilized to evolve the four free parameters. The objective function to be minimized is defined as
\begin{equation}
    -\kappa_m + \beta Im(\pi),
\label{eq:obj5th}
\end{equation}
under the constraint 
\begin{equation}
  \begin{aligned}
\left(W_1, \right. & \left. W_2, \, W_3, \, W_4\right) \in \\
 & \left\{\left.(W_1, W_2, W_3, W_4)\right| W_1 \geq W_2 \geq W_3 \geq W_4, \right. \\
 &\quad\quad\quad\quad\quad\quad\quad\quad \left. \frac{Re(\kappa')}{\kappa}-1 \leq 0.5\% \right\}.
  \end{aligned}
\label{eq:constrained5th}
\end{equation}

In Eq. (\ref{eq:obj5th}), $\beta$ is a hyper-parameter which is used to adjust the dissipation of the scheme. If $\beta$ goes larger, more dissipation is applied to the resulting numerical scheme. Tables \ref{tab:w1w2w3w4} and \ref{tab:w1w2w3w4wang} present the optimized four free parameters for the fifth-order CLS-ID scheme and fifth-order CLS-CD scheme, respectively.

\begin{table}
\caption{Optimized $\left(W_1, W_2, W_3, W_4\right)$ for the fifth-order CLS-ID scheme of Eq. (\ref{eq:clsmywork}) under various $\beta$.}\label{tab:w1w2w3w4}
\begin{ruledtabular}
\begin{tabular}{lcccc}
$\beta$&$W_1$&$W_2$&$W_3$&$W_4$\\
\hline
1.0 & $9.7407\times 10^{-1}$	&   $9.3822 \times 10^{-1}$	&   $1.3219\times 10^{-3}$	&   $1.2388\times 10^{-3}$ \\
0.4 & $9.4334\times 10^{-1}$	&   $6.9128 \times 10^{-1}$	&   $8.3184\times 10^{-4}$	&   $3.3623\times 10^{-4}$ \\
0.3 & $7.7233\times 10^{-1}$	&   $2.0654 \times 10^{-1}$	&   $2.0236\times 10^{-4}$	&   $3.1769\times 10^{-5}$ 
\end{tabular}
\end{ruledtabular}
\end{table}

\begin{table}
\caption{Optimized $\left(W_1, W_2, W_3, W_4\right)$ for the fifth-order CLS-CD scheme of Eq. (\ref{eq:clswang}) under various $\beta$.}\label{tab:w1w2w3w4wang}
\begin{ruledtabular}
\begin{tabular}{lcccc}
$\beta$&$W_1$&$W_2$&$W_3$&$W_4$\\
\hline
1.0 & $9.8925 \times 10^{-1}$& $9.8681\times 10^{-1}$& $7.1976\times 10^{-2}$& $6.8856\times 10^{-2}$\\
0.4 & $9.2056 \times 10^{-1}$& $6.6947\times 10^{-3}$& $3.6870\times 10^{-3}$& $3.6453\times 10^{-3}$\\
0.3 & $8.2832 \times 10^{-1}$& $4.0398\times 10^{-3}$& $3.7802\times 10^{-3}$& $2.8934\times 10^{-3}$\\
\end{tabular}
\end{ruledtabular}
\end{table}

\begin{figure}[htb]
  \centering
    \begin{subfigure}[b]{\columnwidth}
    \includegraphics[width=0.6\columnwidth]{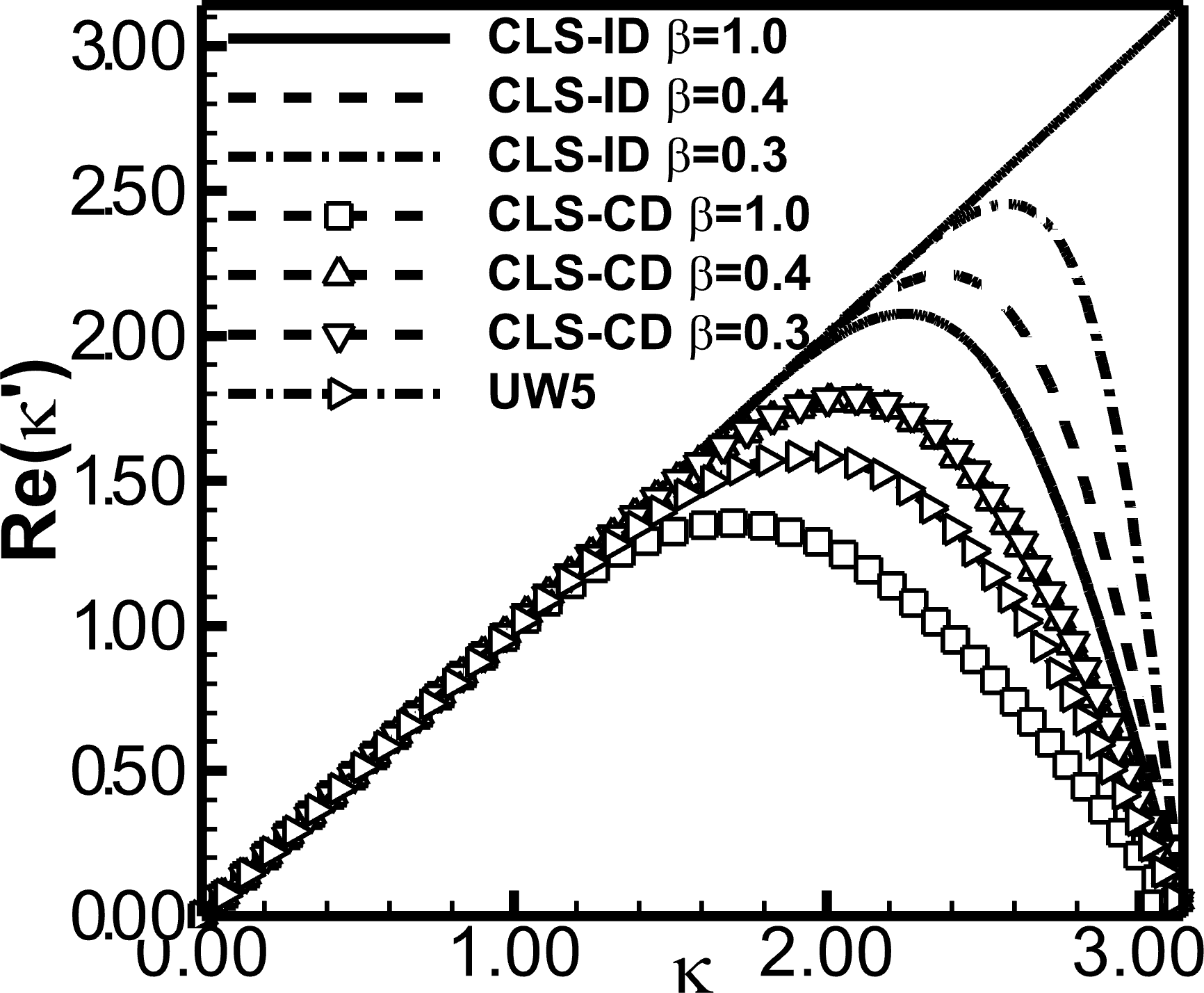}
    \caption{Dispersion error for the fifth-order CLS schemes.}
    \label{fig:dispcls5}
    \end{subfigure}
    \hfill 
    \begin{subfigure}[b]{\columnwidth}
      \includegraphics[width=0.6\columnwidth]{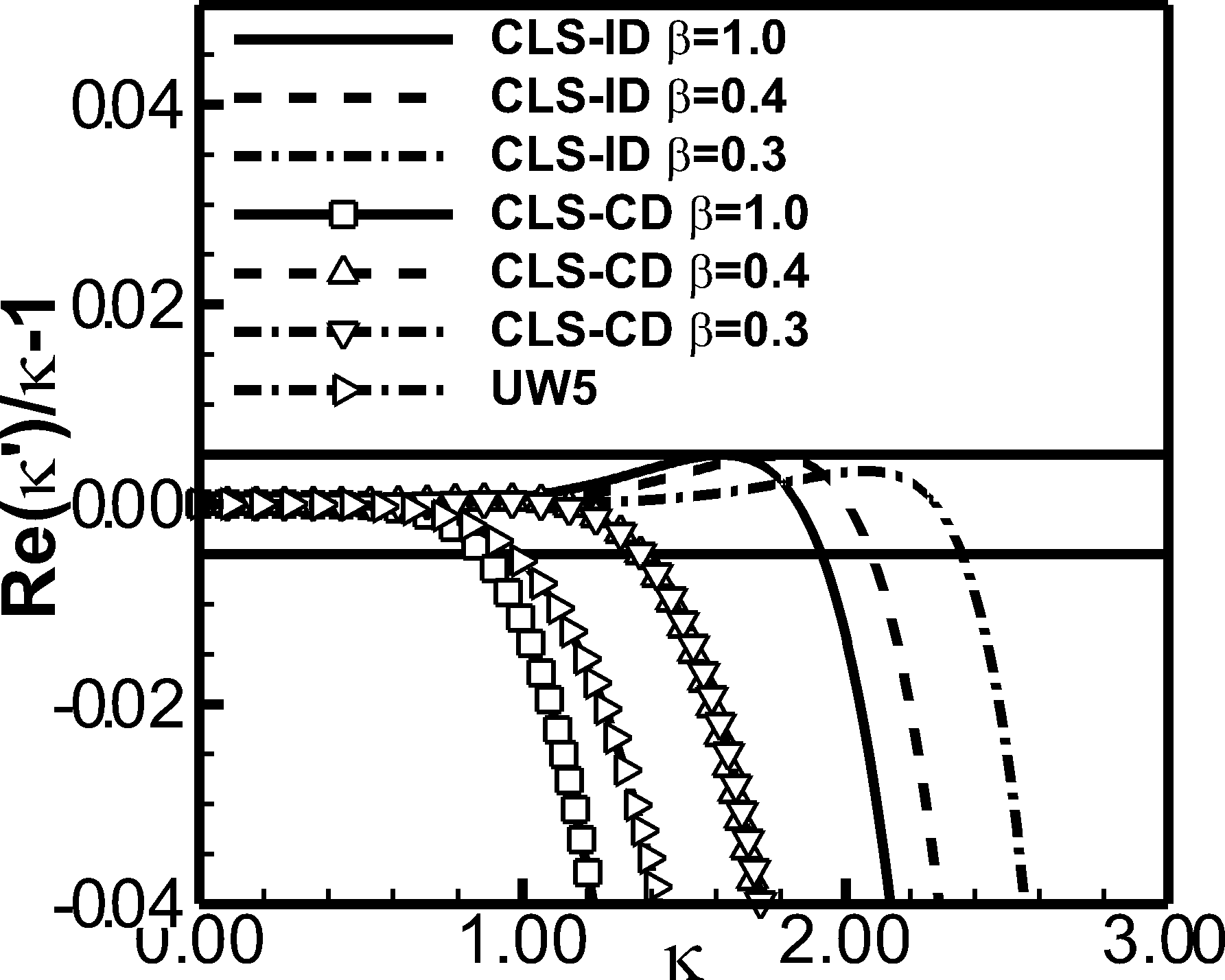}
      \caption{Relative dispersion error for the fifth-order CLS schemes.}
      \label{fig:redispcls5}
    \end{subfigure}
    \hfill
    \begin{subfigure}[b]{\columnwidth}
      \includegraphics[width=0.6\columnwidth]{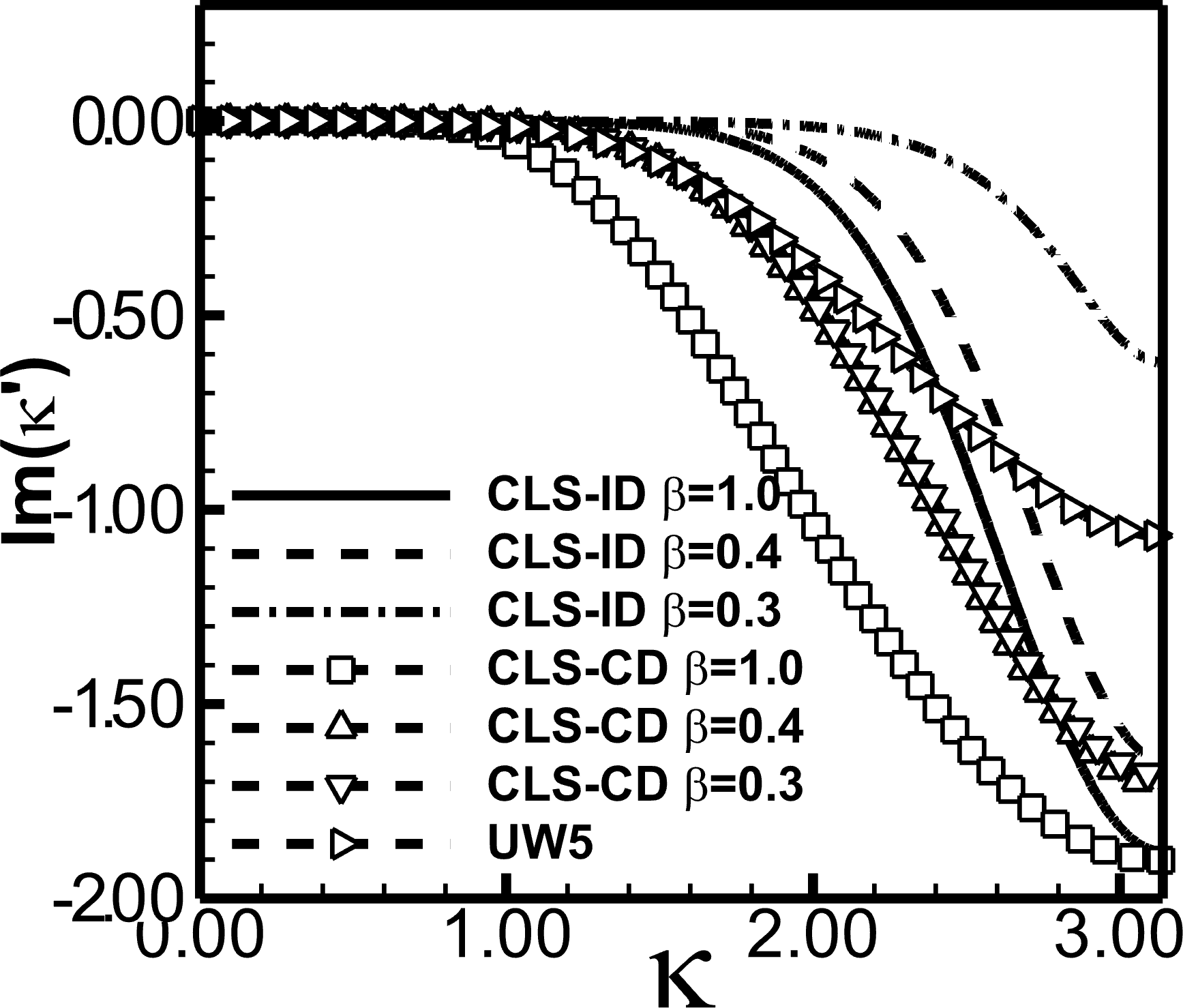}
      \caption{Dissipation error for the fifth-order CLS schemes.}
      \label{fig:disscls5}
    \end{subfigure}
    \caption{\label{fig:dispdisscls5} Spectral properties for the fifth-order CLS schemes.}
\end{figure}

\begin{figure}[htbp]
  \centering
  \includegraphics[width=0.8\columnwidth]{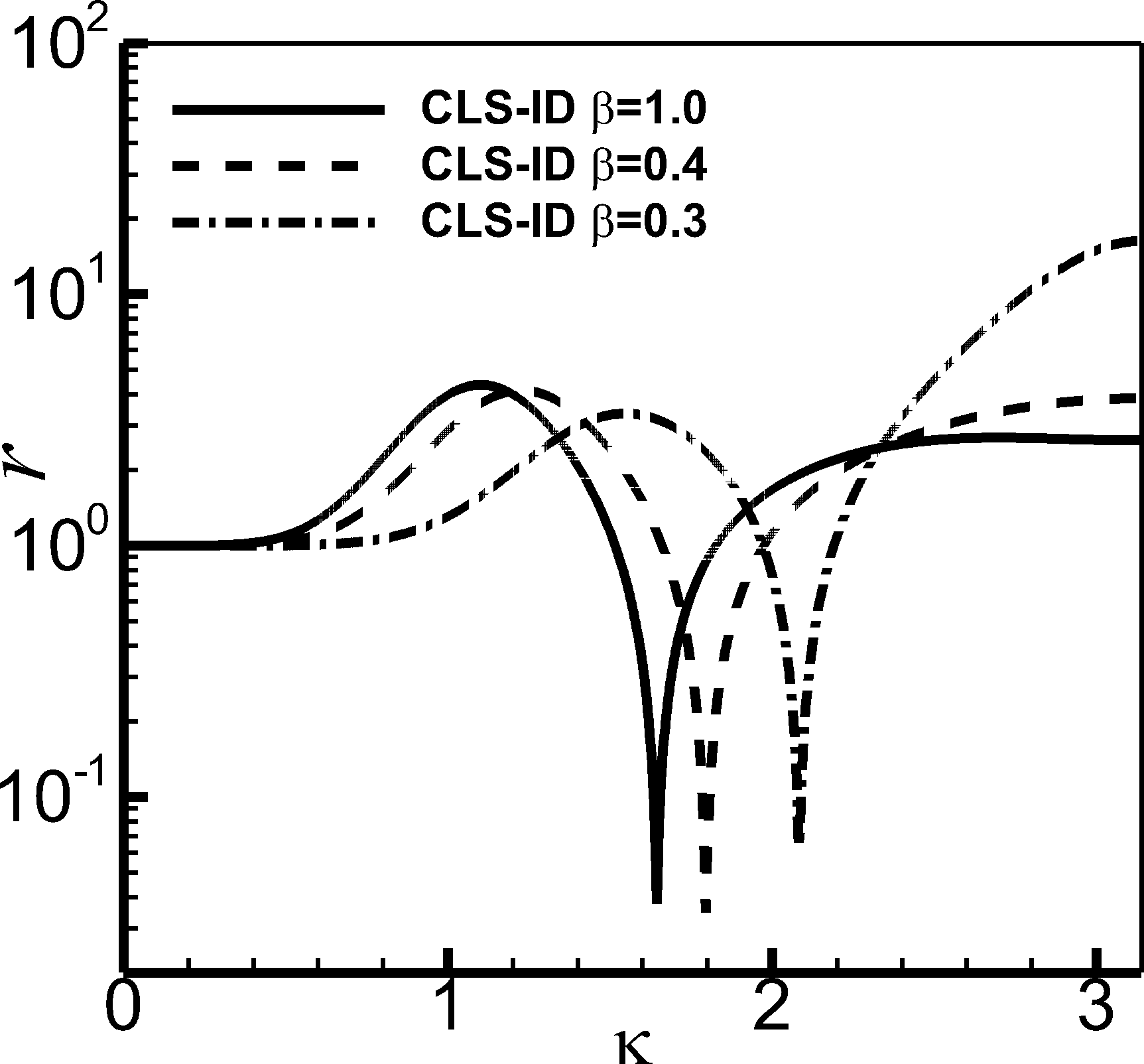}
  \caption{\label{fig:rindexcls5} The $r$-index for the fifth-order CLS-ID scheme.}
\end{figure}

Figure \ref{fig:dispdisscls5} shows the dispersion and dissipation error of the fifth-order CLS schemes. For the fifth-order CLS-ID scheme, the maximum resolved non-dimensional wavenumber can reach as high as 2.0 if only considering the relative dispersion error under the constraint $Re(\kappa')/\kappa-1 < 0.5\%$. It is confirmed again that the dispersion error of the fifth-order CLS-ID scheme is much smaller than the fifth-order CLS-CD scheme.
Figure \ref{fig:rindexcls5} shows the $r$-index for the fifth-order CLS-ID scheme, demonstrating that enough dissipation is applied to the numerical solution with all the three sets of optimized weights.

\section{Hybridization of CLS and CWENO schemes \label{sec:hybrid}}
To capture the discontinuities in the flow field, instead of applying a slop limiter after reconstruction\cite{wang2016compact1,wang2016compact2}, the CLS schemes are hybridized with CWENO schemes\cite{baeza_central_2019} in this work for the first time.
Following the idea of Ren\cite{ren_characteristic-wise_2003}, the linear reconstruction system is modified as
\begin{equation}
  \begin{aligned}
\sigma_i \mathbf{M}_i^{(-1)} \overrightarrow{a}_{i-1}
+\left(\sigma_i \mathbf{M}_i^{(0)} + (1-\sigma_i)\mathbb{I} \right)\overrightarrow{a}_{i} 
+\sigma_i \mathbf{M}_i^{(1)} \overrightarrow{a}_{i+1} \\
\quad  = \sigma_i \overrightarrow{b}_i + (1-\sigma_i)\overrightarrow{a}_i^{\mathrm{CWENO}}, \label{eq:hybrid}
  \end{aligned}
\end{equation}
where $\mathbb{I}$ is an identity matrix and $\sigma_i$ is a shock detector satisfying
\begin{equation}
\sigma_i = \left\{
  \begin{array}{lc}
   O(\left(\Delta x\right)^s),\,\, & s \geq 0\,\,\text{if} \,\,\,\, \Omega_i \,\,\,\,\text{locates inside discontinuity}; \\
   1, & \text{otherwise}.
  \end{array}
\right.
\end{equation}
$\overrightarrow{a}_i^{\mathrm{CWENO}}$ are the coefficients reconstructed by CWENO schemes.

The first shock detector is borrowed from the work of Ren\cite{ren_characteristic-wise_2003}, and is constructed as
\begin{equation}
  \sigma_i = \min\left(1.0, \theta_i'/\theta_c\right)
\label{eq:ren_sigma}
\end{equation}
where 
\begin{equation}
\theta_i' = \min(\theta_{i-1}, \theta_{i}, \theta_{i+1}),
\end{equation}
$\theta_c$ is a critical number below which the control volume is identified as a discontinuous cell and
\begin{equation}
\theta_i = \frac{\left|2 \Delta u_{i-1/2}\Delta u_{i+1/2}\right| + \epsilon}{\left(\Delta u_{i-1/2}\right)^2  + \left(\Delta u_{i+1/2}\right)^2 + \epsilon},
\label{eq:ren_detector_rj}
\end{equation}
where $\Delta u_{i-1/2} = \left(u_i - u_{i-1}\right) \frac{2\Delta x_i}{\Delta x_{i-1} + \Delta x_{i}}$ and $\epsilon$ is a small number avoiding division by zero chosen as
\begin{equation}
\epsilon = \left[\max\left(1\times 10^{-6} \frac{\sum_{m=-1}^{1}{\left|\overline{u}_{i+m}\right|}}{3}, 1\times 10^{-50}\right)\right]^2.
\label{eq:epsilonren}
\end{equation}

\begin{figure}[htbp]
  \centering
  \includegraphics[width=0.8\columnwidth]{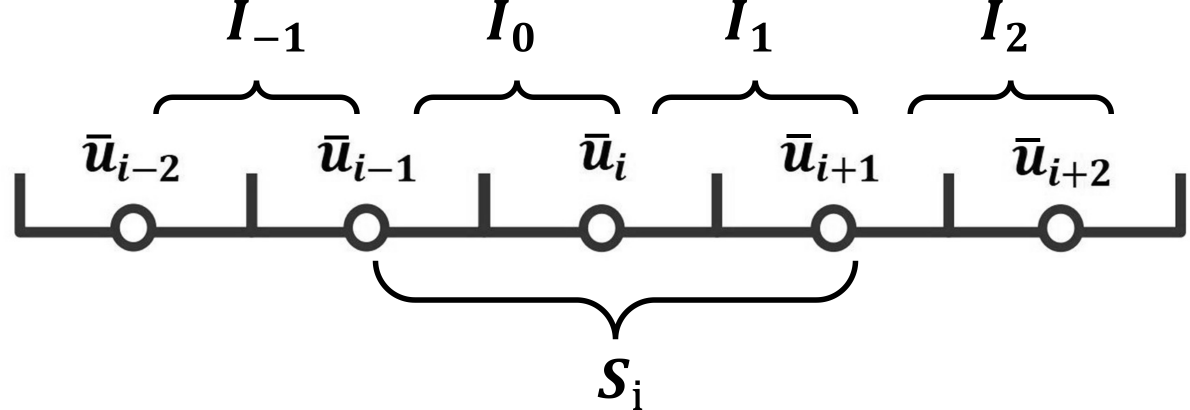}
  \caption{\label{fig:stencil_of_shock_detector} Illustration for the stencils of $\sigma^{\mathrm{Li}}$.}
\end{figure}

Equation (\ref{eq:ren_sigma}) has been validated to be effective \cite{ren_characteristic-wise_2003,wang2015accurate,huang2018high,huang2022adaptive} to recognize most discontinuous cells. However, as pointed out by Baeza\cite{baeza2020efficient}, it is impossible to distinguish between a local extremum and a discontinuity using only three cell-averaged values. Inspired by the work of Baeza \cite{baeza2020efficient}, the second shock detector is proposed in our previous work \cite{PANWCLS}.
Five cell-averaged values are incorporated shown as in Fig. \ref{fig:stencil_of_shock_detector} to calculate the new shock detector as in Eq. (\ref{eq:owenosd}),

\begin{equation}
  \begin{aligned}
  \sigma_i & = \min\left(1.0, \theta_i/\theta_c\right),\\
  \theta_i & = \theta_i^L\cdot \theta_i^R,
  \end{aligned}
  \label{eq:owenosd}
\end{equation}
where $\theta_c$ is a critical value to be calibrated here in this work and
\begin{equation}
  \theta_i^L = \frac{J_L + \epsilon}{J_L + \tau_L + \epsilon},\,\,
  \theta_i^R = \frac{J_R + \epsilon}{J_R + \tau_R + \epsilon},
  \label{eq:sigmalr}
\end{equation}
with
\begin{align}
J_L & = I_0 I_1 + I_{-1} I_{-1},\\
J_R & = I_0 I_1 + I_{2} I_{2},\\
\tau_L & = d_L(I_0+ I_1),\\
\tau_R & = d_R(I_0+ I_1),\\
d_L & = \left(\zeta_1^L \overline{u}_{i+1} + \zeta_0^L \overline{u}_{i} + \zeta_{-1}^L \overline{u}_{i-1} + \zeta_{-2}^L \overline{u}_{i-2}\right)^2,\label{eq:d3ul}\\
d_R & = \left(\zeta_{-1}^R \overline{u}_{i-1} + \zeta_0^R \overline{u}_{i} + \zeta_{1}^R \overline{u}_{i+1} + \zeta_{2}^R \overline{u}_{i+2}\right)^2\label{eq:d3ur},
\end{align}
and
\begin{align}
  I_{-1} & = \left(\frac{2 \lambda_0}{\lambda_{-2} + \lambda_{-1}} \left(\overline{u}_{i-1} - \overline{u}_{i-2}\right)\right)^2,\label{eq:I-1}\\
  I_{0}  & = \left(\frac{2 \lambda_0}{\lambda_{-1} + \lambda_{0}} \left(\overline{u}_{i} - \overline{u}_{i-1}\right)\right)^2,\\
  I_{1}  & = \left(\frac{2 \lambda_0}{\lambda_{0} + \lambda_{1}} \left(\overline{u}_{i+1} - \overline{u}_{i}\right)\right)^2,\\
  I_{2}  & = \left(\frac{2 \lambda_0}{\lambda_{1} + \lambda_{2}} \left(\overline{u}_{i+2} - \overline{u}_{i+1}\right)\right)^2.\label{eq:I2}
\end{align}
In Eq. (\ref{eq:sigmalr}), $\epsilon$ is chosen as
\begin{equation}
\epsilon = \left[\max\left(1\times 10^{-6} \frac{\sum_{m=-1}^{1}{\left|\overline{u}_{i+m}\right|}}{3}, 1\times 10^{-50}\right)\right]^4.
\label{eq:epsilonpan}
\end{equation}
In Eqs. (\ref{eq:I-1}) to (\ref{eq:I2}),
\begin{equation}
\lambda_{m}  = \frac{\Delta x_{i+m}}{\Delta x_i},\,\,m=-2,-1,0,1,2.
\end{equation}
Additionally,
the coefficients of $\zeta^L$ and $\zeta^R$ are defined by
\begin{align}
\zeta_{-2}^L & = -\frac{24}{
  \left(
    \begin{array}{c}
  \left(\lambda _{-2}+\lambda _{-1}\right) \left(\lambda _{-2}+\lambda _{-1}+\lambda _0\right) \\
    \cdot \left(\lambda _{-2}+\lambda _{-1}+\lambda _0+\lambda _1\right)
    \end{array}
    \right)
  },\\
\zeta_{-1}^L&
=\frac{
  \left(
  \begin{array}{c}
  24 \left(\lambda _{-2}^2+\left(3 \lambda _{-1}+2 \lambda _0+\lambda _1\right) \lambda _{-2}\right.\\
  \left. +3 \lambda _{-1}^2+\lambda _0 \left(\lambda _0+\lambda _1\right)+2 \lambda _{-1} \left(2 \lambda _0+\lambda _1\right)\right)
  \end{array}
  \right)
  }{
  \left(
  \begin{array}{c}
    \left(\lambda _{-2}+\lambda _{-1}\right) \left(\lambda _{-1}+\lambda _0\right) \left(\lambda _{-2}+\lambda _{-1}+\lambda _0\right) \\
    \cdot \left(\lambda _{-1}+\lambda _0+\lambda _1\right) \left(\lambda _{-2}+\lambda _{-1}+\lambda _0+\lambda _1\right)
  \end{array}
  \right)
  },\\
\zeta_{0}^L &=-\frac{
  \left(
  \begin{array}{c}
  24 \left(\lambda _{-1}^2+2 \left(2 \lambda _0+\lambda _1\right) \lambda _{-1}+3 \lambda _0^2\right.\\
\left.  +\lambda _1^2+3 \lambda _0 \lambda _1+\lambda _{-2} \left(\lambda _{-1}+2 \lambda _0+\lambda _1\right)\right)
  \end{array}
  \right)
  }{
  \left(
  \begin{array}{c}
    \left(\lambda _{-1}+\lambda _0\right) \left(\lambda _{-2}+\lambda _{-1}+\lambda _0\right) \left(\lambda _0+\lambda _1\right) \\
    \cdot \left(\lambda _{-1}+\lambda _0+\lambda _1\right) \left(\lambda _{-2}+\lambda _{-1}+\lambda _0+\lambda _1\right)
  \end{array}
  \right)
  },\\
\zeta_{1}^L &=\frac{24}{
  \left(
  \begin{array}{c}
  \left(\lambda _0+\lambda _1\right) \left(\lambda _{-1}+\lambda _0+\lambda _1\right) \\
  \cdot \left(\lambda _{-2}+\lambda _{-1}+\lambda _0+\lambda _1\right)
  \end{array}
  \right)
  },\\
  \zeta_{-1}^R & = -\frac{24}{
  \left(
  \begin{array}{c}
    \left(\lambda _{-1}+\lambda _0\right) \left(\lambda _{-1}+\lambda _0+\lambda _1\right) \\
    \cdot\left(\lambda _{-1}+\lambda _0+\lambda _1+\lambda _2\right)
  \end{array}
  \right)
    },\\
  \zeta_{0}^R & =\frac{
  \left(
  \begin{array}{c}
    24 \left(\lambda _{-1}^2+\left(3 \lambda _0+2 \lambda _1+\lambda _2\right) \lambda _{-1}+3 \lambda _0^2 \right.\\
    \left.+\lambda _1 \left(\lambda _1+\lambda _2\right)+2 \lambda _0 \left(2 \lambda _1+\lambda _2\right)\right)
  \end{array}
  \right)
    }{
  \left(
  \begin{array}{c}
      \left(\lambda _{-1}+\lambda _0\right) \left(\lambda _0+\lambda _1\right) \left(\lambda _{-1}+\lambda _0+\lambda _1\right) \\
      \cdot \left(\lambda _0+\lambda _1+\lambda _2\right) \left(\lambda _{-1}+\lambda _0+\lambda _1+\lambda _2\right)
  \end{array}
  \right)
    },\\
\zeta_{1}^R & =  -\frac{
  \left(
  \begin{array}{c}
  24 \left(\lambda _0^2+2 \left(2 \lambda _1+\lambda _2\right) \lambda _0+3 \lambda _1^2+\lambda _2^2\right.\\
\left.+3 \lambda _1 \lambda _2+\lambda _{-1} \left(\lambda _0+2 \lambda _1+\lambda _2\right)\right)
  \end{array}
  \right)
  }{
  \left(
  \begin{array}{c}
    \left(\lambda _0+\lambda _1\right) \left(\lambda _{-1}+\lambda _0+\lambda _1\right) \left(\lambda _1+\lambda _2\right) \\
    \cdot \left(\lambda _0+\lambda _1+\lambda _2\right) \left(\lambda _{-1}+\lambda _0+\lambda _1+\lambda _2\right)
  \end{array}
  \right)
  },\\
\zeta_{2}^R & =  \frac{24}{
  \left(
  \begin{array}{c}
  \left(\lambda _1+\lambda _2\right) \left(\lambda _0+\lambda _1+\lambda _2\right) \\
  \cdot \left(\lambda _{-1}+\lambda _0+\lambda _1+\lambda _2\right)
  \end{array}
  \right)
  }.
\end{align}
In smooth regions, $d_L$ in Eq. (\ref{eq:d3ul}) and $d_R$ in Eq. (\ref{eq:d3ur}) are \cite{baeza2020efficient,PANWCLS}
\begin{align}
d_L & = \left(\frac{\partial^3u}{\partial x^3}\right)\left(\Delta x\right)^6 + O(\left(\Delta x\right)^8),\\
d_R & = \left(\frac{\partial^3u}{\partial x^3}\right)\left(\Delta x\right)^6 + O(\left(\Delta x\right)^8).
\end{align}

To better distinguish between the two shock detectors, a superscript $\left(\cdot\right)^{\mathrm{Ren}}$ is added to the variables defined as in Eqs. (\ref{eq:ren_sigma}) to (\ref{eq:ren_detector_rj}), e.g., $\sigma_i^{\mathrm{Ren}}$; and a superscript $\left(\cdot \right)^{\mathrm{Li}}$ is added to the variables defined in  Eqs. (\ref{eq:owenosd}) to (\ref{eq:sigmalr}), e.g., $\sigma_i^{\mathrm{Li}}$.

\begin{assumption}
  \label{assump}
  $u(x)$ that is to be reconstructed has a critical point of order $k$, $k \in \left\{0, 1\right\}$; 
  At least one of $I_0$ and $I_1$ are of order $O(\left(\Delta x\right)^{2m}), m \geq 1$;
  If $S_i = { \left\{ \overline{u}_{i-1}, \overline{u}_{i}, \overline{u}_{i+1} \right\}
  }$ contains discontinuity, $I_{-1}$ and $I_{2}$ are not of order $O(1)$ at the same time.
\end{assumption}
With Assumption \ref{assump}, the scale of $\theta_i^{\mathrm{Li}}$ has been proved \cite{PANWCLS} to have

\begin{theorem}\label{theorem:1}\cite{PANWCLS}
  Suppose  Assumption \ref{assump} holds,
  then the shock detector $\theta_i^{\mathrm{Li}}$ in Eq. (\ref{eq:owenosd}) satisfies the following equations
  \begin{equation}
  \theta_i^{\mathrm{Li}} = \left\{
    \begin{array}{ll}
      1+O((\Delta x)^{2m}) + O(\epsilon), & \,\text{if}\,\,S_i\,\,\text{is smooth}, m \geq 1,\\
      O((\Delta x)^{2m}) + O(\epsilon), & 
      \begin{array}{l}
      \text{if}\,\,S_i\,\,\text{includes discon-}\\
      \text{tinuity}, m \geq 1.\\
      \end{array}
    \end{array}
  \right.
  \end{equation}
\end{theorem}

\begin{remark}
The small number $\epsilon$ influences the asymptotic behavior of the shock detector \cite{baeza_central_2019}. However, this is not the focus of the present paper. In this work, a scale-invariant formula is chosen as in Eqs. (\ref{eq:epsilonren}) and (\ref{eq:epsilonpan}). And for the fair of comparison, $1\times 10^{-6}$ are utilized as the coefficients in both Eq. (\ref{eq:epsilonren}) and Eq. (\ref{eq:epsilonpan}).
\end{remark}

\begin{figure}[htbp]
  \centering
  \includegraphics[width=0.8\columnwidth]{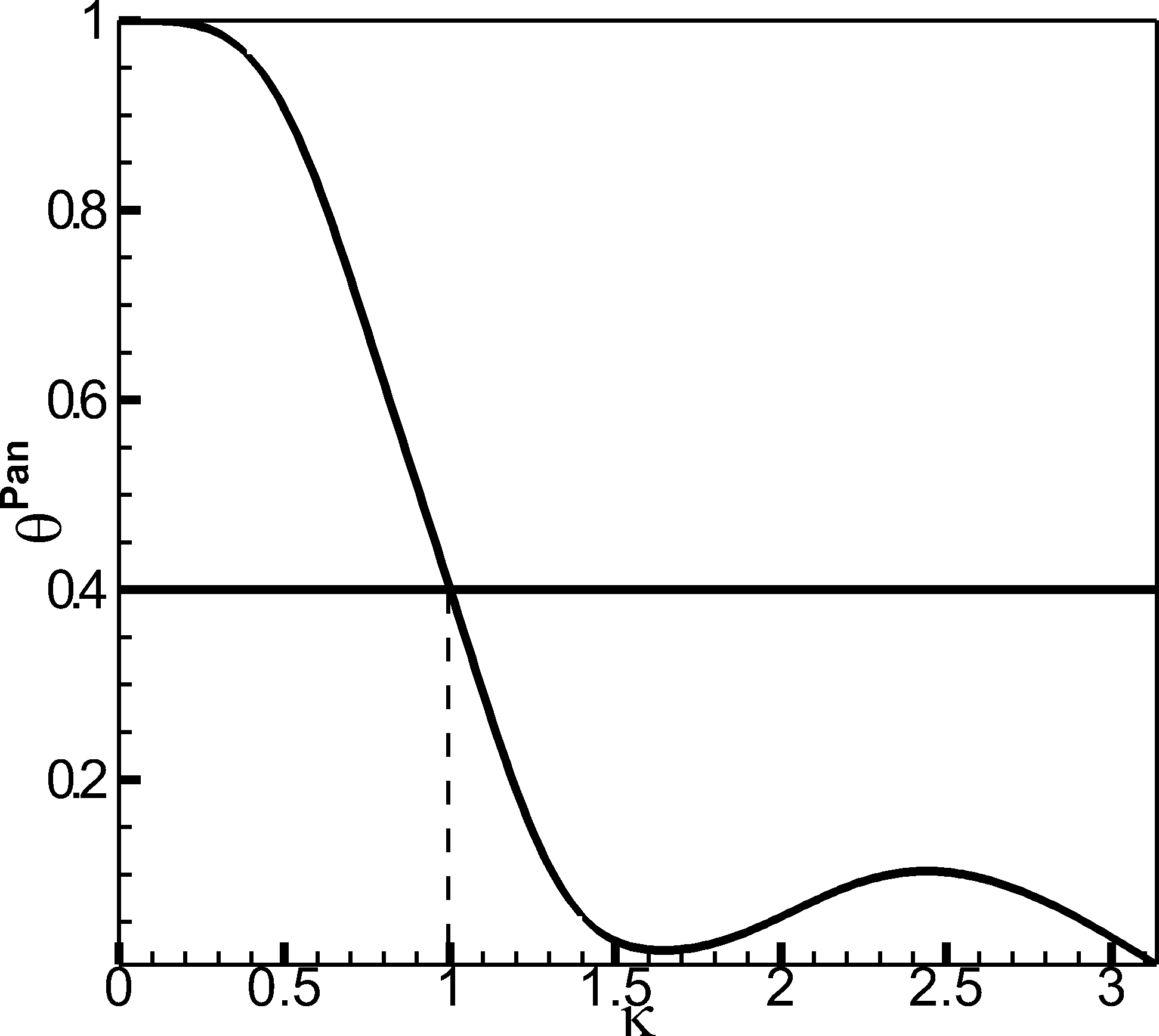}
  \caption{The averaged value of shock detector $\theta^{\mathrm{Li}}$ with function $u(x) = \sin(2\pi k x)$, $k = 1,2,3, \cdots, N/2$.
  \label{fig:sigmapanvk}}
\end{figure}

Taking $u(x) = \sin(2\pi kx)$ as the function to be reconstructed in the computational domain $\Omega  = [0,1]$ and $k = 1,2,3,\cdots, N/2$, the averaged value of $\theta_i^{\mathrm{Li}}$ versus the non-dimensional wavenumber $\kappa = 2 \pi k/N$ is shown as Fig. (\ref{fig:sigmapanvk}). When $\kappa = 1$, $\theta_i^{\mathrm{Li}}$ has the value of approximately $0.4$; when $\kappa = \pi$, $\theta_i^{\mathrm{Li}} = 0.005155$. And $\theta_i^{\mathrm{Li}}$ equals approximately 0.02 at the first local minimum and approximately 0.1 at the local maximum near $\kappa=2.5$. Figure \ref{fig:sigmapanvk} demonstrates that $\theta_i^{\mathrm{Li}}$ can accurately recognize the smooth regions where $\kappa < 1$ when $\theta_c \leq 0.4$.

\section{Extension to Navier-Stokes equations}
The 2D Navier-Stokes equations are
\begin{equation}
\frac{\partial U}{\partial t} + \frac{\partial \left(F-F_{\nu}\right)}{\partial x} + \frac{\partial \left(G-G_{\nu}\right)}{\partial y} = 0,
\label{eq:ns}
\end{equation}
where
\begin{align}
U = \left[\rho, \rho u_1, \rho u_2, \rho E\right]^T.
\end{align}
The inviscid fluxes are
\begin{align}
F & = \left[\rho u_1, \rho u_1^2 + p, \rho u_1 u_2, \rho H u_1\right]^T,\\
G & = \left[\rho u_2, \rho u_1 u_2, \rho u_2^2 + p, \rho H u_2\right]^T,
\end{align}
and the viscous fluxes are
\begin{align}
F_{\nu} & = \left[0, \tau_{11}, \tau_{21}, u_1 \tau_{11} + u_2\tau_{12} + k_h\frac{\partial T}{\partial x}\right]^T,\label{eq:visflux1}\\
G_{\nu} & = \left[0, \tau_{12}, \tau_{22}, u_1 \tau_{21} + u_2\tau_{22} + k_h\frac{\partial T}{\partial y}\right]^T.\label{eq:visflux2}
\end{align}
$\rho$, $u_1$, $u_2$, $p$, $H$ and $E$ are the density, $x$-velocity, $y$-velocity, pressure, total enthalpy and total energy for the fluid, respectively.
\begin{align}
E & = e + \frac{1}{2}(u_1^2+u_2^2),\\
H & = e + \frac{1}{2}(u_1^2+u_2^2) + \frac{p}{\rho},
\end{align}
where $e$ is the specific internal energy. In this paper, the perfect gas is considered and $e = \frac{p}{(\gamma-1)\rho}$, where $\gamma$ is the ratio of specific heat capacity.
$\tau_{ij}$ are the viscous stress tensor defined as
\begin{equation}
  \tau_{ij} = 2\mu S_{ij} -\frac{2}{3}S_{kk}\delta_{ij},
\end{equation}
where $\delta_{ij}$ is the Kronecker delta function,
\begin{equation}
  S_{ij} = \frac{1}{2}\left(\frac{\partial u_j}{\partial x_i} + \frac{\partial u_i}{\partial x_j} \right),
\end{equation}
and $\mu$ is the viscosity coefficient. $k_h$ in Eqs. (\ref{eq:visflux1}) and (\ref{eq:visflux2}) is the conductivity coefficient and relate with $\mu$ through the Prandtl number as
\begin{equation}
  \mathrm{Pr} = \frac{c_p \mu}{k_h},
\end{equation}
where $c_p$ is the specific heat capability at constant pressure.

\begin{figure}[htbp]
  \centering
  \includegraphics[width=0.6\columnwidth]{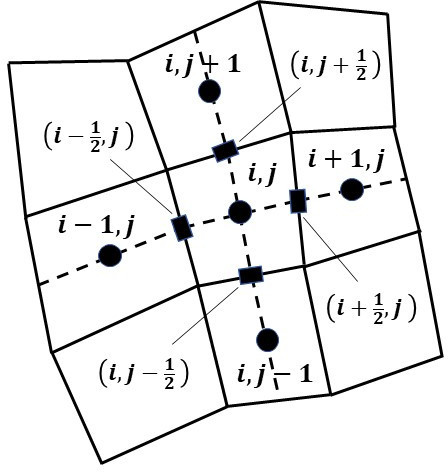}
  \caption{\label{fig:2dillusion} Illustration of the dimension-by-dimension reconstruction way on 2D curvilinear grids\cite{huang2018high}.}
\end{figure}

In this paper, the 2D Navier-Stokes equations are solved in a dimension-by-dimension way shown as in Fig. \ref{fig:2dillusion}.

In each direction, the hybrid CLS-CWENO schemes are utilized to obtain piece-wise smooth polynomials for each cell, from which the physical variables at the center of cell interfaces are calculated.  A common shock detector $\sigma_i$ calculated through density and pressure as
\begin{equation}
\sigma_i = \min(\sigma_{i}(\rho), \sigma_{i}(p))
\end{equation}
is utilized for all the conservative variables.
Approximate Roe Riemann solver with h-correction \cite{sanders1998multidimensional} is used to obtain the universal numerical inviscid flux at a certain interface.

As to the viscous flux, firstly, a second-order cell averaged derivatives are obtained by Gauss formula
\begin{equation}
  \begin{aligned}
\overline{\left[\nabla U\right]}_{i,j} & = 
  \frac{1}{\Omega_{ij}}\int_{\partial \Omega_{ij}}{U\cdot \vec{n}\mathrm{d}S} \\
   & = 
  \frac{1}{\Omega_{ij}}\left(
     U_{i+1/2,j} \cdot \vec{n}_{i+1/2,j} S_{i+1/2,j}\right. \\
  &\quad\quad\quad \left.  +U_{i-1/2,j} \cdot \vec{n}_{i-1/2,j} S_{i-1/2,j} \right.\\
  &\quad\quad\quad   +U_{i,j+1/2} \cdot \vec{n}_{i,j+1/2} S_{i,j+1/2}\\
  &\quad\quad\quad  \left. +U_{i,j-1/2} \cdot \vec{n}_{i,j-1/2} S_{i,j-1/2}
    \right), \\
  \end{aligned}
\end{equation}
where $\vec{n}$ and $S$ are the outward unit normal vector and length of the interface, respectively. 
\begin{equation}
  \begin{aligned}
  U_{i\pm 1/2,j} = \frac{1}{2}(U_{i \pm 1/2,j}^L + U_{i \pm 1/2,j}^R),\\
  U_{i,j\pm 1/2} = \frac{1}{2}(U_{i ,j\pm 1/2}^L + U_{i ,j\pm 1/2}^R).
  \end{aligned}
\end{equation}
The variables with superscript $L$ and $R$ are the reconstructed point values by the hybrid CLS-CWENO schemes.
Then the gradients of variables at interfaces are interpolated by, taking the face $(i+1/2,j)$ as an example, 
\begin{equation}
  \begin{aligned}
  \left[\nabla U\right]_{i+1/2,j} = \frac{
  1 }{
    \left(\Delta x\right)_{i,j} + 
    \left(\Delta x\right)_{i+1,j}
  } \left(
    \left(\Delta x\right)_{i+1,j}\cdot \overline{\left[\nabla U\right]}_{i,j} \right. \\
    \left. \quad\quad +\left(\Delta x\right)_{i,j}\cdot \overline{\left[\nabla U\right]}_{i+1,j} 
\right) \\
+ \frac{\vec{n}_{i+1/2,j}}{
    \left(\Delta x\right)_{i,j} + 
    \left(\Delta x\right)_{i+1,j}
} \cdot \left(U_{i+1/2,j}^R - U_{i+1/2,j}^L\right).
  \end{aligned}
\end{equation}
Eventually, a Runge-Kutta (RK) method is utilized to discretize Eqs. (\ref{eq:ns}) in temporal direction.

\section{Numerical examples \label{sec:examples}}
If not specifically emphasized, the time discretization utilized in this section is the third-order SSP-RK method\cite{gottlieb2011strong}.

\subsection{Calibration of $\theta_c$}
Firstly, a linear problem with convection equation and a nonlinear problem with Euler equations are utilized to check the influence of the critical number $\theta_c$ in the two different shock detectors, i.e., $\sigma^{\mathrm{Ren}}$ and $\sigma^{\mathrm{Li}}$.
In this section, the coefficients for the third-order scheme are the ones optimized by $\kappa_c = 1.0$ as in Tab. \ref{tab:w1w2}; the coefficients for the fifth-order scheme are the ones optimized by $\beta = 1.0$ as in Tab. \ref{tab:w1w2w3w4}.

\subsubsection{Linear problem with Gaussian-square-triangle-ellipse waves\label{sec:gste}}
The initial condition is
\begin{equation}
  \begin{aligned}
  &u_0(x)  = \\
  &\,\, \left\{
  \begin{array}{ll}
    \begin{array}{lr}
    \frac{1}{6}\left(
      G(x,\beta,z-\delta) 
      +G(x,\beta, \right.\\
      \quad\left. z+\delta) +4G(x,\beta,z)
    \right), 
    \end{array}
    & 0.2 \leq x \leq 0.4,\\
    1, & 0.6 \leq x \leq 0.8,\\
    1-\left|10(x-1.1)\right|,& 1.0 \leq x \leq 1.2,\\
    \begin{array}{lr}
    \frac{1}{6}\left(
      F(x,\alpha,a-\delta)
      +F(x,\alpha, \right.\\
      \quad \left.a+\delta)+4 F(x,\alpha,a)
    \right) ,
    \end{array}
     & 1.4 \leq x \leq 1.6,\\
    0,& \text{elsewise},
  \end{array}
  \right.
\end{aligned}
  \label{eq:gste}
\end{equation}

\noindent where 
\begin{align}
G(x,\beta,z) & = e^{\beta(x-1-z)^2},\\
F(x,\alpha,a) & = \sqrt{\max(1-\alpha^2(x-1-a^2), 0)},
\end{align}
$z = -0.7$, $\delta = 0.005$, $\beta = \frac{\log_{10} 2}{36\delta^2}$, $\alpha = 10$ and $a = 0.5$.

The computational domain is $\Omega = [0,2]$ with periodic boundary condition. The number of cells is $N = 400$, the simulation end time is $t_{end}=2$ and the CFL number is $0.5$.

\begin{figure}[!htbp]
  \centering
    \begin{subfigure}[b]{\columnwidth}
    \includegraphics[width=0.6\columnwidth]{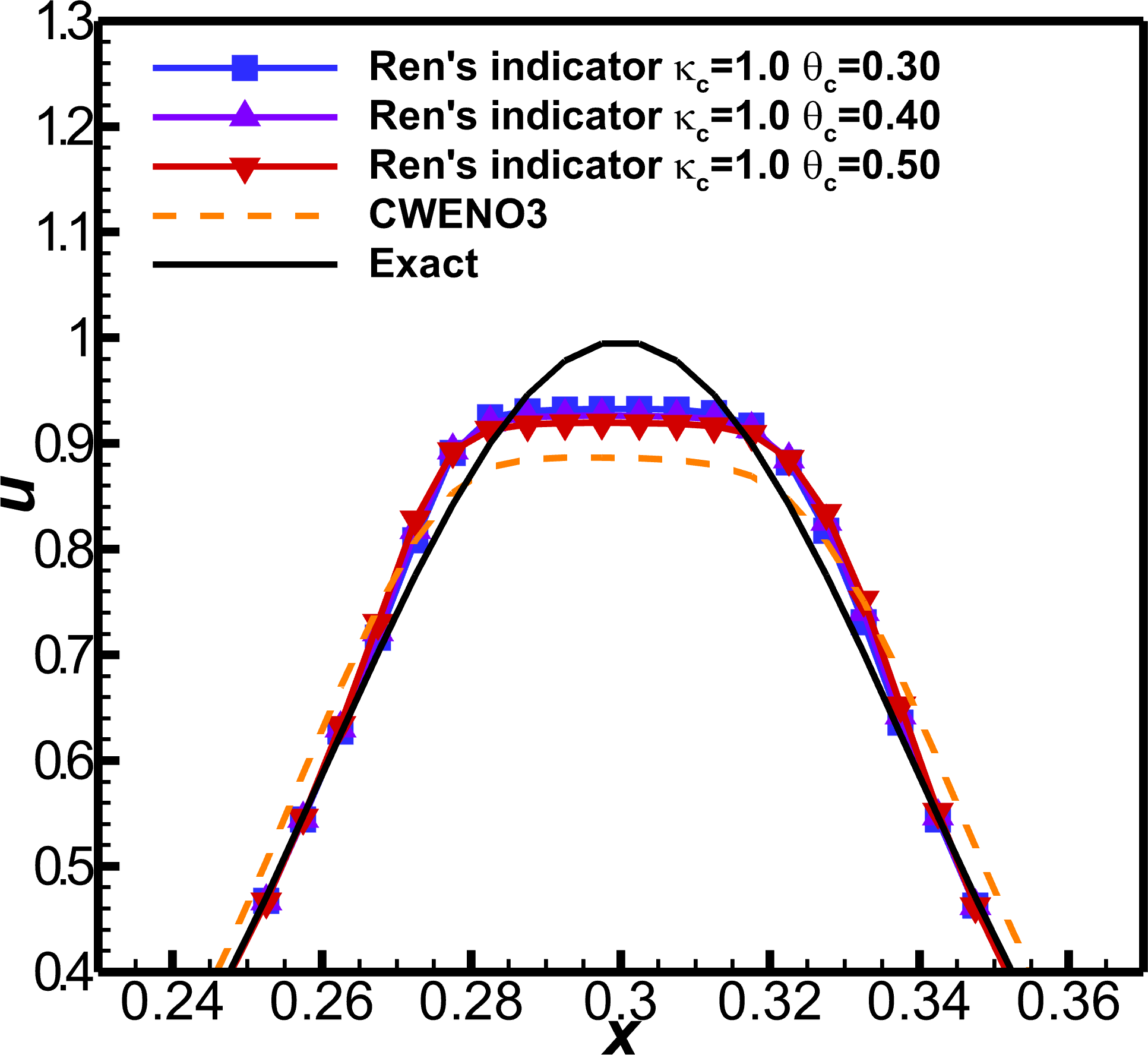}
    \caption{\label{fig:clsweno3_critical_value_compare_gste_ren_1}Close view of Gaussian wave.}
    \end{subfigure}
    \hfill
    \begin{subfigure}[b]{\columnwidth}
    \includegraphics[width=0.6\columnwidth]{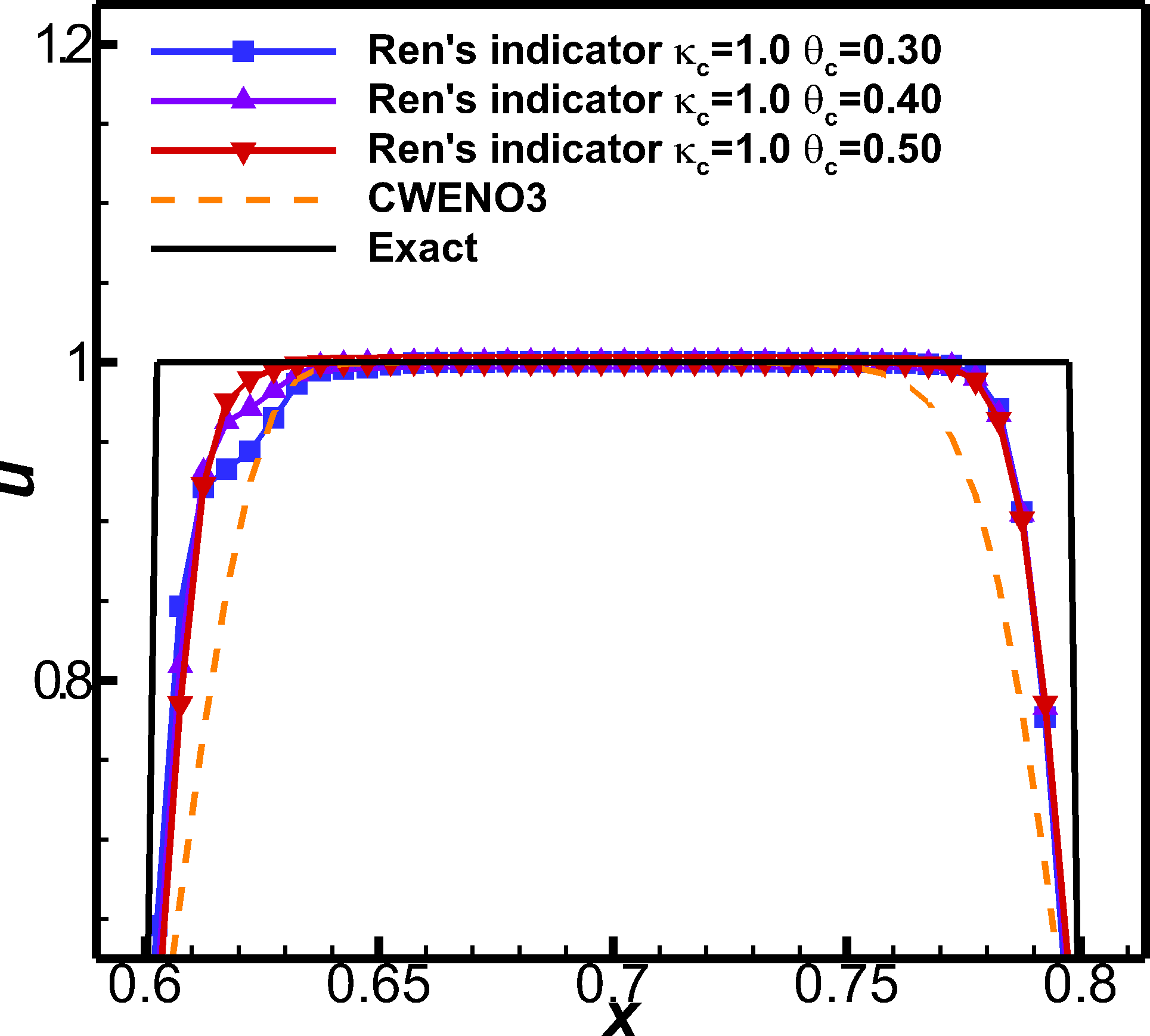}
    \caption{\label{fig:clsweno3_critical_value_compare_gste_ren_2}Close view of square wave.}
    \end{subfigure}
    \begin{subfigure}[b]{\columnwidth}
    \includegraphics[width=0.6\columnwidth]{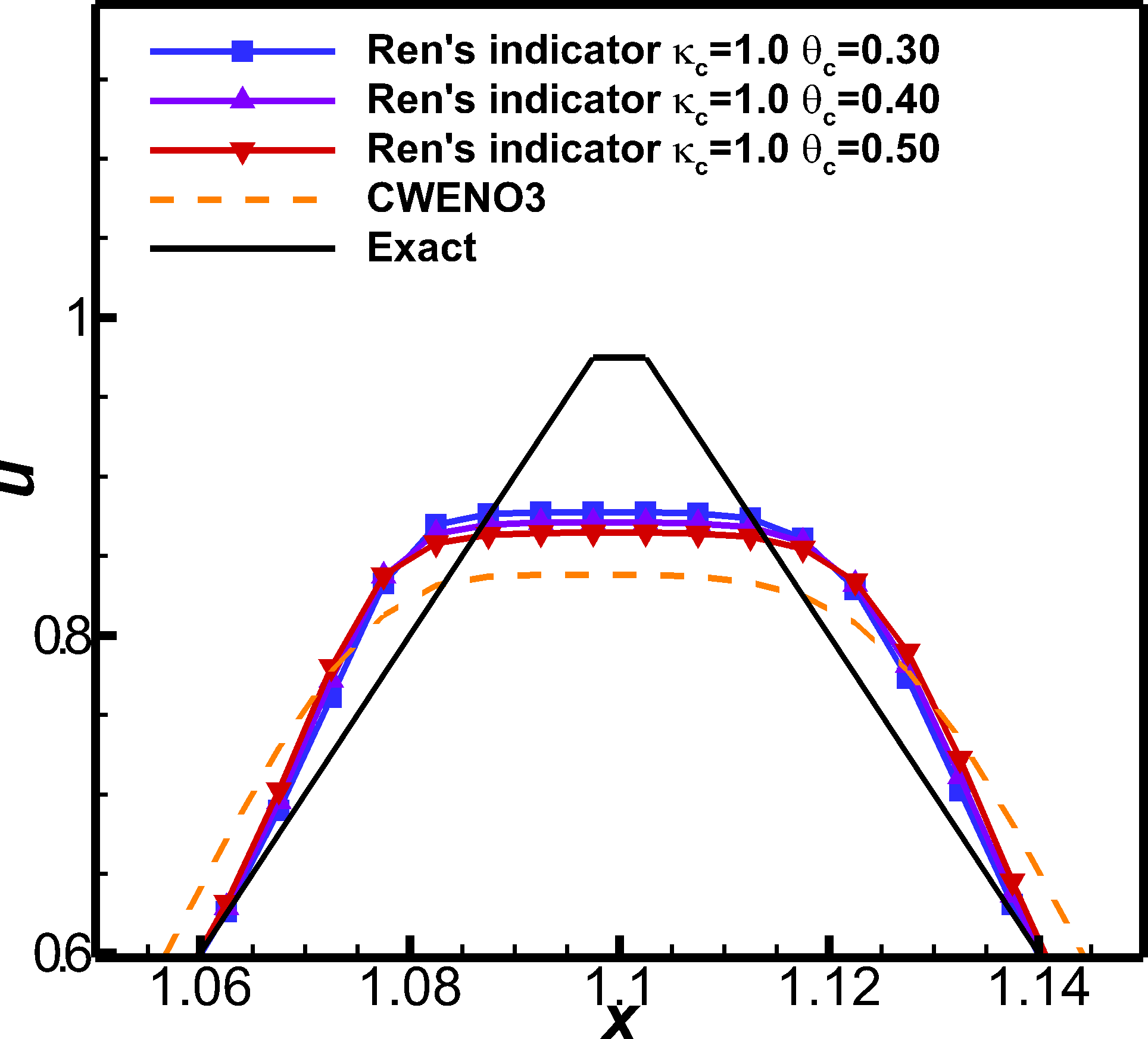}
    \caption{\label{fig:clsweno3_critical_value_compare_gste_ren_3}Close view of triangle wave.}
    \end{subfigure}
    \hfill
    \begin{subfigure}[b]{\columnwidth}
    \includegraphics[width=0.6\columnwidth]{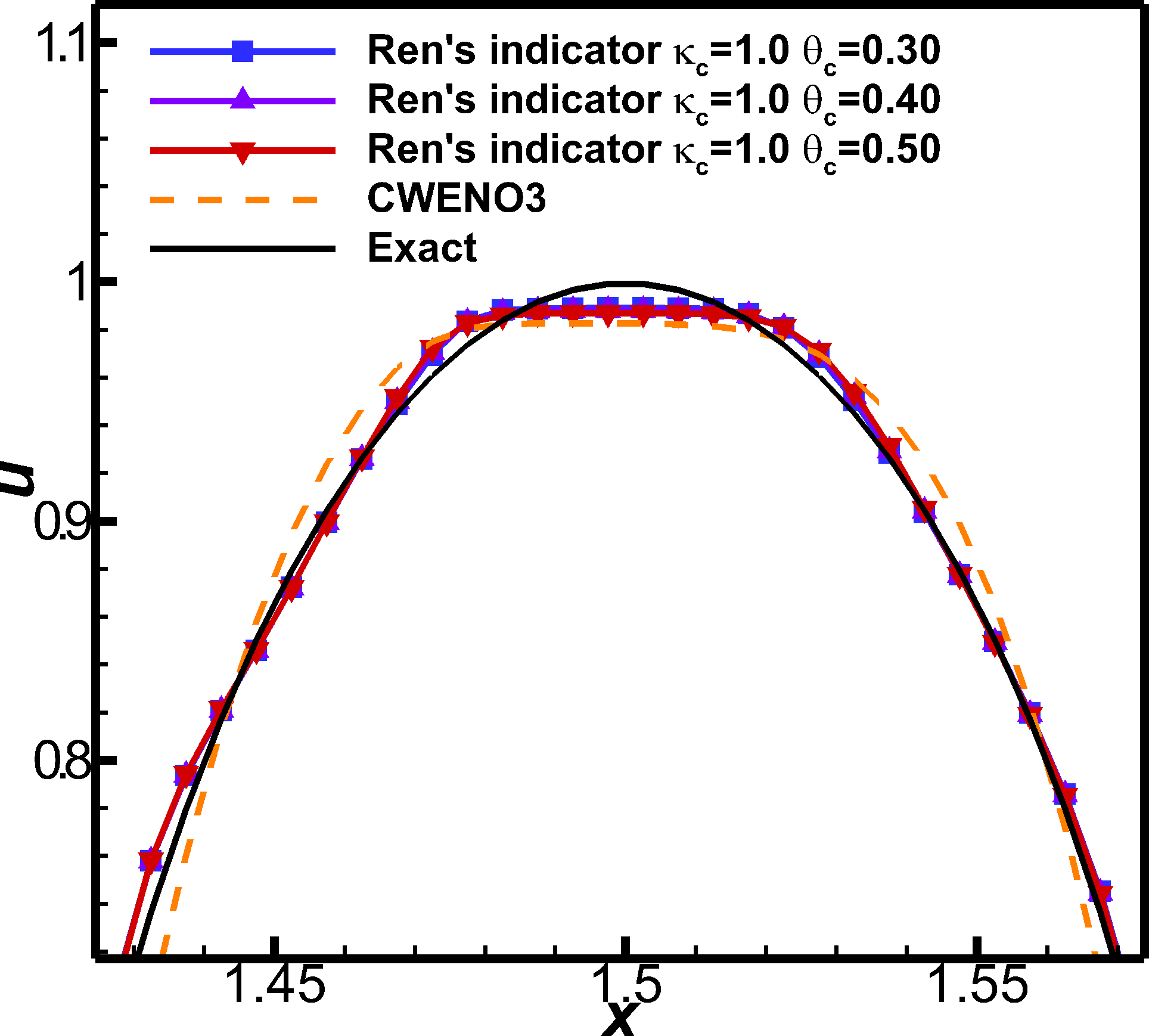}
    \caption{\label{fig:clsweno3_critical_value_compare_gste_ren_4}Close view of ellipse wave.}
    \end{subfigure}
    \caption{\label{fig:clsweno3_critical_value_compare_gste_ren} Linear convection problem with Gaussian-square-triangle-ellipse waves solved by the third-order hybrid CLS-CWENO scheme. Shock detector is $\sigma^{\mathrm{Ren}}$.}
\end{figure}

\begin{figure}[!htbp]
  \centering
    \begin{subfigure}[b]{\columnwidth}
    \includegraphics[width=0.6\columnwidth]{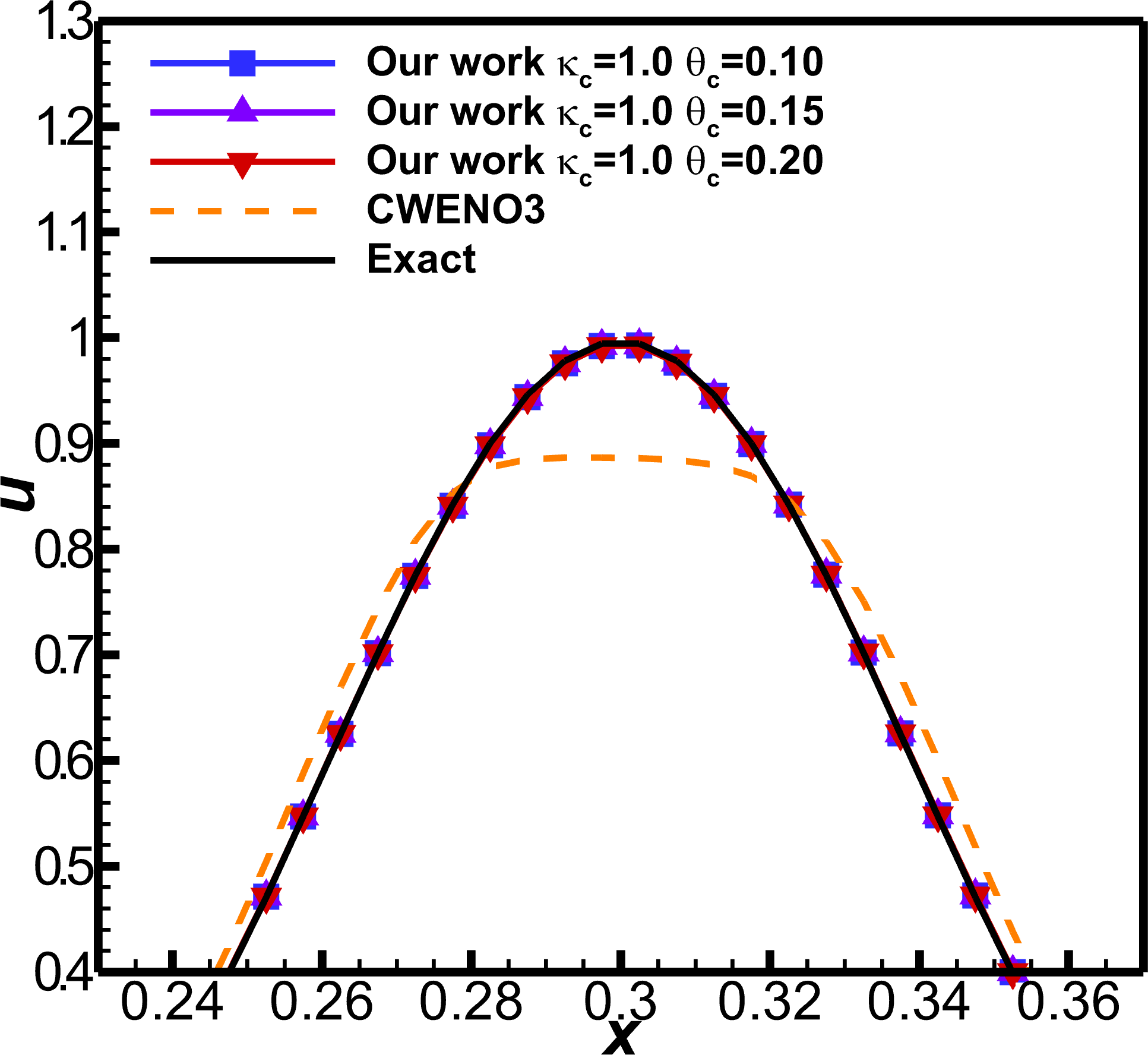}
    \caption{\label{fig:clsweno3_critical_value_compare_gste_our_work_1}Close view of Gaussian wave.}
    \end{subfigure}
    \hfill
    \begin{subfigure}[b]{\columnwidth}
    \includegraphics[width=0.6\columnwidth]{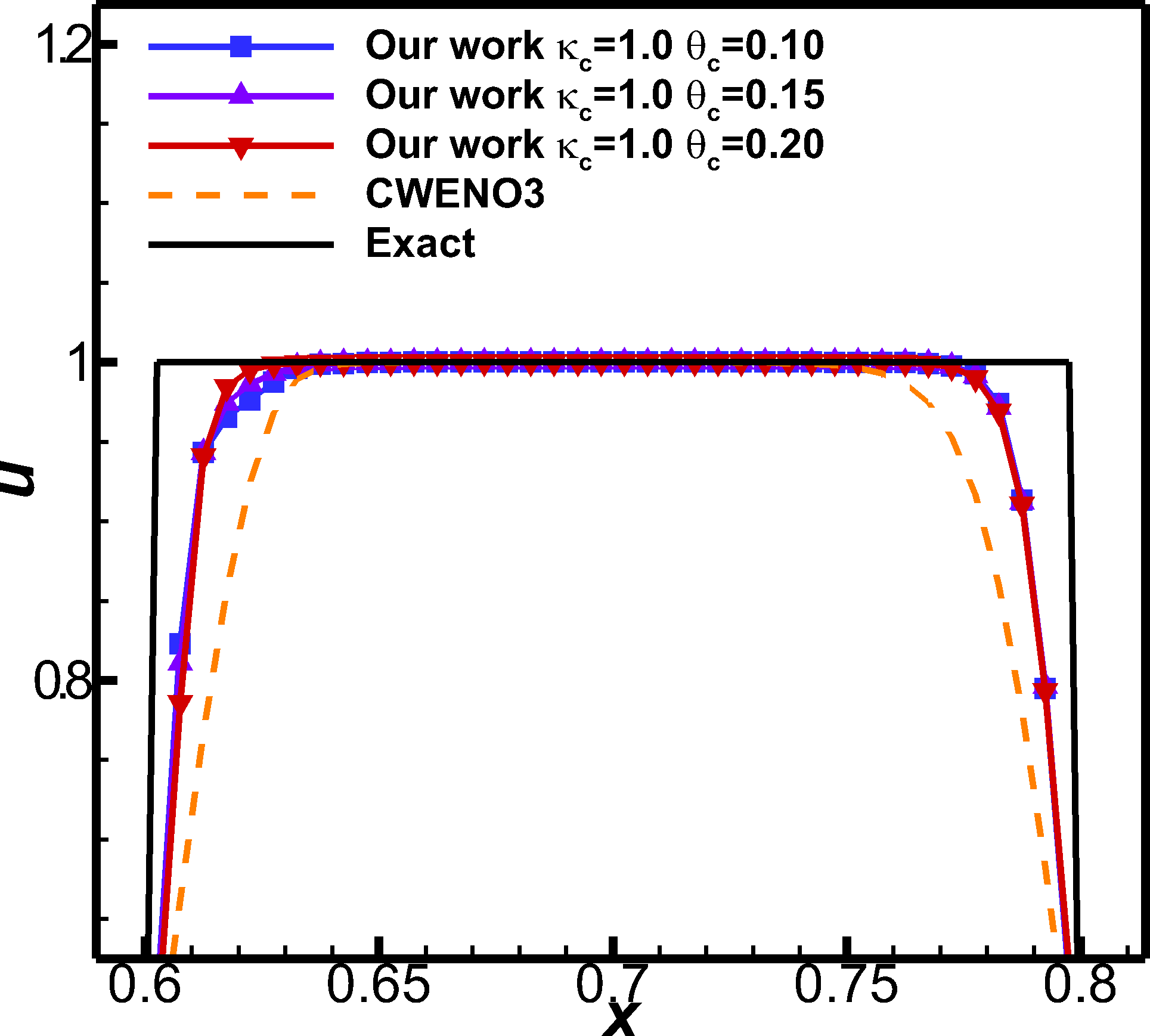}
    \caption{\label{fig:clsweno3_critical_value_compare_gste_our_work_2}Close view of square wave.}
    \end{subfigure}
    \begin{subfigure}[b]{\columnwidth}
    \includegraphics[width=0.6\columnwidth]{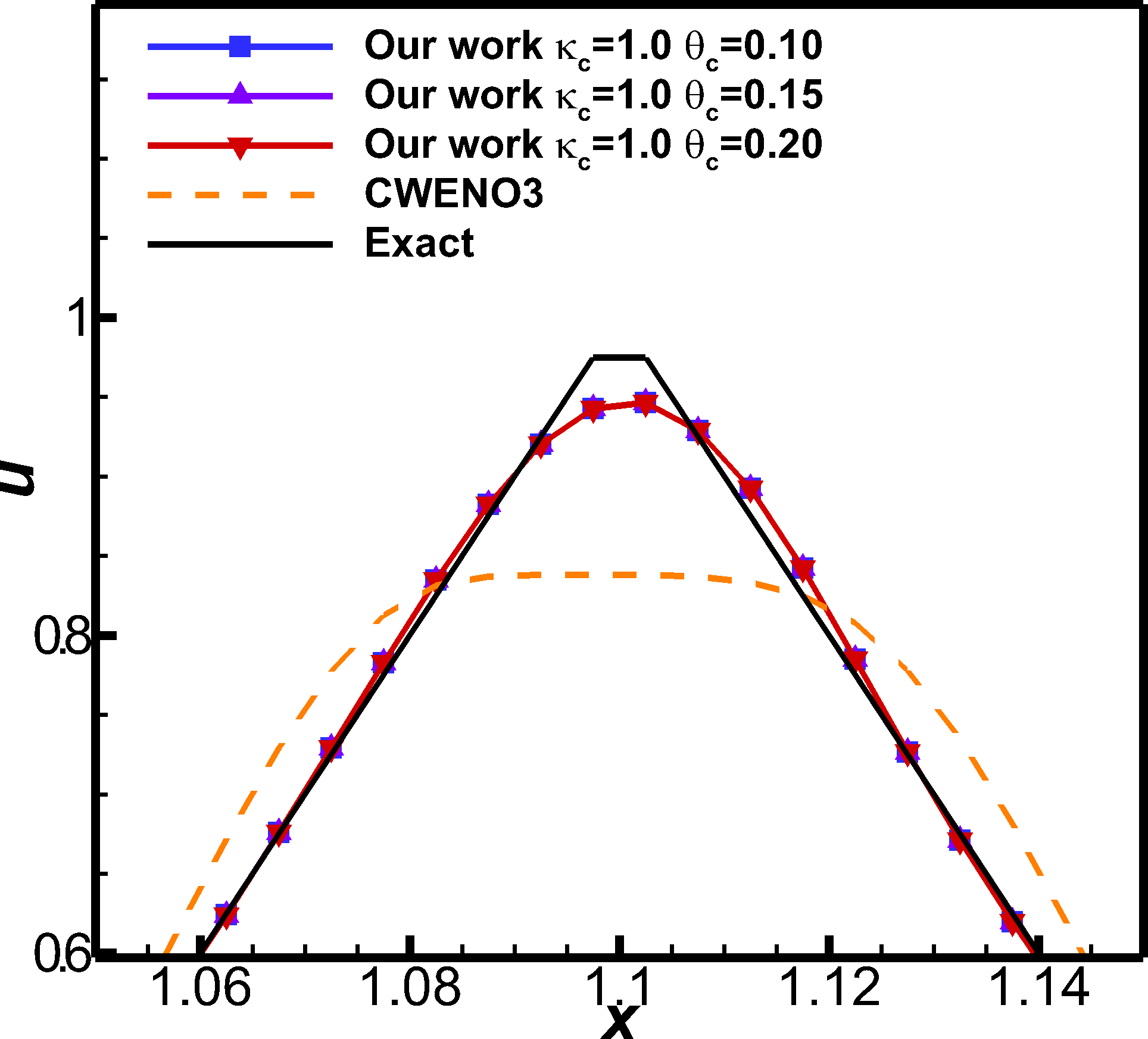}
    \caption{\label{fig:clsweno3_critical_value_compare_gste_our_work_3}Close view of triangle wave.}
    \end{subfigure}
    \hfill
    \begin{subfigure}[b]{\columnwidth}
    \includegraphics[width=0.6\columnwidth]{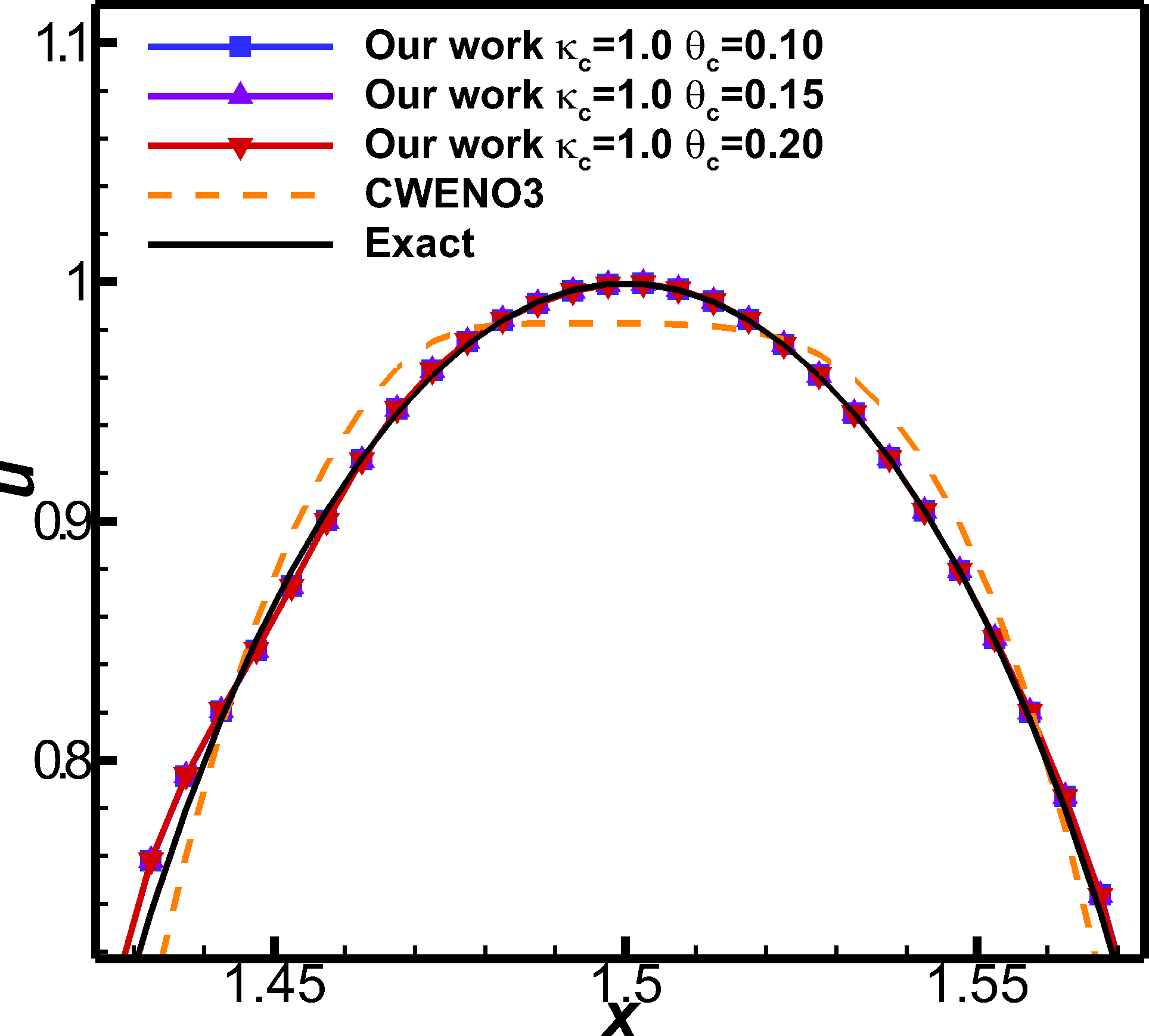}
    \caption{\label{fig:clsweno3_critical_value_compare_gste_our_work_4}Close view of ellipse wave.}
    \end{subfigure}
    \caption{\label{fig:clsweno3_critical_value_compare_gste_our_work} Linear convection problem with Gaussian-square-triangle-ellipse waves solved by the third-order hybrid CLS-CWENO scheme. Shock detector is $\sigma^{\mathrm{Li}}$.}
\end{figure}

Figures \ref{fig:clsweno3_critical_value_compare_gste_ren} and \ref{fig:clsweno3_critical_value_compare_gste_our_work} show the results of the third-order hybrid CLS-CWENO scheme using the shock detector $\sigma^{\mathrm{Ren}}$ and $\sigma^{\mathrm{Li}}$, respectively, with different critical number $\theta_c$. According to Fig. \ref{fig:clsweno3_critical_value_compare_gste_ren}, for the third-order hybrid CLS-CWENO scheme with $\sigma^{\mathrm{Ren}}$, all the three values of $\theta_c = 0.3$, $0.4$ and $0.5$ can eliminate the oscillations near square wave and capture the Gaussian, triangle and ellipse waves smoothly. In addition, as shown in Fig. \ref{fig:clsweno3_critical_value_compare_gste_ren_2}, for third-order hybrid CLS-CWENO scheme with $\sigma^{\mathrm{Ren}}$, $\theta_c = 0.5$ produces the smoothest result around the square wave. As shown in Figs. \ref{fig:clsweno3_critical_value_compare_gste_ren_1}, \ref{fig:clsweno3_critical_value_compare_gste_ren_3} and \ref{fig:clsweno3_critical_value_compare_gste_ren_4}, the smooth extrema are dissipated due to the inability of $\sigma^{\mathrm{Ren}}$ to distinguish between the local discontinuities and local extrema. 
On the other hand, as shown in Fig. \ref{fig:clsweno3_critical_value_compare_gste_our_work}, the third-order CLS-CWENO scheme using $\sigma^{\mathrm{Li}}$ with $\theta_c = 0.10$, $0.15$ and $0.20$ all give oscillation-free results and the extrema of Gaussian, triangle and ellipse waves are correctly identified. According to Fig. \ref{fig:clsweno3_critical_value_compare_gste_our_work_2}, $\theta_c = 0.20$ gives the smoothest result around the discontinuity of the square wave for the third-order hybrid CLS-CWENO scheme using $\sigma^{\mathrm{Li}}$.

\begin{figure}[!htbp]
  \centering
    \begin{subfigure}[b]{\columnwidth}
    \includegraphics[width=0.6\columnwidth]{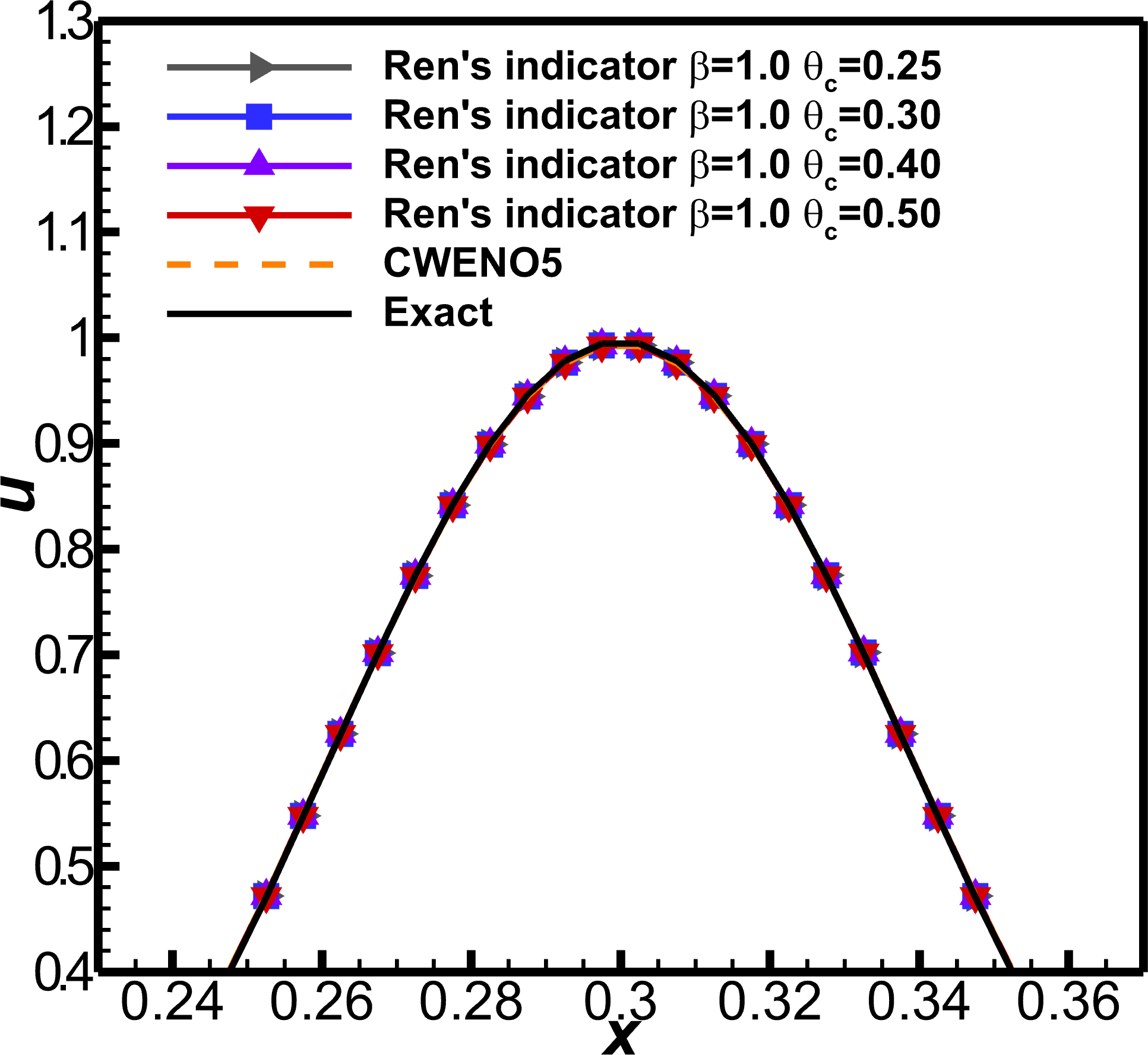}
    \caption{\label{fig:clsweno5_critical_value_compare_gste_ren_1}Close view of Gaussian wave.}
    \end{subfigure}
    \hfill
    \begin{subfigure}[b]{\columnwidth}
    \includegraphics[width=0.6\columnwidth]{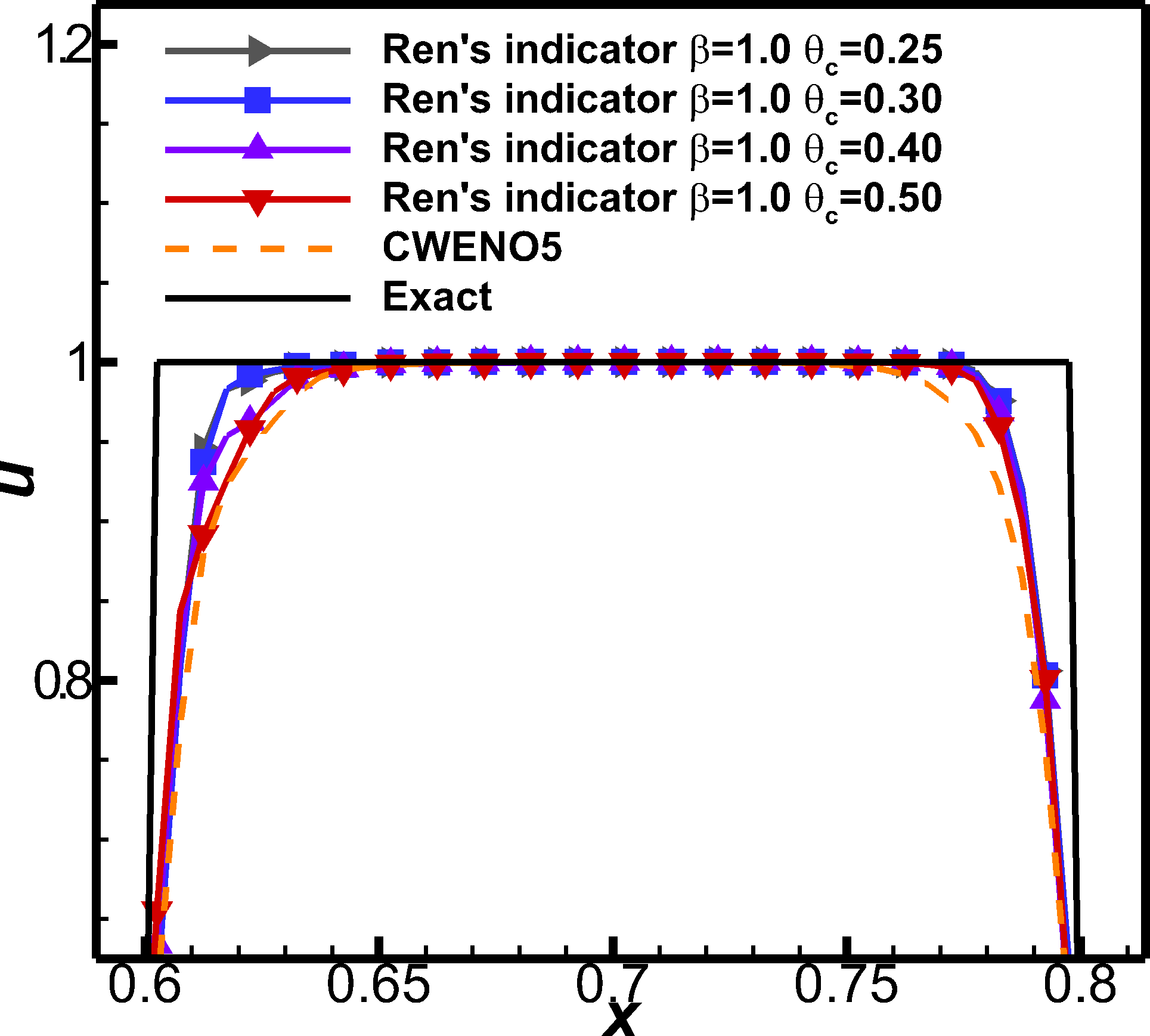}
    \caption{\label{fig:clsweno5_critical_value_compare_gste_ren_2}Close view of square wave.}
    \end{subfigure}
    \begin{subfigure}[b]{\columnwidth}
    \includegraphics[width=0.6\columnwidth]{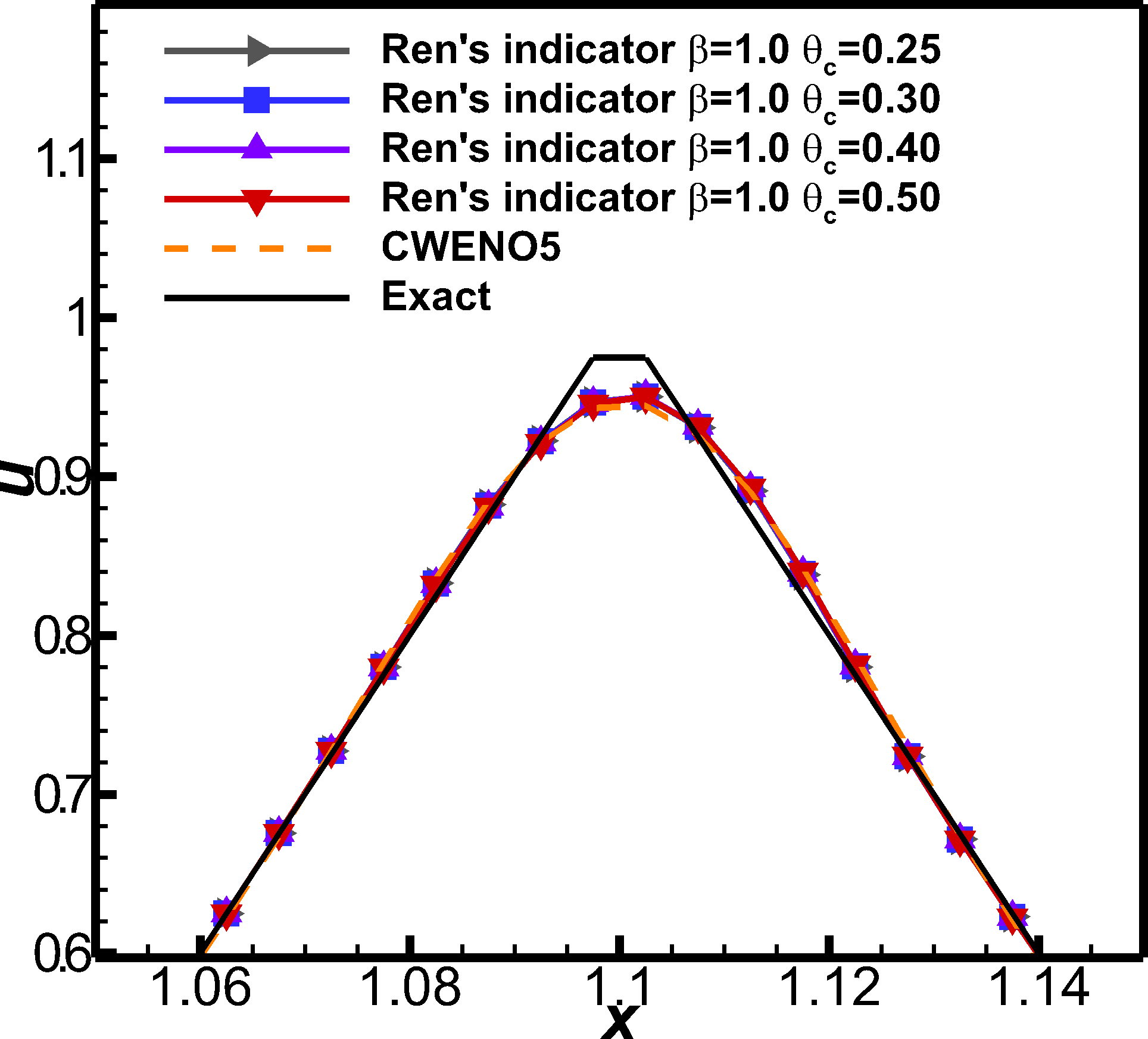}
    \caption{\label{fig:clsweno5_critical_value_compare_gste_ren_3}Close view of triangle wave.}
    \end{subfigure}
    \hfill
    \begin{subfigure}[b]{\columnwidth}
    \includegraphics[width=0.6\columnwidth]{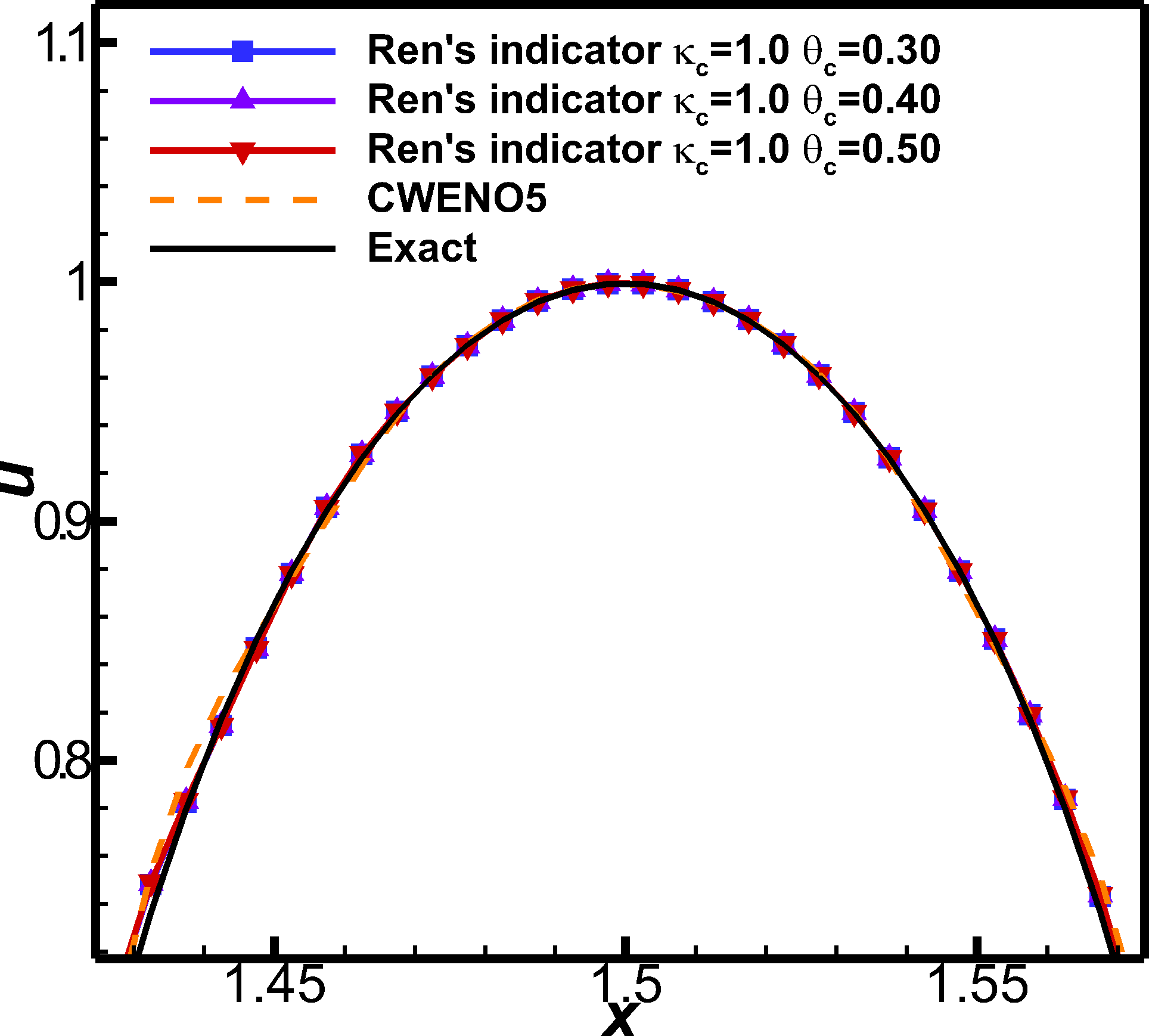}
    \caption{\label{fig:clsweno5_critical_value_compare_gste_ren_4}Close view of ellipse wave.}
    \end{subfigure}
    \caption{\label{fig:clsweno5_critical_value_compare_gste_ren} Linear convection problem with Gaussian-square-triangle-ellipse waves solved by the fifth-order hybrid CLS-CWENO scheme. Shock detector is $\sigma^{\mathrm{Ren}}$.}
\end{figure}

\begin{figure}[!htbp]
  \centering
    \begin{subfigure}[b]{\columnwidth}
    \includegraphics[width=0.6\columnwidth]{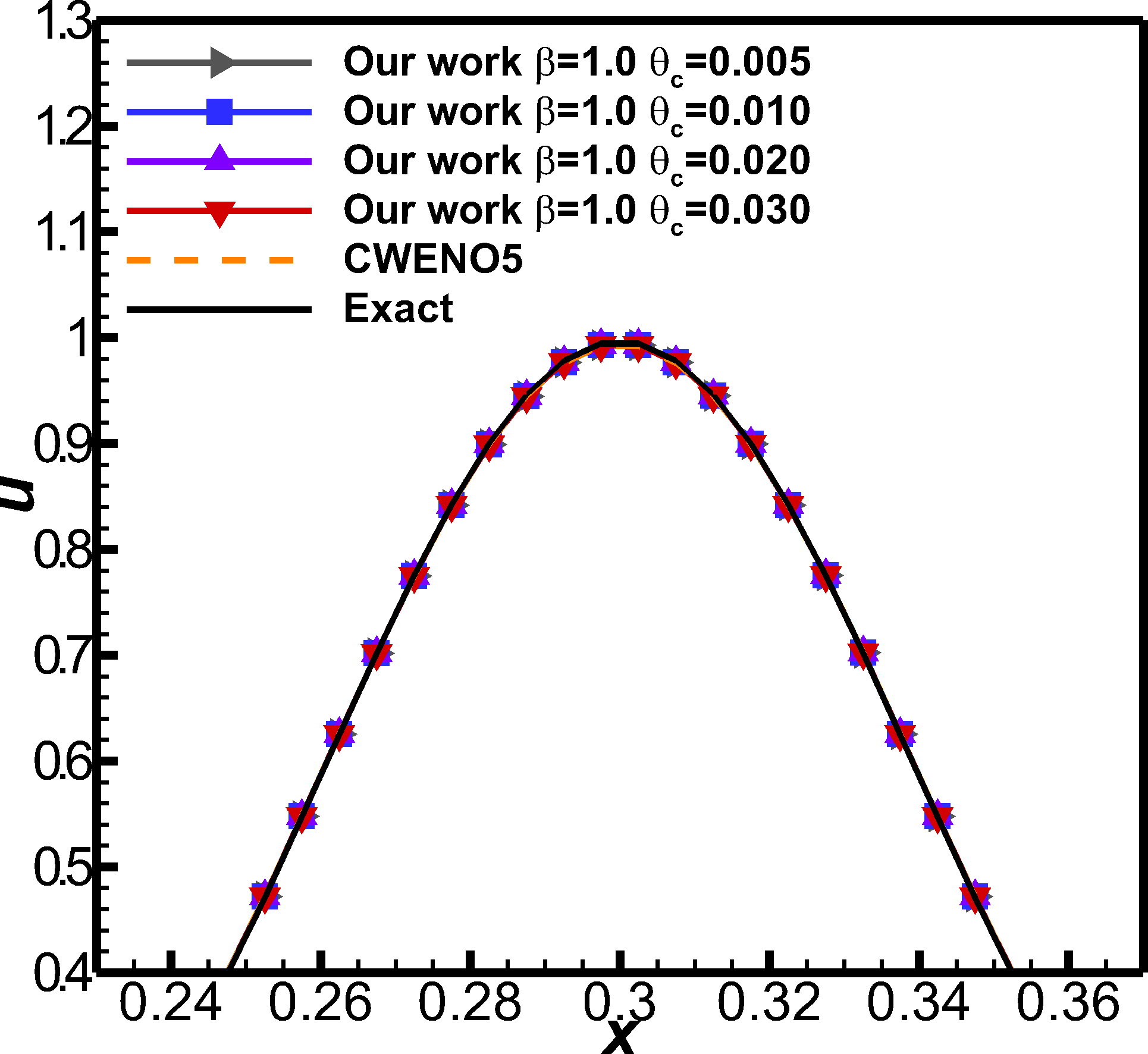}
    \caption{\label{fig:clsweno5_critical_value_compare_gste_our_work_1}Close view of Gaussian wave.}
    \end{subfigure}
    \hfill
    \begin{subfigure}[b]{\columnwidth}
    \includegraphics[width=0.6\columnwidth]{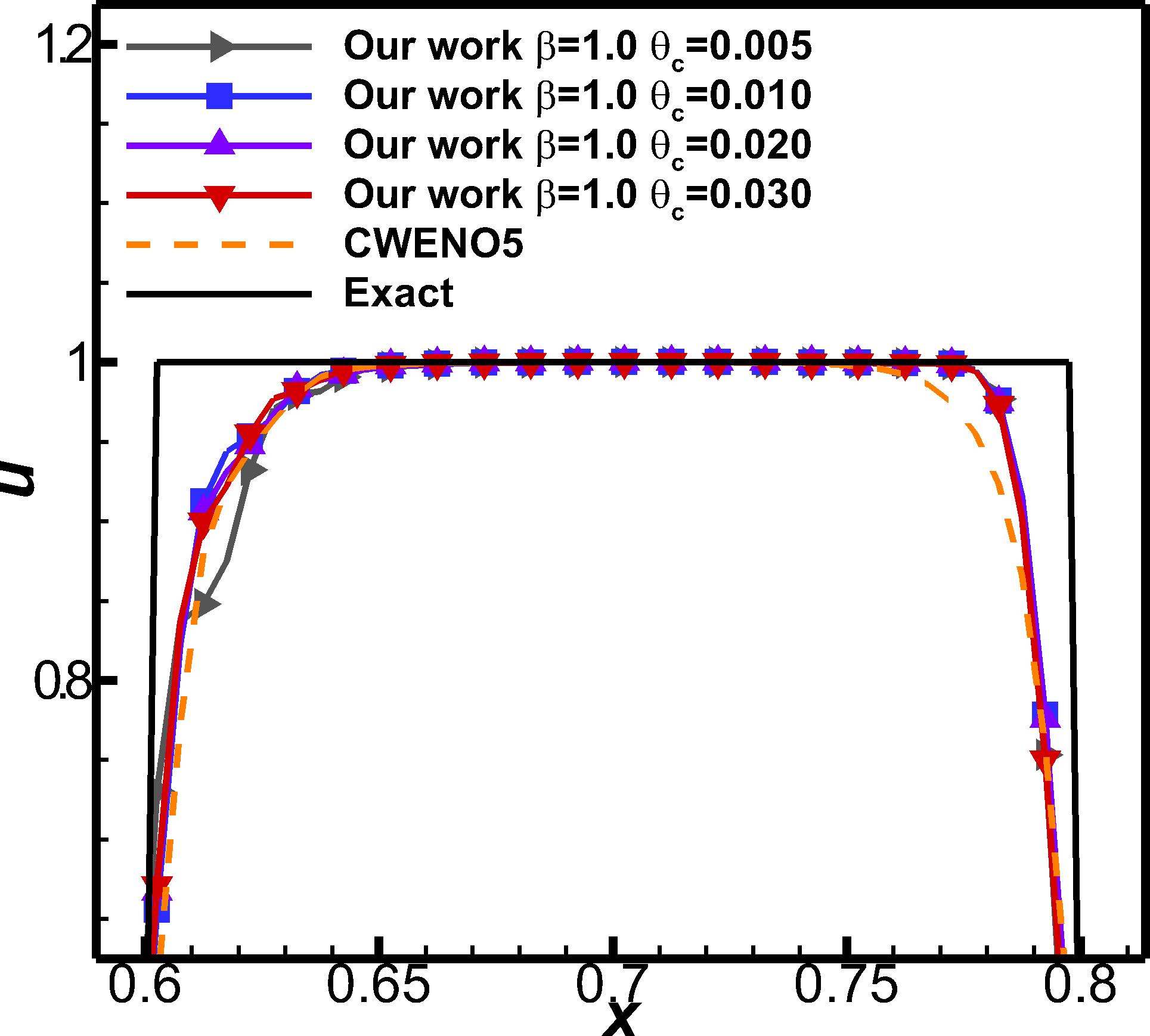}
    \caption{\label{fig:clsweno5_critical_value_compare_gste_our_work_2}Close view of square wave.}
    \end{subfigure}
    \begin{subfigure}[b]{\columnwidth}
    \includegraphics[width=0.6\columnwidth]{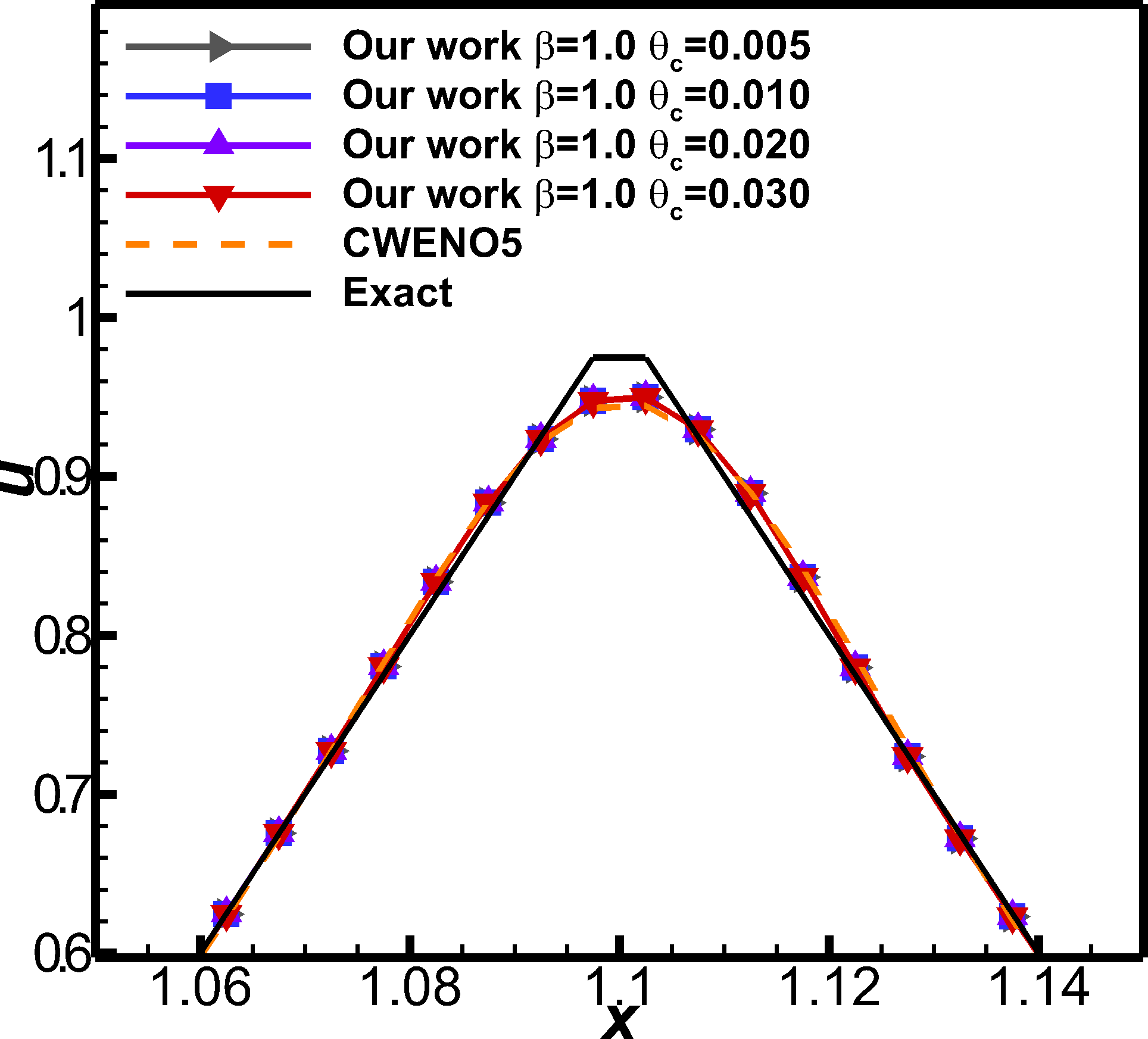}
    \caption{\label{fig:clsweno5_critical_value_compare_gste_our_work_3}Close view of triangle wave.}
    \end{subfigure}
    \hfill
    \begin{subfigure}[b]{\columnwidth}
    \includegraphics[width=0.6\columnwidth]{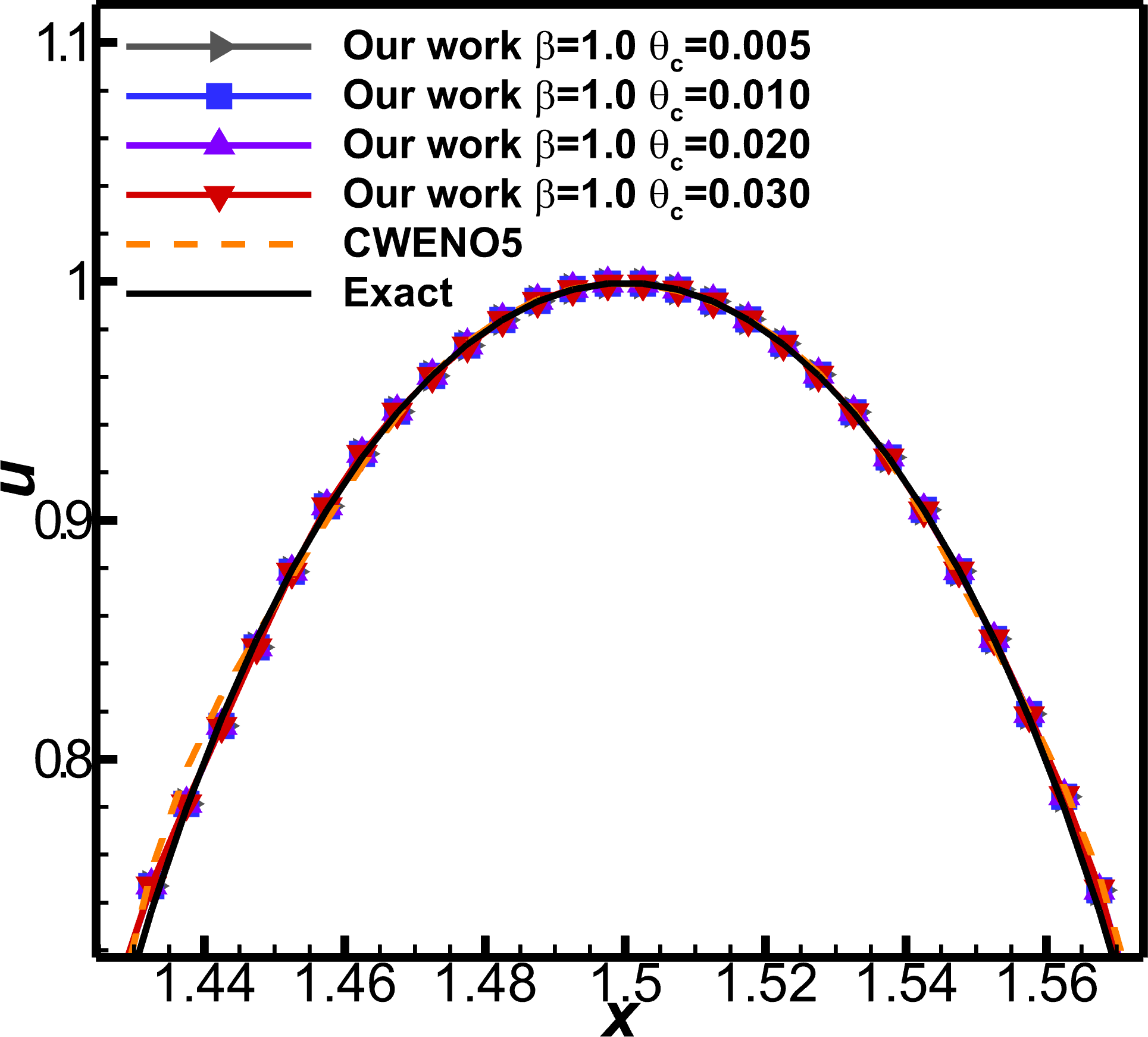}
    \caption{\label{fig:clsweno5_critical_value_compare_gste_our_work_4}Close view of ellipse wave.}
    \end{subfigure}
    \caption{\label{fig:clsweno5_critical_value_compare_gste_our_work} Linear convection problem with Gaussian-square-triangle-ellipse waves solved by the fifth-order hybrid CLS-CWENO scheme. Shock detector is $\sigma^{\mathrm{Li}}$.}
\end{figure}

Figures \ref{fig:clsweno5_critical_value_compare_gste_ren} and \ref{fig:clsweno5_critical_value_compare_gste_our_work} show the results of the fifth-order hybrid CLS-CWENO scheme using $\sigma^{\mathrm{Ren}}$ and $\sigma^{\mathrm{Li}}$, respectively. All the figures present oscillation-free results around the Gaussian, square, triangle and ellipse waves. Additionally, the fifth-order hybrid CLS-CWENO scheme using $\sigma^{\mathrm{Ren}}$ with $\theta_c=0.3$ gives the best result according to Fig. \ref{fig:clsweno5_critical_value_compare_gste_ren_2} and the fifth-order hybrid CLS-CWENO scheme using $\sigma^{\mathrm{Li}}$ with $\theta = 0.03$ are the smoothest around the square shown as in Fig. \ref{fig:clsweno5_critical_value_compare_gste_our_work_2}. The fifth-order hybrid CLS-CWENO schemes are able to capture the extrema whichever $\sigma^{\mathrm{Ren}}$ or $\sigma^{\mathrm{Li}}$ is utilized due to the superiority of the underlying fifth-order CWENO scheme than the third-order CWENO scheme.

\subsubsection{Lax shock tube problem\cite{lax}\label{sec:calibration_lax}}
Lax shock tube problem with the following initial conditions is utilized to test the performance of $\theta_c$ around discontinuities for the proposed hybrid schemes,
\begin{equation}
  \left[\rho, u, p\right] = \left\{
  \begin{array}{ll}
    0.445, 0.698, 3.528, & \text{if}\,\,0\leq x\le 0.5,\\
    0.5, 0, 0.571, & \text{if}\,\,0.5 \leq x\leq 1.0.\\
  \end{array}
  \right.
\end{equation}
The computational domain is $\Omega = [0,1]$ discretized by $200$ uniform cells. The final simulation time is $t_{end} = 0.1$ and the CFL number is 0.5.
\begin{figure}[!htbp]
  \centering
    \begin{subfigure}[b]{\columnwidth}
    \includegraphics[width=0.6\columnwidth]{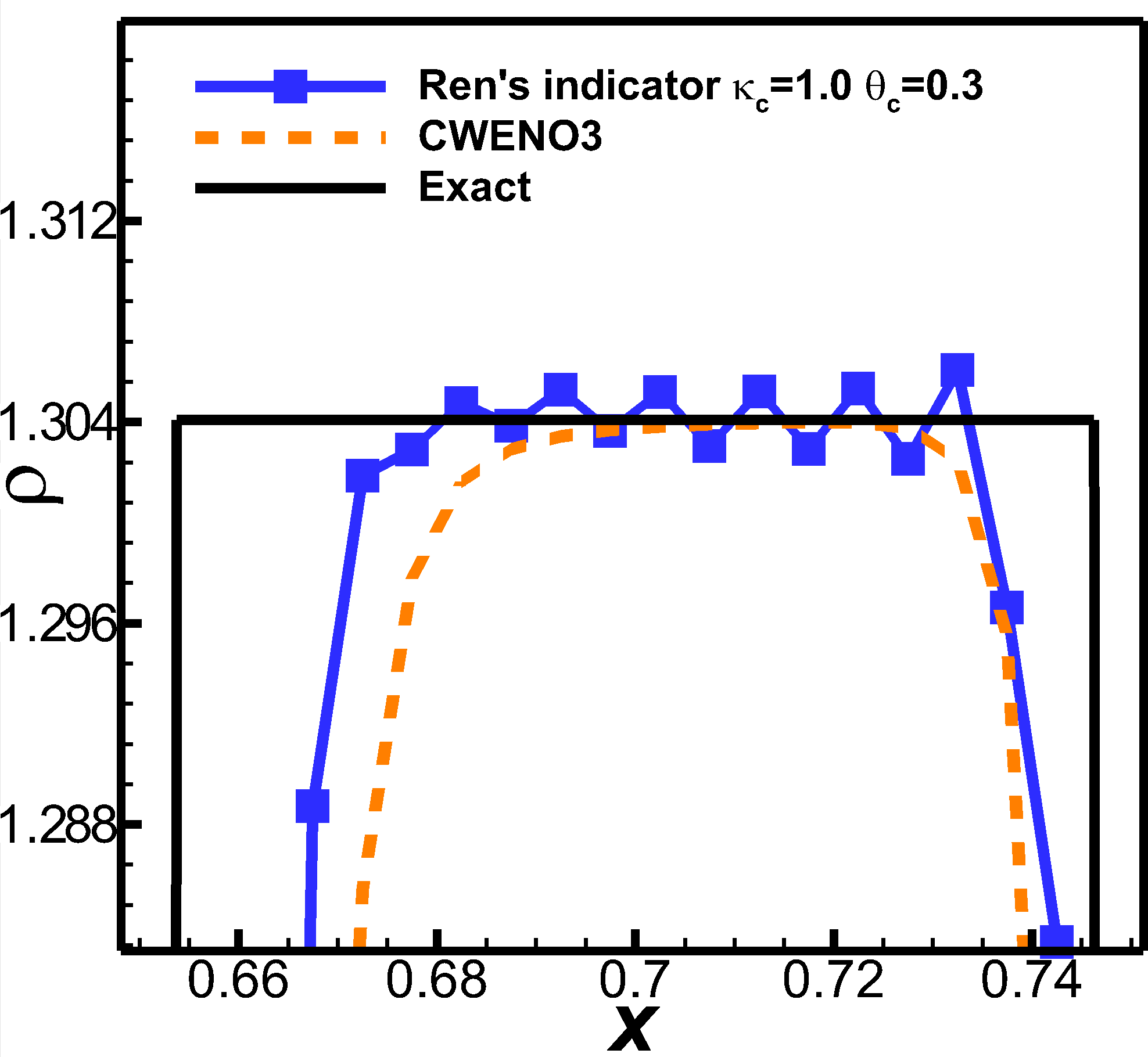}
    \caption{\label{fig:clsweno3_critical_value_compare_lax_ren_1}$\theta_c = 0.30$.}
    \end{subfigure}
    \begin{subfigure}[b]{\columnwidth}
    \includegraphics[width=0.6\columnwidth]{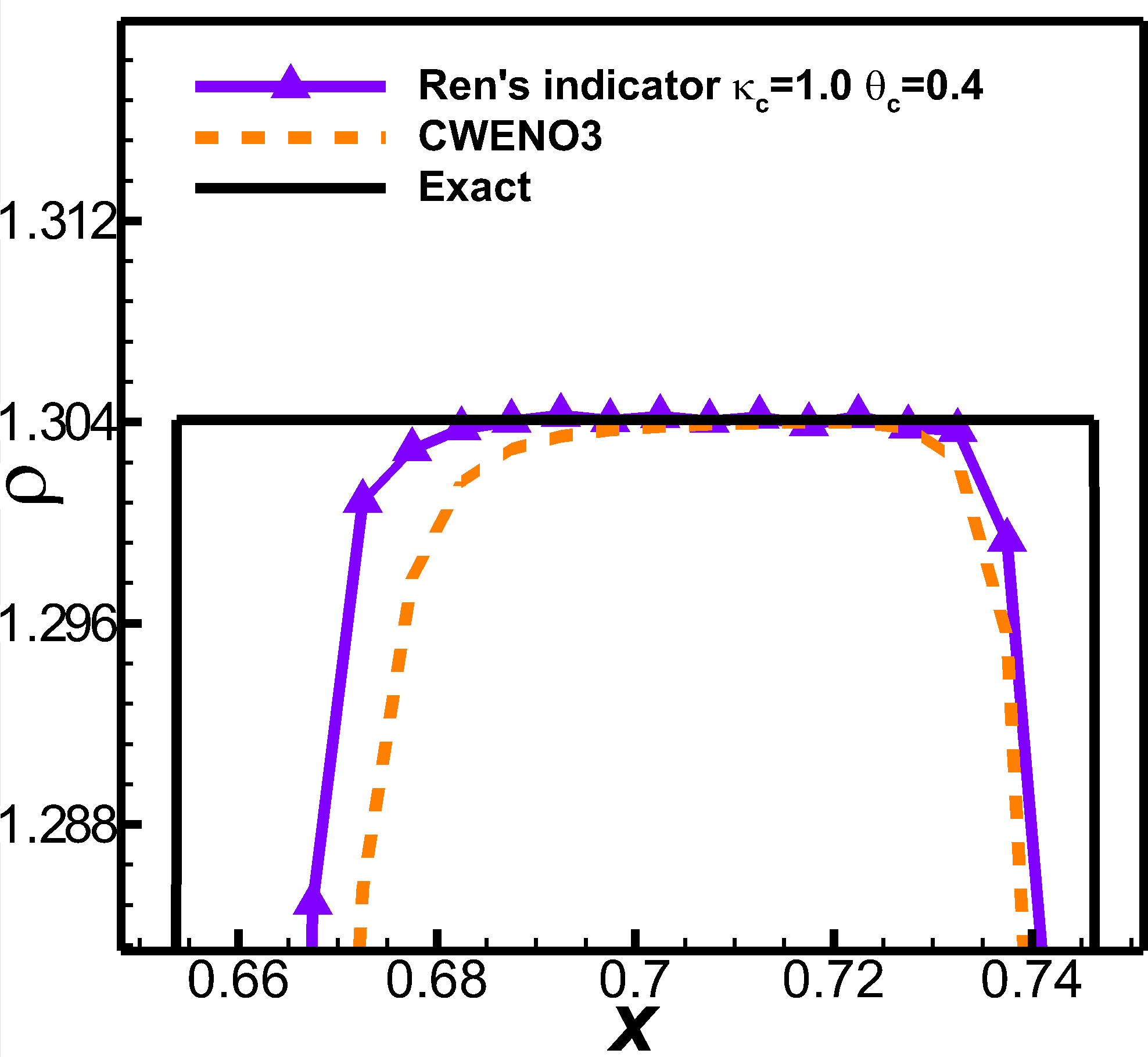}
    \caption{\label{fig:clsweno3_critical_value_compare_lax_ren_2}$\theta_c = 0.40$.}
    \end{subfigure}
    \begin{subfigure}[b]{\columnwidth}
    \includegraphics[width=0.6\columnwidth]{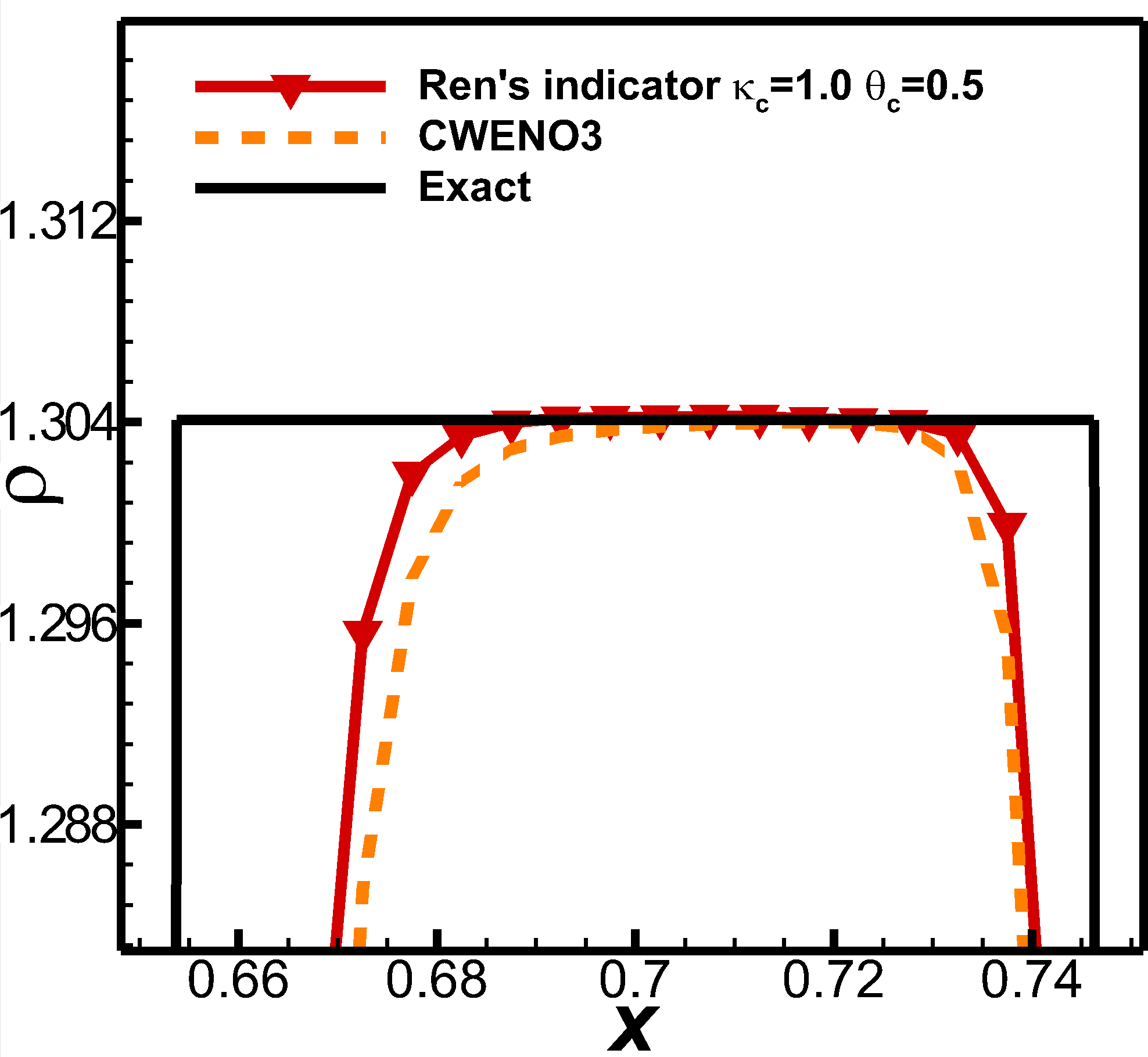}
    \caption{\label{fig:clsweno3_critical_value_compare_lax_ren_3}$\theta_c = 0.50$.}
    \end{subfigure}
    \caption{\label{fig:clsweno3_critical_value_compare_lax_ren} Lax shock tube problem solved by the third-order hybrid CLS-CWENO scheme. Shock detector is $\sigma^{\mathrm{Ren}}$.}
\end{figure}

\begin{figure}[!htbp]
  \centering
    \begin{subfigure}[b]{\columnwidth}
    \includegraphics[width=0.6\columnwidth]{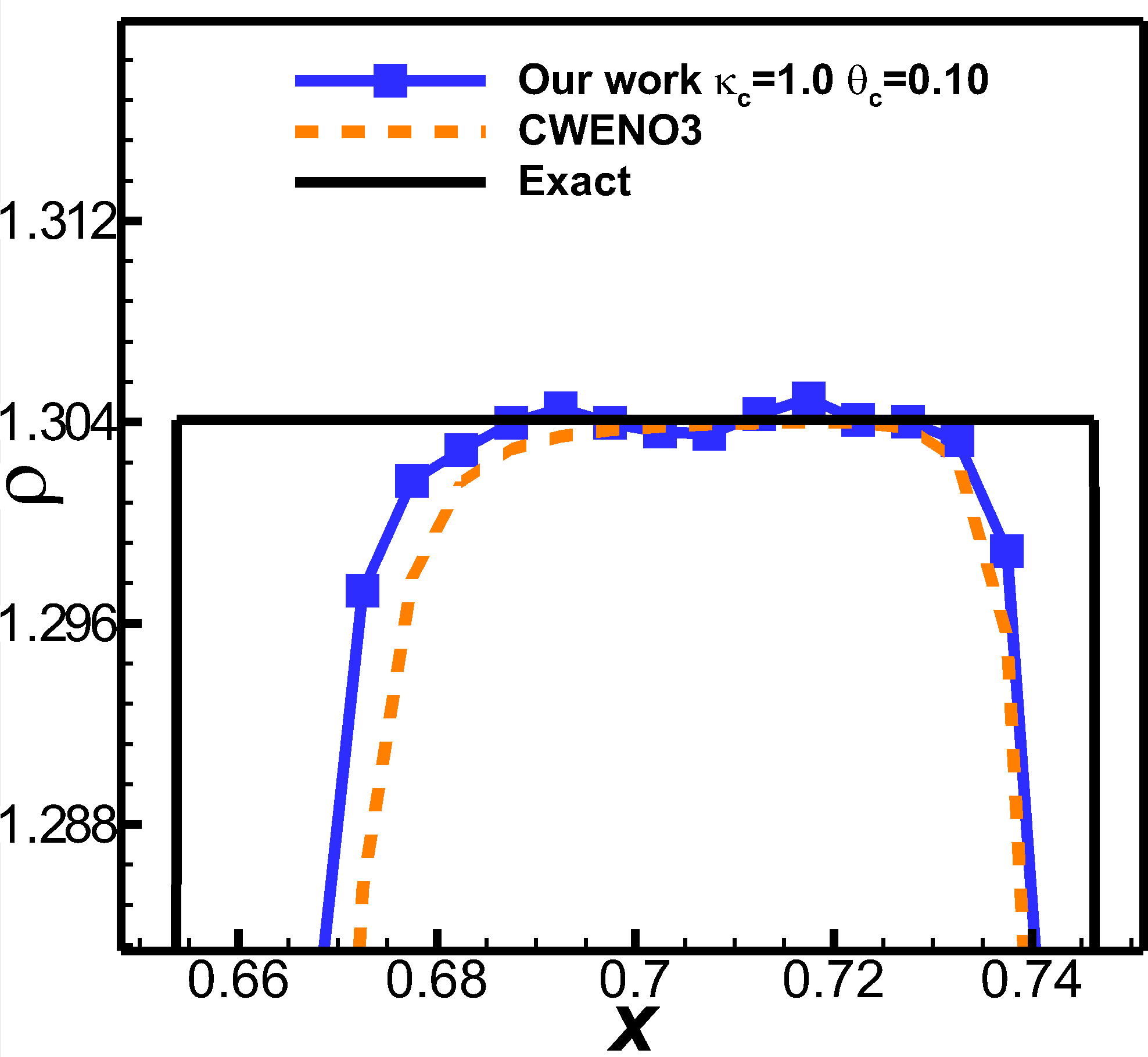}
    \caption{\label{fig:clsweno3_critical_value_compare_lax_our_work_1}$\theta_c = 0.10$.}
    \end{subfigure}
    \begin{subfigure}[b]{\columnwidth}
    \includegraphics[width=0.6\columnwidth]{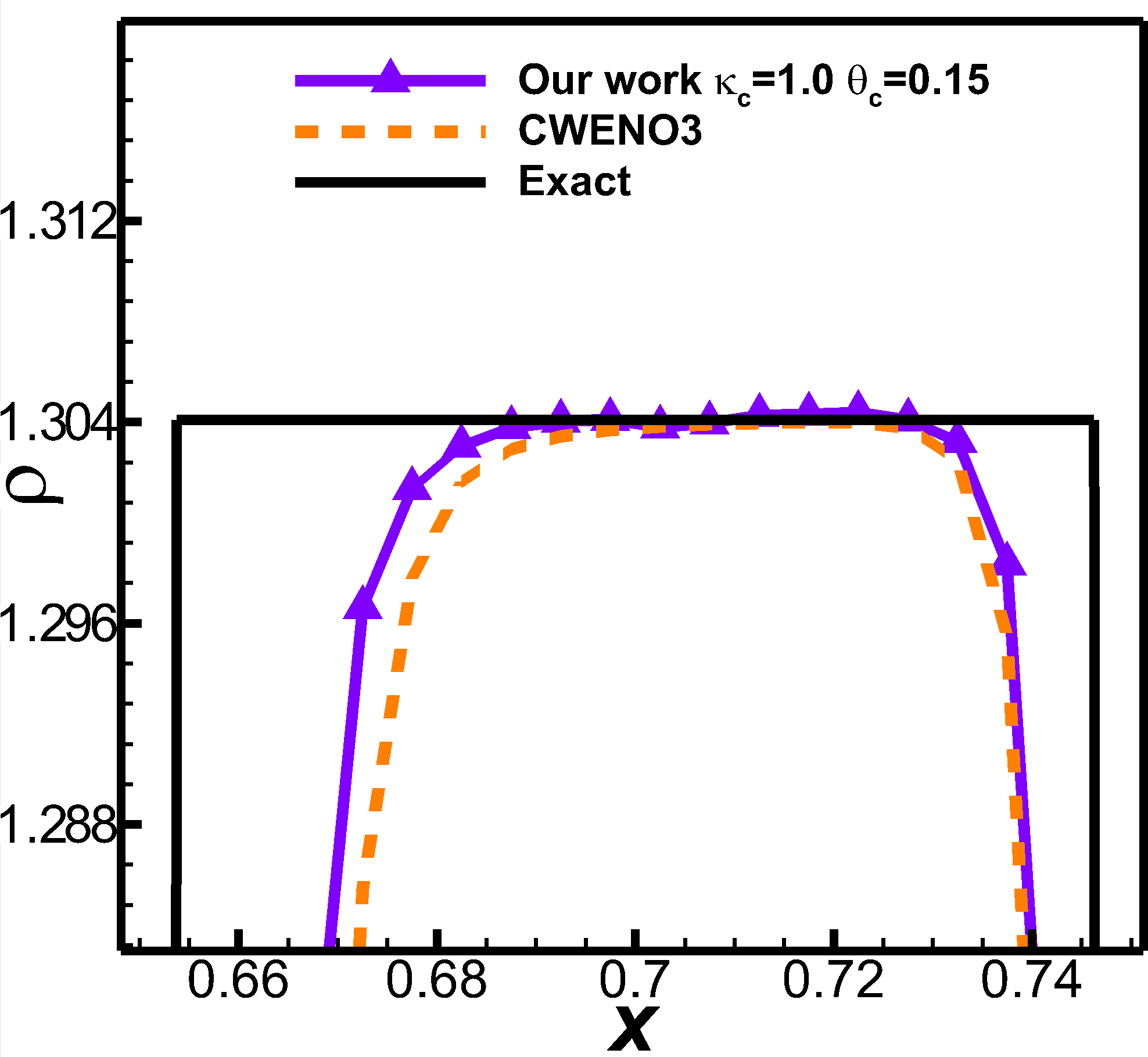}
    \caption{\label{fig:clsweno3_critical_value_compare_lax_our_work_2}$\theta_c = 0.15$.}
    \end{subfigure}
    \begin{subfigure}[b]{\columnwidth}
    \includegraphics[width=0.6\columnwidth]{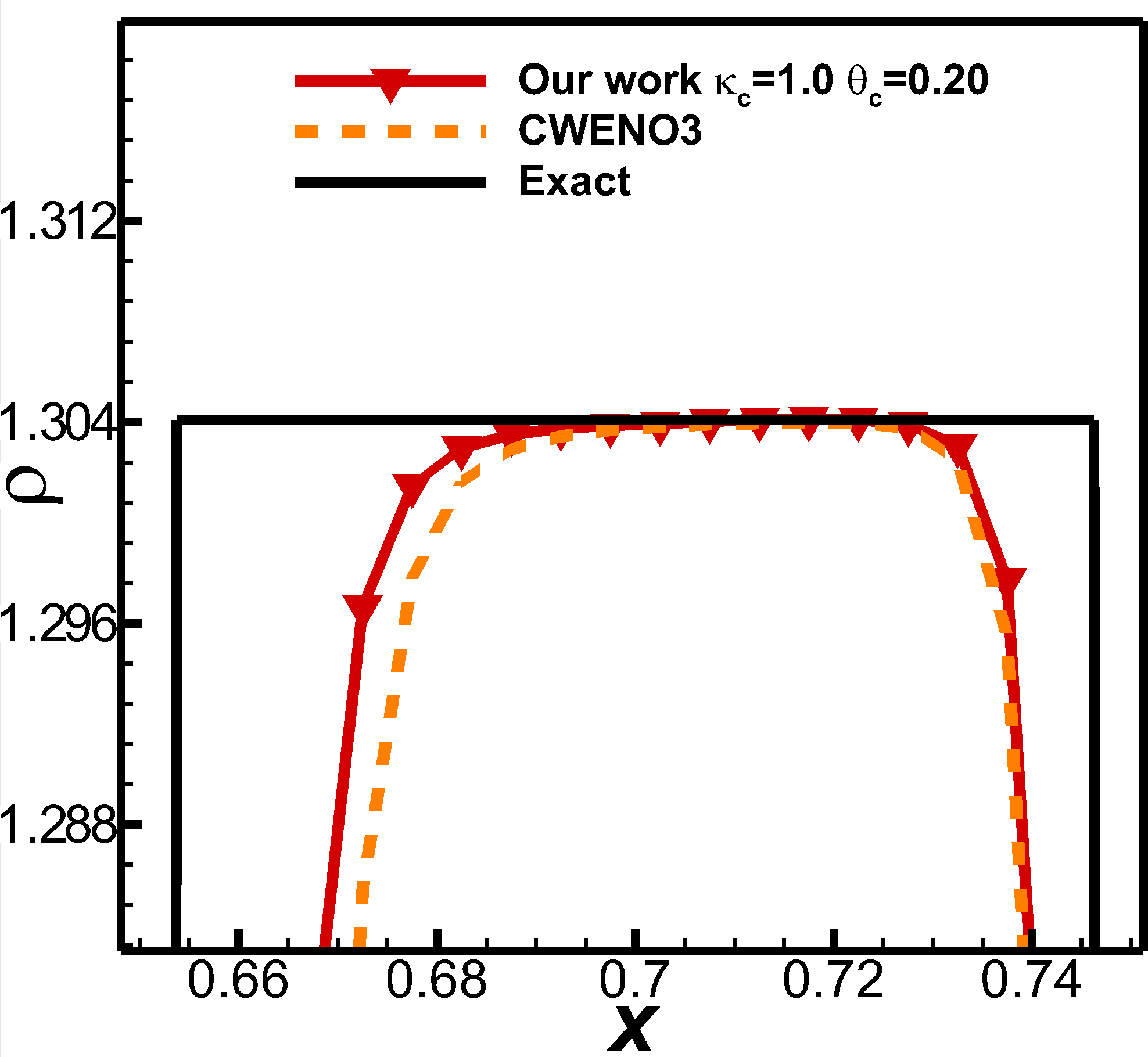}
    \caption{\label{fig:clsweno3_critical_value_compare_lax_our_work_3}$\theta_c = 0.20$.}
    \end{subfigure}
    \caption{\label{fig:clsweno3_critical_value_compare_lax_our_work} Lax shock tube problem solved by the third-order hybrid CLS-CWENO scheme. Shock detector is $\sigma^{\mathrm{Li}}$.}
\end{figure}

As shown in Fig. \ref{fig:clsweno3_critical_value_compare_lax_ren}, the third-order CLS-CWENO scheme utilizing $\sigma^{\mathrm{Ren}}$ with $\theta_c = 0.3$ produces zig-zag oscillations after the shock while $\sigma^{\mathrm{Ren}}$ with $\theta_c = 0.4$ and $0.5$ produces smooth results.
On the other hand, as shown in Fig. \ref{fig:clsweno3_critical_value_compare_lax_our_work}, for the third-order hybrid CLS-CWENO scheme utilizing $\sigma^{\mathrm{Li}}$, the results of all three values of $\theta_c$ are acceptable and $\theta_c = 0.20$ gives the least oscillating result.

\begin{figure}[!htbp]
  \centering
    \begin{subfigure}[b]{\columnwidth}
    \includegraphics[width=0.6\columnwidth]{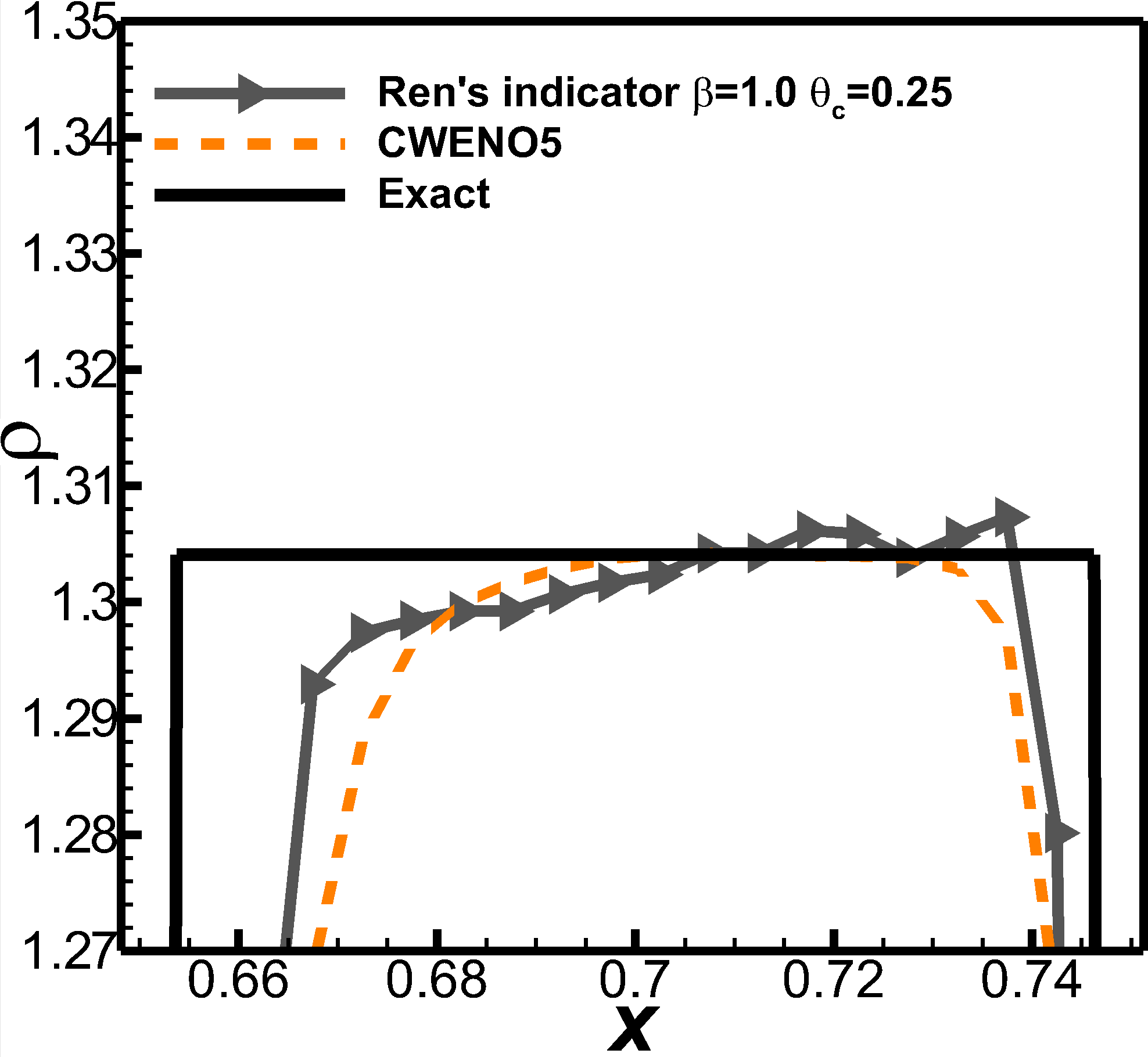}
    \caption{\label{fig:clsweno5_critical_value_compare_lax_ren_1}$\theta_c = 0.25$.}
    \end{subfigure}
    \begin{subfigure}[b]{\columnwidth}
    \includegraphics[width=0.6\columnwidth]{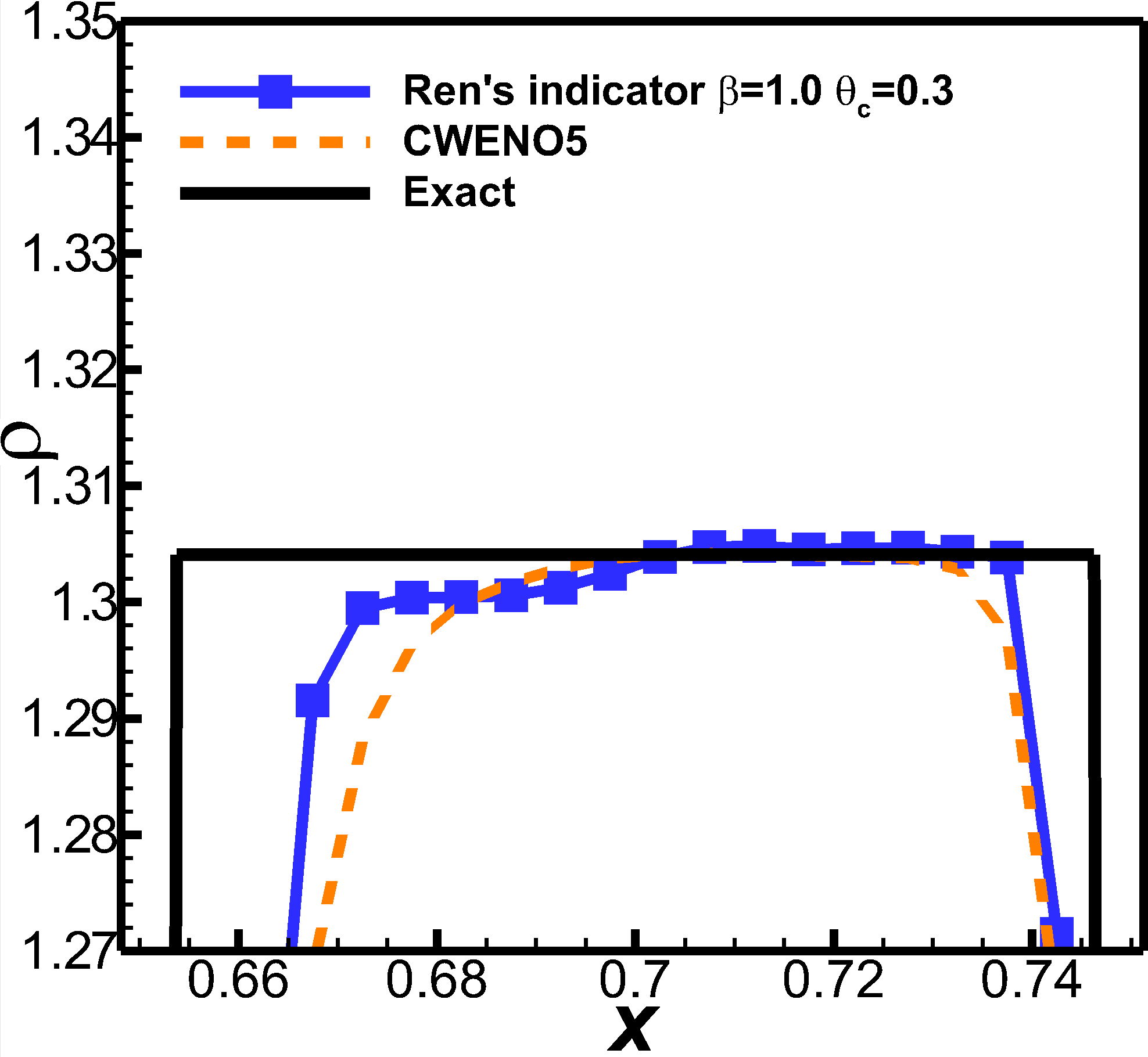}
    \caption{\label{fig:clsweno5_critical_value_compare_lax_ren_2}$\theta_c = 0.3$.}
    \end{subfigure}
    \begin{subfigure}[b]{\columnwidth}
    \includegraphics[width=0.6\columnwidth]{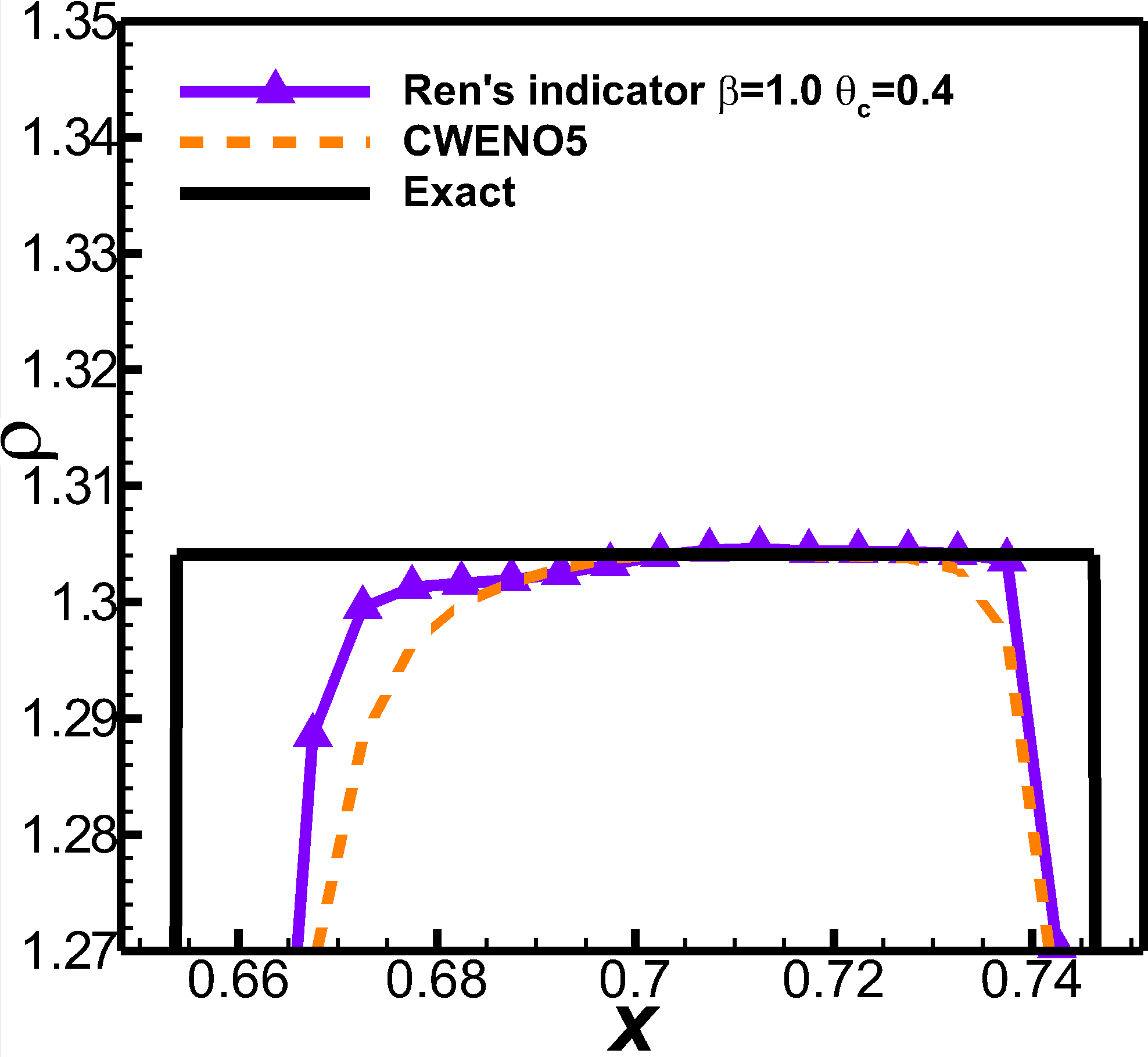}
    \caption{\label{fig:clsweno5_critical_value_compare_lax_ren_3}$\theta_c = 0.4$.}
    \end{subfigure}
    \begin{subfigure}[b]{\columnwidth}
    \includegraphics[width=0.6\columnwidth]{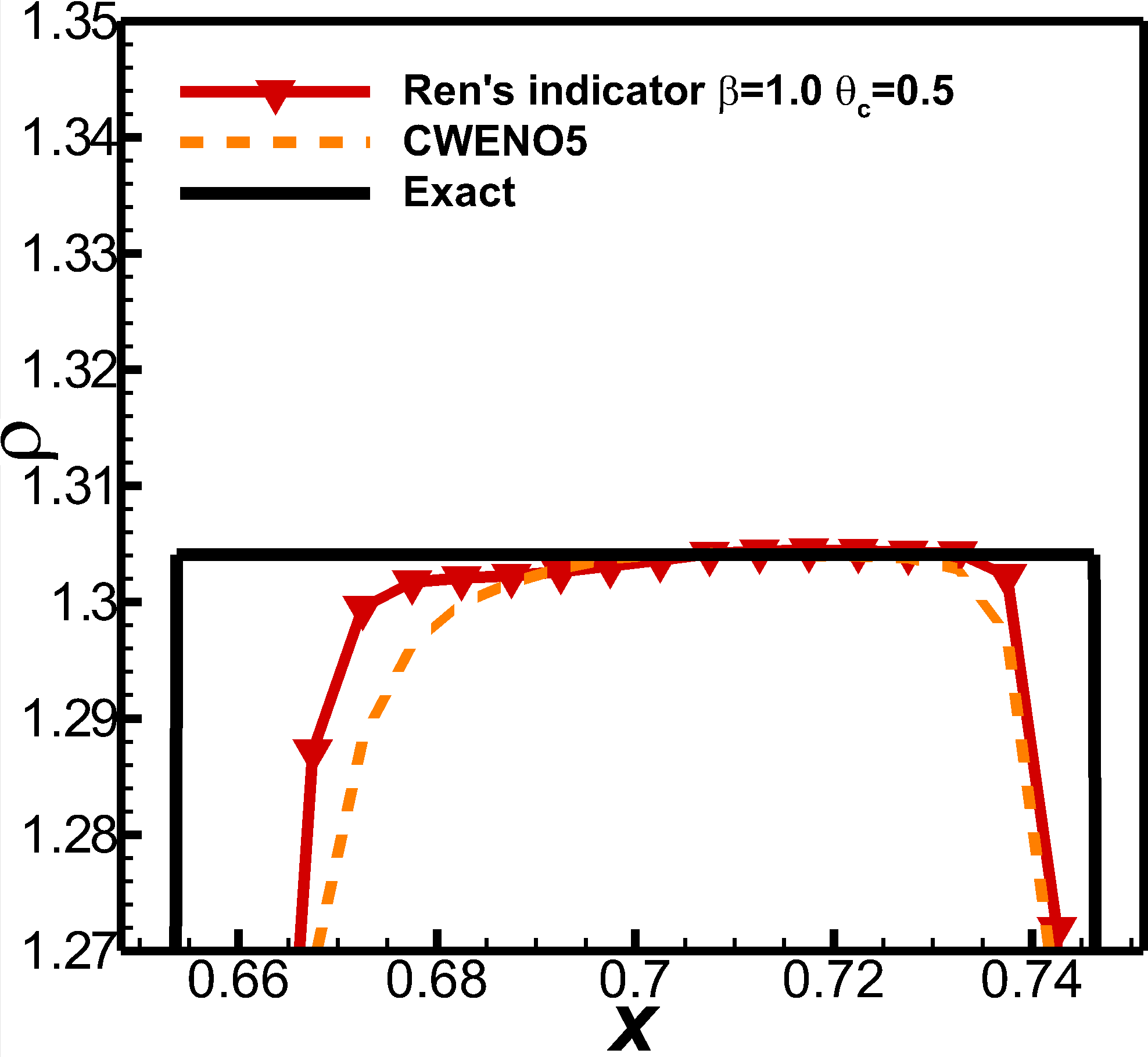}
    \caption{\label{fig:clsweno5_critical_value_compare_lax_ren_4}$\theta_c = 0.5$.}
    \end{subfigure}
    \caption{\label{fig:clsweno5_critical_value_compare_lax_ren} Lax shock tube problem solved by the fifth-order hybrid CLS-CWENO scheme. Shock detector is $\sigma^{\mathrm{Ren}}$.}
\end{figure}

\begin{figure}[!htbp]
  \centering
    \begin{subfigure}[b]{\columnwidth}
    \includegraphics[width=0.6\columnwidth]{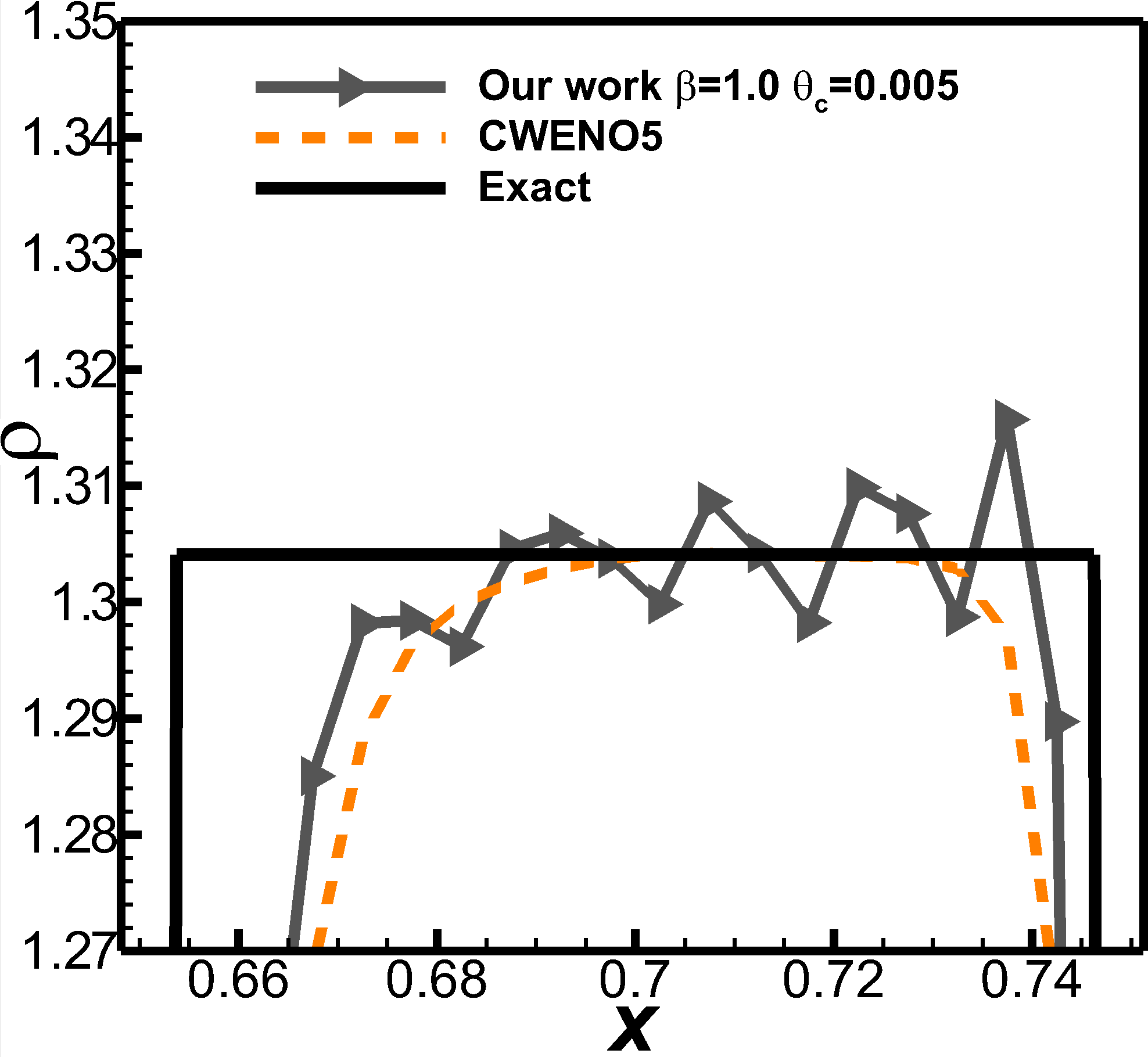}
    \caption{\label{fig:clsweno5_critical_value_compare_lax_our_work_1}$\theta_c = 0.005$.}
    \end{subfigure}
    \begin{subfigure}[b]{\columnwidth}
    \includegraphics[width=0.6\columnwidth]{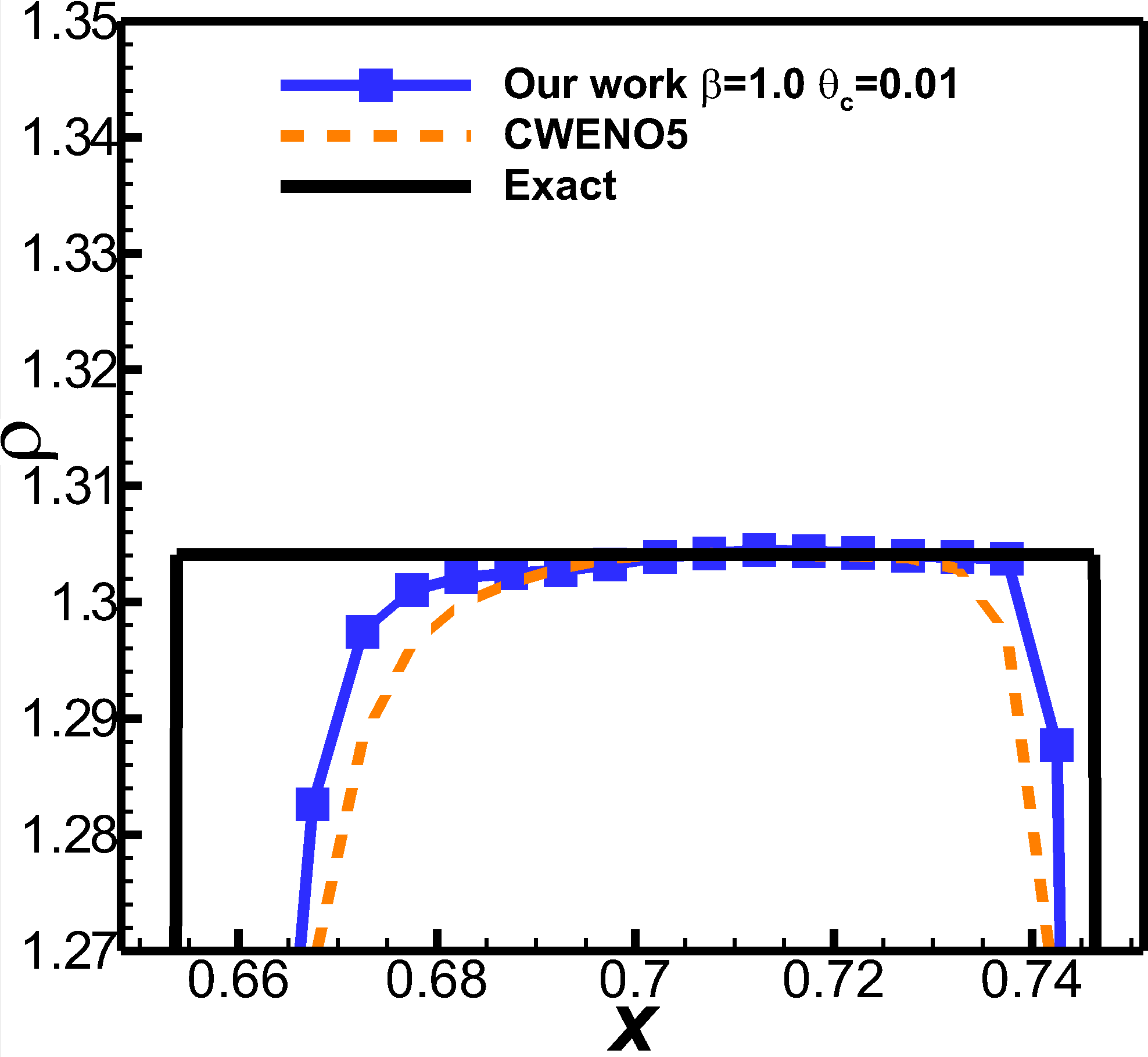}
    \caption{\label{fig:clsweno5_critical_value_compare_lax_our_work_2}$\theta_c = 0.010$.}
    \end{subfigure}
    \begin{subfigure}[b]{\columnwidth}
    \includegraphics[width=0.6\columnwidth]{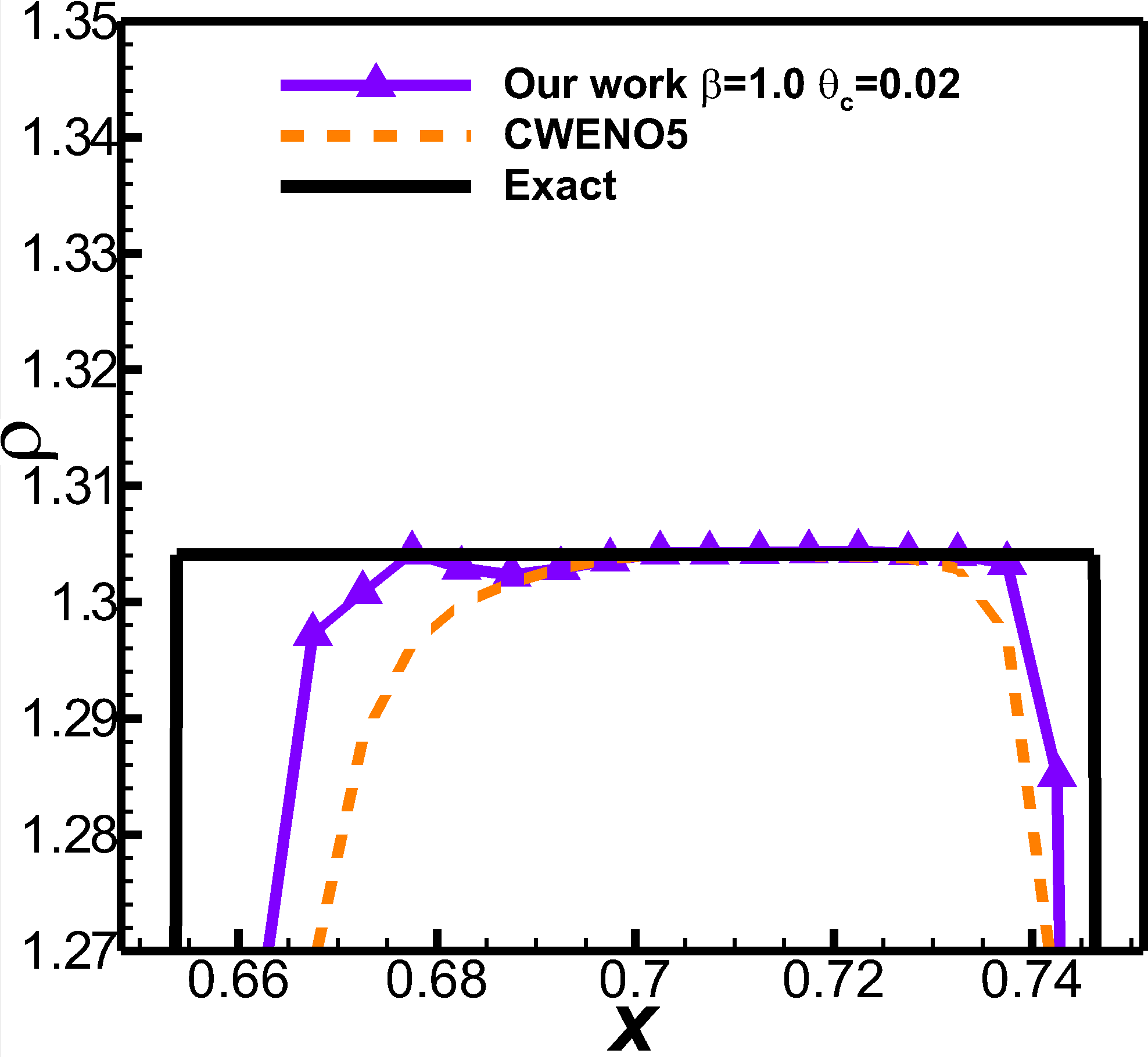}
    \caption{\label{fig:clsweno5_critical_value_compare_lax_our_work_3}$\theta_c = 0.020$.}
    \end{subfigure}
    \begin{subfigure}[b]{\columnwidth}
    \includegraphics[width=0.6\columnwidth]{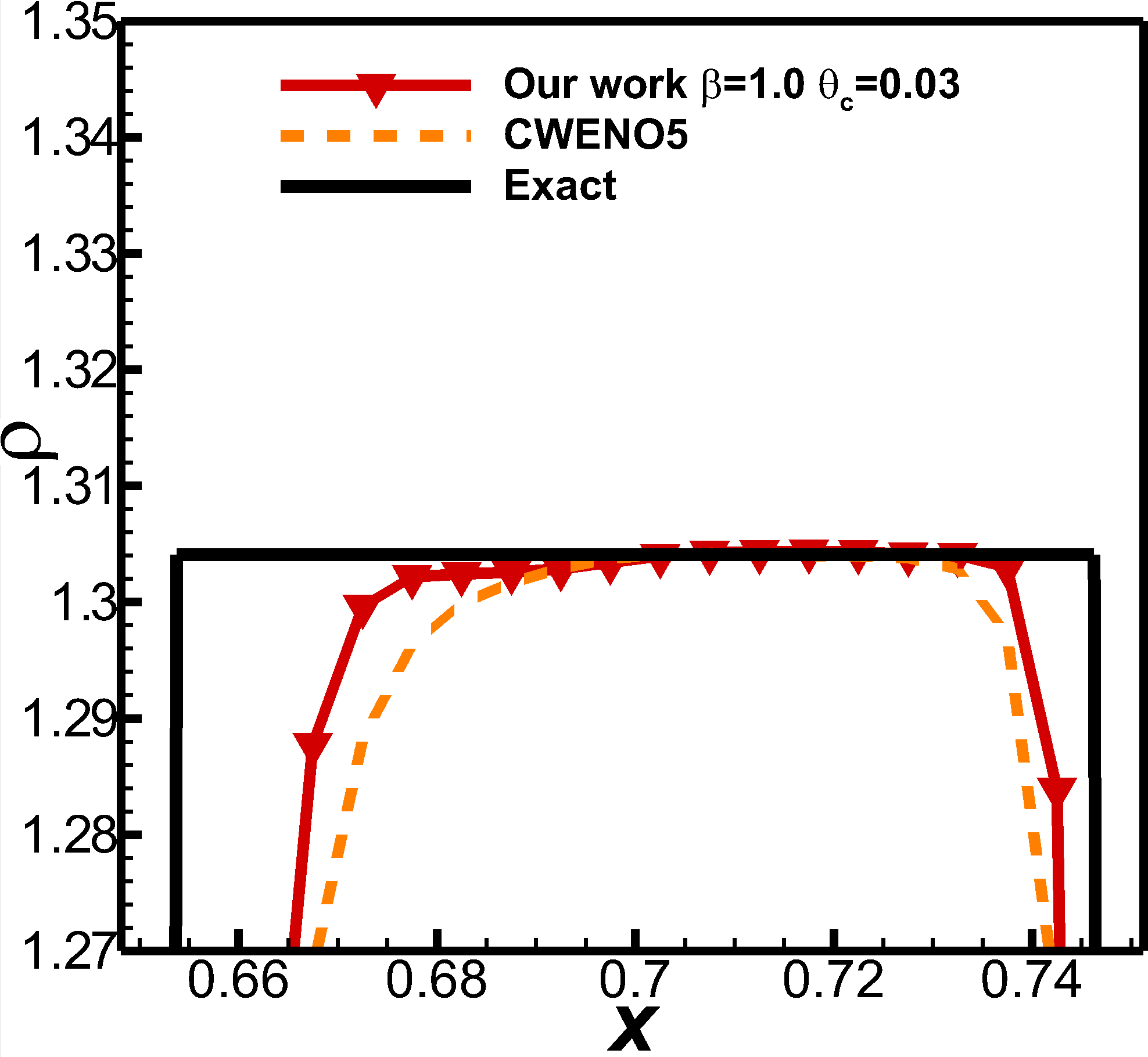}
    \caption{\label{fig:clsweno5_critical_value_compare_lax_our_work_4}$\theta_c = 0.030$.}
    \end{subfigure}
    \caption{\label{fig:clsweno5_critical_value_compare_lax_our_work} Lax shock tube problem solved by the fifth-order hybrid CLS-CWENO scheme. Shock detector is $\sigma^{\mathrm{Li}}$.}
\end{figure}

Figures \ref{fig:clsweno5_critical_value_compare_lax_ren} and \ref{fig:clsweno5_critical_value_compare_lax_our_work} plot the results for the fifth-order hybrid CLS-CWENO scheme. With the increasing of $\theta_c$, the scheme utilizing both $\sigma^{\mathrm{Ren}}$ and $\sigma^{\mathrm{Li}}$ produces oscillation-free results for the Lax shock tube problem.

Concluding from the results of both the linear convection equation in Sec. \ref{sec:gste} and the nonlinear Euler equations in Sec. \ref{sec:calibration_lax} from Figs. \ref{fig:clsweno3_critical_value_compare_gste_ren}-\ref{fig:clsweno5_critical_value_compare_lax_our_work}, the critical $\theta_c$ for the third- and fifth-order hybrid CLS-CWENO scheme is chosen as $0.5$ when utilizing $\sigma^{\mathrm{Ren}}$ hereafter in this paper; the critical $\theta_c$ is chosen as $0.2$ for the third-order hybrid CLS-CWENO scheme and $0.03$ for the fifth-order hybrid CLS-CWENO scheme when utilizing $\sigma^{\mathrm{Li}}$.

\subsection{Linear problem with smooth initial condition}
In this section, the linear convection equation with smooth initial conditions of $u_0(x) = \sin(2\pi x)$ and $u_0(x) = \sin^2(2\pi x)$  is utilized to test the accuracy of the hybrid CLS-CWENO schemes. The computational domain is $\Omega = [0,1]$ with periodic boundary condition and the simulation end time is $t_{end} = 1.0$. Third- and fifth-order RK schemes are utilized to discretize the linear equation in temporal direction for the third- and fifth-order hybrid CLS-CWENO schemes, respectively. The Courant number is chosen as 0.5.
\begin{figure}[!h]
  \centering
    \begin{subfigure}[b]{\columnwidth}
    \includegraphics[width=0.8\columnwidth]{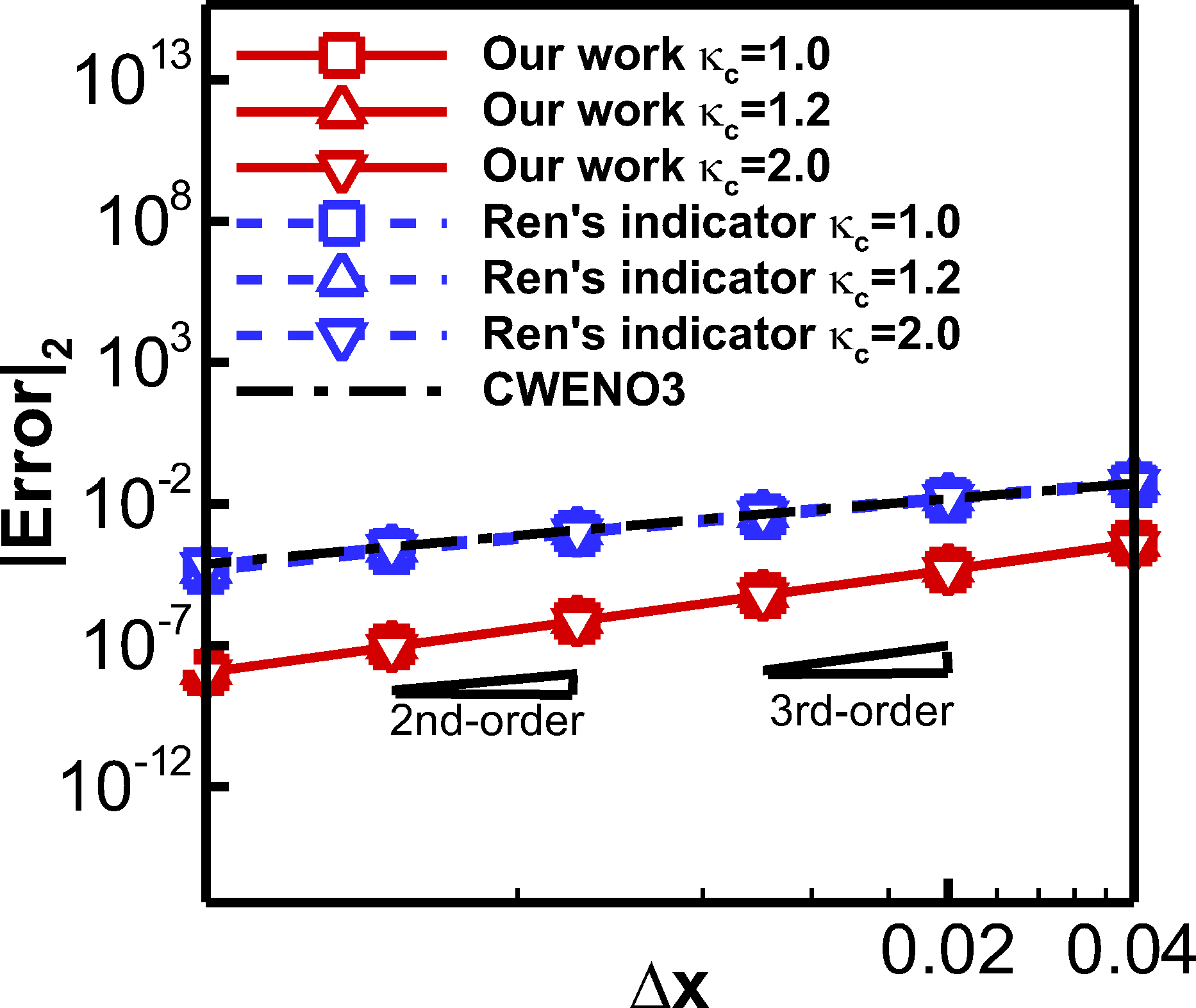}
    \caption{$u_0 = \sin(2\pi x)$.\label{fig:accusin1_3rd}}
    \end{subfigure}
    \begin{subfigure}[b]{\columnwidth}
    \includegraphics[width=0.8\columnwidth]{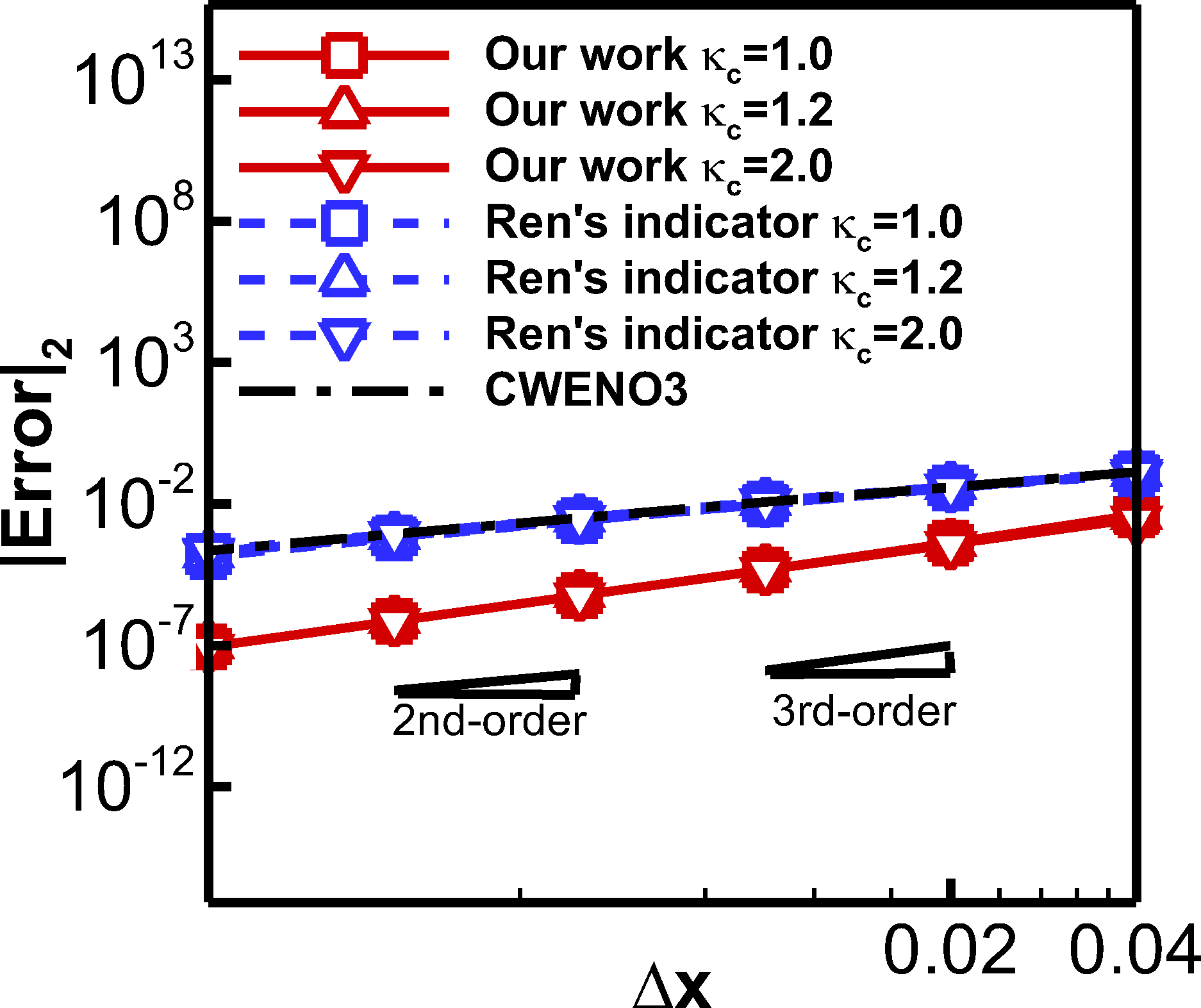}
    \caption{$u_0 = \sin^2(2\pi x)$.\label{fig:accusin2_3rd}}
    \end{subfigure}
    \caption{\label{fig:clsweno_accuracy_3rd} Accuracy test for the third-order scheme. $\theta_c=0.2$ for $\sigma^{\mathrm{Li}}$ and $\theta_c=0.5$ for $\sigma^{\mathrm{Ren}}$.}
\end{figure}

\begin{figure}
  \centering
    \begin{subfigure}[b]{\columnwidth}
    \includegraphics[width=0.8\columnwidth]{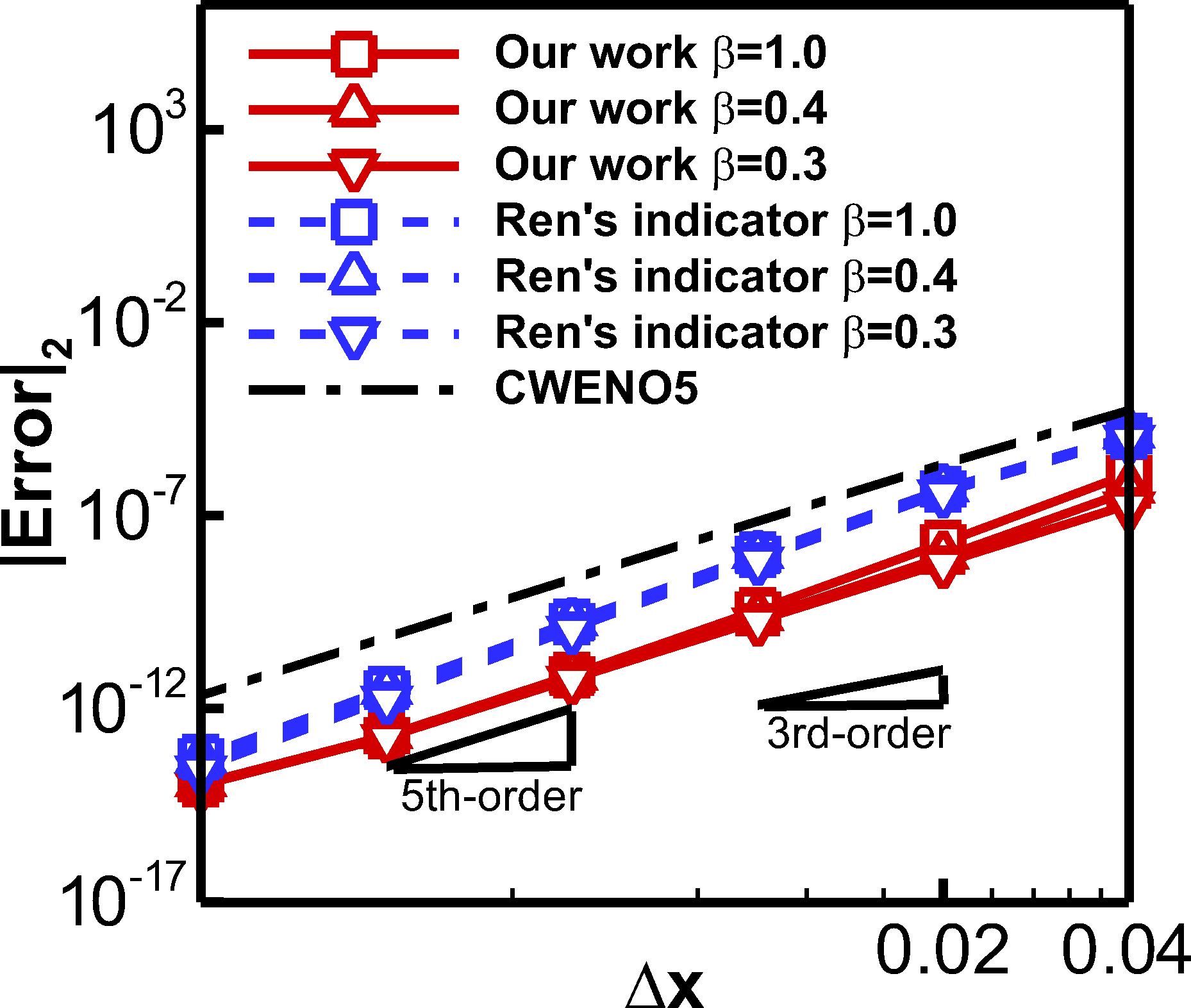}
    \caption{$u_0 = \sin(2\pi x)$.\label{fig:accusin1_5th}}
    \end{subfigure}
    \begin{subfigure}[b]{\columnwidth}
    \includegraphics[width=0.8\columnwidth]{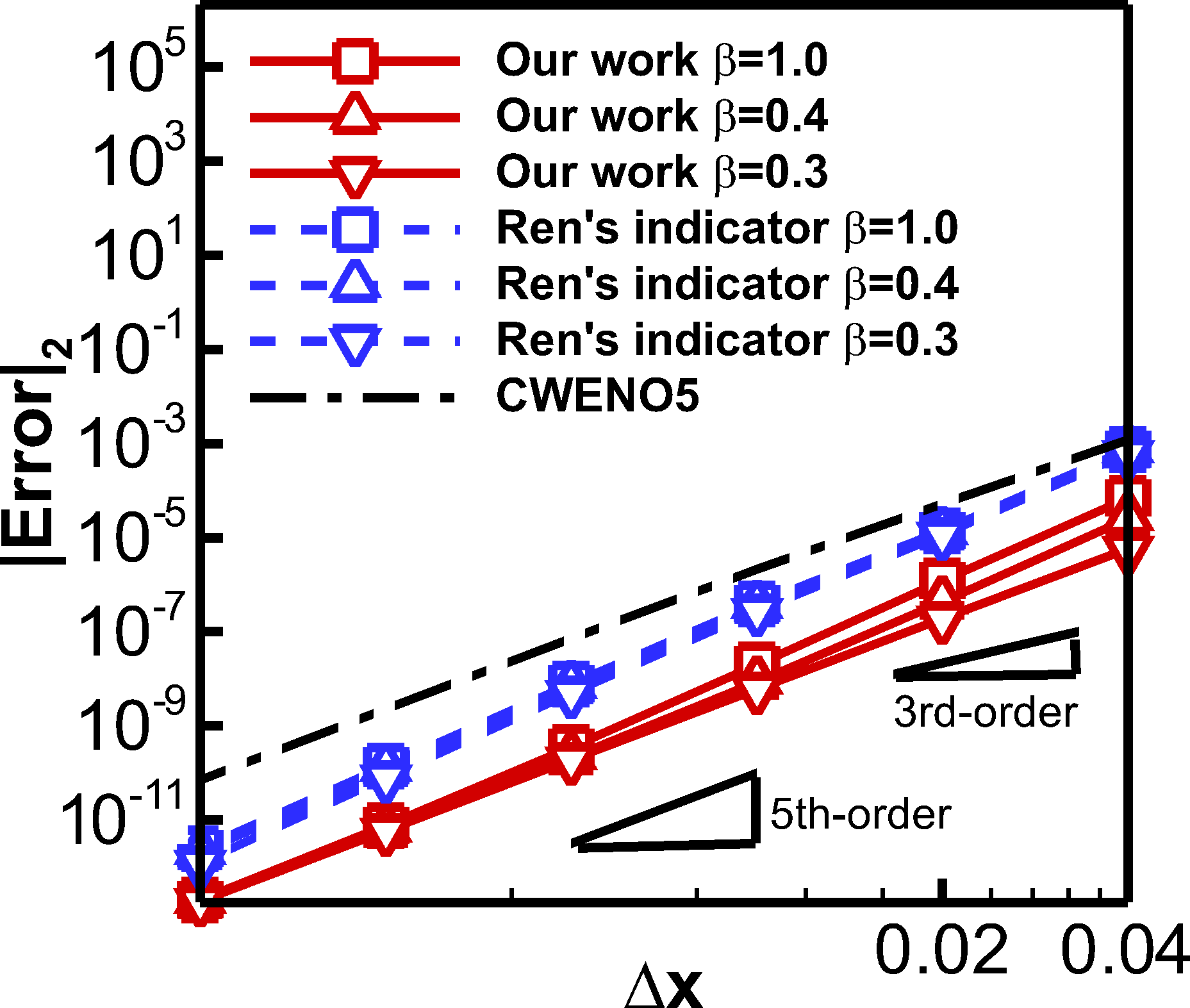}
    \caption{$u_0 = \sin^2(2\pi x)$.\label{fig:accusin2_5th}}
    \end{subfigure}
    \caption{\label{fig:clsweno_accuracy_5th} Accuracy test for the fifth-order scheme. $\theta_c=0.03$ for $\sigma^{\mathrm{Li}}$ and $\theta_c=0.5$ for $\sigma^{\mathrm{Ren}}$.}
\end{figure}

Figures \ref{fig:clsweno_accuracy_3rd}  and \ref{fig:clsweno_accuracy_5th} plot the $L_2$-norm of the error for the hybrid CLS-CWENO schemes with different optimized coefficients listed in Tabs. \ref{tab:w1w2} and \ref{tab:w1w2w3w4}, respectively.
As Figs. \ref{fig:clsweno_accuracy_3rd} and \ref{fig:clsweno_accuracy_5th} illustrate, the error of the hybrid CLS-CWENO schemes with shock detector $\sigma^{\mathrm{Li}}$ is of orders smaller than the error of pure CWENO schemes, demonstrating the high-accuracy of the proposed schemes.
In Figs. \ref{fig:accusin1_3rd} and \ref{fig:accusin2_3rd}, the converging rate for the third-order hybrid CLS-CWENO scheme with shock detector $\sigma^{\mathrm{Li}}$ is about $3$, which confirms the capability of $\sigma^{\mathrm{Li}}$ in identifying the first-order extrema.
In Figs. \ref{fig:accusin1_5th} and \ref{fig:accusin2_5th}, the converging rate for the fifth-order hybrid CLS-CWENO scheme can reach about $5$ with both shock detectors. And for the fifth-order scheme, the errors using coefficients optimized by different $\beta$ is ordered as $\left|E\right|_{\beta = 0.3} \le \left|E\right|_{\beta = 0.4} < \left|E\right|_{\beta = 1.0} $ on coarse grids, which is in accordance with the spectral analysis in Sec. \ref{sec:optimization}.

\subsection{Propagation of broadband sound waves}
This case describes the propagation of a sound wave packet which contains acoustic turbulent structures with various length scales. The initial condition is
\begin{align}
  p(x, 0) & = p_0\left( 1+ \epsilon \sum_{k=1}^{N/2}{\left[E_p(k)\right]^{0.5} \sin(2\pi k(x+\psi_k))}\right),\\
  \rho(x,0) & = \rho_0 \left(\frac{p(x,0)}{p_0}\right)^{1/\gamma},\\
  u(x,0) &= u_0 + \frac{2}{\gamma-1}\left(c(x,0)-c_0\right),
\end{align}
where 
\begin{equation}
  E_p(k) = \left(k/k_0\right)^4e^{-2(k/k_0)^2}
\end{equation}
is the energy spectrum reaching its maximum at $k=k_0$. $p_0 = 1$, $u_0 = 1$ and $\rho_0 = 1$. The computational domain is $\Omega = [0,1]$ discretized by 128 uniform cells and $\psi_k,\,k=1,2,3,\cdots,N/2$ are random numbers ranging from 0 to 1. $\epsilon$ is a small parameter which determines the intensity of the acoustic turbulence and is chosen as 0.001 in the simulation. $c = \sqrt{\gamma p/\rho}$ is the speed of sound and $\gamma=1.4$. Periodic boundary conditions are imposed at the two ends. The problem is solved until $t_{end} = 1.0$ with Courant number 0.5.
\begin{figure}[!htbp]
  \centering
    \begin{subfigure}[b]{\columnwidth}
    \includegraphics[width=0.61\columnwidth]{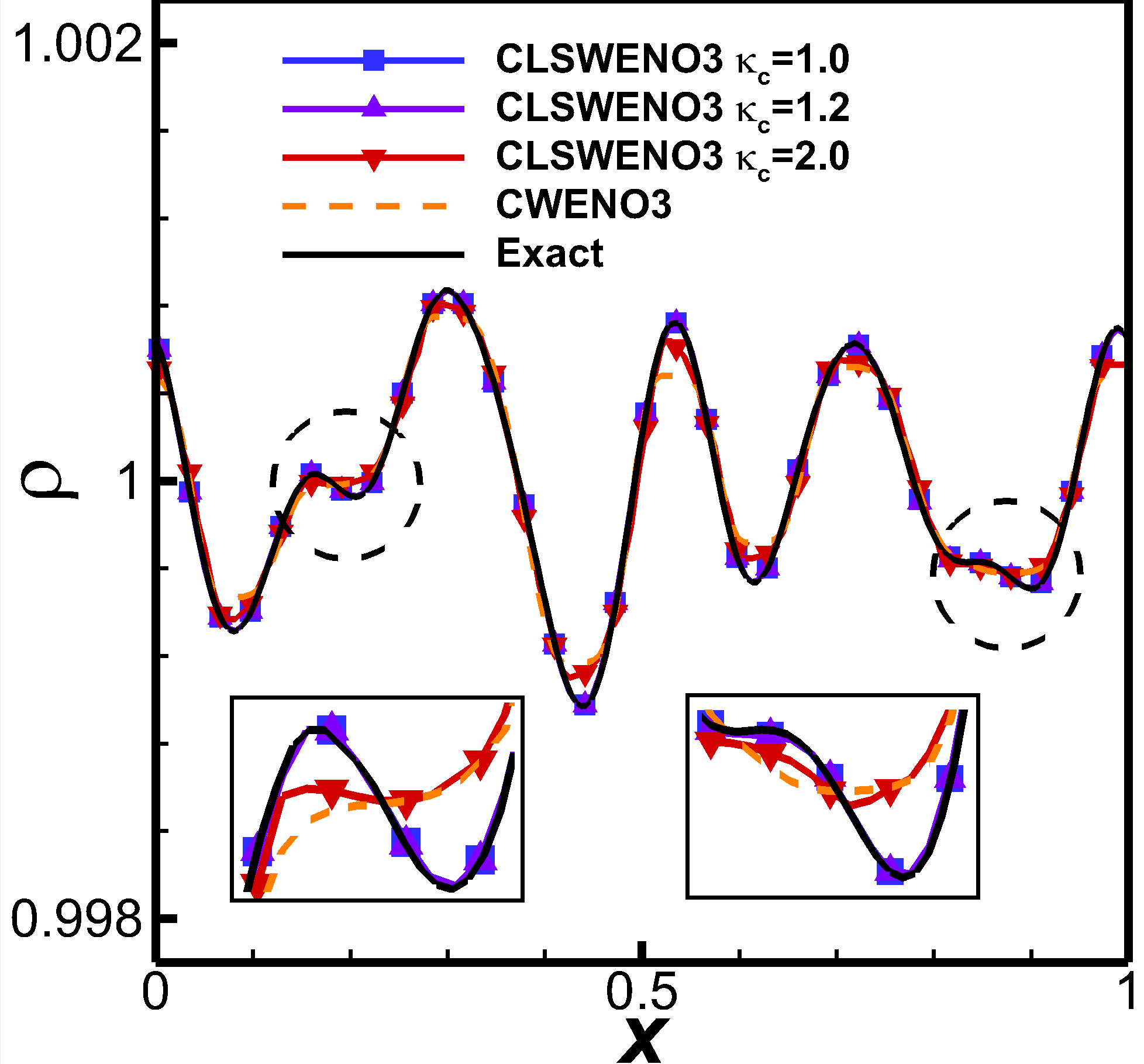}
    \caption{With coefficients optimized by varying $\kappa_c$ with $\sigma^{\mathrm{Li}}$.}
    \end{subfigure}
    \begin{subfigure}[b]{\columnwidth}
    \includegraphics[width=0.61\columnwidth]{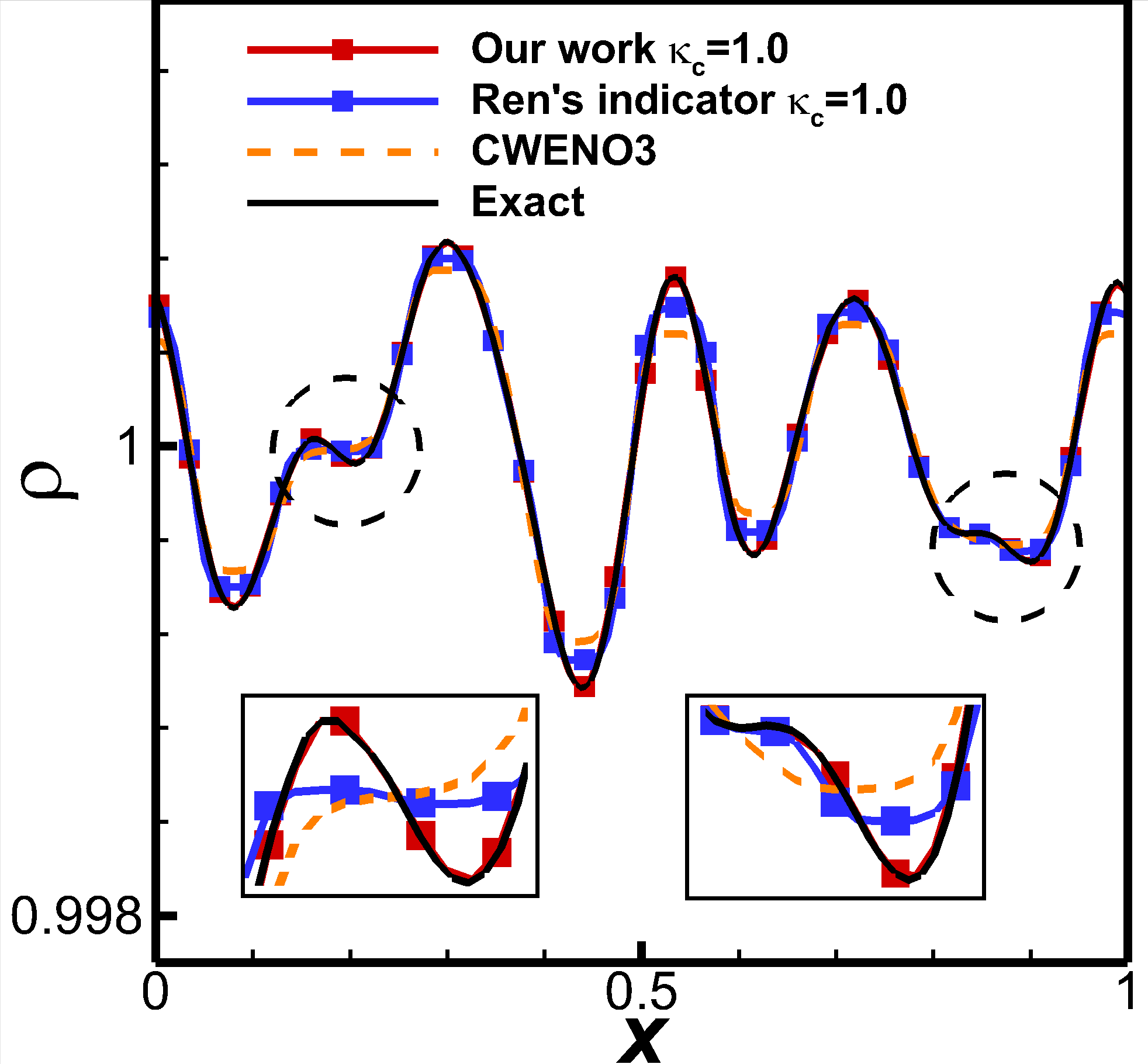}
    \caption{With coefficients optmized by $\kappa_c = 1.0$.}
    \end{subfigure}
    \begin{subfigure}[b]{\columnwidth}
    \includegraphics[width=0.61\columnwidth]{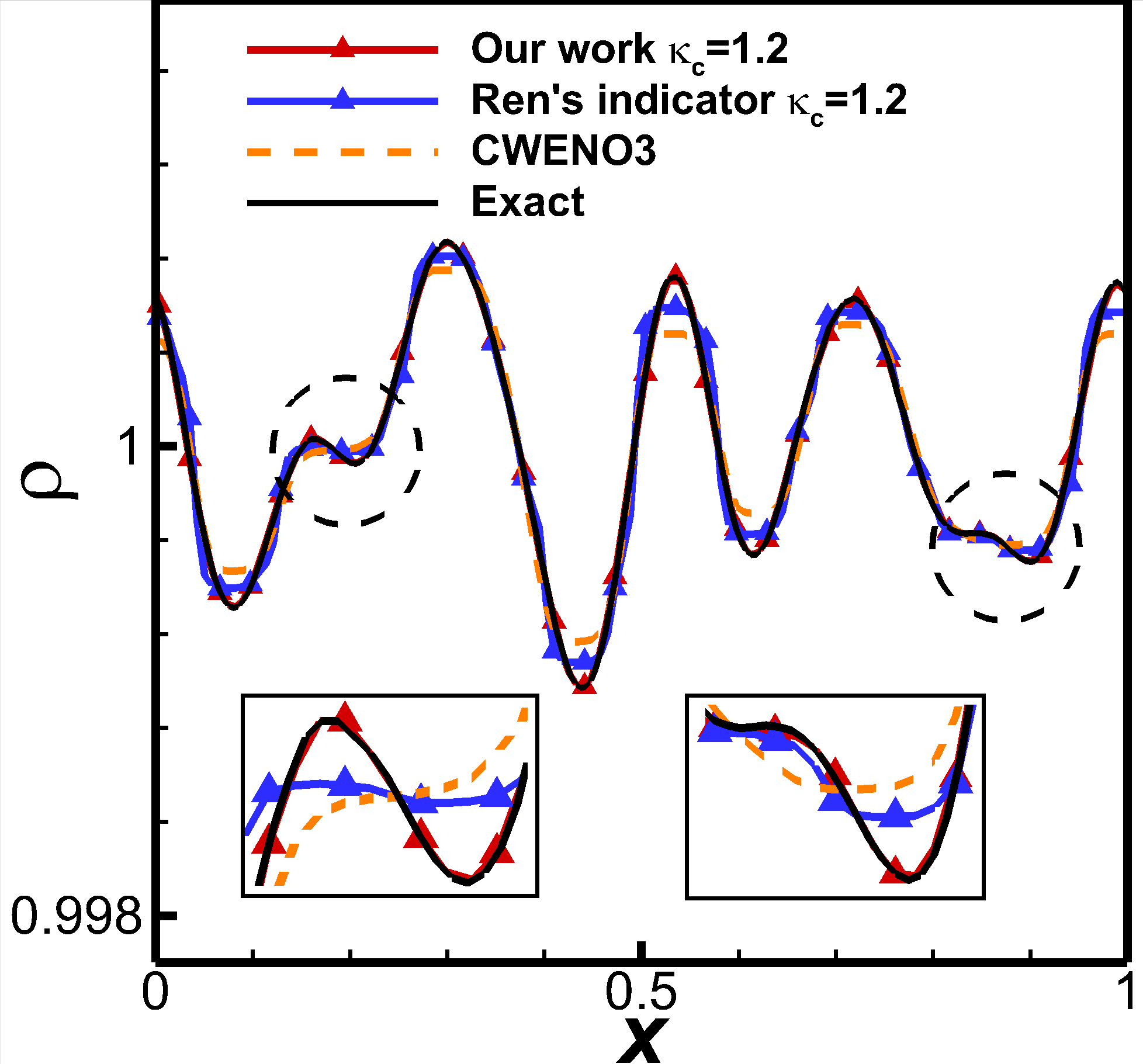}
    \caption{With coefficients optmized by $\kappa_c = 1.2$.}
    \end{subfigure}
    \begin{subfigure}[b]{\columnwidth}
    \includegraphics[width=0.61\columnwidth]{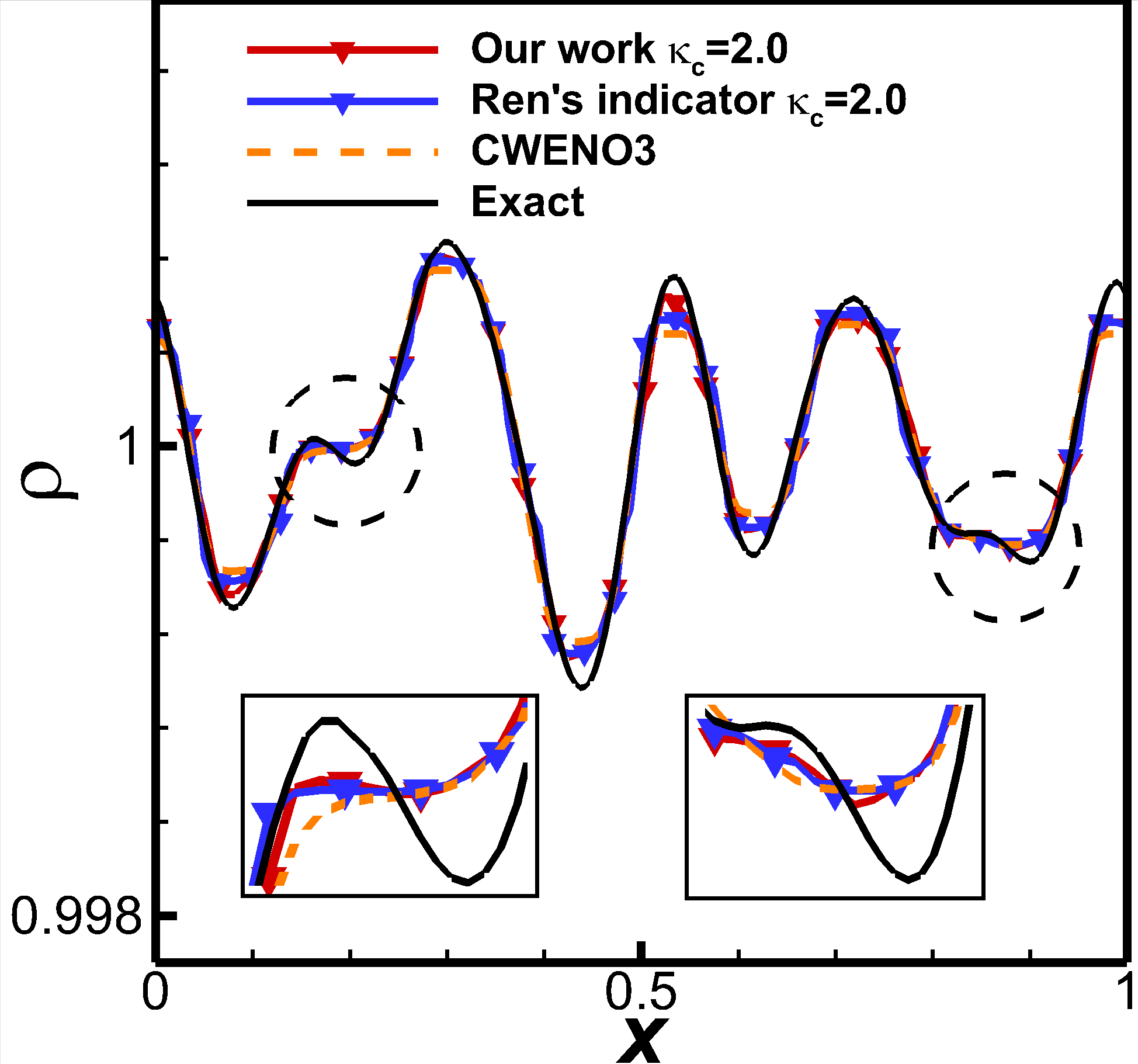}
    \caption{With coefficients optmized by $\kappa_c = 2.0$.}
    \end{subfigure}
    \caption{\label{fig:bwp_k4_compare_3rd} Results for the propagation of broadband sound waves with $k_0 = 4$ by the third-order hybrid CLS-CWENO scheme. }
\end{figure}

\begin{figure}[!htbp]
  \centering
    \begin{subfigure}[b]{\columnwidth}
    \includegraphics[width=0.61\columnwidth]{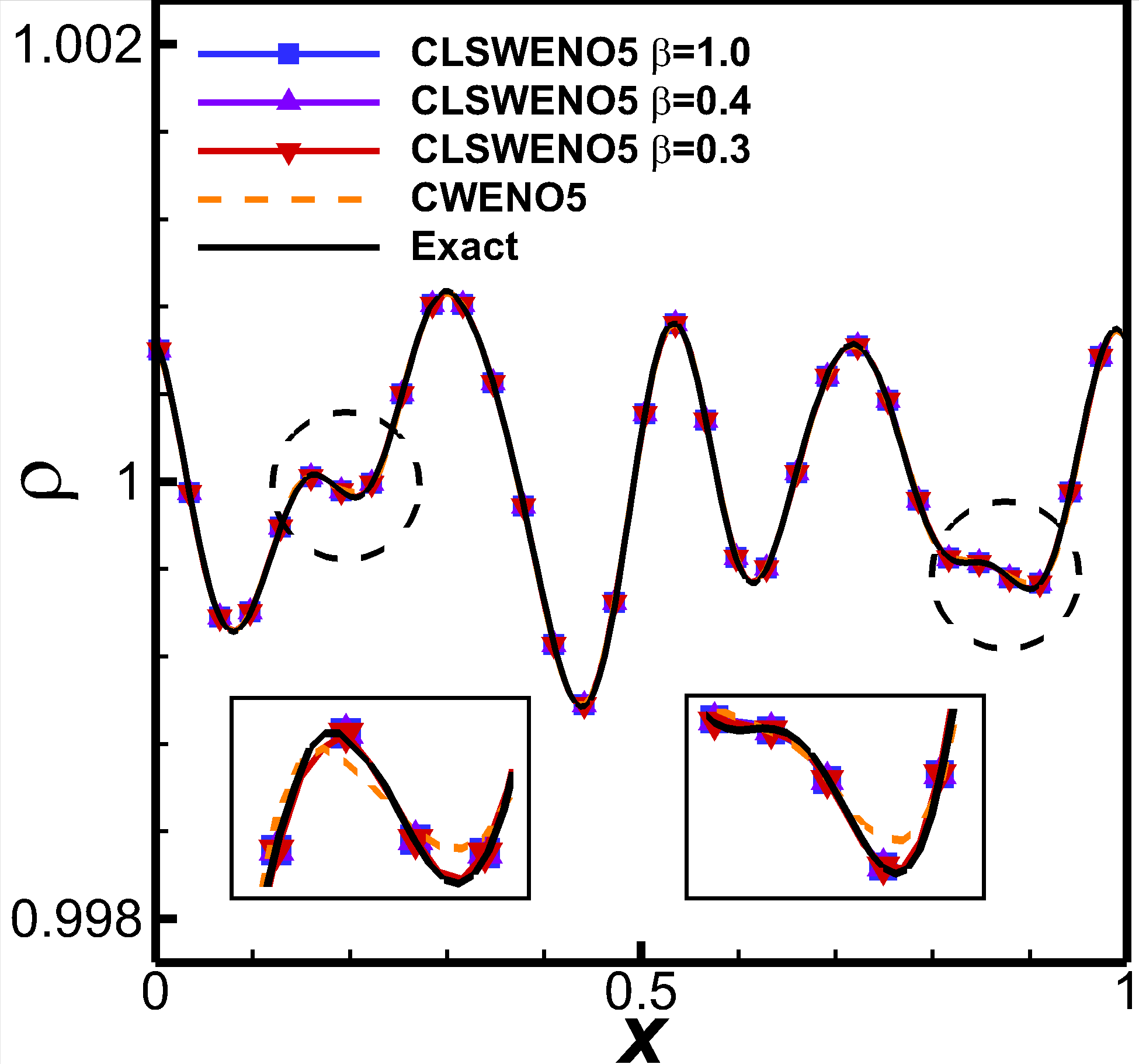}
    \caption{With coefficients optimized by varying $\beta$ with $\sigma^{\mathrm{Li}}$.}
    \end{subfigure}
    \begin{subfigure}[b]{\columnwidth}
    \includegraphics[width=0.61\columnwidth]{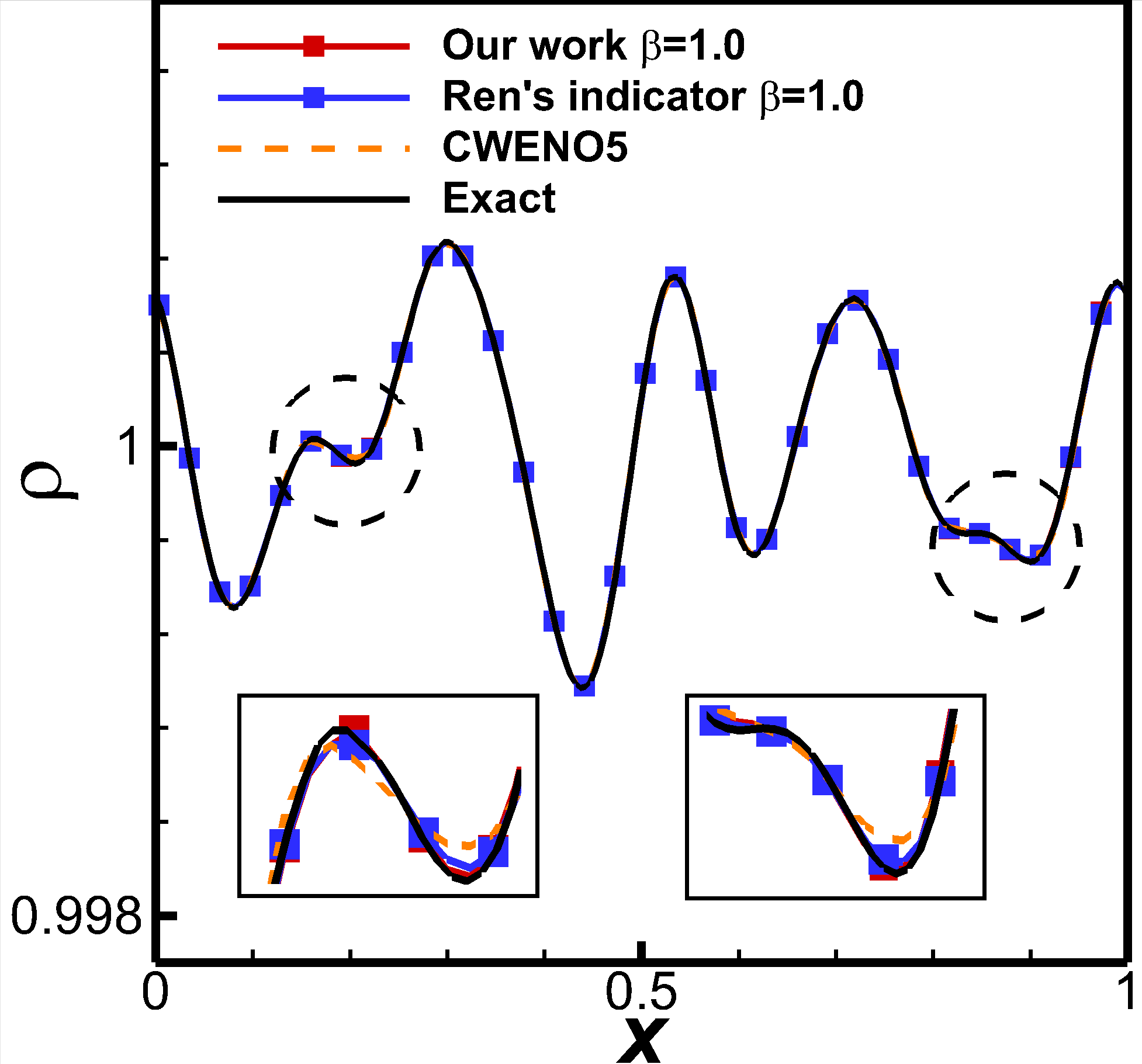}
    \caption{With coefficients optmized by $\beta = 1.0$.}
    \end{subfigure}
    \begin{subfigure}[b]{\columnwidth}
    \includegraphics[width=0.61\columnwidth]{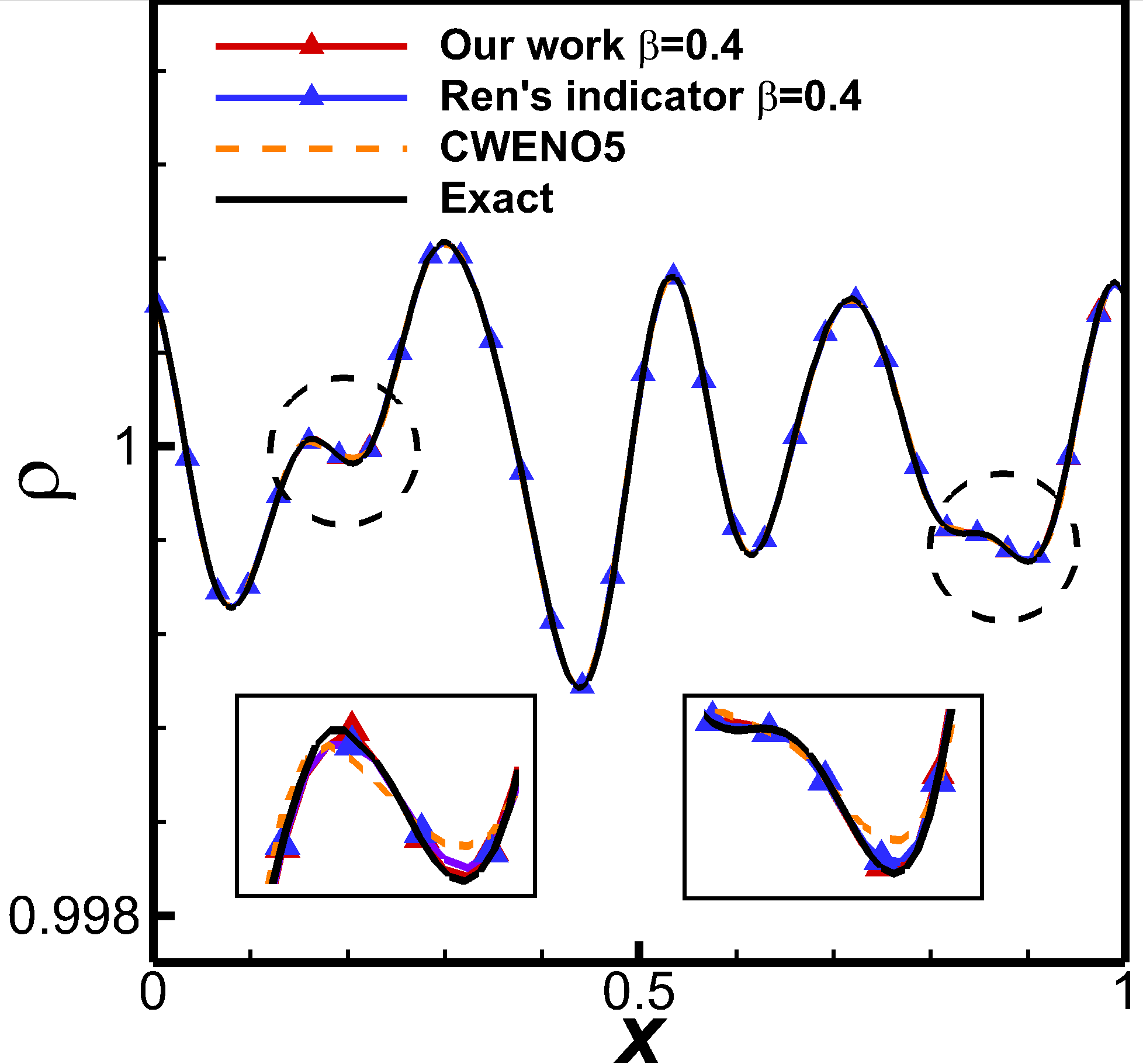}
    \caption{With coefficients optmized by $\beta = 0.4$.}
    \end{subfigure}
    \begin{subfigure}[b]{\columnwidth}
    \includegraphics[width=0.61\columnwidth]{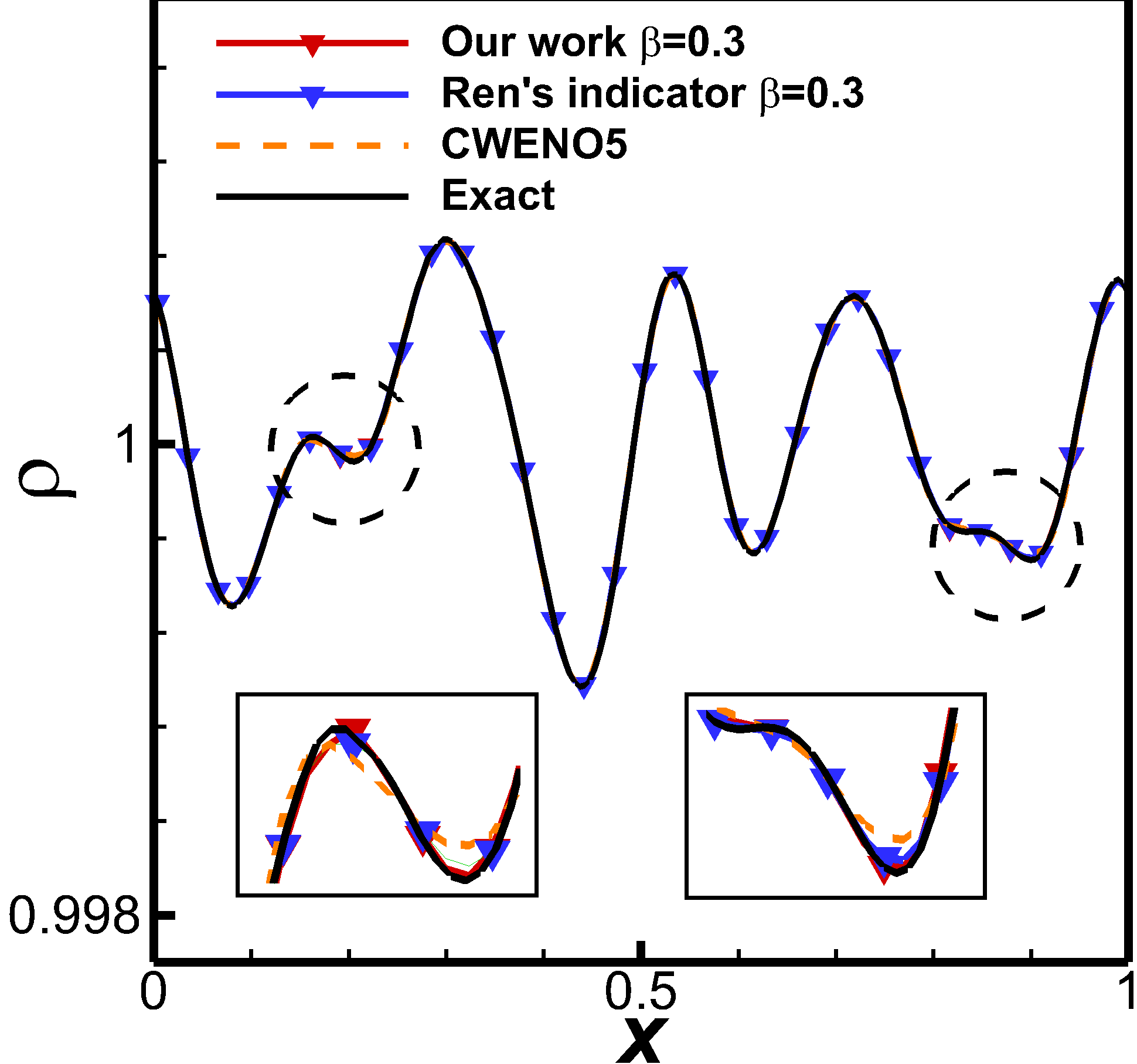}
    \caption{With coefficients optmized by $\beta = 0.3$.}
    \end{subfigure}
    \caption{\label{fig:bwp_k4_compare_5th} Results for the propagation of broadband sound waves with $k_0 = 4$ by the fifth-order hybrid CLS-CWENO scheme. }
\end{figure}

\begin{figure}[!htbp]
  \centering
    \begin{subfigure}[b]{\columnwidth}
    \includegraphics[width=0.61\columnwidth]{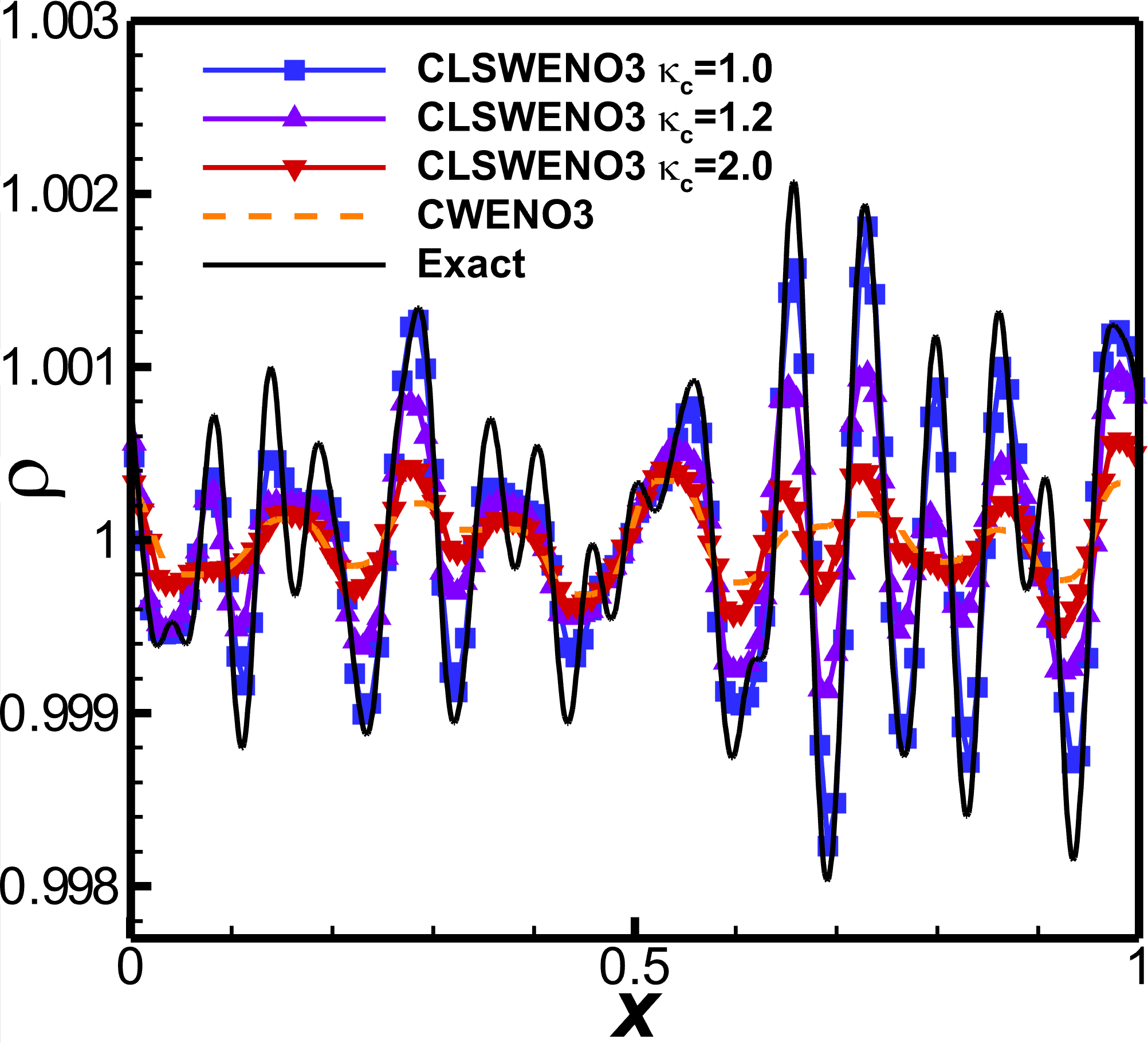}
    \caption{With coefficients optimized by varying $\kappa_c$ with $\sigma^{\mathrm{Li}}$.}
    \end{subfigure}
    \begin{subfigure}[b]{\columnwidth}
    \includegraphics[width=0.61\columnwidth]{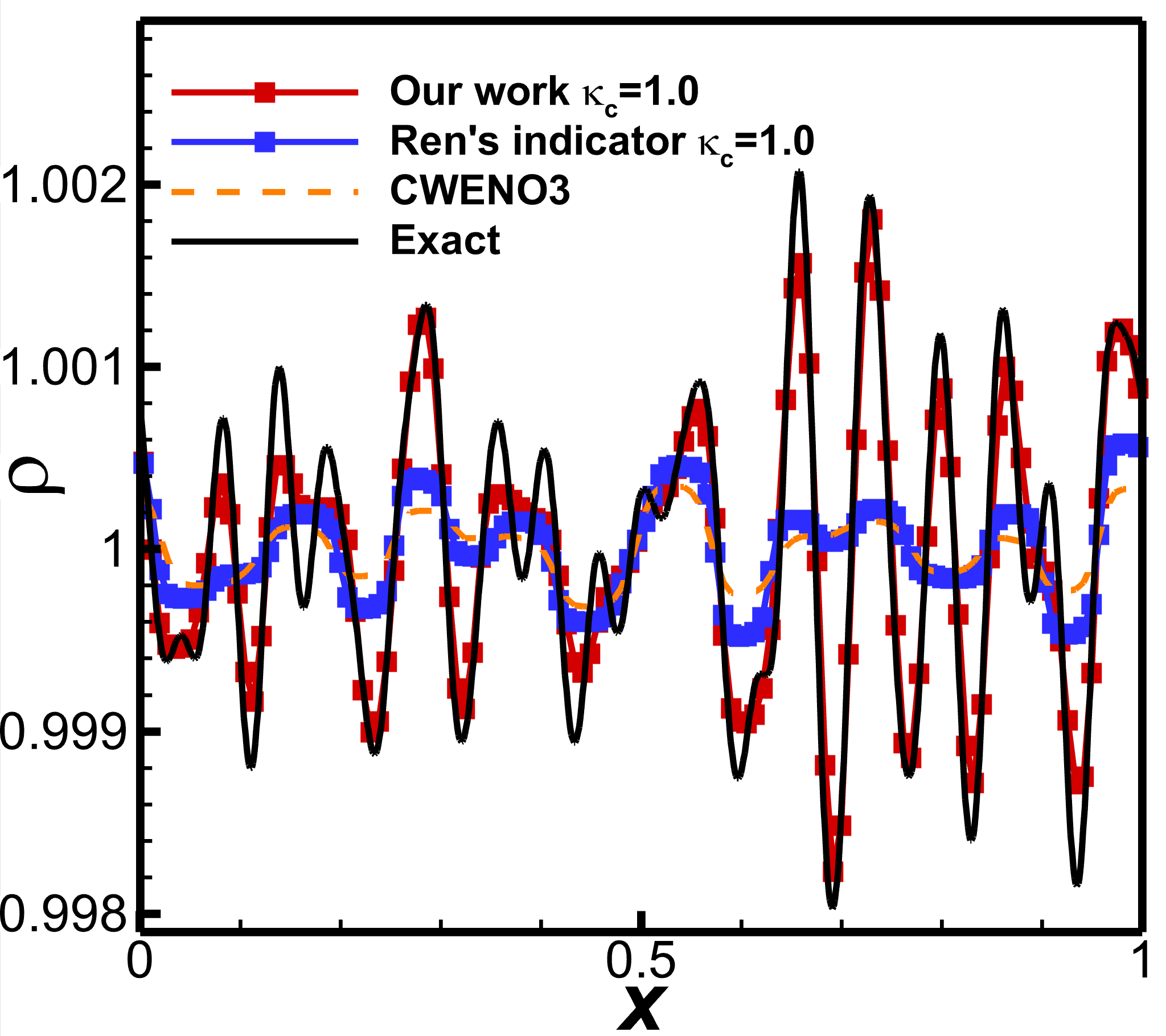}
    \caption{With coefficients optmized by $\kappa_c = 1.0$.}
    \end{subfigure}
    \begin{subfigure}[b]{\columnwidth}
    \includegraphics[width=0.61\columnwidth]{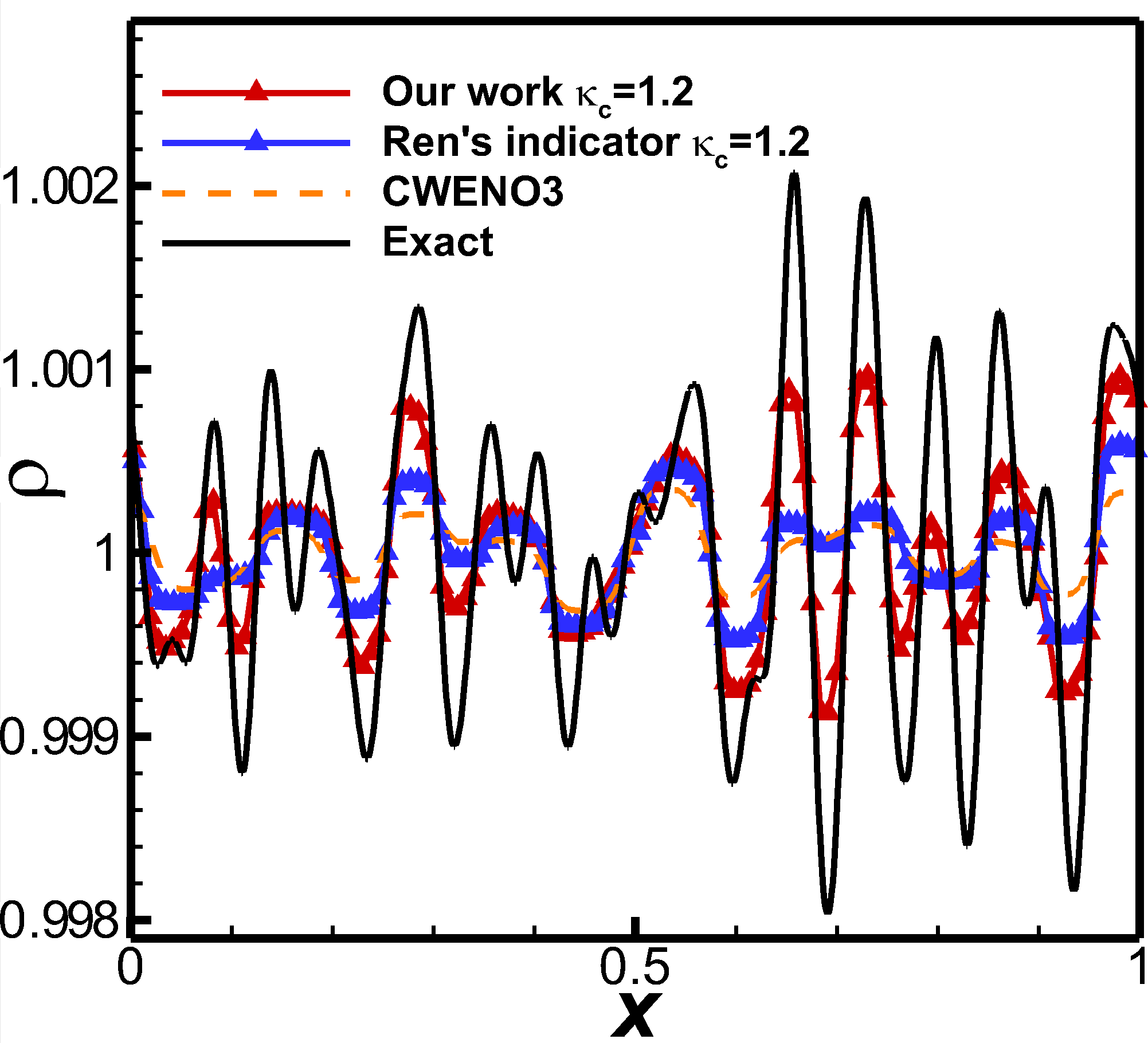}
    \caption{With coefficients optmized by $\kappa_c = 1.2$.}
    \end{subfigure}
    \begin{subfigure}[b]{\columnwidth}
    \includegraphics[width=0.61\columnwidth]{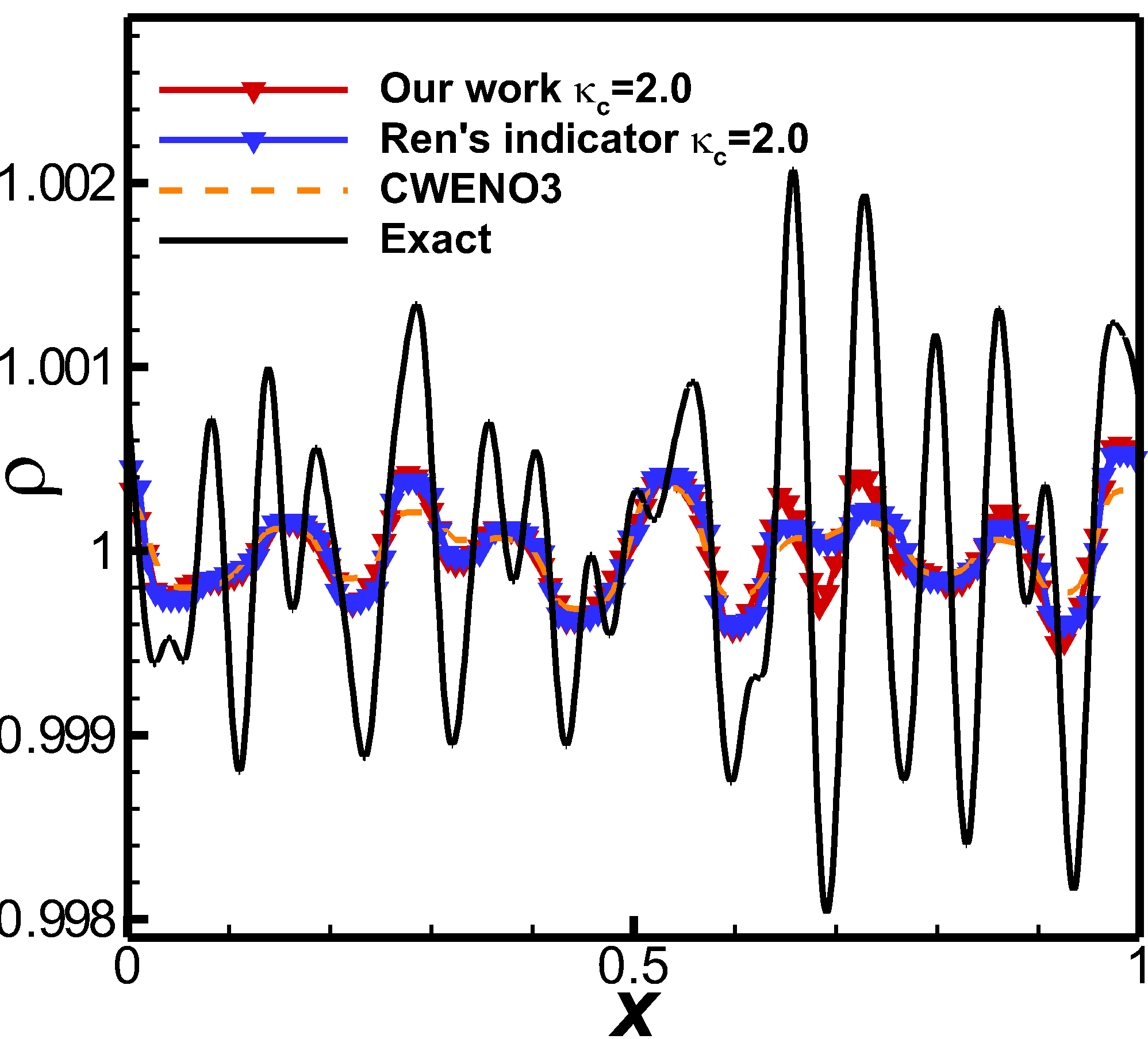}
    \caption{With coefficients optmized by $\kappa_c = 2.0$.}
    \end{subfigure}
    \caption{\label{fig:bwp_k12_compare_3rd} Results for the propagation of broadband sound waves with $k_0 = 12$ by the third-order hybrid CLS-CWENO scheme. }
\end{figure}

\begin{figure}[!htbp]
  \centering
    \begin{subfigure}[b]{\columnwidth}
    \includegraphics[width=0.61\columnwidth]{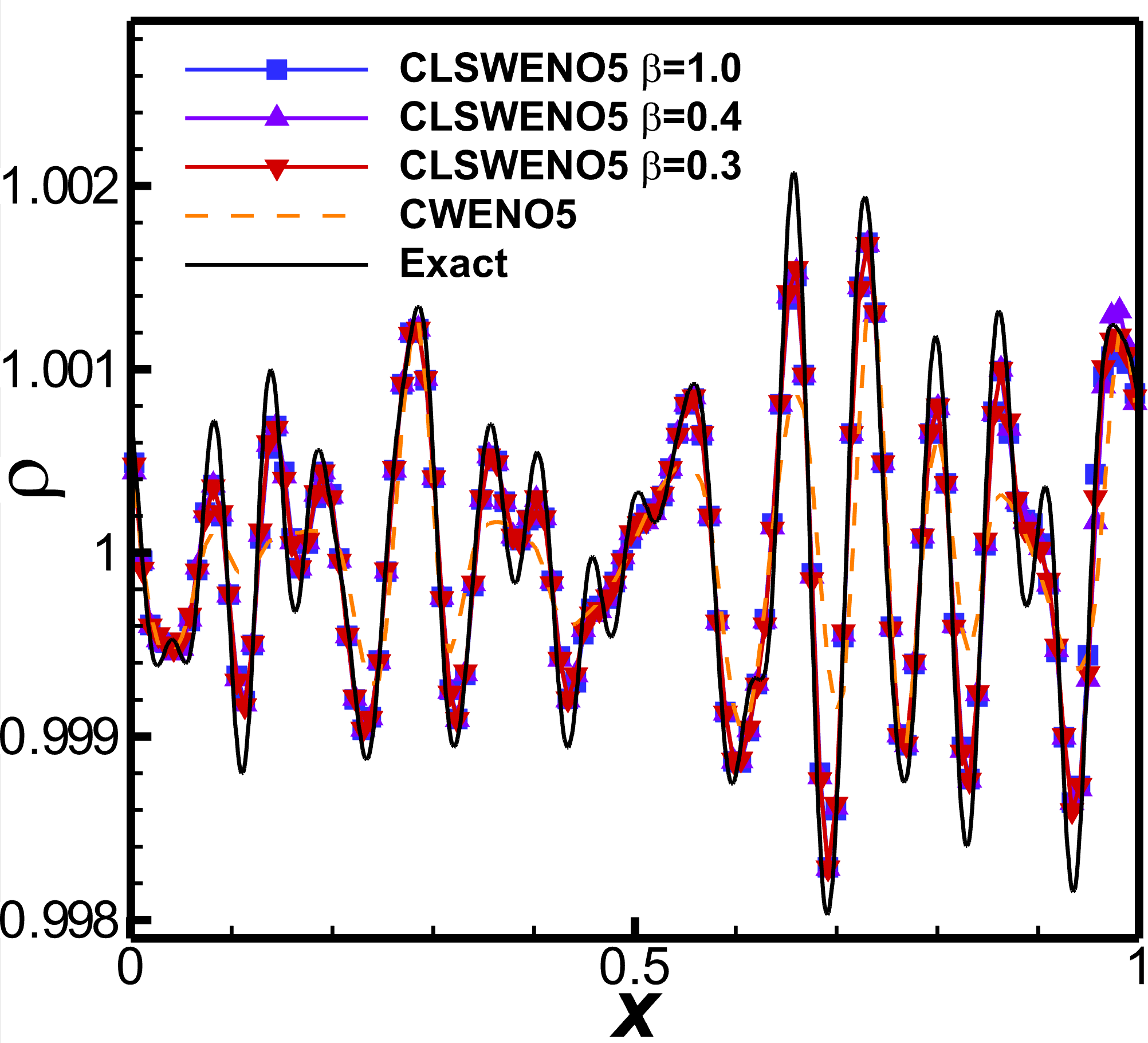}
    \caption{With coefficients optimized by varying $\beta$ with $\sigma^{\mathrm{Li}}$.}
    \end{subfigure}
    \begin{subfigure}[b]{\columnwidth}
    \includegraphics[width=0.61\columnwidth]{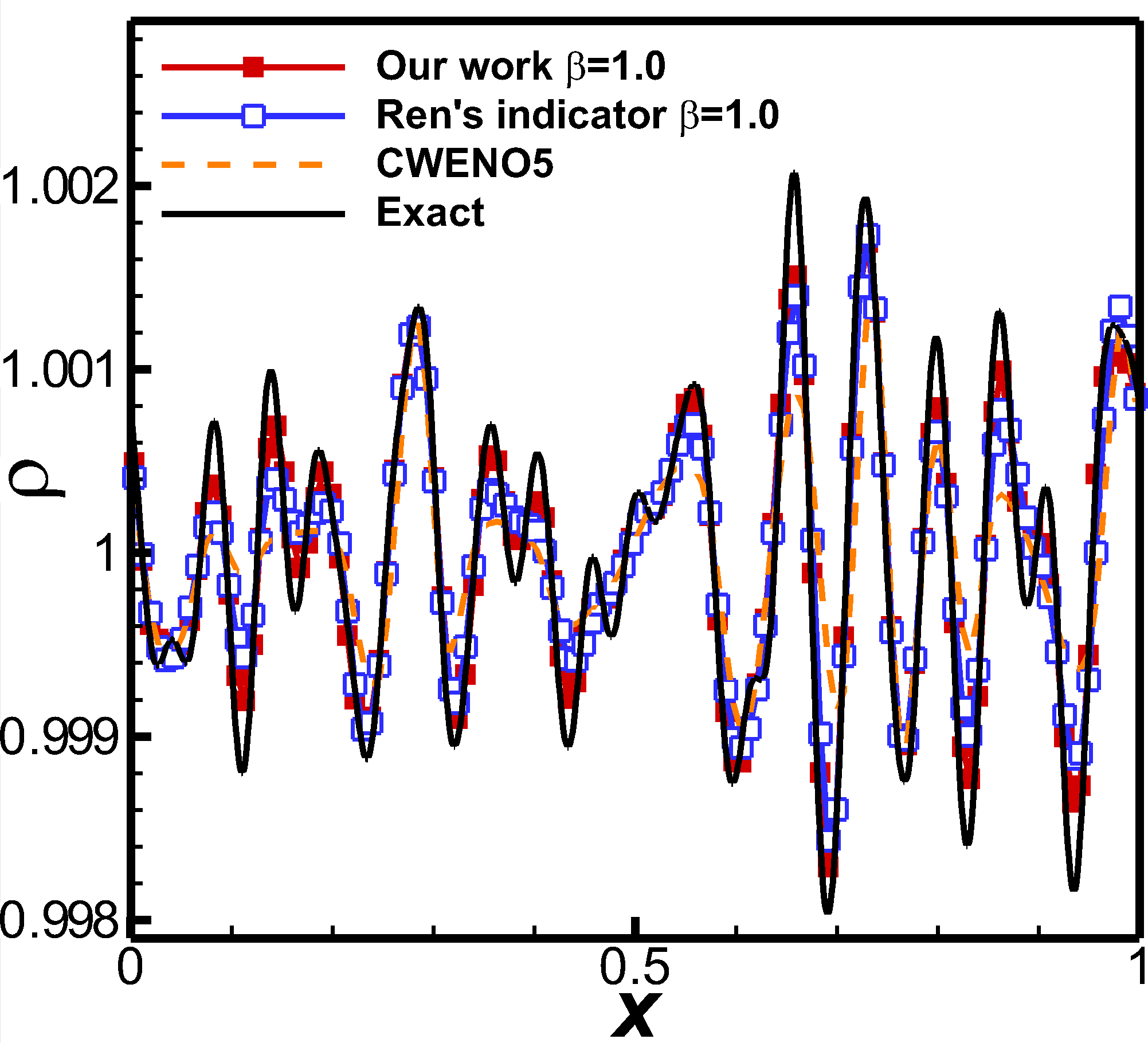}
    \caption{With coefficients optmized by $\beta = 1.0$.}
    \end{subfigure}
    \begin{subfigure}[b]{\columnwidth}
    \includegraphics[width=0.61\columnwidth]{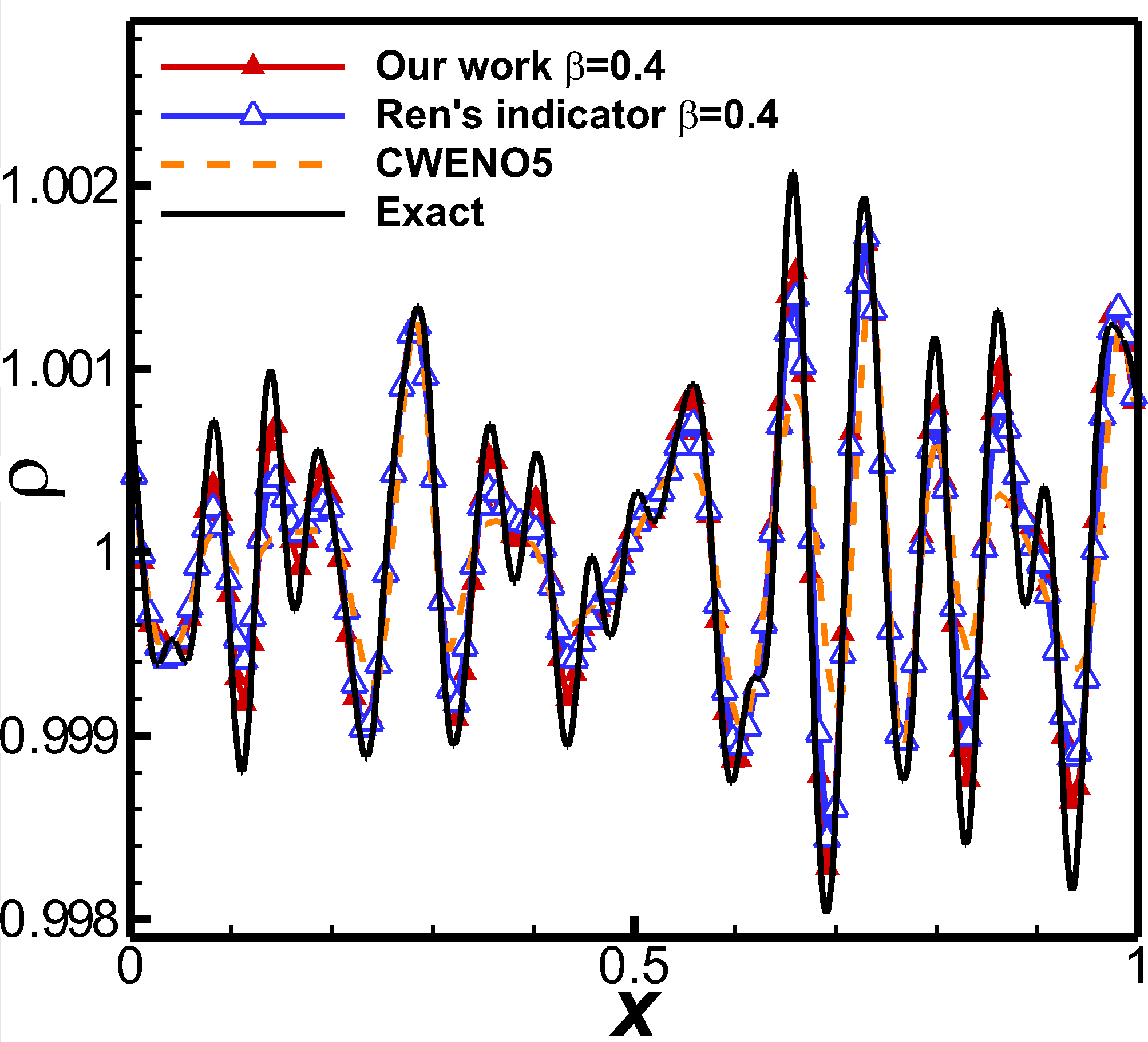}
    \caption{With coefficients optmized by $\beta = 0.4$.}
    \end{subfigure}
    \begin{subfigure}[b]{\columnwidth}
    \includegraphics[width=0.61\columnwidth]{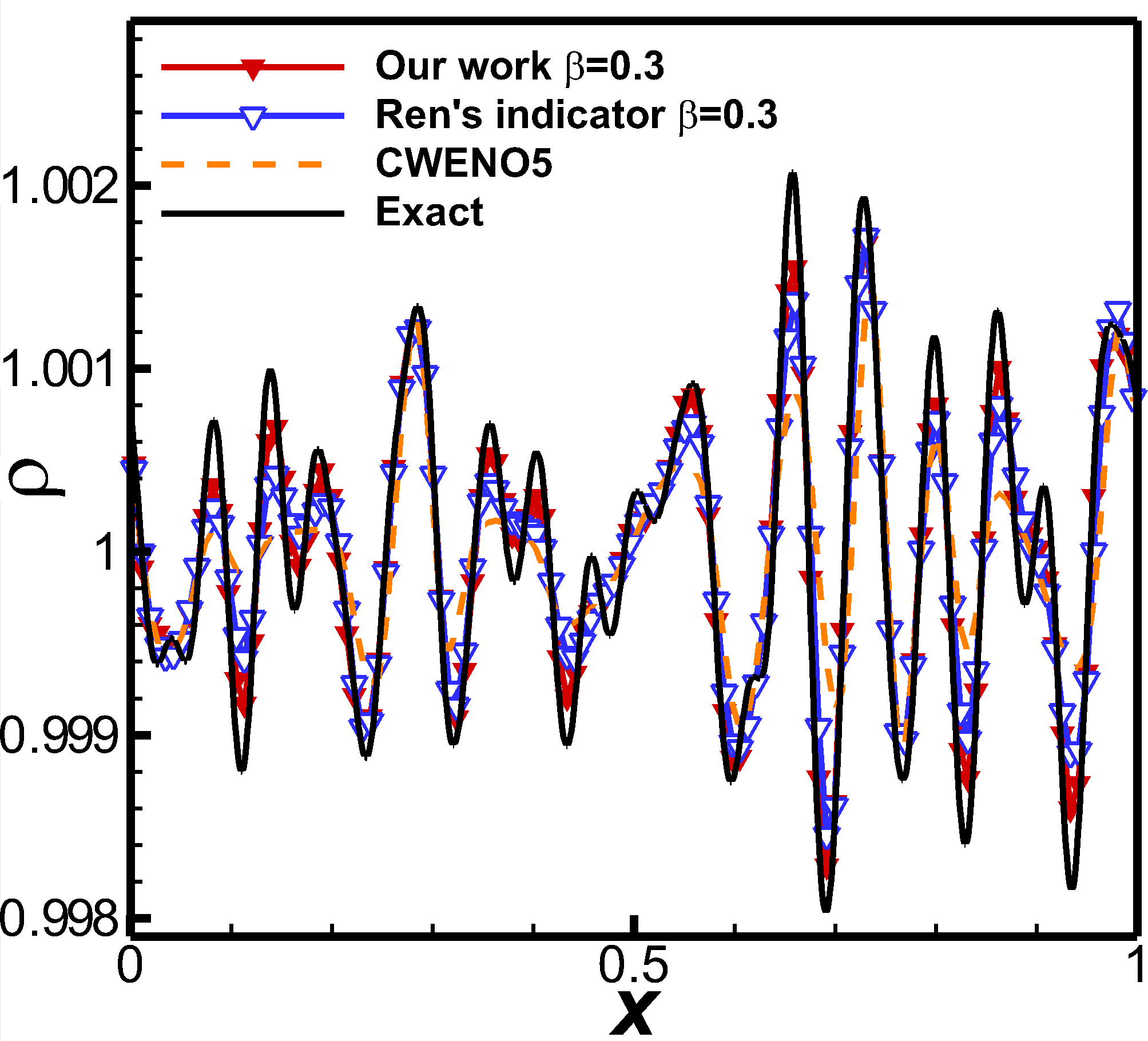}
    \caption{With coefficients optmized by $\beta = 0.3$.}
    \end{subfigure}
    \caption{\label{fig:bwp_k12_compare_5th} Results for the propagation of broadband sound waves with $k_0 = 12$ by the fifth-order hybrid CLS-CWENO scheme. }
\end{figure}

Figures \ref{fig:bwp_k4_compare_3rd} and \ref{fig:bwp_k4_compare_5th} show the results of the hybrid CLS-CWENO schemes with $k_0 = 4$. For the third-order scheme with proposed shock detector, extrema are well-resolved as shown in Fig. \ref{fig:bwp_k4_compare_3rd}. However, there exists order degradation when utilizing $\sigma^{\mathrm{Ren}}$ which cannot identify the smooth extrema. For the fifth-order hybrid scheme, all the flow structures are resolved whichever shock detector is utilized.

When increasing the wavenumber $k_0$ to 12, the results are shown as in Figs. \ref{fig:bwp_k12_compare_3rd} and \ref{fig:bwp_k12_compare_5th}. For the third-order scheme, an interesting phenomenon is observed: although the coefficients optimized by $\kappa_c = 1.0$ have the largest dissipation in high-wave-number region, the hybrid scheme with coefficients optimized by $\kappa_c = 1.0$ performs best. In addition, Fig. \ref{fig:bwp_k12_compare_3rd} confirms the superiority of the proposed $\sigma^{\mathrm{Li}}$ in identifying local extrema than $\sigma^{\mathrm{Ren}}$. As shown in Fig. \ref{fig:bwp_k12_compare_5th}, flow structures with different scales are all resolved with the fifth-order hybrid CLS-CWENO schemes.
\subsection{Lax shock tube problem}
\begin{figure}[!htbp]
  \centering
    \begin{subfigure}[b]{\columnwidth}
    \includegraphics[width=0.61\columnwidth]{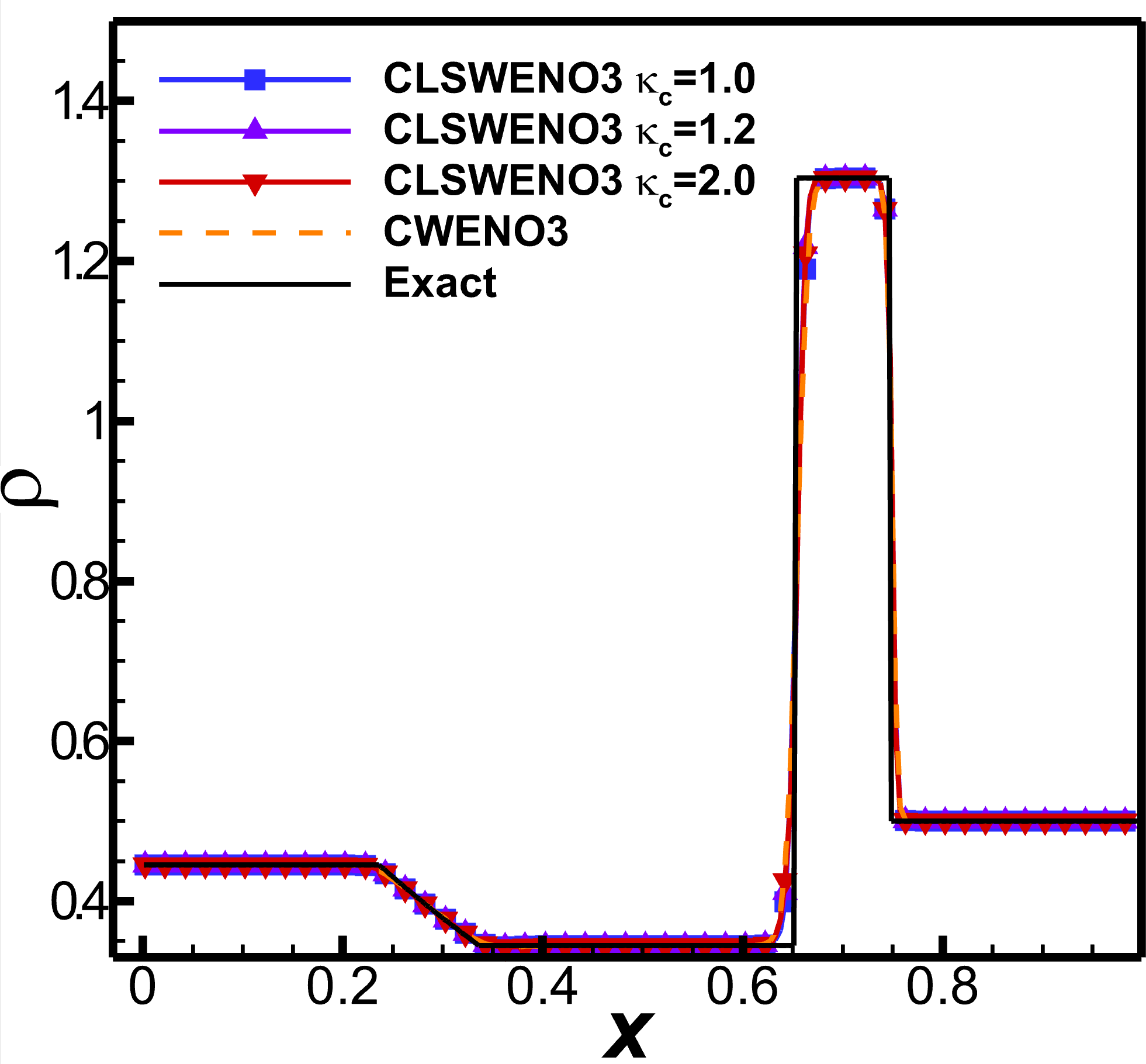}
    \caption{Density.\label{fig:lax_3rd_diff_beta_density}}
    \end{subfigure}
    \begin{subfigure}[b]{\columnwidth}
    \includegraphics[width=0.61\columnwidth]{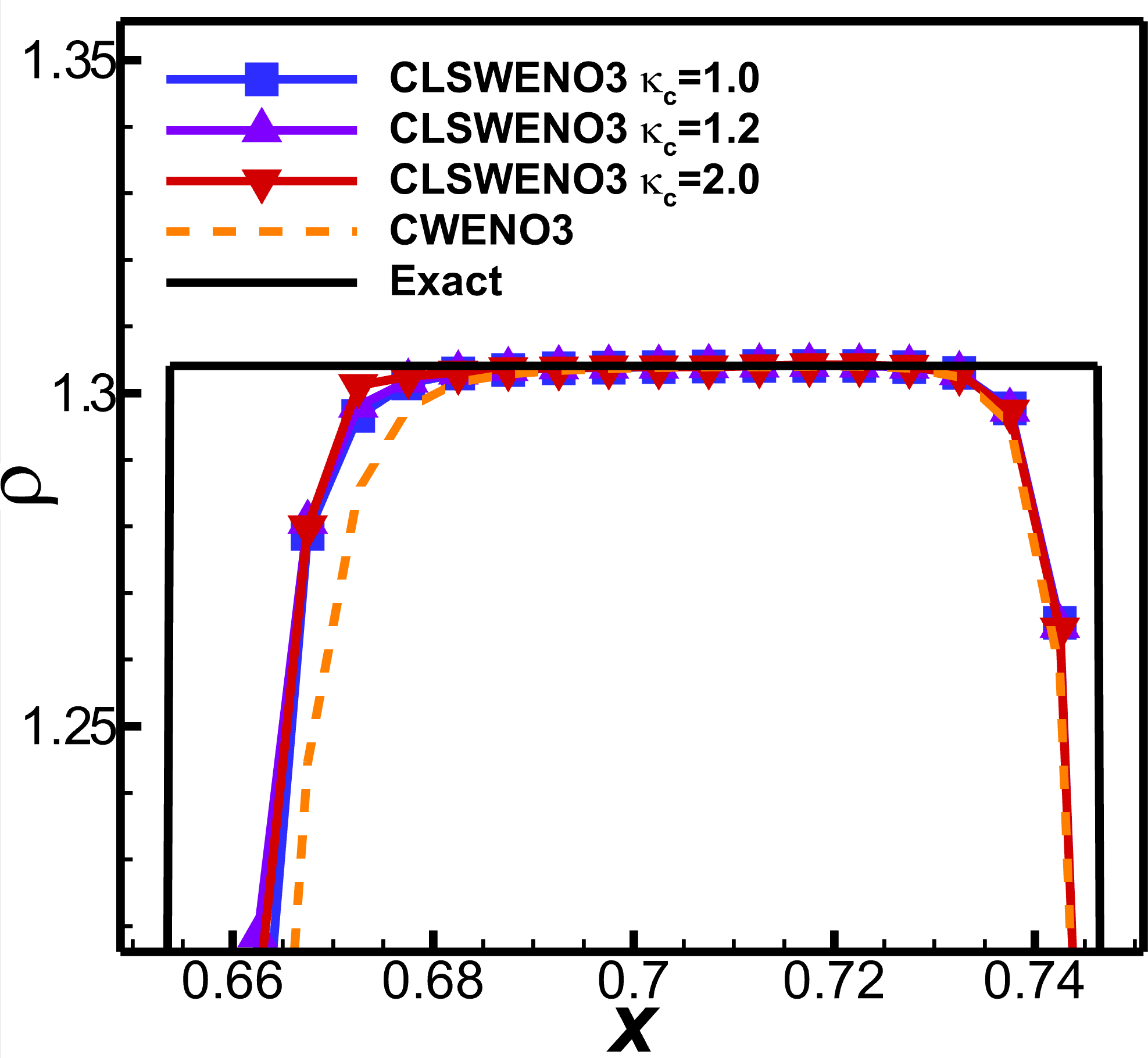}
    \caption{Close view of density.\label{fig:lax_3rd_diff_beta_density_cv}}
    \end{subfigure}
    \begin{subfigure}[b]{\columnwidth}
    \includegraphics[width=0.61\columnwidth]{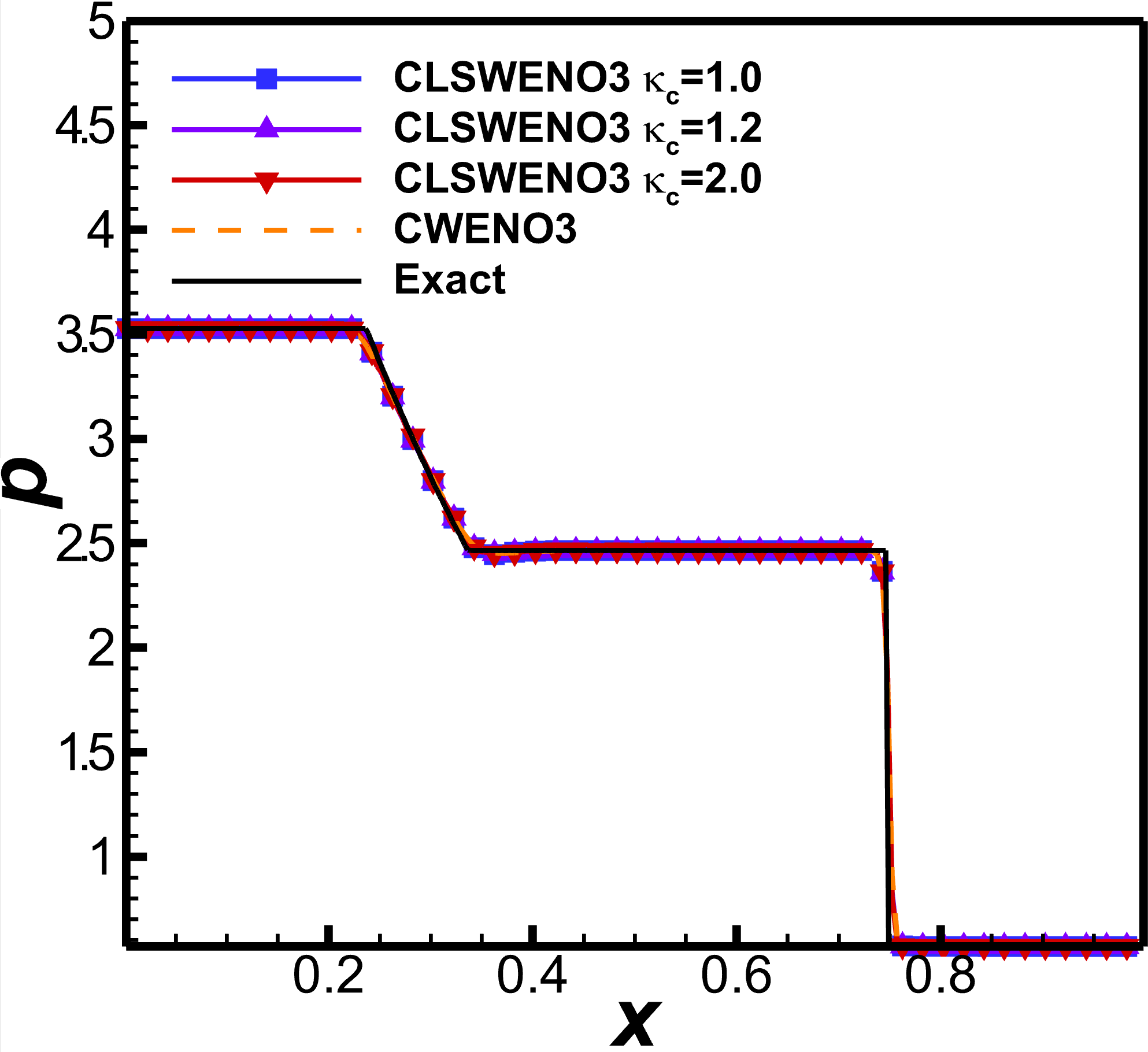}
    \caption{Pressure.\label{fig:lax_3rd_diff_beta_pre}}
    \end{subfigure}
    \caption{\label{fig:lax_3rd_diff_beta} Results of the Lax shock tube problem with third-order scheme utilizing $\sigma^{\mathrm{Li}}$ with $\theta_c=0.20$.}
\end{figure}

\begin{figure}[!htbp]
  \centering
    \begin{subfigure}[b]{\columnwidth}
    \includegraphics[width=0.61\columnwidth]{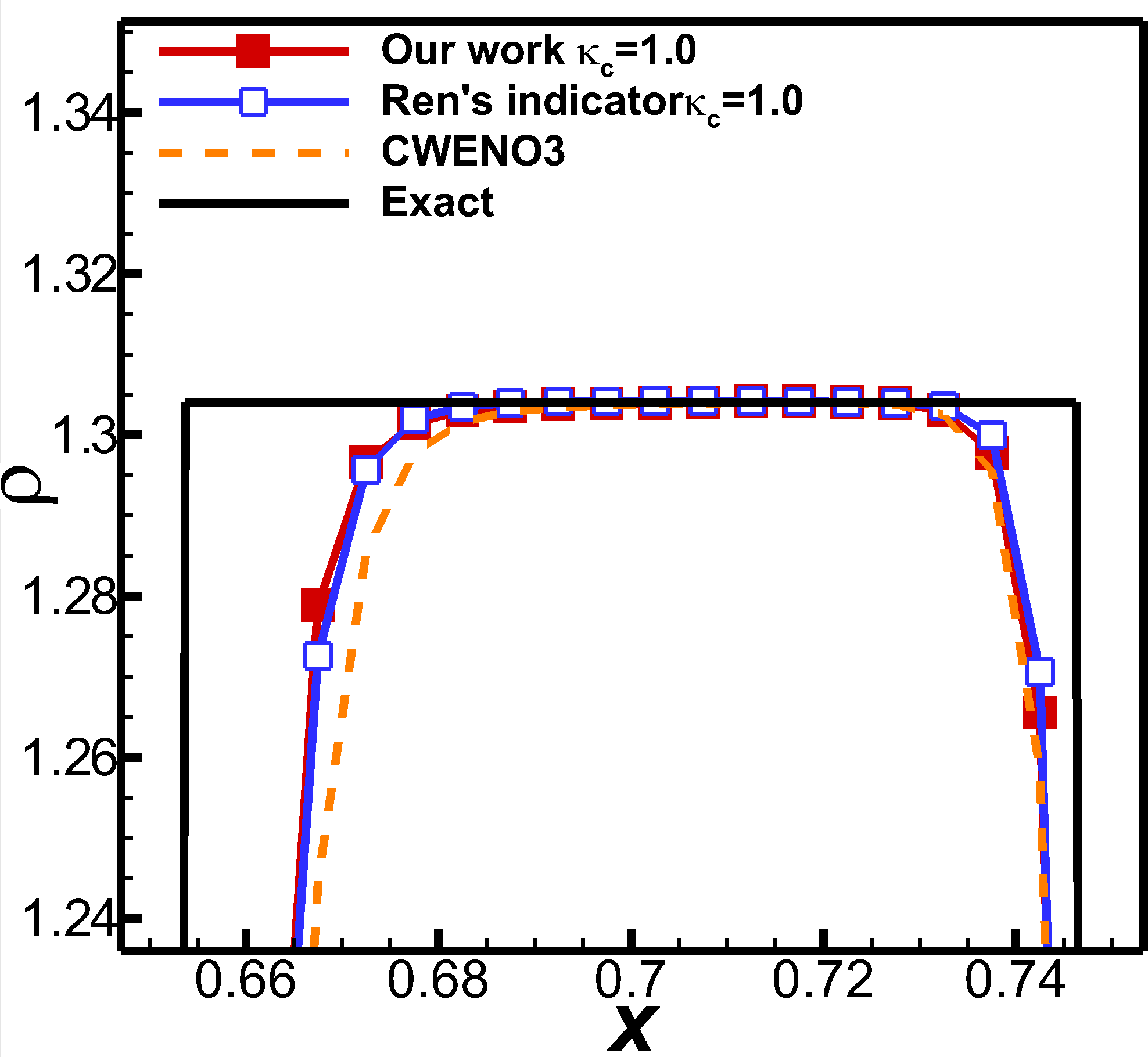}
    \caption{With coefficients optimized by $\kappa = 1.0$.\label{fig:lax_3rd_compare_with_ren_1.0}}
    \end{subfigure}
    \begin{subfigure}[b]{\columnwidth}
    \includegraphics[width=0.61\columnwidth]{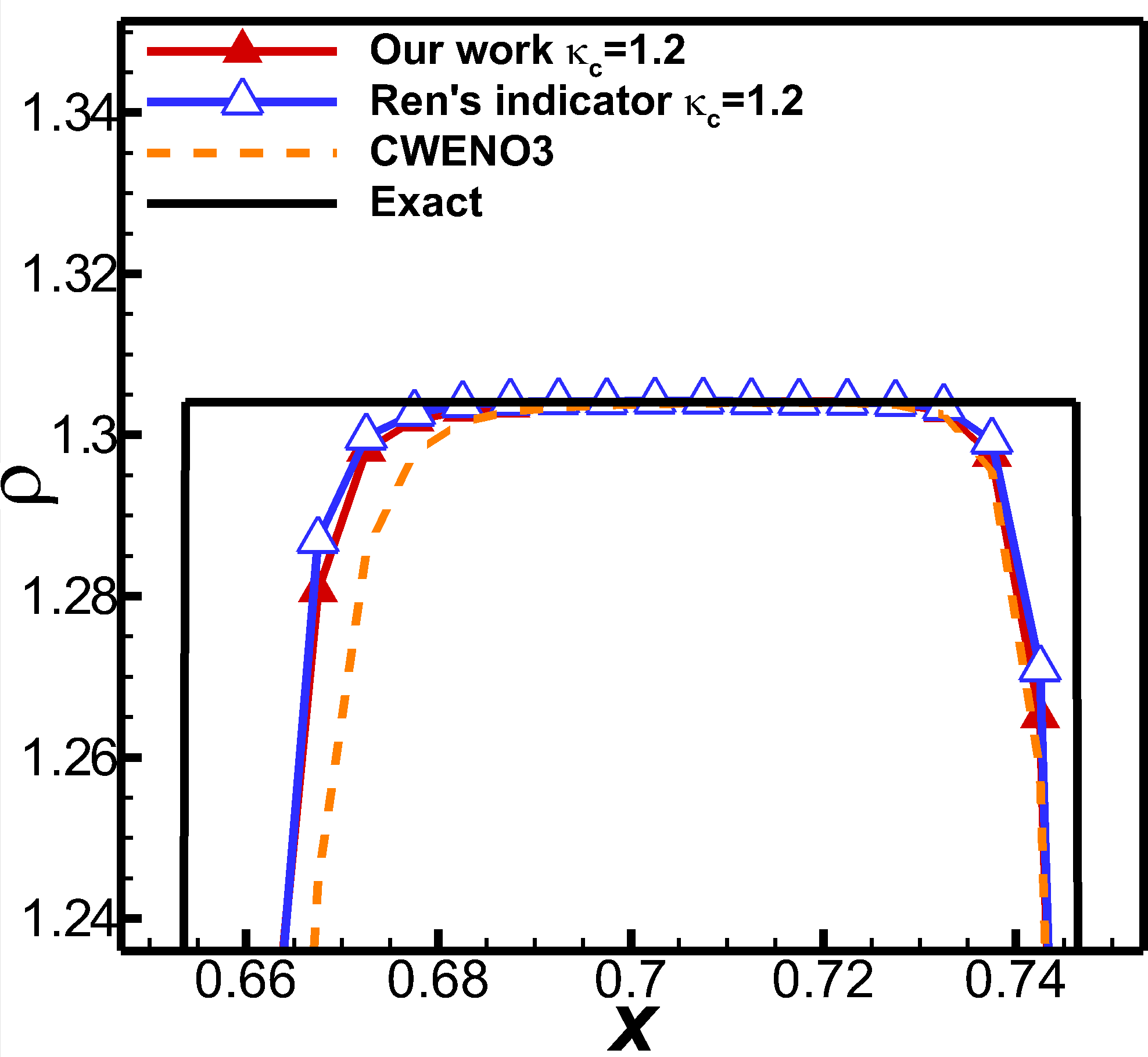}
    \caption{With coefficients optimized by $\kappa = 1.2$.\label{fig:lax_3rd_compare_with_ren_1.2}}
    \end{subfigure}
    \begin{subfigure}[b]{\columnwidth}
    \includegraphics[width=0.61\columnwidth]{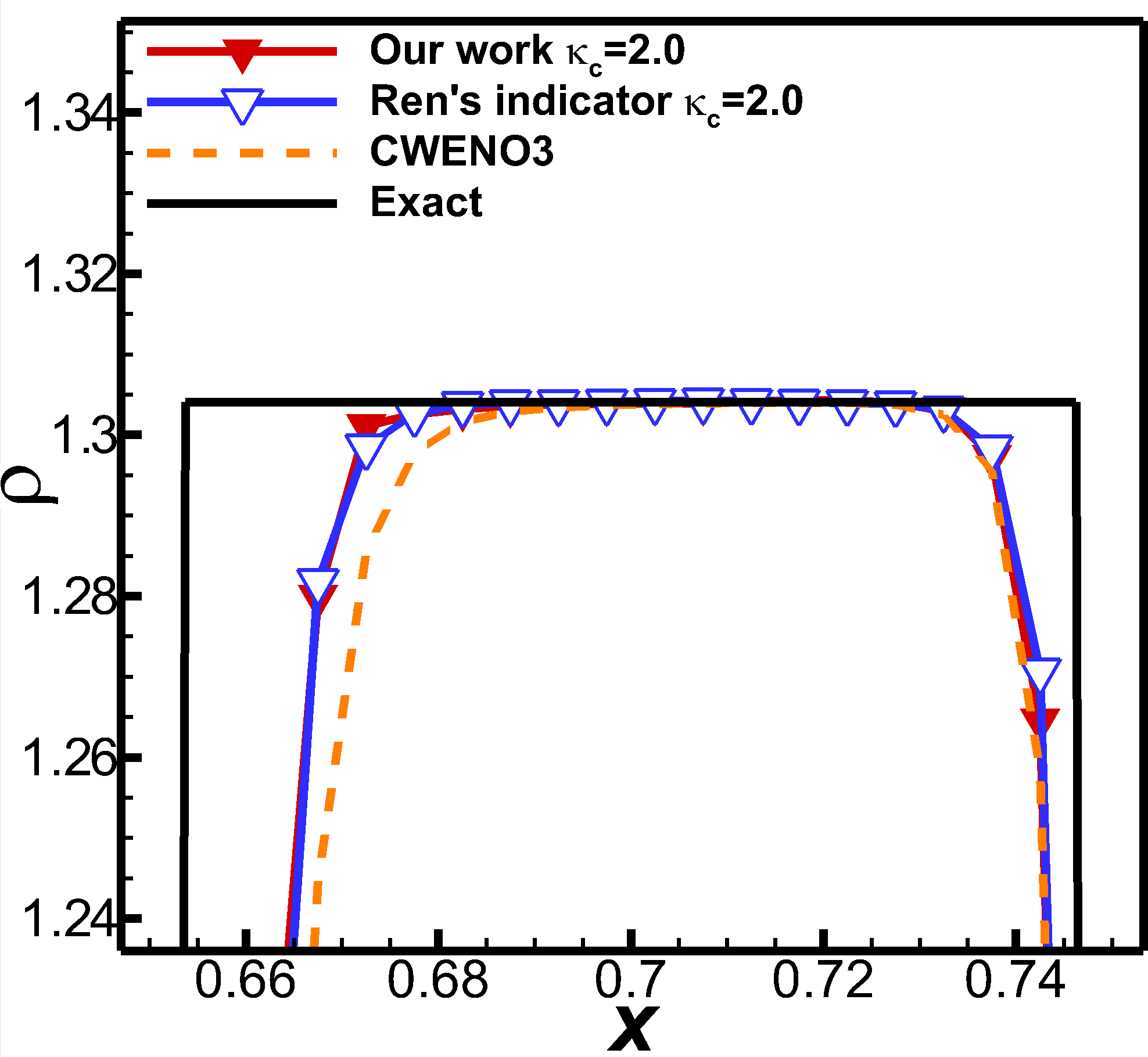}
    \caption{With coefficients optimized by $\kappa = 2.0$.\label{fig:lax_3rd_compare_with_ren_2.0}}
    \end{subfigure}
    \caption{\label{fig:lax_diff_sd_3rd} Comparison of the shock detector for the Lax shock tube problem. Third-order hybrid CLS-CWENO scheme. $\theta_c=0.20$ for $\sigma^{\mathrm{Li}}$ and $\theta_c=0.50$ for $\sigma^{\mathrm{Ren}}$.}
\end{figure}

\begin{figure}[!htbp]
  \centering
    \begin{subfigure}[b]{\columnwidth}
    \includegraphics[width=0.61\columnwidth]{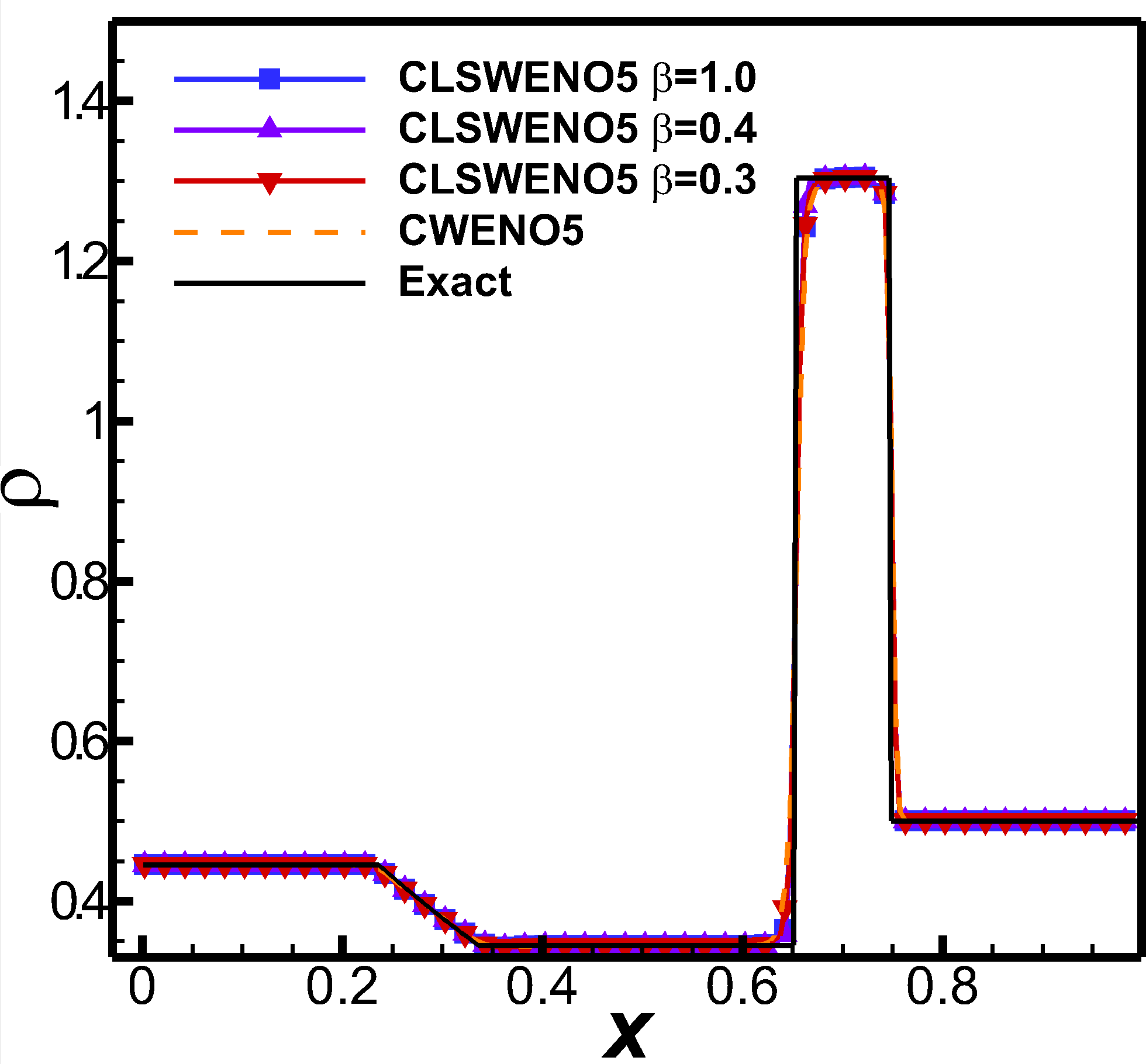}
    \caption{Density.\label{fig:lax_5th_diff_beta_density}}
    \end{subfigure}
    \begin{subfigure}[b]{\columnwidth}
    \includegraphics[width=0.61\columnwidth]{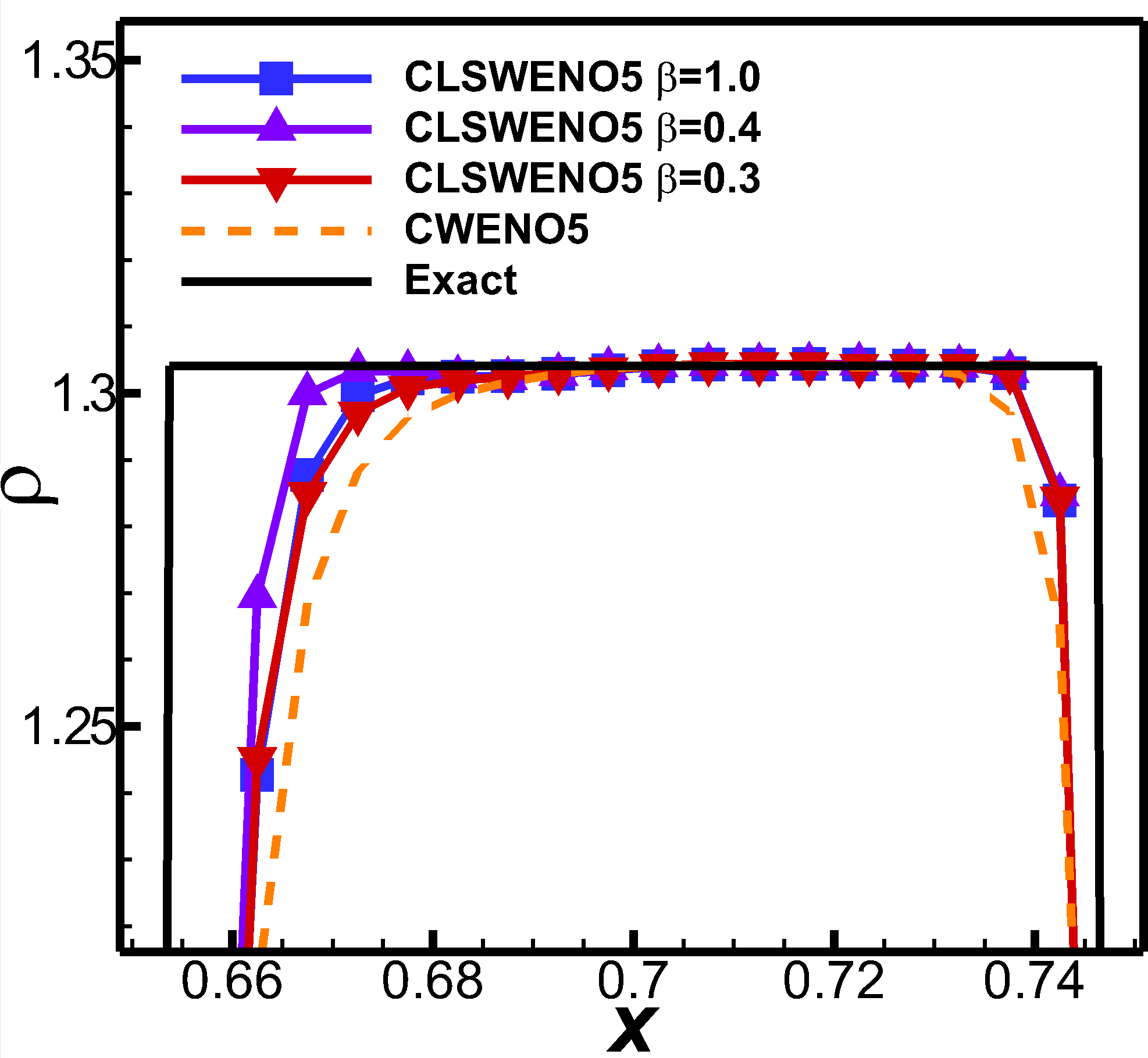}
    \caption{Close view of density.\label{fig:lax_5th_diff_beta_density_cv}}
    \end{subfigure}
    \begin{subfigure}[b]{\columnwidth}
    \includegraphics[width=0.61\columnwidth]{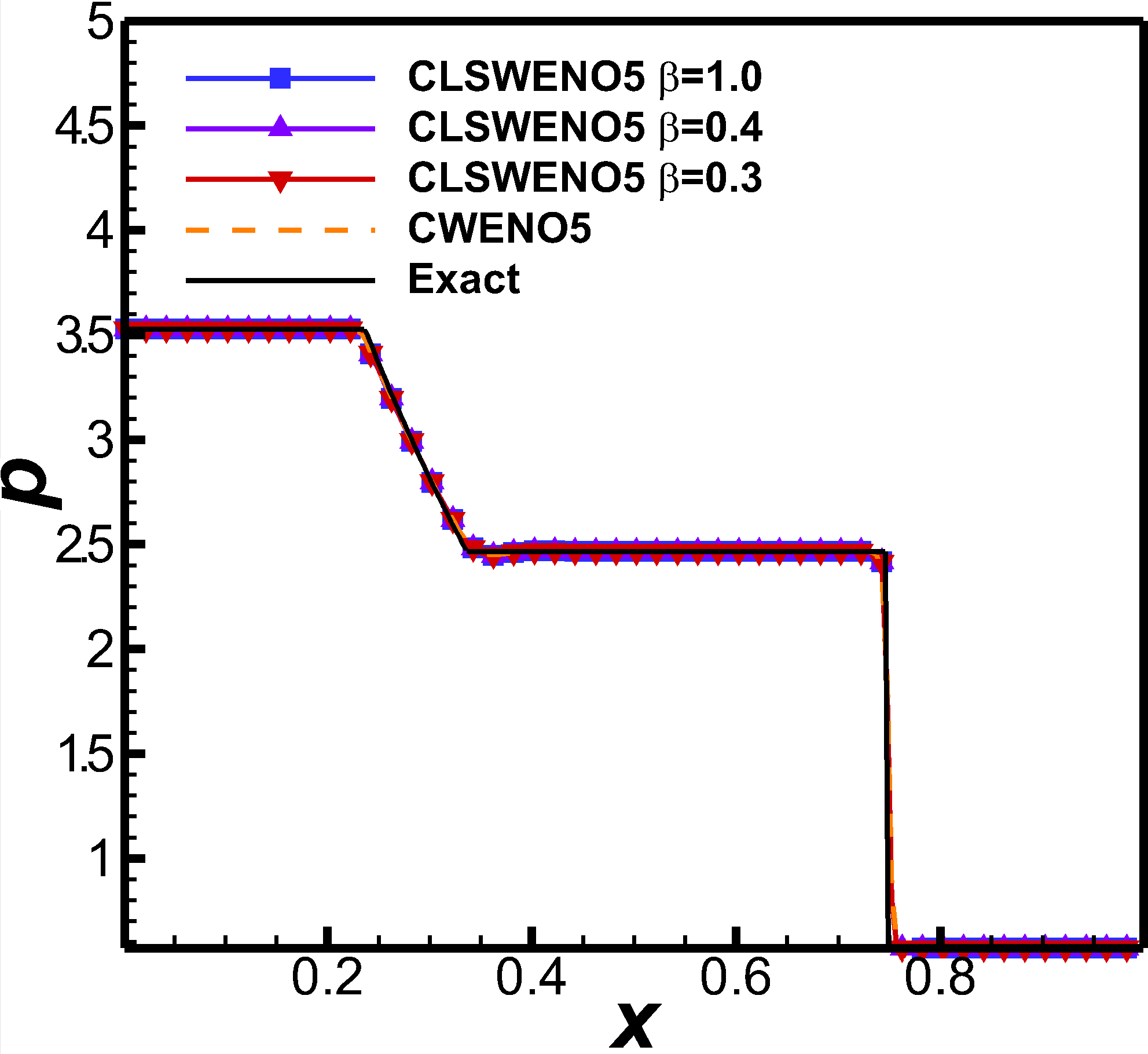}
    \caption{Pressure.\label{fig:lax_5th_diff_beta_pre}}
    \end{subfigure}
    \caption{\label{fig:lax_5th_diff_beta} Results of the Lax shock tube problem with fifth-order scheme utilizing $\sigma^{\mathrm{Li}}$ with $\theta_c=0.03$.}
\end{figure}

\begin{figure}[!htbp]
  \centering
    \begin{subfigure}[b]{\columnwidth}
    \includegraphics[width=0.61\columnwidth]{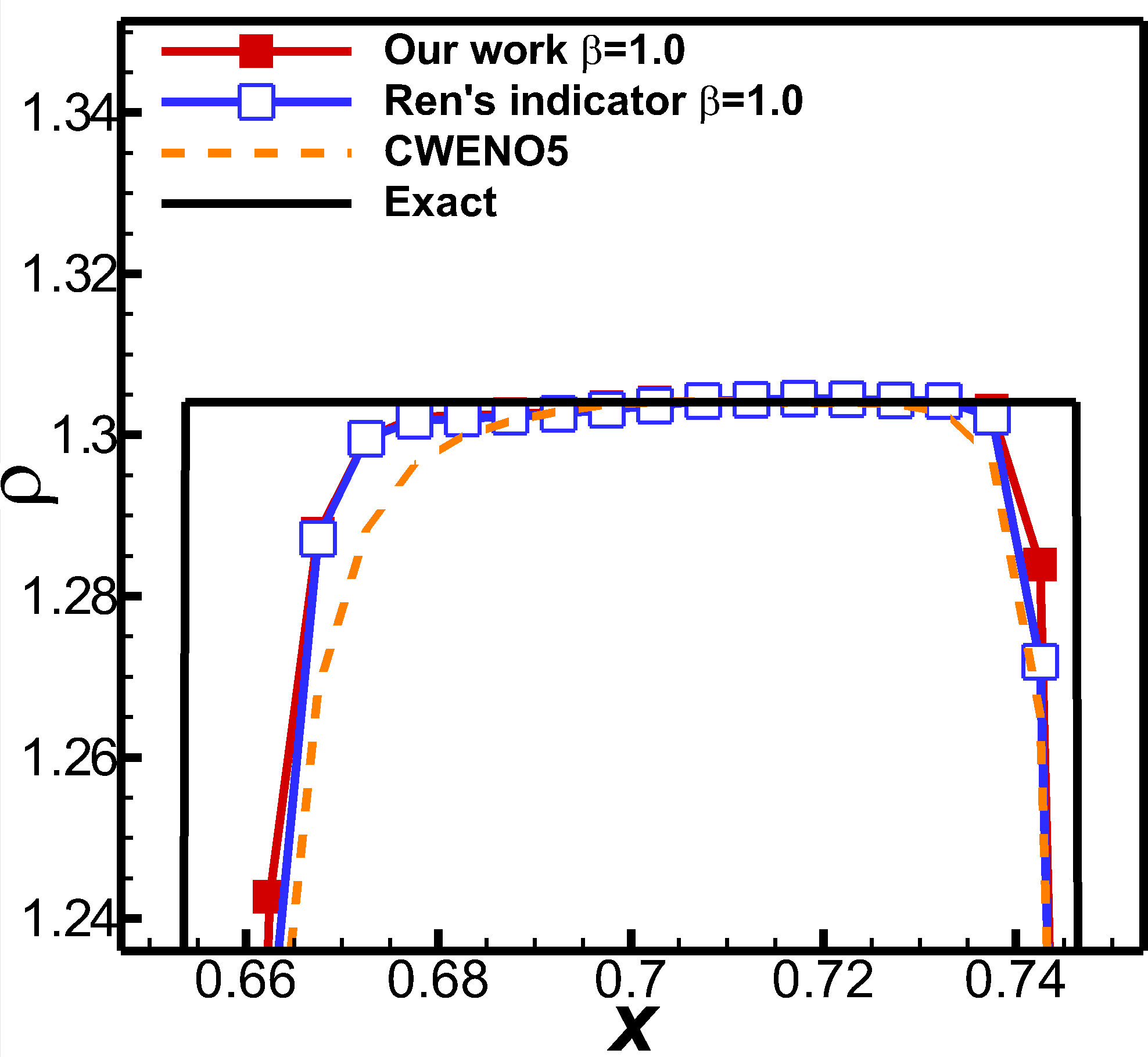}
    \caption{With coefficients optimized by $\beta = 1.0$.\label{fig:lax_5th_compare_with_ren_1.0}}
    \end{subfigure}
    \begin{subfigure}[b]{\columnwidth}
    \includegraphics[width=0.61\columnwidth]{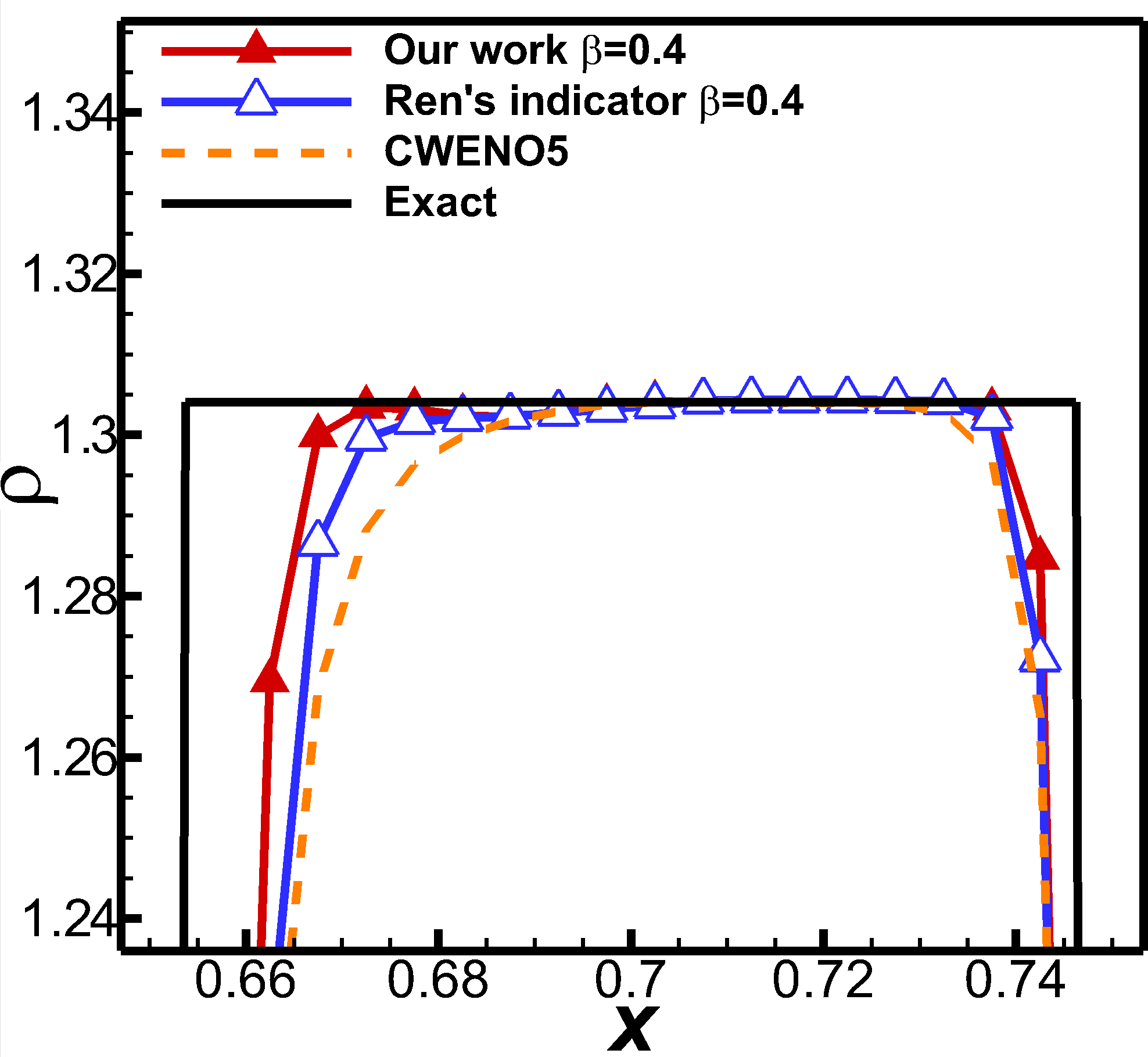}
    \caption{With coefficients optimized by $\beta = 0.4$.\label{fig:lax_5th_compare_with_ren_1.2}}
    \end{subfigure}
    \begin{subfigure}[b]{\columnwidth}
    \includegraphics[width=0.61\columnwidth]{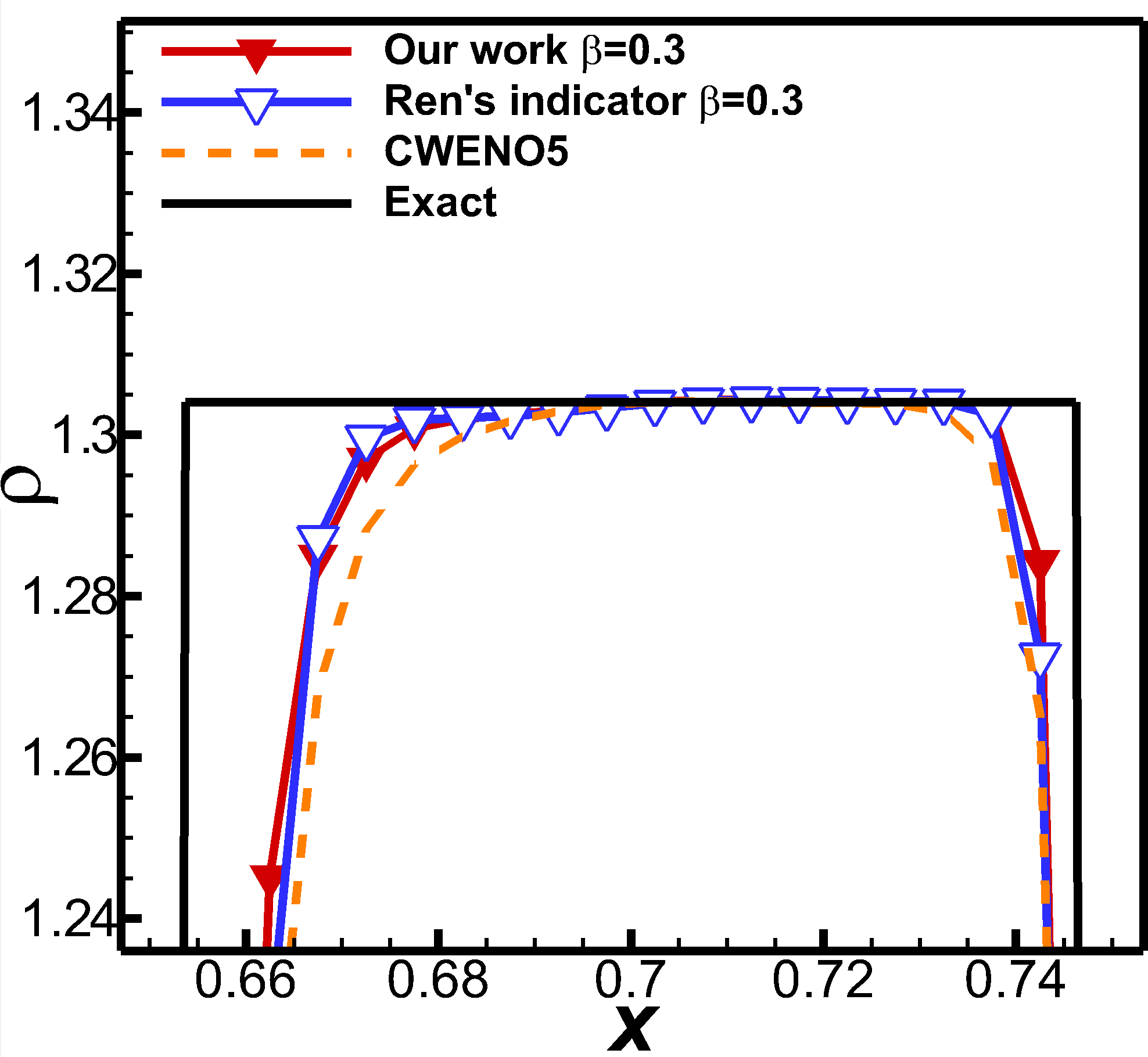}
    \caption{With coefficients optimized by $\beta = 0.3$.\label{fig:lax_5th_compare_with_ren_2.0}}
    \end{subfigure}
    \caption{\label{fig:lax_diff_sd_5th} Comparison of the shock detector for the Lax shock tube problem. Fifth-order hybrid CLS-CWENO scheme. $\theta_c=0.03$ for $\sigma^{\mathrm{Li}}$ and $\theta_c=0.50$ for $\sigma^{\mathrm{Ren}}$.}
\end{figure}

The results of the Lax shock tube problem with different optimized coefficients are shown from Figs. \ref{fig:lax_3rd_diff_beta} to \ref{fig:lax_diff_sd_5th}. The simulation setup is not replicated here, and readers are referred to Sec. \ref{sec:calibration_lax}.

Figure \ref{fig:lax_3rd_diff_beta} shows the results of the third-order hybrid CLS-CWENO schemes using shock detector $\sigma^{\mathrm{Li}}$ with $\theta_c = 0.20$. The density and pressure are captured without oscillations. With the increase of $\kappa_c$, the discontinuities are captured with steeper gradients.

Figure \ref{fig:lax_diff_sd_3rd} compares the performance between the proposed shock detector $\sigma^{\mathrm{Li}}$ and $\sigma^{\mathrm{Ren}}$ for the third-order scheme near the contact and shock discontinuities. With the coefficients optimized by $\kappa_c = 1.0$
and $2.0$, $\sigma^{\mathrm{Li}}$ resolves the discontinuities sharper than $\sigma^{\mathrm{Ren}}$; with coefficients optimized by $\kappa_c = 1.2$, $\sigma^{\mathrm{Ren}}$ performs better than $\sigma^{\mathrm{Li}}$. In conclusion, as to the recognition of discontinuities, the performance of the two shock detectors is very close to each other and $\sigma^{\mathrm{Li}}$ performs slightly better in this case.

Figure \ref{fig:lax_5th_diff_beta} shows the results of the fifth-order hybrid CLS-CWENO schemes using shock detector $\sigma^{\mathrm{Li}}$ with $\theta_c = 0.03$. No oscillation is observed and the coefficients optimized by $\beta=0.4$ resolve the contact discontinuity with steepest gradients.

Figures \ref{fig:lax_diff_sd_5th} compare the performance between the proposed shock detector $\sigma^{\mathrm{Li}}$ and $\sigma^{\mathrm{Ren}}$ for the fifth-order scheme near the contact and shock discontinuities.  For the fifth-order scheme, the performance of $\sigma^{\mathrm{Li}}$ is generally better than $\sigma^{\mathrm{Ren}}$ near discontinuities.

\subsection{Sod shock tube problem}

\begin{figure}[!htbp]
  \centering
    \begin{subfigure}[b]{\columnwidth}
    \includegraphics[width=0.61\columnwidth]{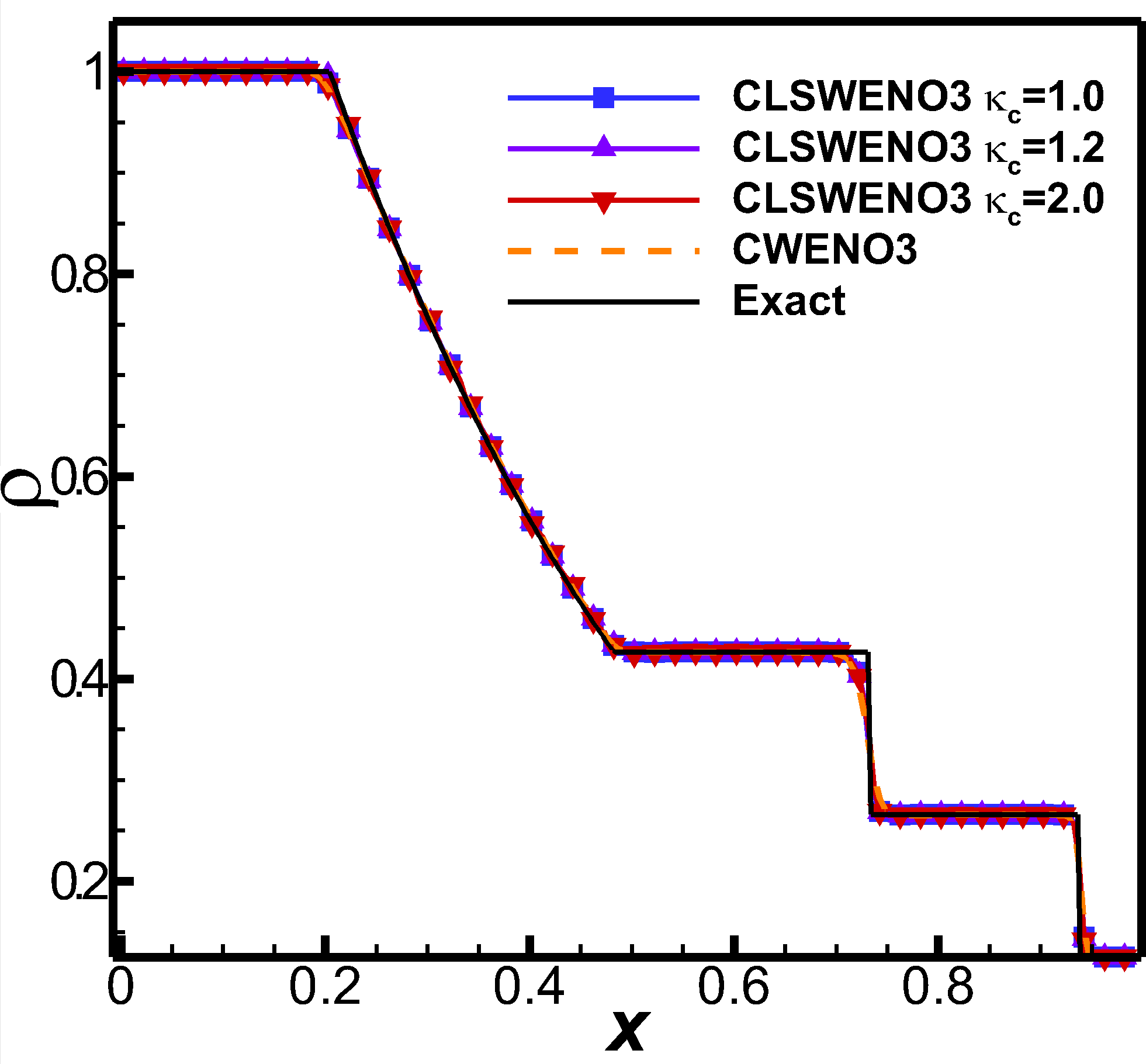}
    \caption{Density.\label{fig:sod_3rd_diff_kappa_density}}
    \end{subfigure}
    \begin{subfigure}[b]{\columnwidth}
    \includegraphics[width=0.61\columnwidth]{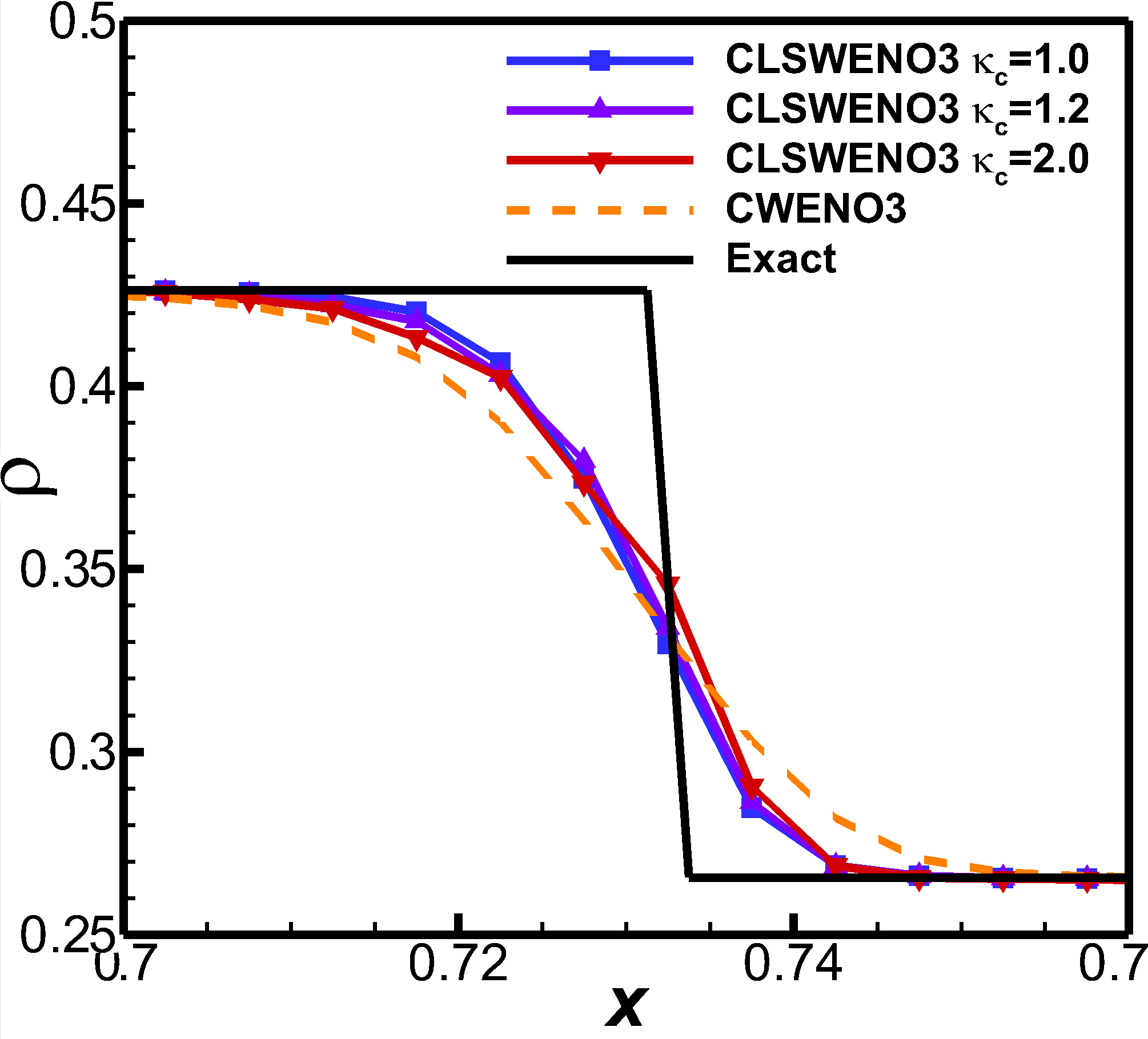}
    \caption{Close view of density.\label{fig:sod_3rd_diff_kappa_density_enlarged}}
    \end{subfigure}
    \begin{subfigure}[b]{\columnwidth}
    \includegraphics[width=0.61\columnwidth]{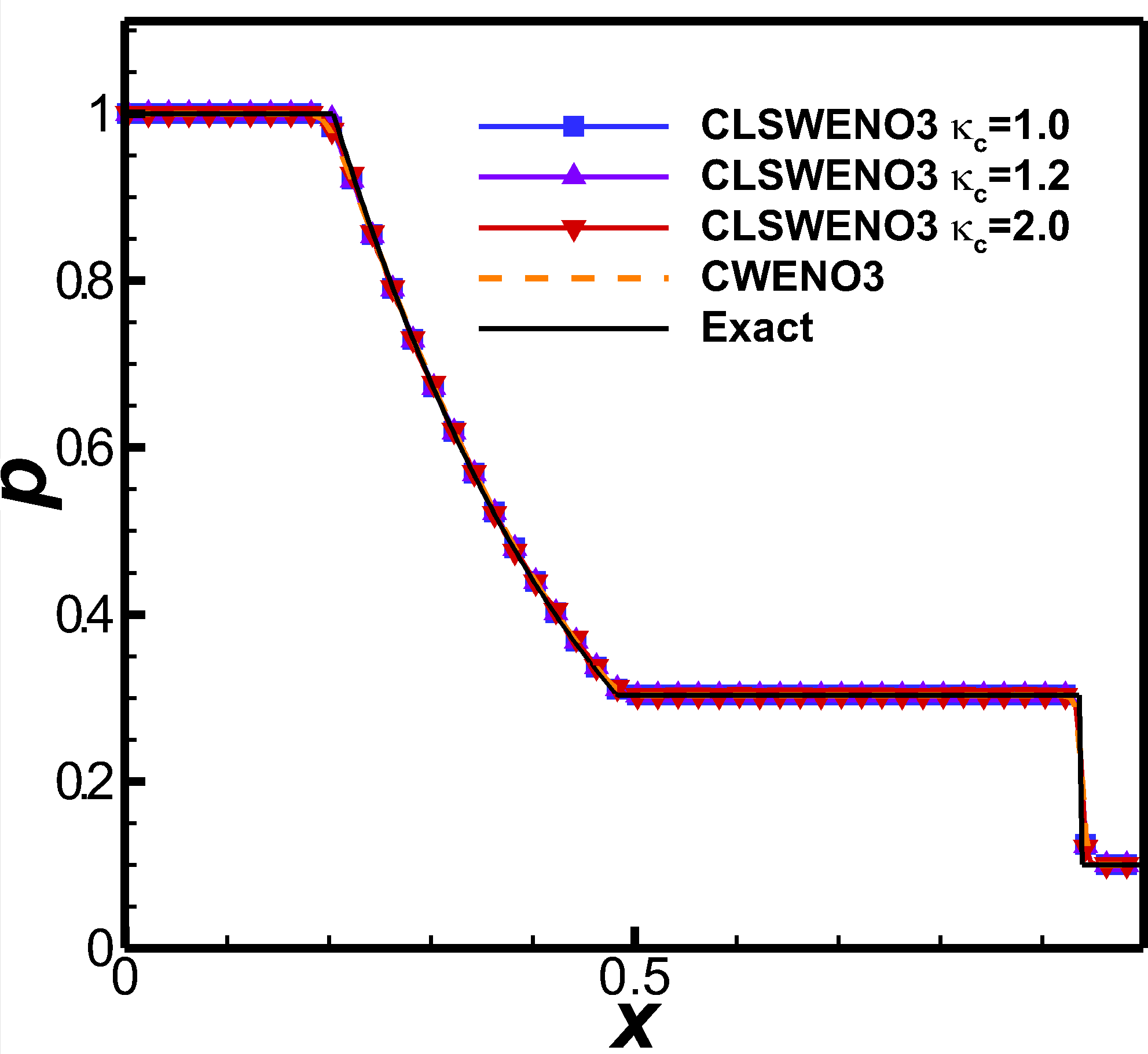}
    \caption{Pressure.\label{fig:sod_3rd_diff_kappa_pressure}}
    \end{subfigure}
    \caption{\label{fig:sod_3rd} Sod shock tube problem with third-order hybrid CLS-CWENO scheme.}
\end{figure}

\begin{figure}[!htbp]
  \centering
    \begin{subfigure}[b]{\columnwidth}
    \includegraphics[width=0.61\columnwidth]{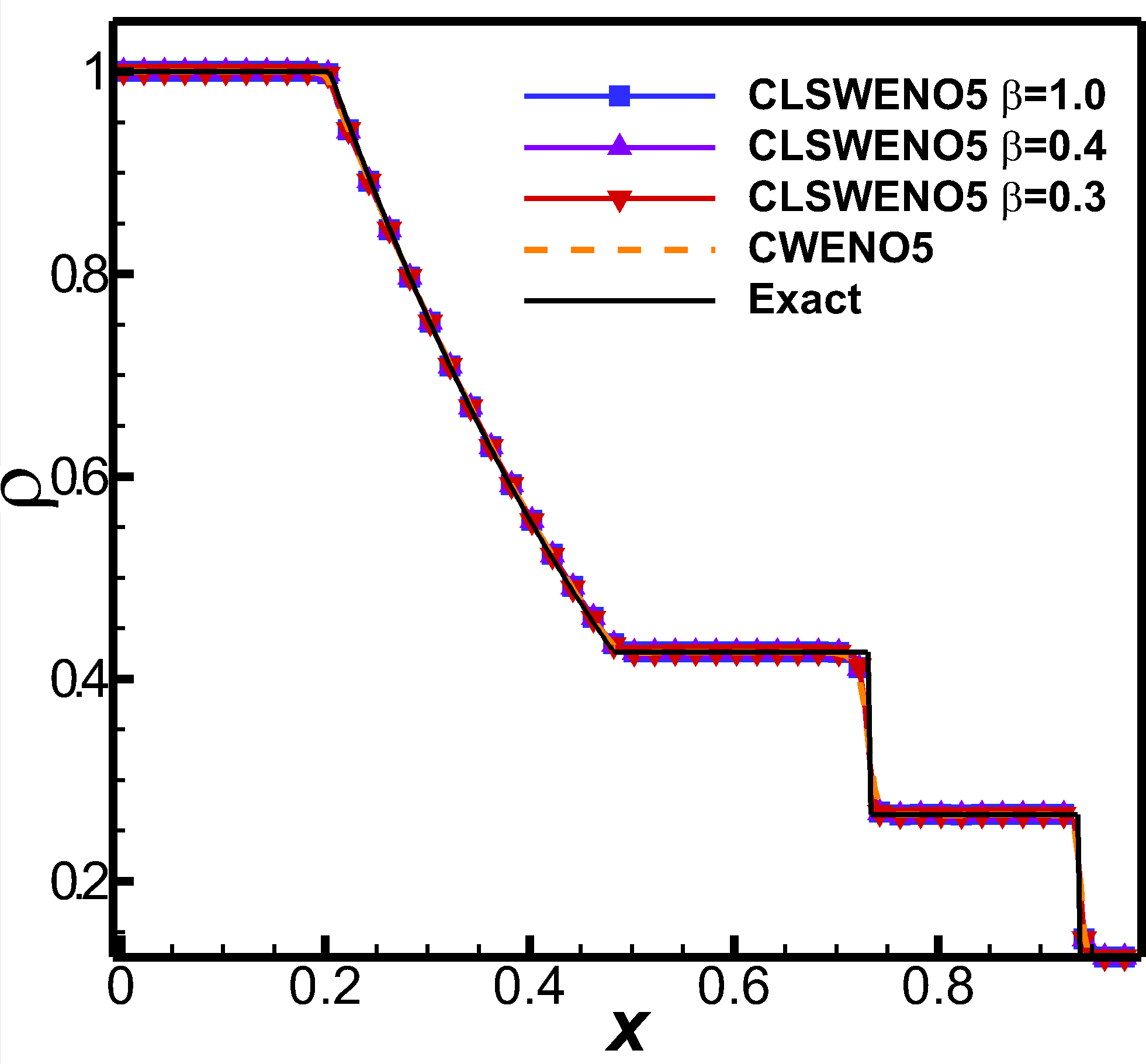}
    \caption{Density.\label{fig:sod_5th_diff_kappa_density}}
    \end{subfigure}
    \begin{subfigure}[b]{\columnwidth}
    \includegraphics[width=0.61\columnwidth]{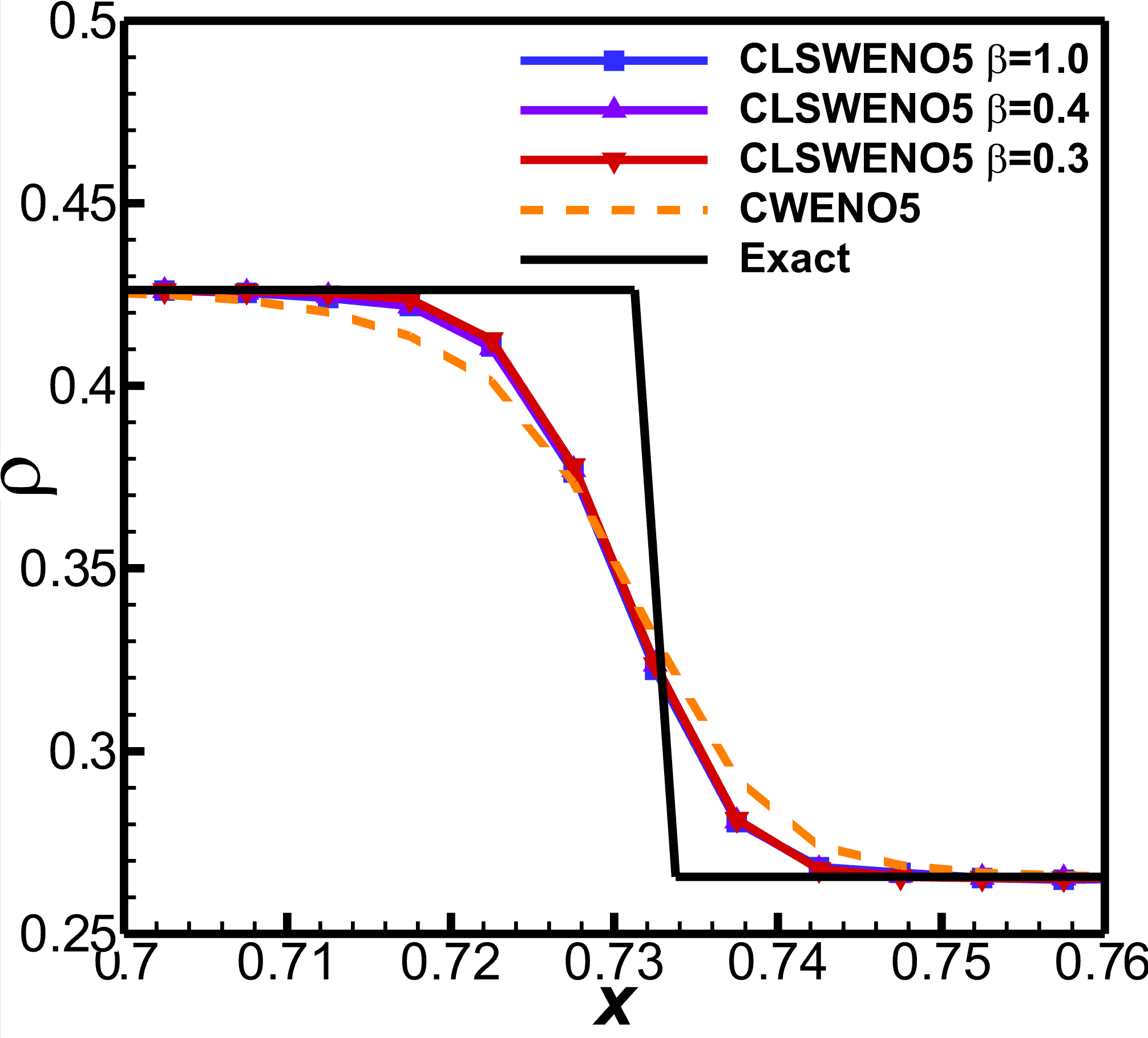}
    \caption{Close view of density.\label{fig:sod_5th_diff_kappa_density_enlarged}}
    \end{subfigure}
    \begin{subfigure}[b]{\columnwidth}
    \includegraphics[width=0.61\columnwidth]{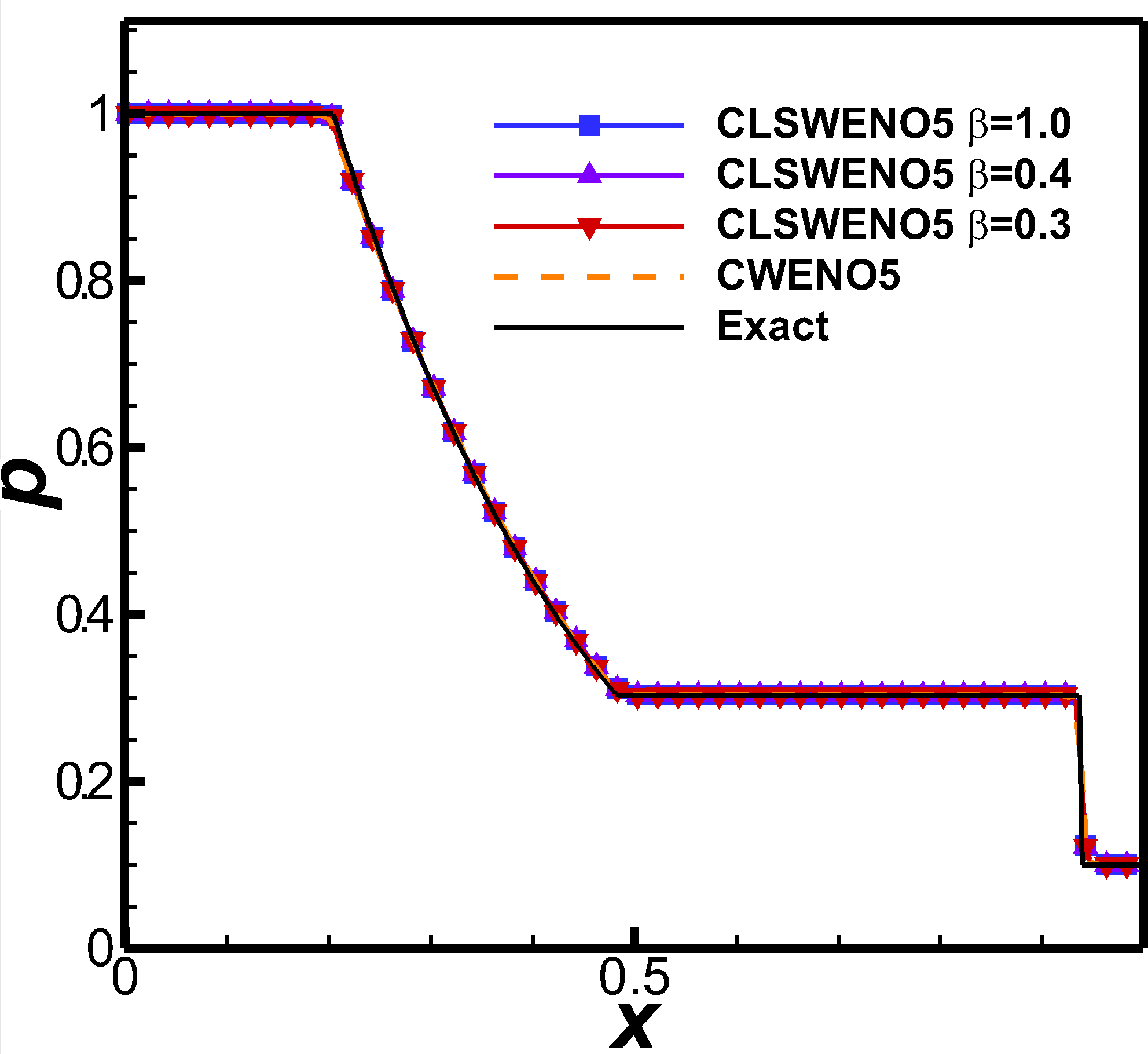}
    \caption{Pressure.\label{fig:sod_5th_diff_kappa_pressure}}
    \end{subfigure}
    \caption{\label{fig:sod_5th} Sod shock tube problem with fifth-order hybrid CLS-CWENO scheme.}
\end{figure}

\begin{figure}[!htbp]
  \centering
    \begin{subfigure}[b]{\columnwidth}
    \includegraphics[width=0.61\columnwidth]{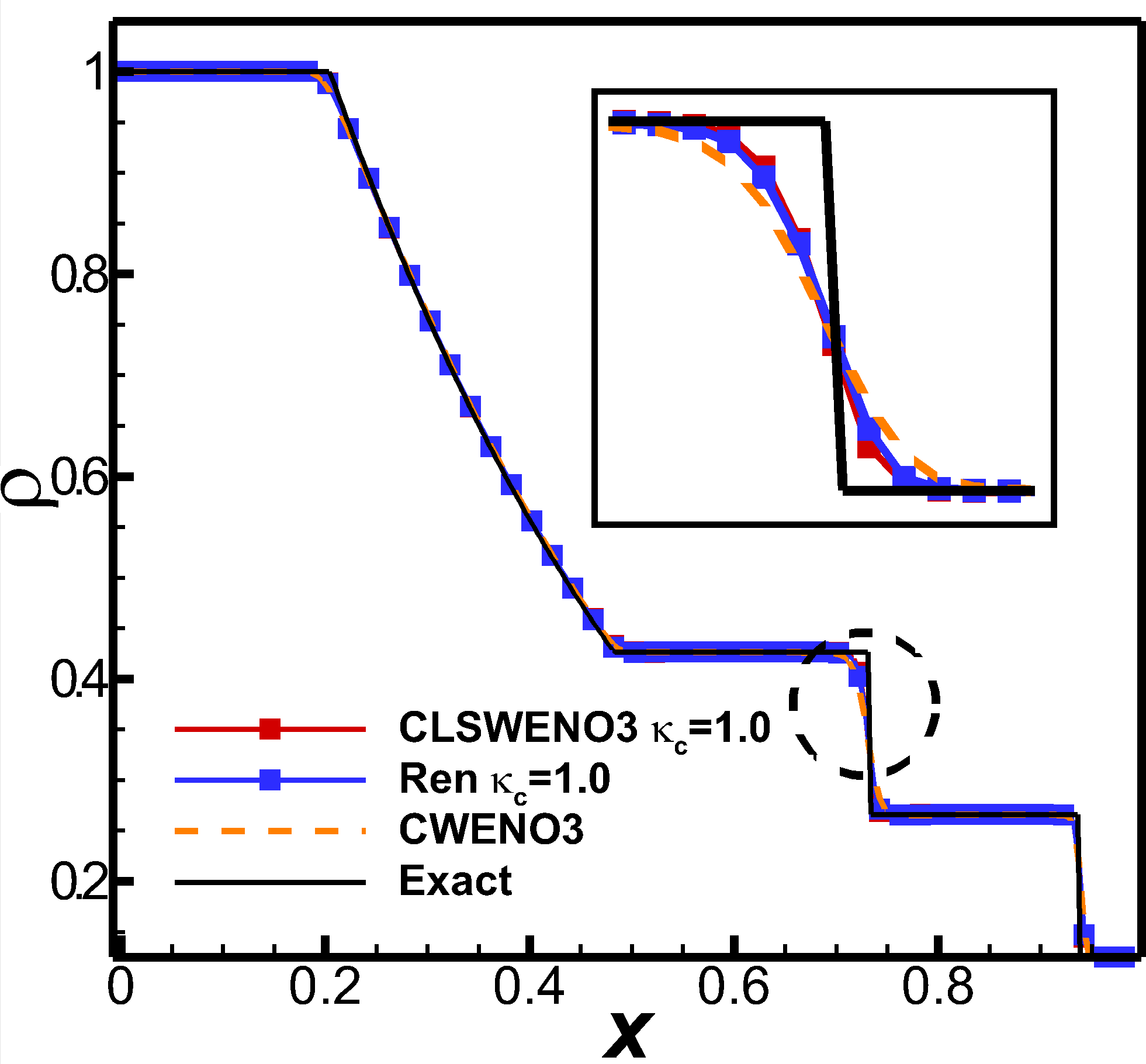}
    \caption{With coefficients optimized by $\kappa = 1.0$.\label{fig:sod_3rd_compare_with_ren_1.0}}
    \end{subfigure}
    \begin{subfigure}[b]{\columnwidth}
    \includegraphics[width=0.61\columnwidth]{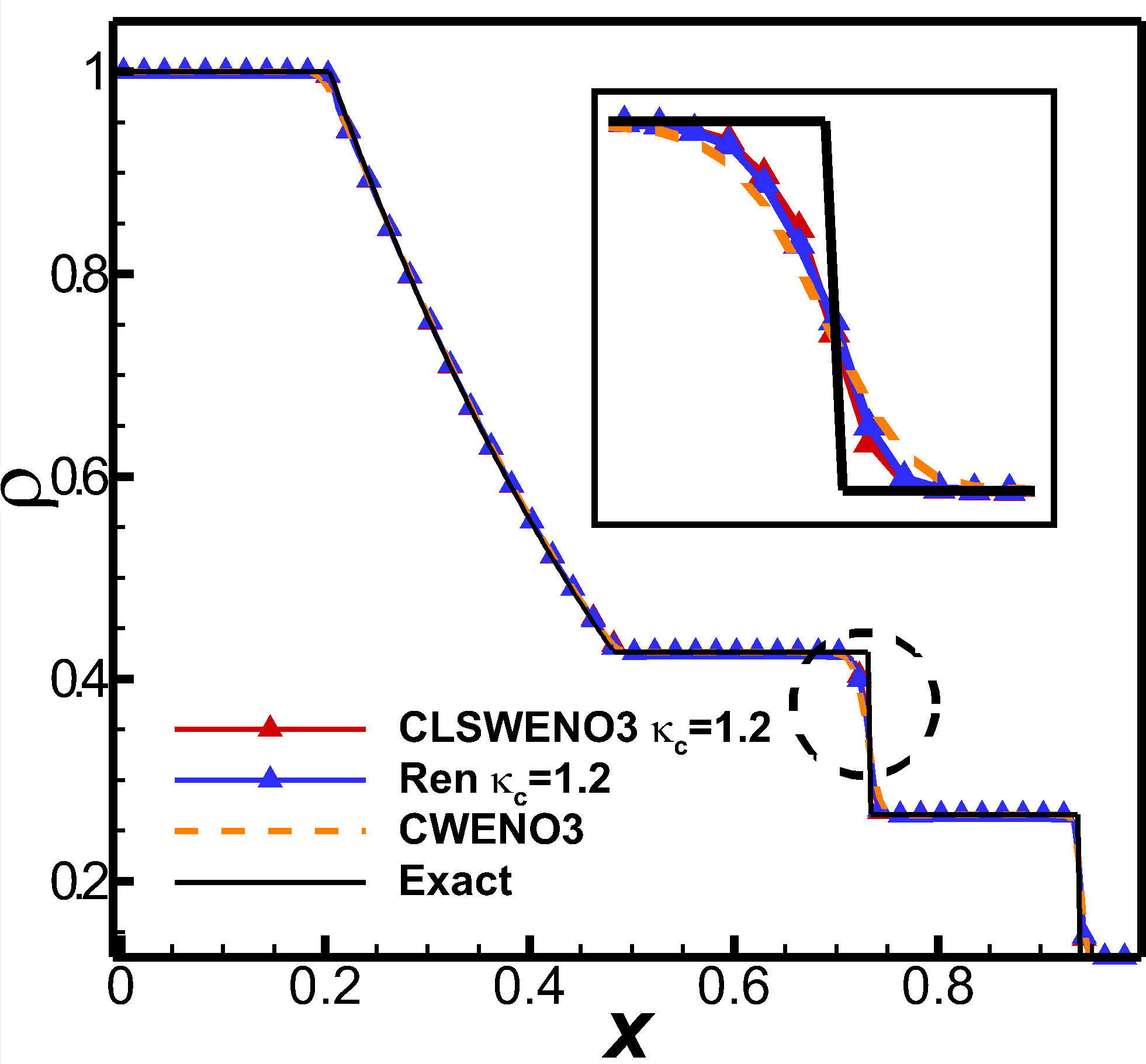}
    \caption{With coefficients optimized by $\kappa = 1.2$.\label{fig:sod_3rd_compare_with_ren_1.2}}
    \end{subfigure}
    \begin{subfigure}[b]{\columnwidth}
    \includegraphics[width=0.61\columnwidth]{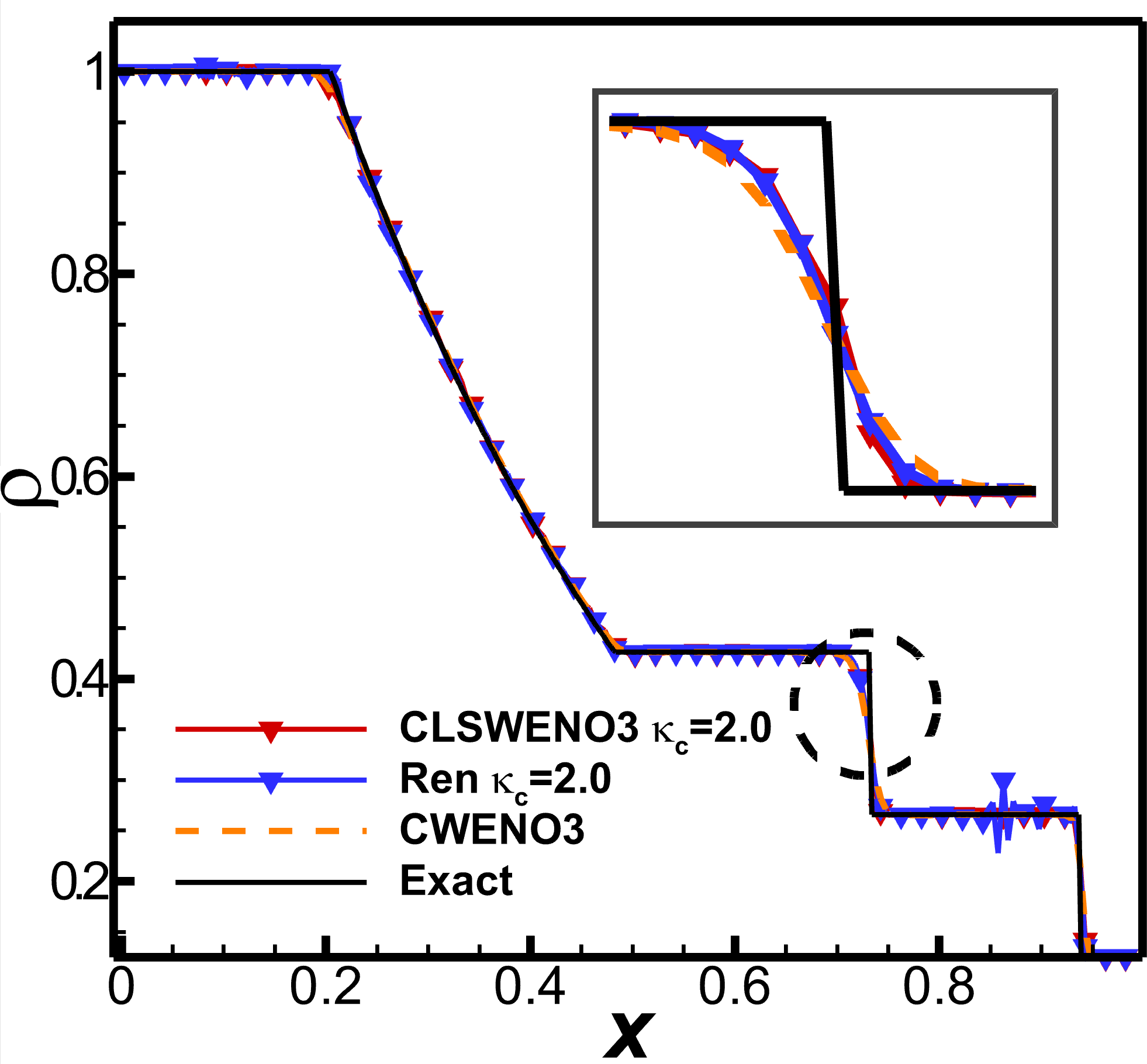}
    \caption{With coefficients optimized by $\kappa = 2.0$.\label{fig:sod_3rd_compare_with_ren_2.0}}
    \end{subfigure}
    \caption{\label{fig:sod_diff_sd_3rd} Comparison of the shock detector for the Sod shock tube problem. Third-order hybrid CLS-CWENO scheme. $\theta_c=0.20$ for $\sigma^{\mathrm{Li}}$ and $\theta_c=0.50$ for $\sigma^{\mathrm{Ren}}$.}
\end{figure}

\begin{figure}[!htbp]
  \centering
    \begin{subfigure}[b]{\columnwidth}
    \includegraphics[width=0.61\columnwidth]{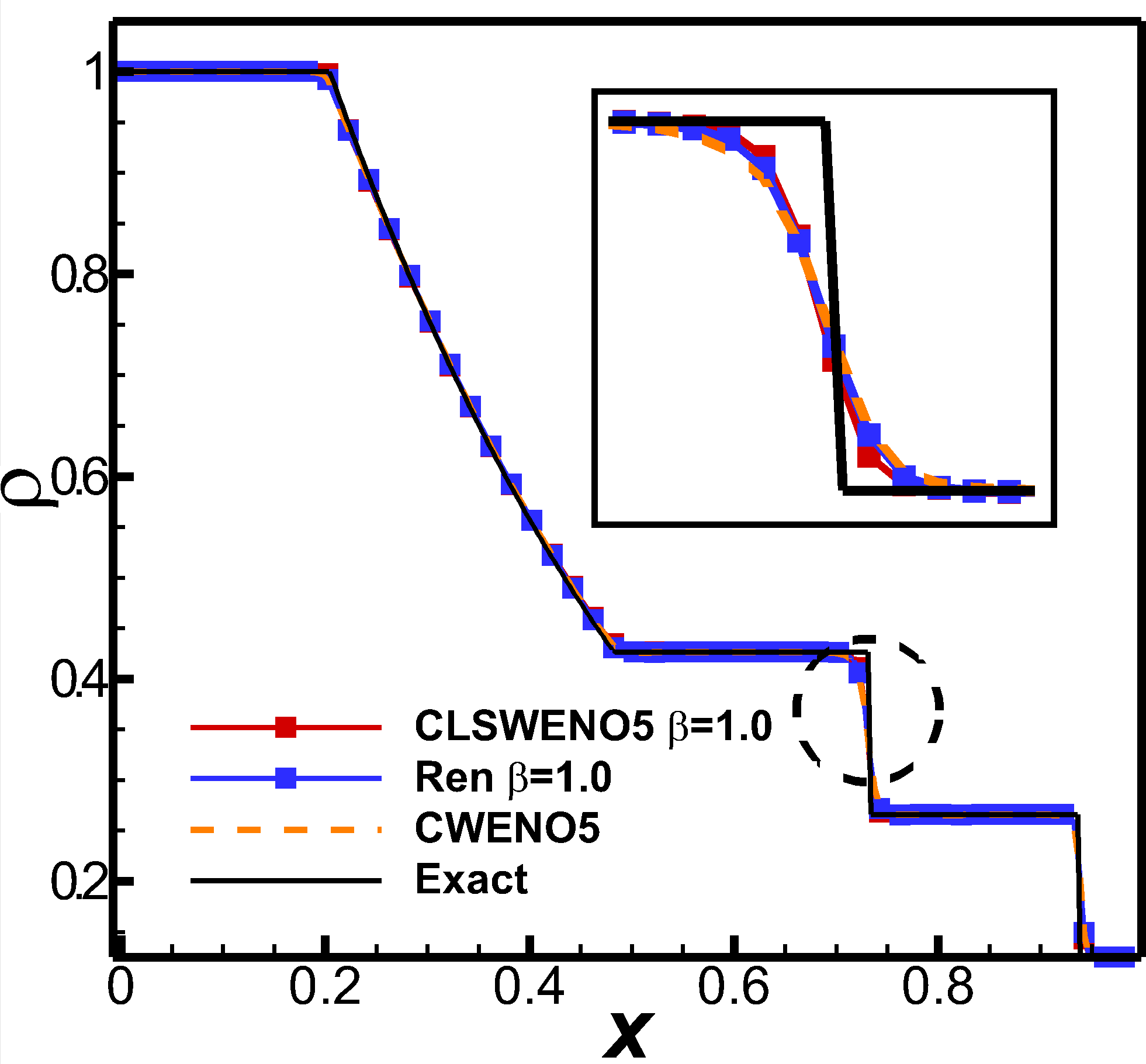}
    \caption{With coefficients optimized by $\beta = 1.0$.\label{fig:sod_5th_compare_with_ren_1.0}}
    \end{subfigure}
    \begin{subfigure}[b]{\columnwidth}
    \includegraphics[width=0.61\columnwidth]{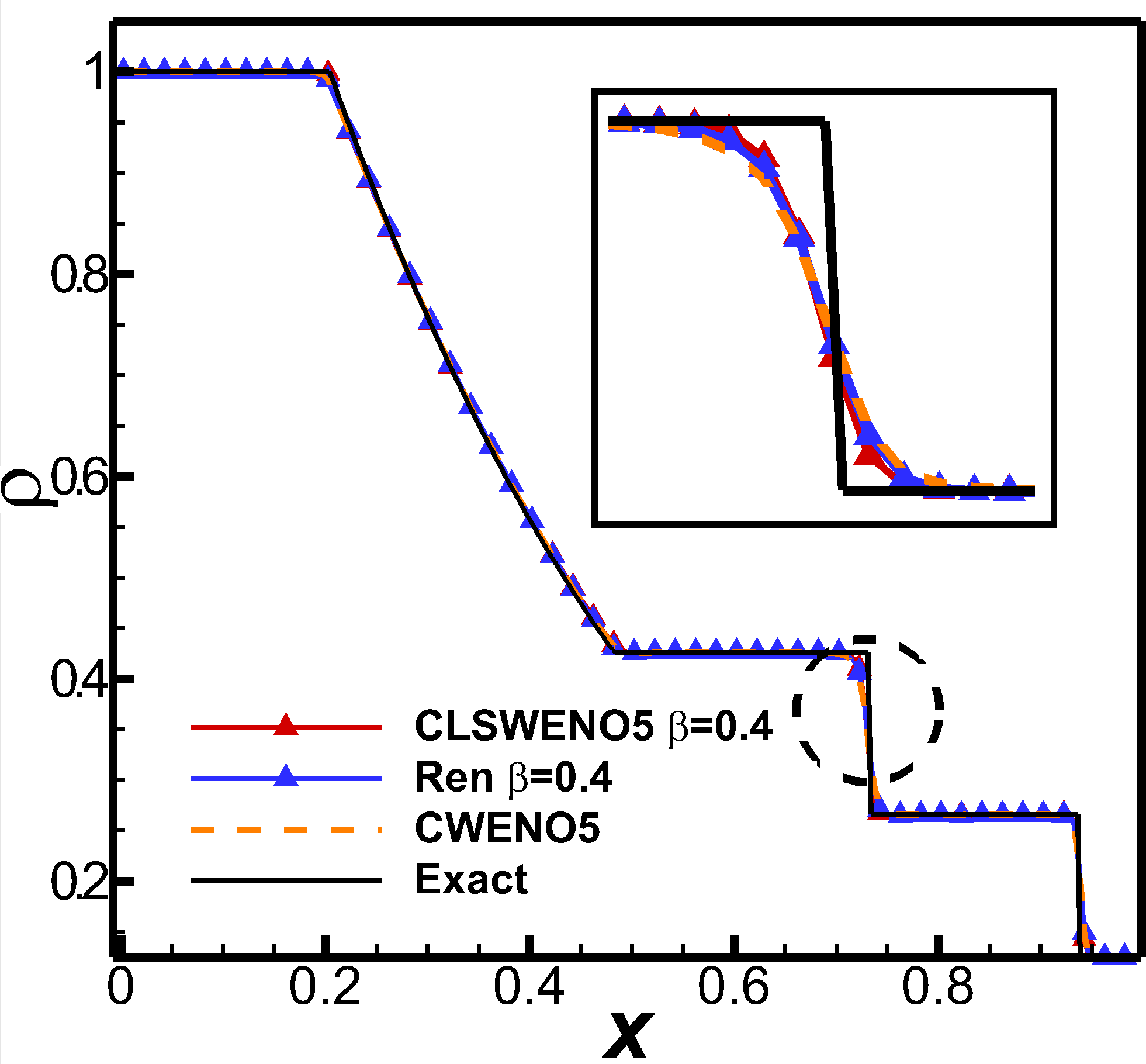}
    \caption{With coefficients optimized by $\beta = 0.4$.\label{fig:sod_5th_compare_with_ren_1.2}}
    \end{subfigure}
    \begin{subfigure}[b]{\columnwidth}
    \includegraphics[width=0.61\columnwidth]{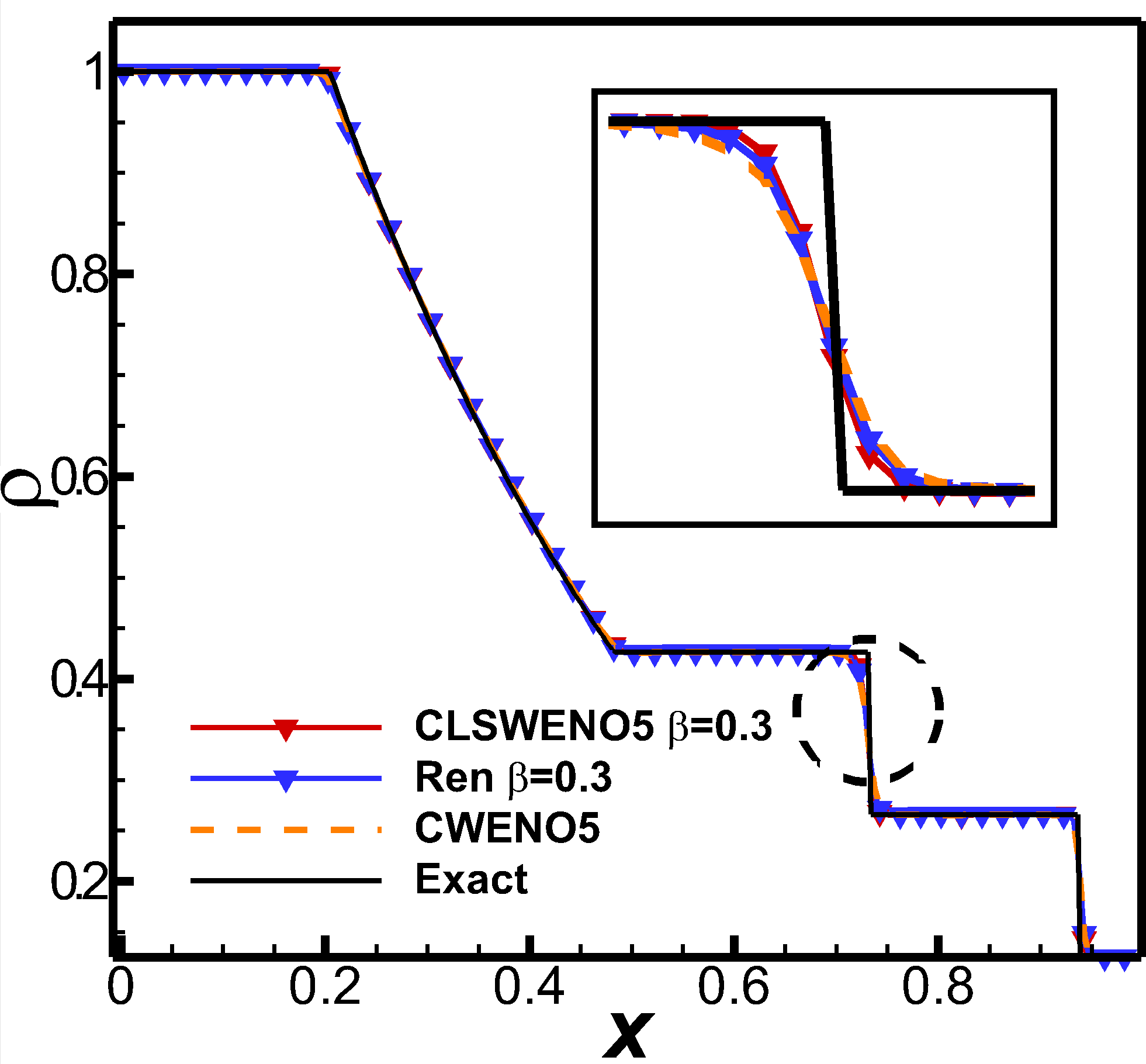}
    \caption{With coefficients optimized by $\beta = 0.3$.\label{fig:sod_5th_compare_with_ren_2.0}}
    \end{subfigure}
    \caption{\label{fig:sod_diff_sd_5th} Comparison of the shock detector for the Sod shock tube problem. Fifth-order hybrid CLS-CWENO scheme. $\theta_c=0.03$ for $\sigma^{\mathrm{Li}}$ and $\theta_c=0.50$ for $\sigma^{\mathrm{Ren}}$.}
\end{figure}

The Sod shock tube problem is another popular test case in validating the performance of shock-capturing schemes. The initial condition is
\begin{equation}
  \left[\rho, u, p\right] = \left\{
  \begin{array}{ll}
    1, 0, 1, & \text{if}\,\,0\leq x\le 0.5,\\
    0.125, 0, 0.1, & \text{if}\,\,0.5 \leq x\leq 1.0.\\
  \end{array}
  \right.
\end{equation}
The computational domain of $\Omega = [0,1]$ consists of 200 uniform cells. The final simulation time is $t_{end} = 0.25$ and the CFL number is 0.5 for the simulation.
Figures \ref{fig:sod_3rd} and \ref{fig:sod_5th} show the results of the hybrid CLS-CWENO schemes using the proposed shock detector $\sigma^{\mathrm{Li}}$.

The profile of density and pressure confirms the oscillation-free shock capturing capability of the proposed method.
Figures \ref{fig:sod_diff_sd_3rd} and \ref{fig:sod_diff_sd_5th} compare the performance of the two different shock detectors. It can be concluded that the hybrid scheme coupled with $\sigma^{\mathrm{Li}}$ captures discontinuity with higher resolution than the scheme coupled with $\sigma^{\mathrm{Ren}}$. In addition, as shown in Fig. \ref{fig:sod_3rd_compare_with_ren_2.0}, oscillation is triggered for the third-order hybrid CLS-CWENO scheme utilizing $\sigma^{\mathrm{Ren}}$ when the dissipation is too small with coefficients optimized by $\kappa_c = 2.0$.
However, the hybrid CLS-CWENO scheme coupled with $\sigma^{\mathrm{Li}}$ can still simulate the problem without oscillations with the same coefficients.

\subsection{Shu-Osher problem}
The initial condition for the Shu-Osher problem\cite{shu1989efficient} is
\begin{equation}
  \left[\rho, u, p\right] = \left\{
  \begin{array}{ll}
    3.857143, 2.629369, 10.333333, & \text{if}\,\,0\leq x\le 1,\\
    1+0.2\sin(5x), 0, 1, & \text{if}\,\,1 \leq x\leq 10.\\
  \end{array}
  \right.
\end{equation}
The computational domain is $[0,10]$ discretized by 200 uniform control volumes. The simulation end time is $t_{end}=1.8$ with Courant number of 0.5.

\begin{figure}[!htbp]
  \centering
    \begin{subfigure}[b]{\columnwidth}
    \includegraphics[width=0.61\columnwidth]{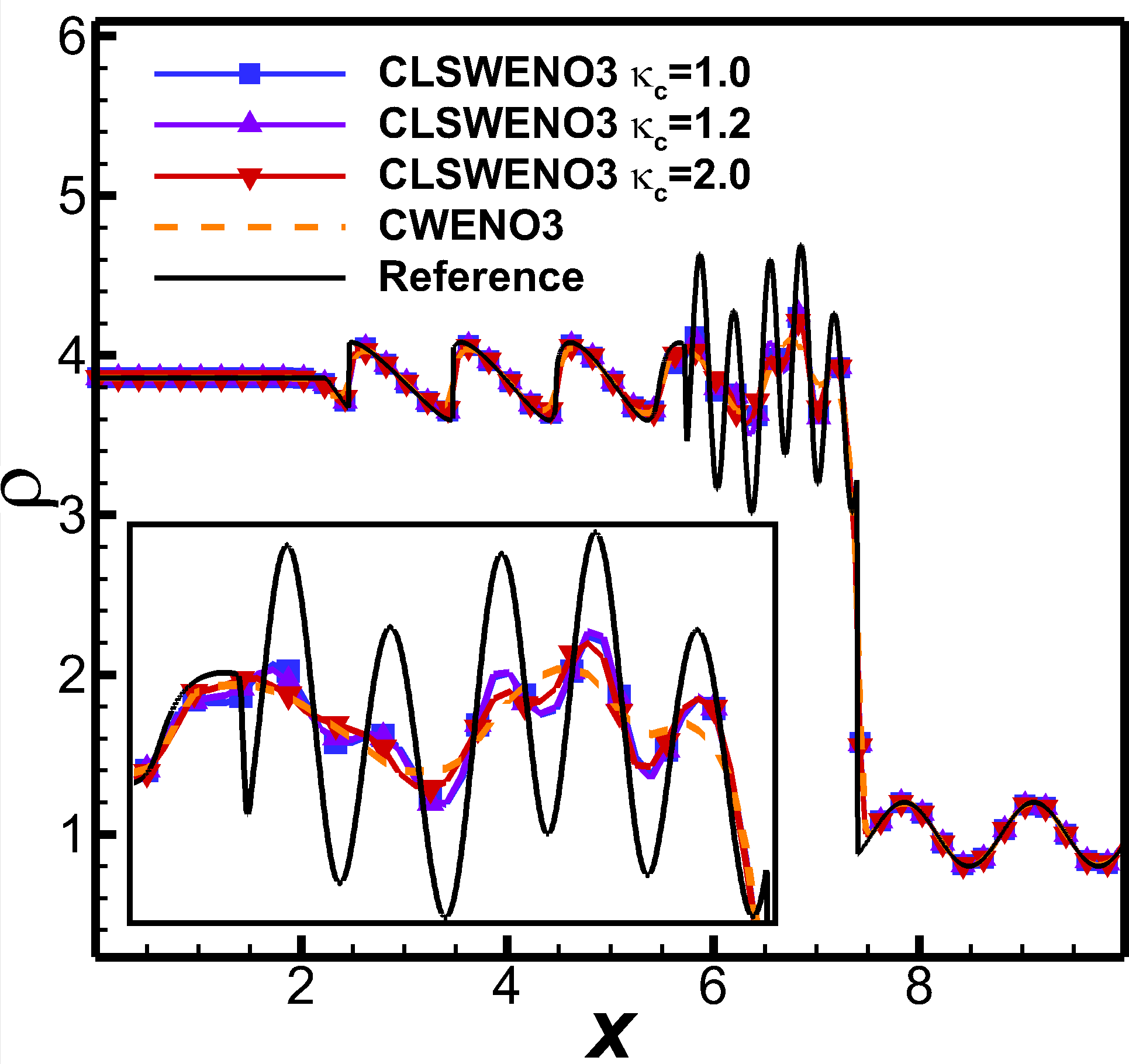}
    \caption{Third-order scheme.\label{fig:so_3rd}}
    \end{subfigure}
    \begin{subfigure}[b]{\columnwidth}
    \includegraphics[width=0.61\columnwidth]{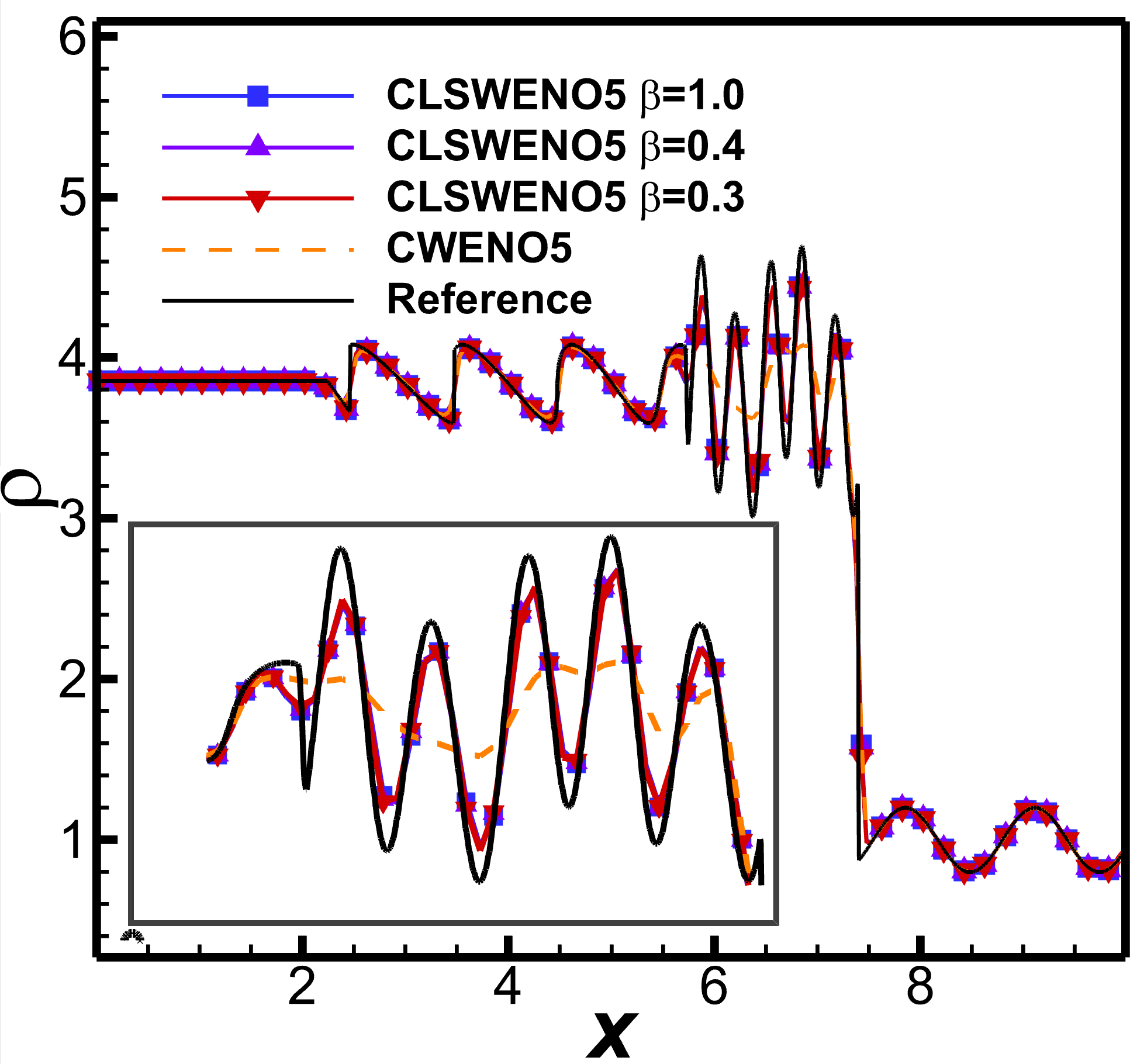}
    \caption{Fifth-order scheme.\label{fig:so_5th}}
    \end{subfigure}
    \caption{\label{fig:so_density} Density distribution for the Shu-Osher problem.}
\end{figure}

The right moving shock interacts with a disturbance of density and generates multiscale flow structures.
The density distribution of the Shu-Osher problem with third- and fifth-order hybrid schemes utilizing shock detector $\sigma^{\mathrm{Li}}$ are shown in Fig. \ref{fig:so_density}, where the reference result is simulated by the fifth-order CWENO scheme using 4000 cells with third-order SSP-RK method. For the third-order scheme, the coefficients optimized by $\kappa_c = 1.2$ perform best. The flow structures with coefficients optimized by $\kappa_c = 2.0$ are dissipated more than the results with coefficients optimized by $\kappa_c = 1.2$ even though the dispersion error of the coefficient optimized by $\kappa_c = 2.0$ is much smaller than the coefficients optimized by $\kappa_c = 1.2$. This phenomenon confirms the necessity of enough dissipation in high-wave-number region. For the fifth-order scheme, all the three sets of coefficients can resolve the multiscale structures and capture the discontinuities robustly.

\begin{figure}[!htbp]
  \centering
    \begin{subfigure}[b]{\columnwidth}
    \includegraphics[width=0.61\columnwidth]{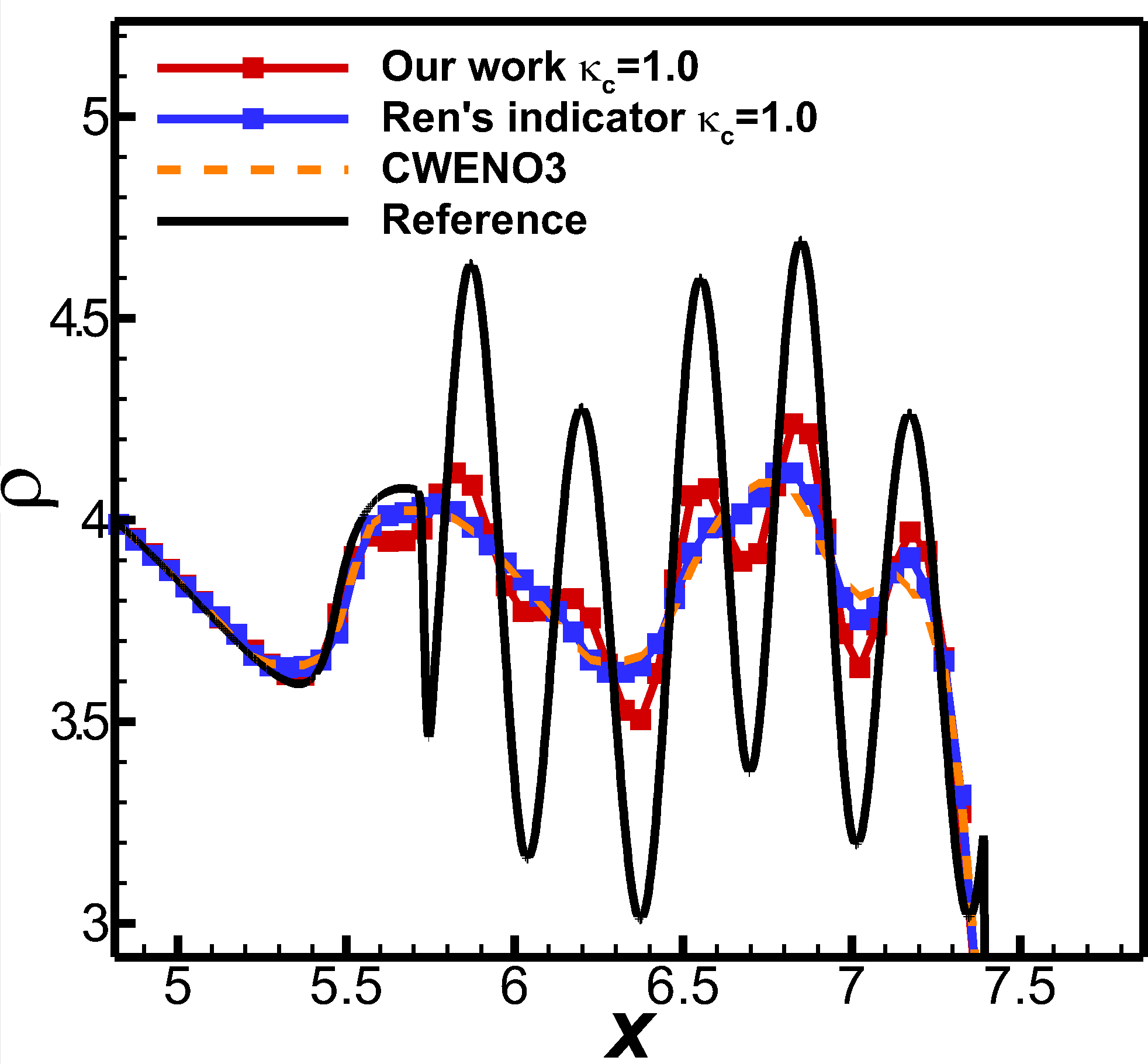}
    \caption{With coefficients optimized by $\kappa = 1.0$.\label{fig:shuosher_3rd_compare_with_ren_1.0}}
    \end{subfigure}
    \begin{subfigure}[b]{\columnwidth}
    \includegraphics[width=0.61\columnwidth]{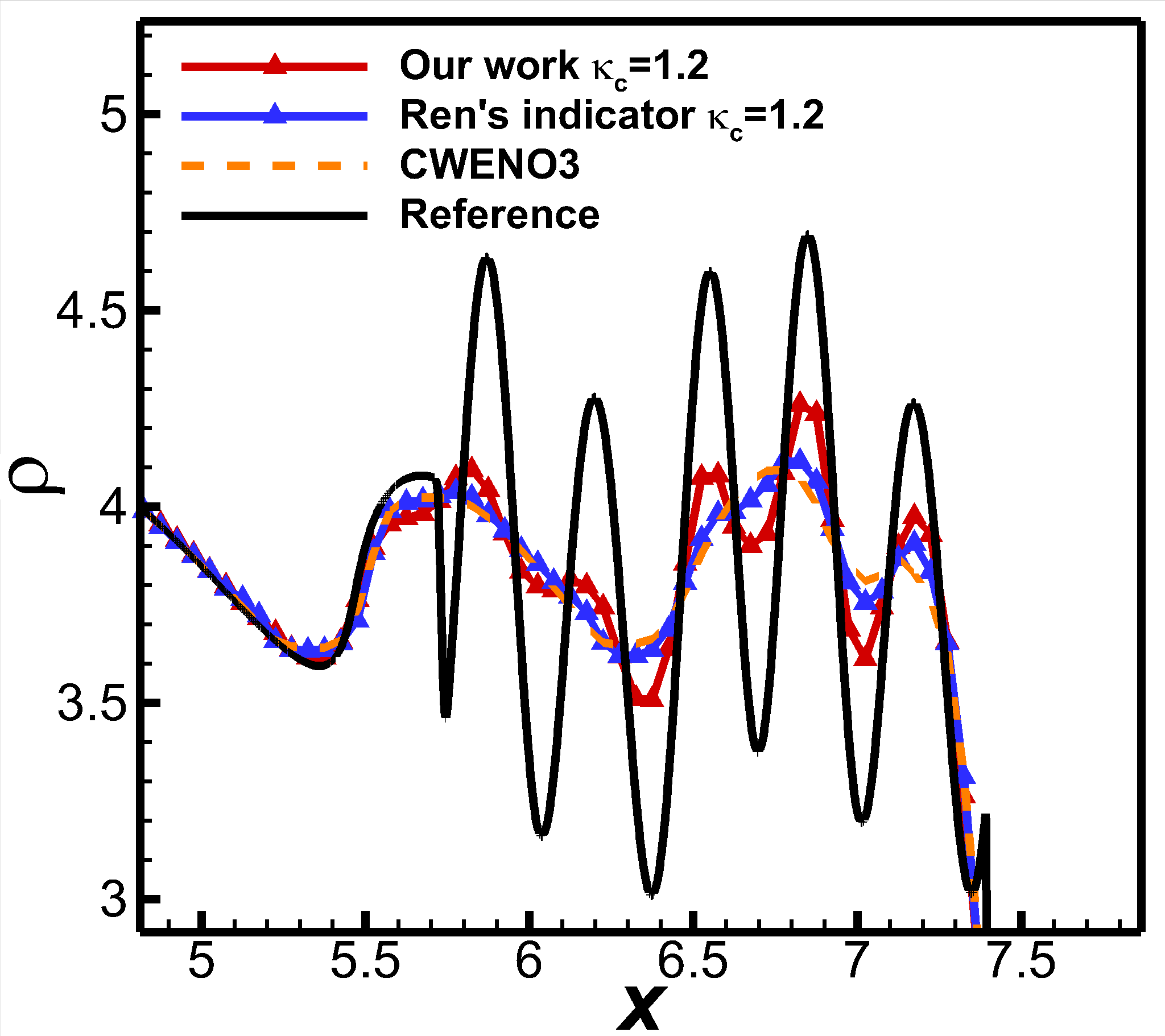}
    \caption{With coefficients optimized by $\kappa = 1.2$.\label{fig:shuosher_3rd_compare_with_ren_1.2}}
    \end{subfigure}
    \begin{subfigure}[b]{\columnwidth}
    \includegraphics[width=0.61\columnwidth]{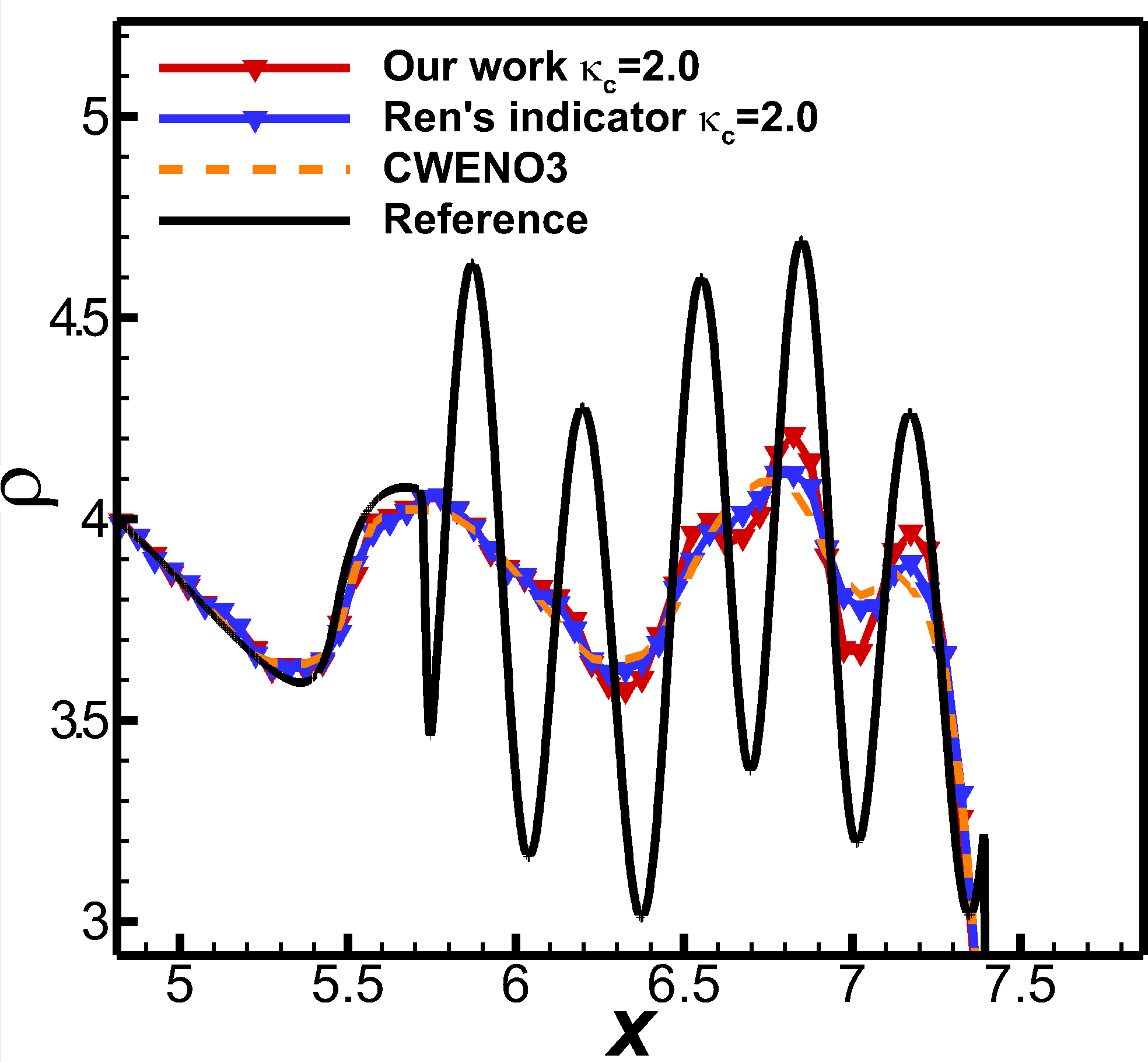}
    \caption{With coefficients optimized by $\kappa = 2.0$.\label{fig:shuosher_3rd_compare_with_ren_2.0}}
    \end{subfigure}
    \caption{\label{fig:shuosher_diff_sd_3rd} Comparison of the shock detector for the Shu-Osher shock tube problem. Third-order hybrid CLS-CWENO scheme. $\theta_c=0.20$ for $\sigma^{\mathrm{Li}}$ and $\theta_c=0.50$ for $\sigma^{\mathrm{Ren}}$.}
\end{figure}

\begin{figure}[!htbp]
  \centering
    \begin{subfigure}[b]{\columnwidth}
    \includegraphics[width=0.61\columnwidth]{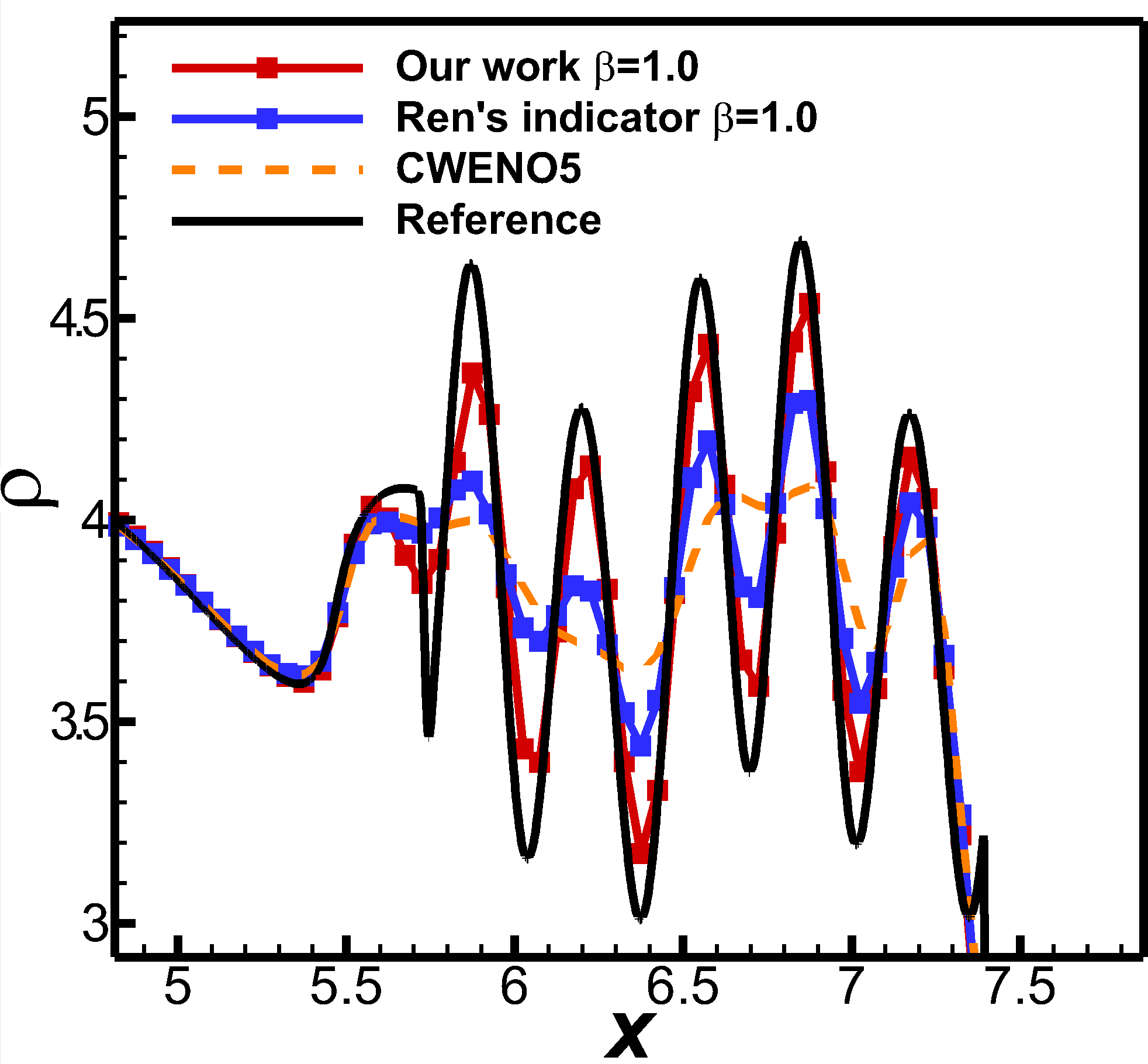}
    \caption{With coefficients optimized by $\beta = 1.0$.\label{fig:shuosher_5th_compare_with_ren_1.0}}
    \end{subfigure}
    \begin{subfigure}[b]{\columnwidth}
    \includegraphics[width=0.61\columnwidth]{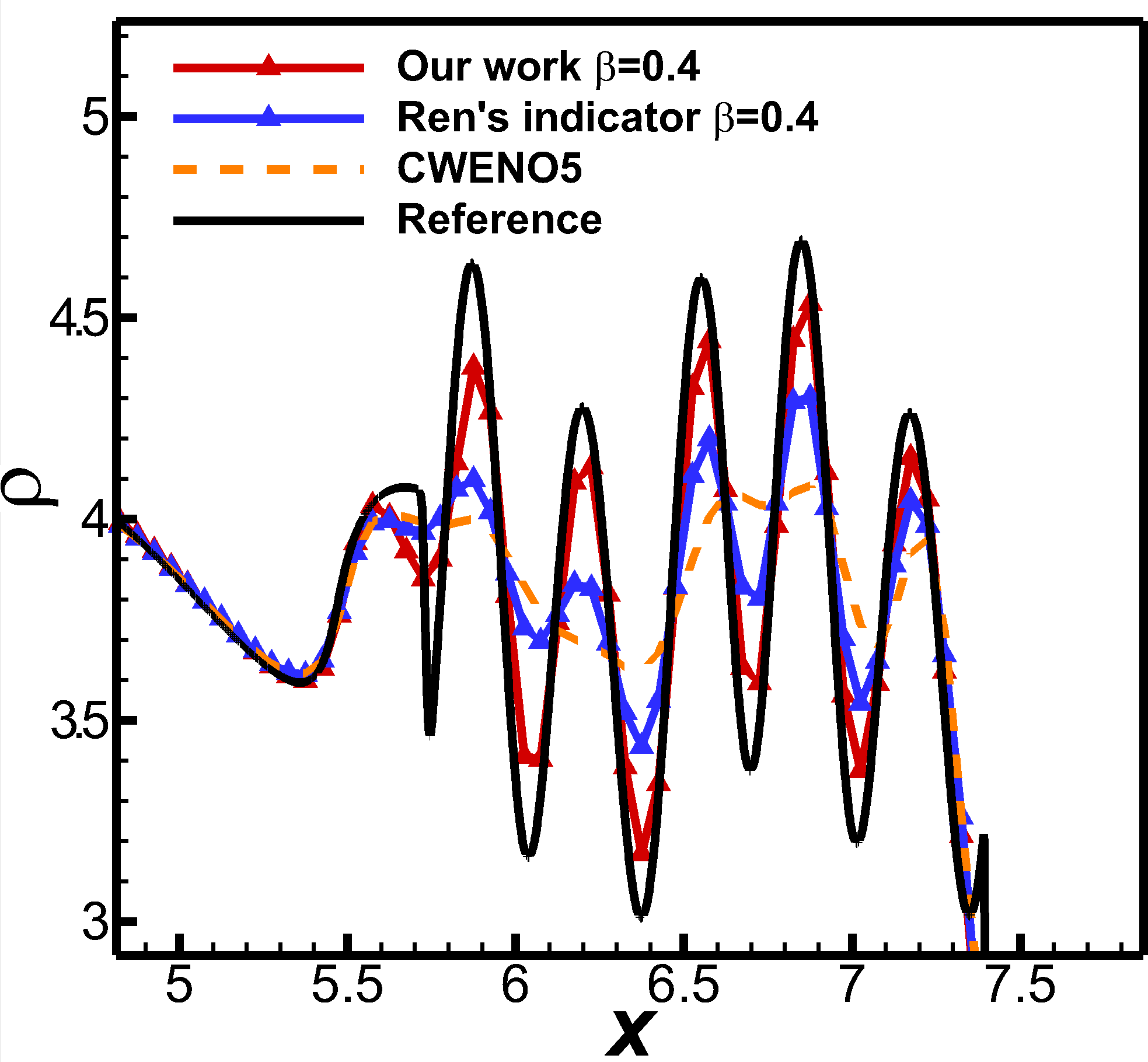}
    \caption{With coefficients optimized by $\beta = 0.4$.\label{fig:shuosher_5th_compare_with_ren_1.2}}
    \end{subfigure}
    \begin{subfigure}[b]{\columnwidth}
    \includegraphics[width=0.61\columnwidth]{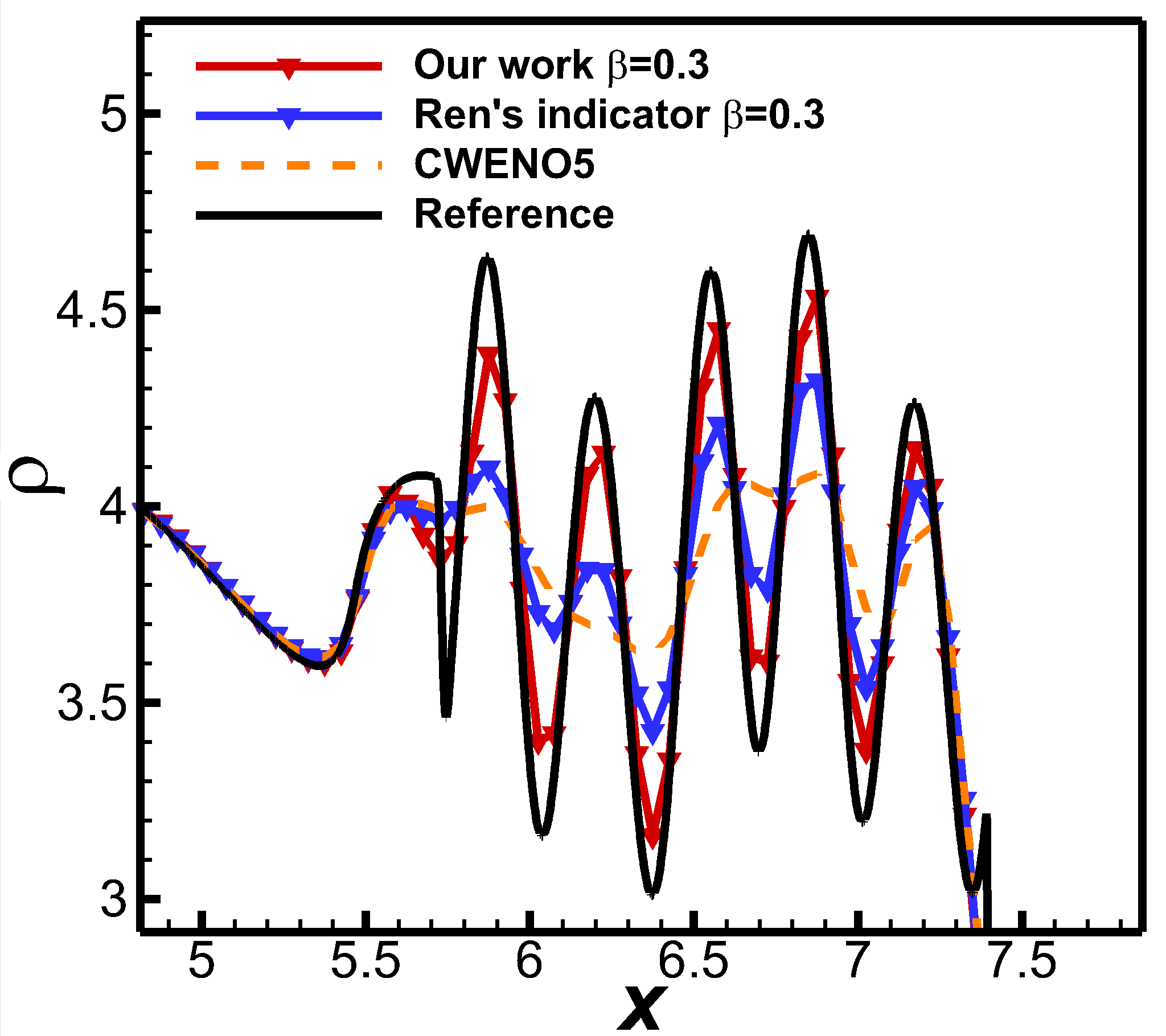}
    \caption{With coefficients optimized by $\beta = 0.3$.\label{fig:shuosher_5th_compare_with_ren_2.0}}
    \end{subfigure}
    \caption{\label{fig:shuosher_diff_sd_5th} Comparison of the shock detector for the Shu-Osher shock tube problem. Fifth-order hybrid CLS-CWENO scheme. $\theta_c=0.03$ for $\sigma^{\mathrm{Li}}$ and $\theta_c=0.50$ for $\sigma^{\mathrm{Ren}}$.}
\end{figure}

Figures \ref{fig:shuosher_diff_sd_3rd} and \ref{fig:shuosher_diff_sd_5th} compare the results of the hybrid schemes with two different shock detectors. As shown in Fig. \ref{fig:shuosher_diff_sd_3rd}, the third-order hybrid CLS-CWENO scheme with proposed shock detector $\sigma^{\mathrm{Li}}$ is successful in identifying the local extrema, while the resolution of hybrid CLS-CWENO scheme coupled with $\sigma^{\mathrm{Ren}}$ is almost identical to the third-order CWENO scheme in this case. For the fifth-order scheme, the proposed shock detector is also superior in improving the resolution of multiscale flow structures.

\subsection{Interacting blast waves\cite{woodward_numerical_1984}}
Euler equations with the following initial condition is solved,
\begin{equation}
  \left[\rho, u, p\right] = \left\{
  \begin{array}{ll}
    1, 0, 1000, & \text{if}\,\,0\leq x\le 0.1,\\
    1, 0, 0.01, & \text{if}\,\,0.1\leq x\le 0.9,\\
    1, 0, 100, & \text{if}\,\,0.9\leq x\le 1.\\
  \end{array}
  \right.
\end{equation}
The computational domain is $[0,1]$ with refective boundary conditions on the two ends. The final simulation time is $t_{end} = 0.038 $ with Courant number of 0.5.

\begin{figure}[!htbp]
  \centering
    \begin{subfigure}[b]{\columnwidth}
    \includegraphics[width=0.61\columnwidth]{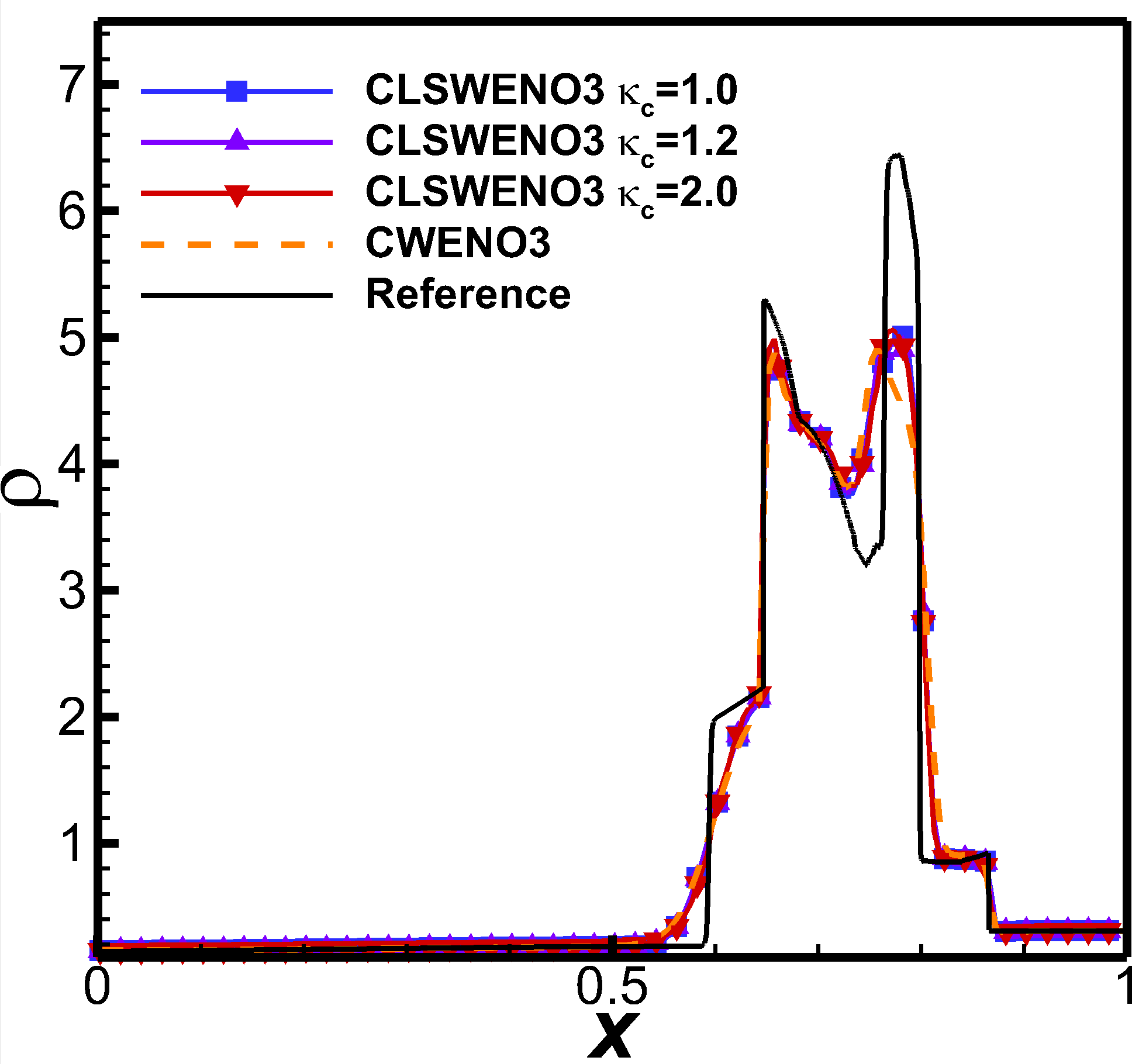}
    \caption{Density. Third-order scheme.\label{fig:ibw_3rd}}
    \end{subfigure}
    \begin{subfigure}[b]{\columnwidth}
    \includegraphics[width=0.61\columnwidth]{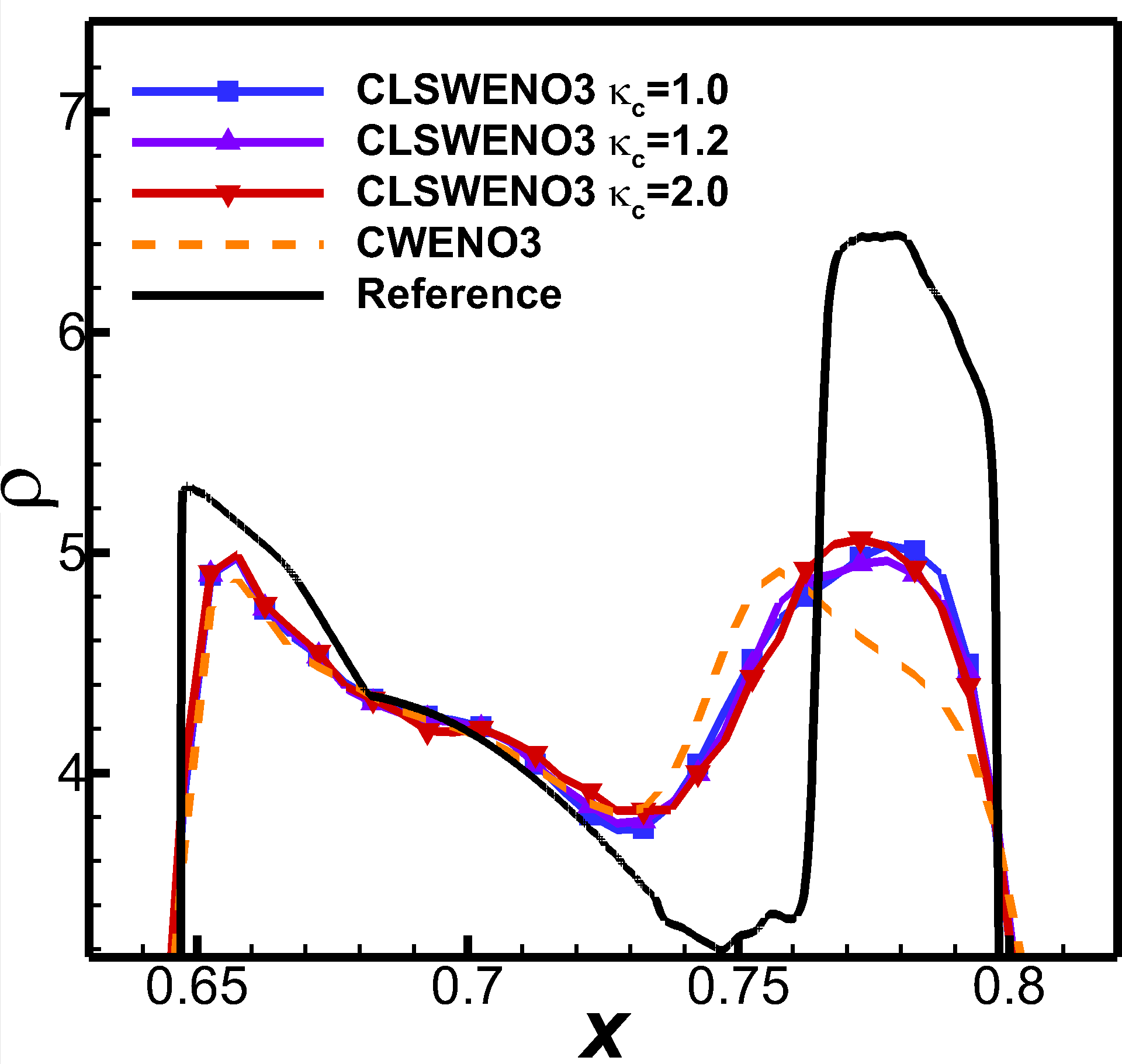}
    \caption{Close view of density. Third-order scheme.\label{fig:ibw_3rd_cv}}
    \end{subfigure}
    \begin{subfigure}[b]{\columnwidth}
    \includegraphics[width=0.61\columnwidth]{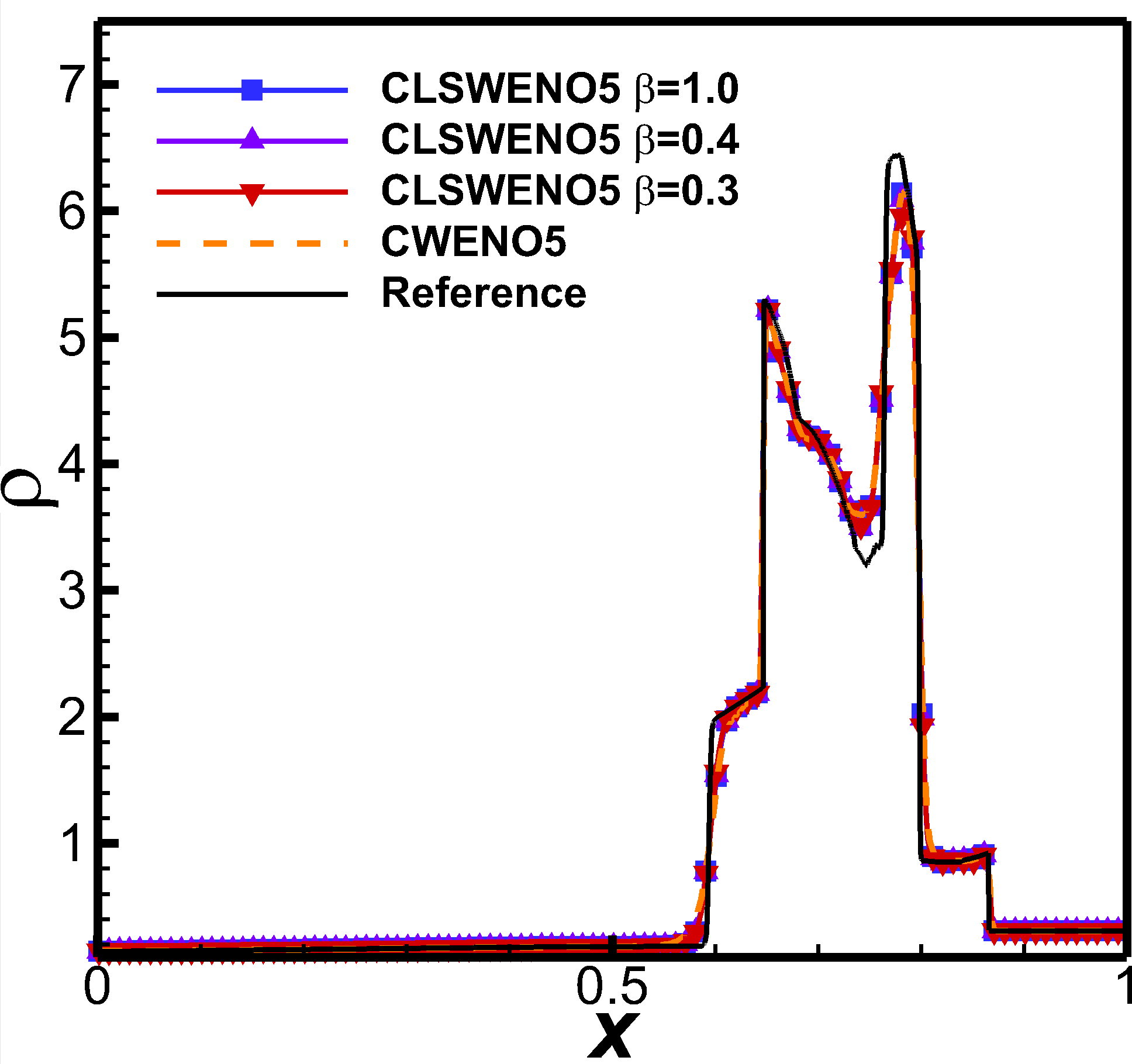}
    \caption{Density. Fifth-order scheme.\label{fig:ibw_5th}}
    \end{subfigure}
    \begin{subfigure}[b]{\columnwidth}
    \includegraphics[width=0.61\columnwidth]{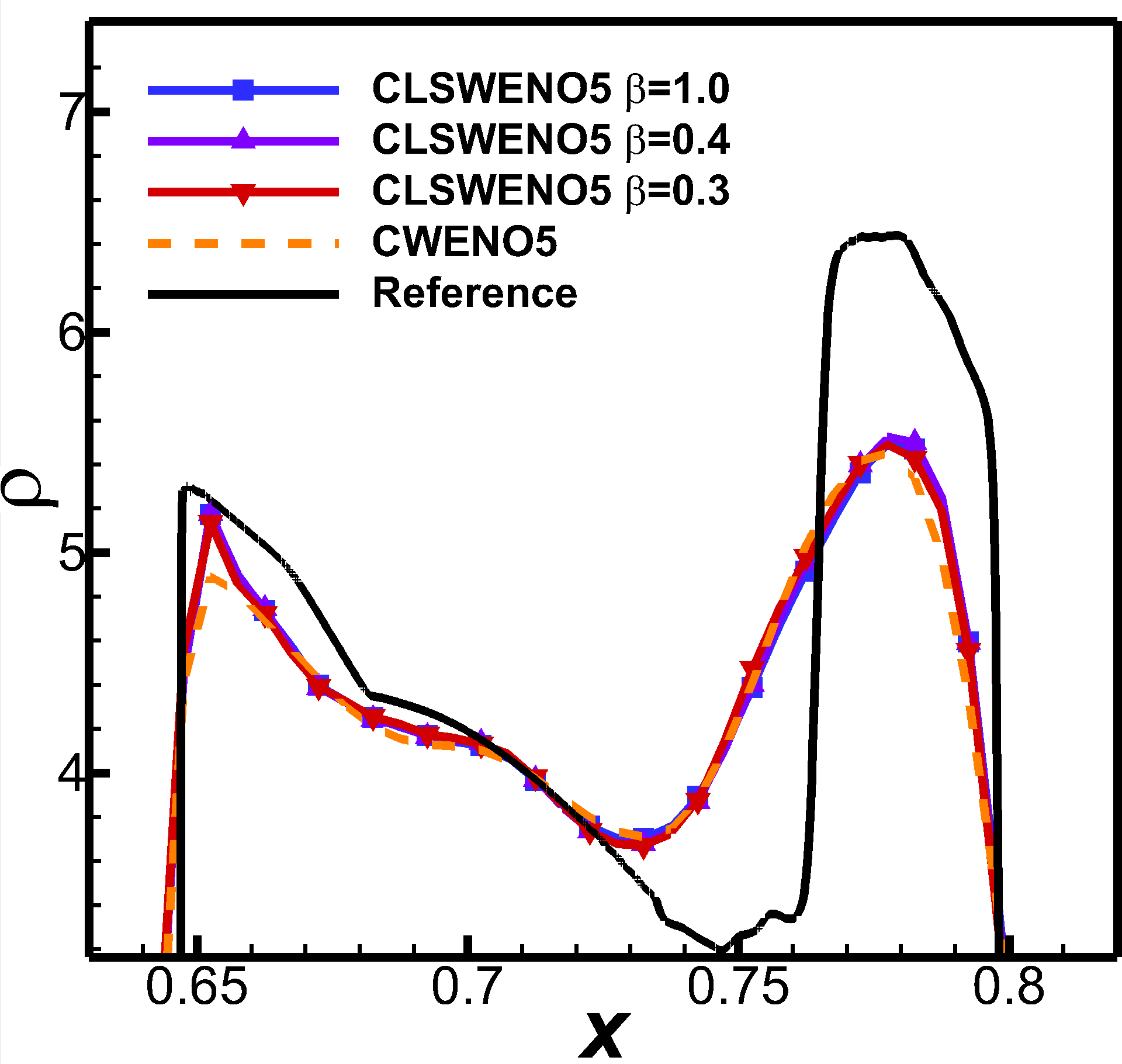}
    \caption{Close view of density. Fifth-order scheme.\label{fig:ibw_5th_cv}}
    \end{subfigure}
    \caption{\label{fig:ibw_density} Density distribution for the two interacting blast wave problem with 200 uniform cells.}
\end{figure}

\begin{figure}[!htbp]
  \centering
    \begin{subfigure}[b]{\columnwidth}
    \includegraphics[width=0.61\columnwidth]{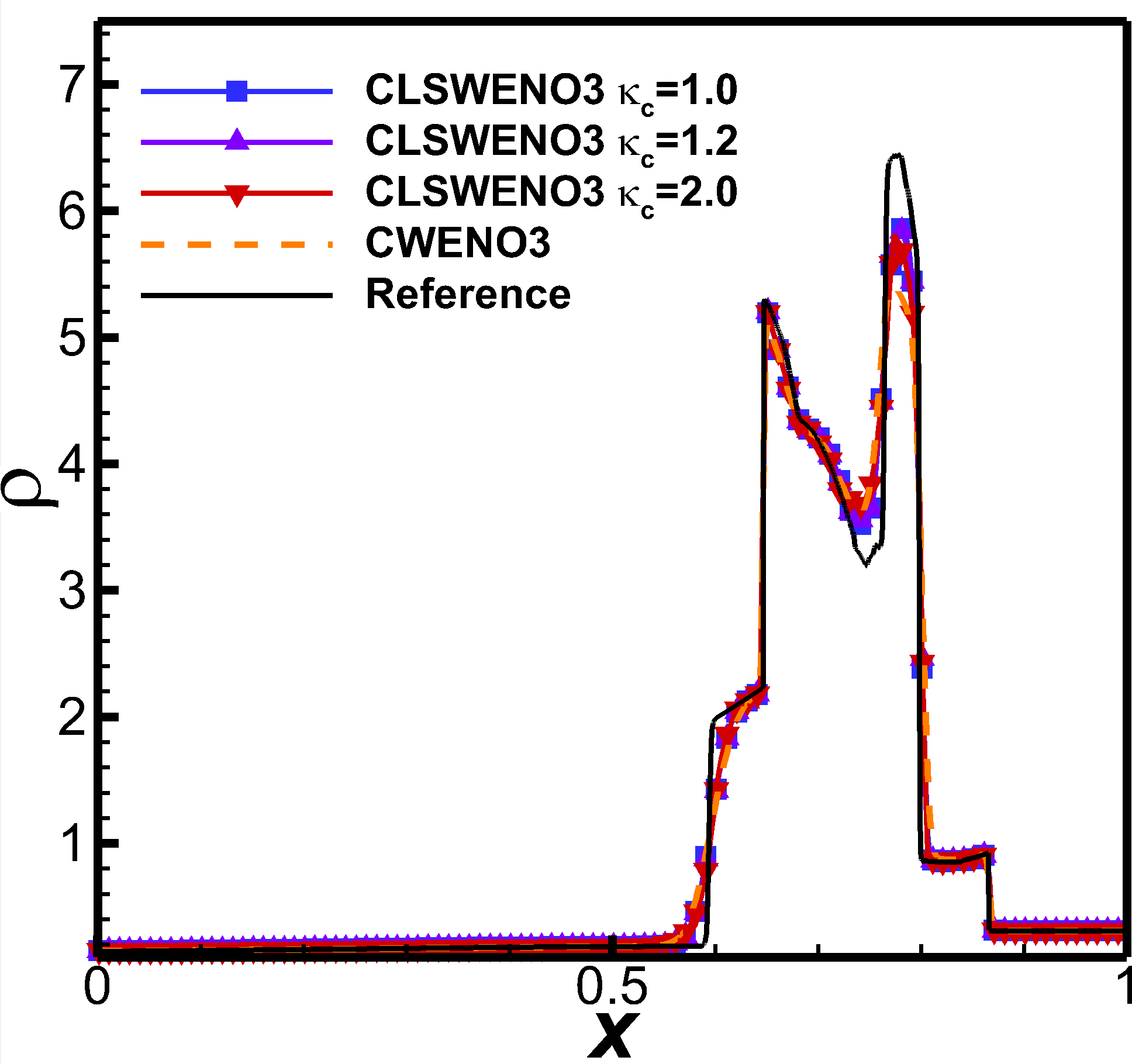}
    \caption{Density. Third-order scheme.\label{fig:ibw_3rd_N400}}
    \end{subfigure}
    \begin{subfigure}[b]{\columnwidth}
    \includegraphics[width=0.61\columnwidth]{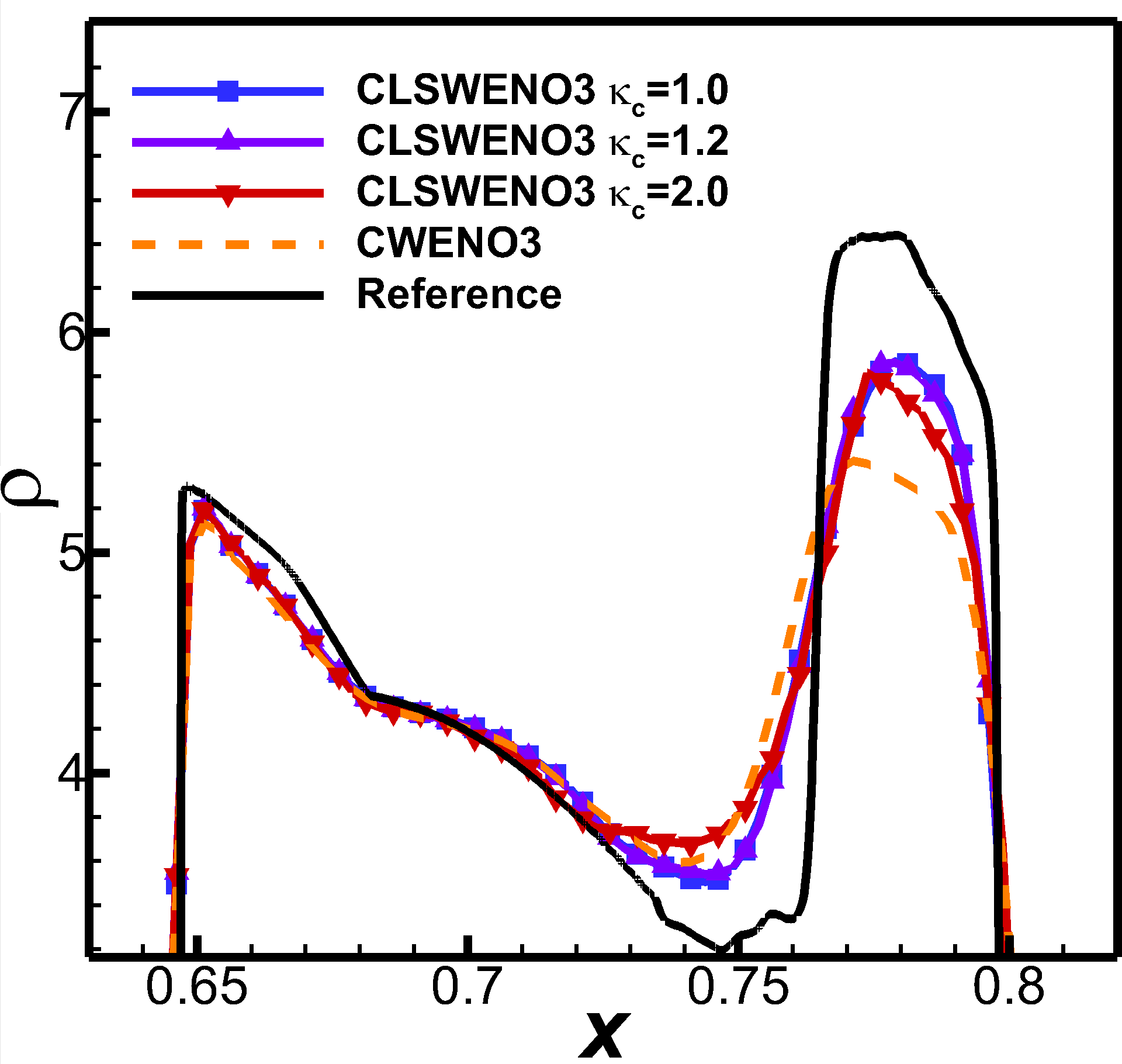}
    \caption{Close view of density. Third-order scheme.\label{fig:ibw_3rd_cv_N400}}
    \end{subfigure}
    \begin{subfigure}[b]{\columnwidth}
    \includegraphics[width=0.61\columnwidth]{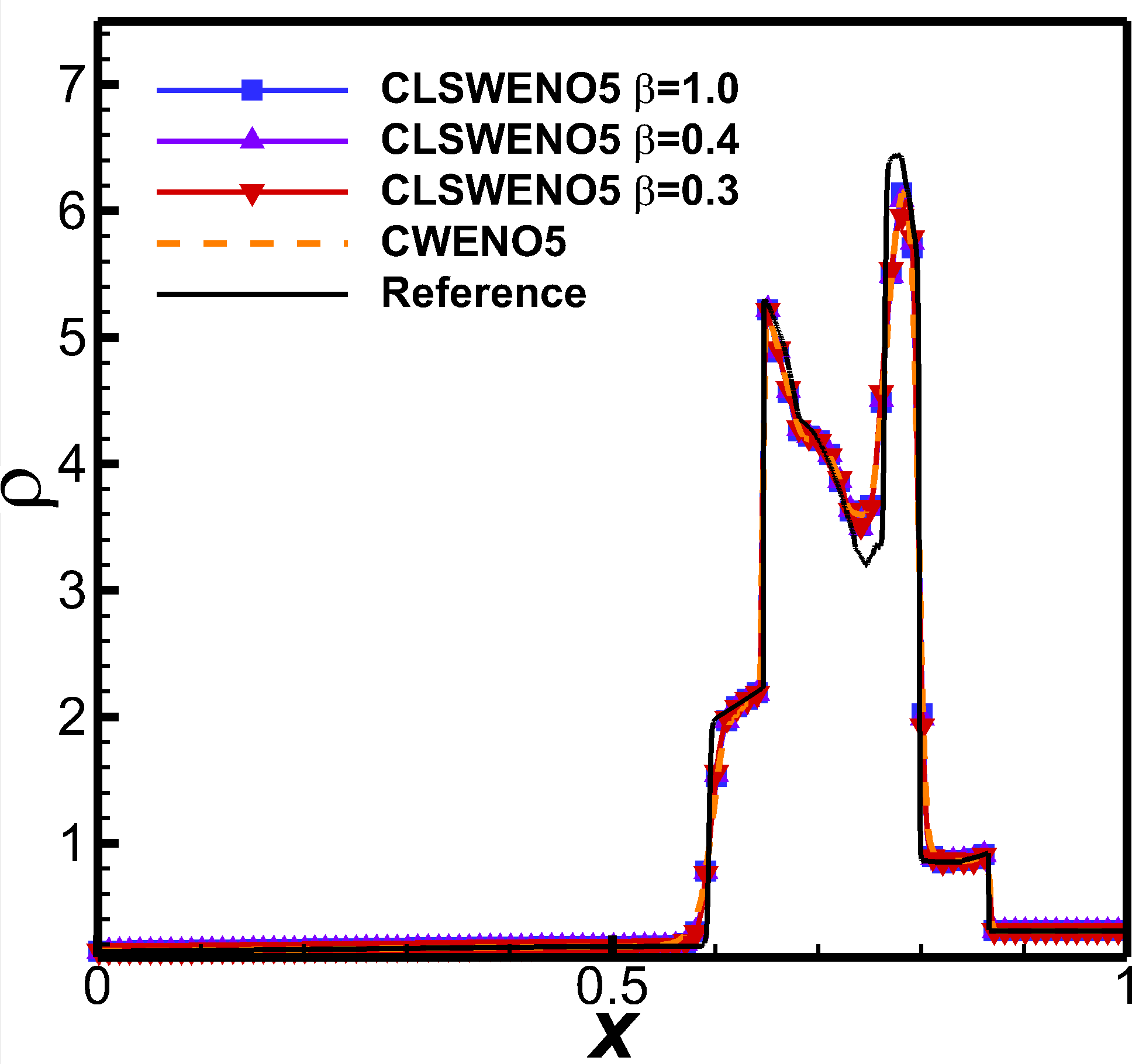}
    \caption{Density. Fifth-order scheme.\label{fig:ibw_5th_N400}}
    \end{subfigure}
    \begin{subfigure}[b]{\columnwidth}
    \includegraphics[width=0.61\columnwidth]{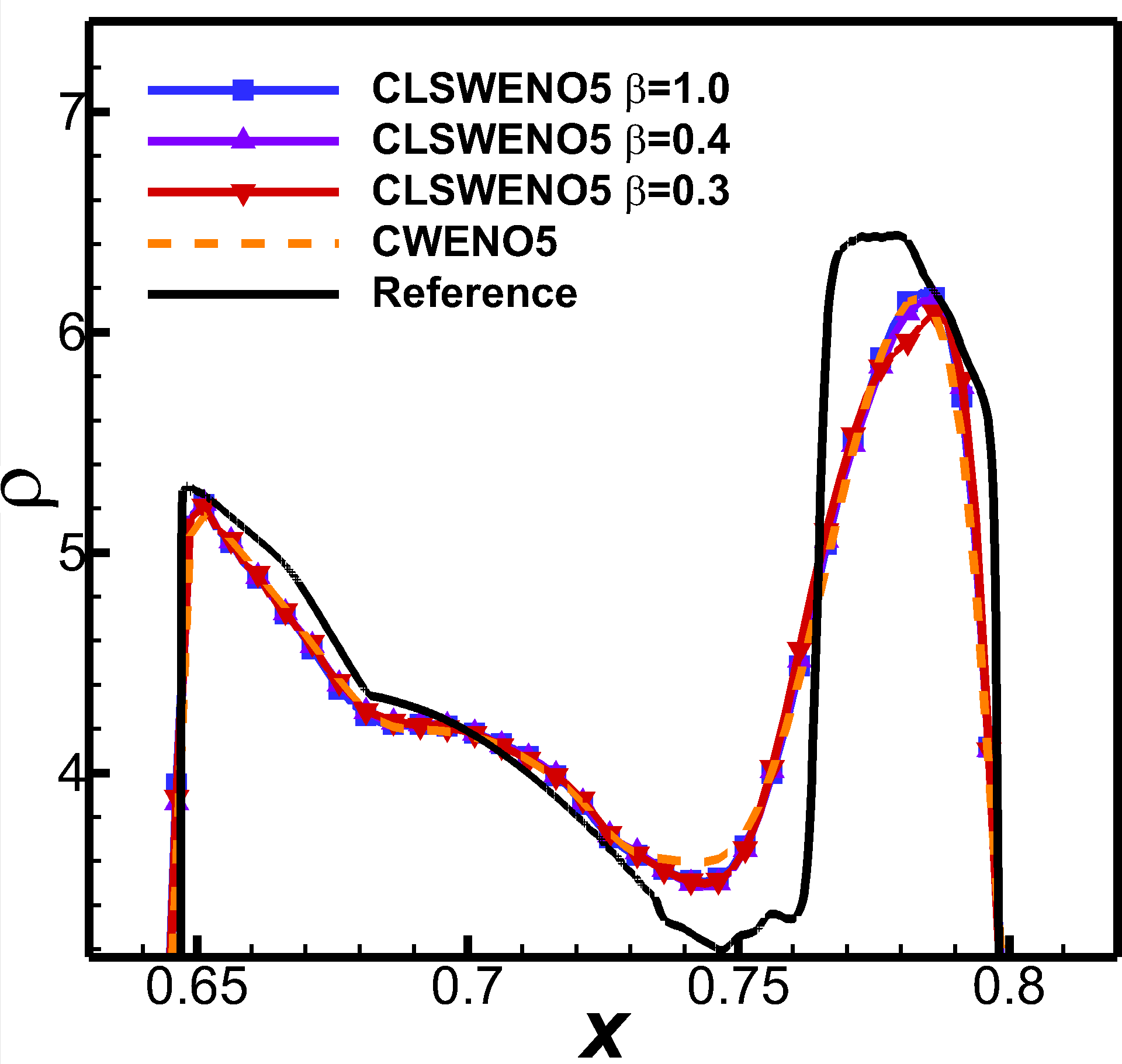}
    \caption{Close view of density. Fifth-order scheme.\label{fig:ibw_5th_cv_N400}}
    \end{subfigure}
    \caption{\label{fig:ibw_density_N400} Density distribution for the two interacting blast wave problem with 400 uniform cells.}
\end{figure}

Figures \ref{fig:ibw_density} and \ref{fig:ibw_density_N400} show the results for the third- and fifth-order schemes utilizing the proposed shock detector $\sigma^{\mathrm{Li}}$ with 200 and 400 uniform cells, respectively. As shown, the  shocks are captured without oscillation. The resolution of the hybrid CLS-CWENO schemes is improved when compared to the CWENO schemes. For the third-order hybrid CLS-CWENO scheme, the coefficients optimized by $\kappa_c = 1.0$ and $1.2$ outperforms the coefficients optimized by $\kappa_c = 2.0$, especially as shown in Fig. \ref{fig:ibw_3rd_cv_N400}.

Until now, the performance of different optimized coefficients is thoroughly tested through 1D linear and nonlinear cases. To sum up, for the third-order scheme, the coefficients optimized by $\kappa_c = 1.0$ is the best taking considerations of both the stability around discontinuities and the resolution in smooth areas; for the fifth-order scheme, the performance of the optimized three sets of coefficients are comparative. For the brevity of the paper, coefficients optimized by $\kappa_c = 1.0$ is utilized for the third-order hybrid CLS-CWENO scheme and coefficients optimized by $\beta=0.4$ is adopted for the fifth-order hybrid CLS-CWENO scheme in the remainder of 2D numerical examples. Additionally, only the results with the proposed shock detector $\sigma^{\mathrm{Li}}$ are presented hereafter.

\subsection{Isentropic vortex problem\label{sec:iv}}
In this problem, an isentropic disturbance is added to a mean flow and the initial condition is
\begin{equation}
  \begin{aligned}
u &= 1-\frac{5}{2\pi}e^{0.5(1-r^2)} y,\\
v &= 1+\frac{5}{2\pi}e^{0.5(1-r^2)} x,\\
T &= 1-\frac{25(\Gamma-1)}{8\gamma \pi^2}e^{1-r^2},\\
p&=\rho^{\gamma}.
  \end{aligned}
\end{equation}

\begin{figure}[!htbp]
  \centering
    \includegraphics[width=0.8\columnwidth]{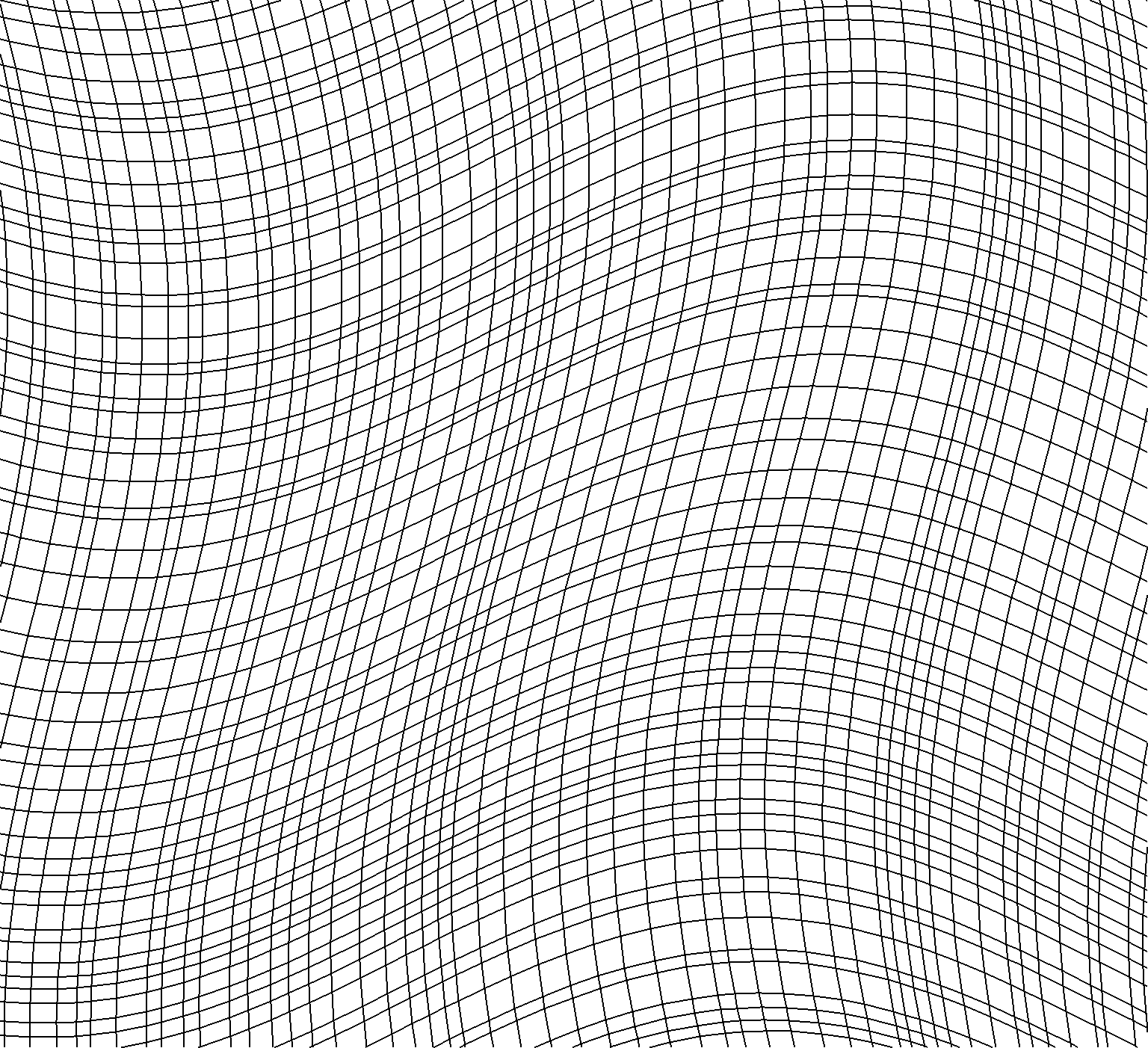}
    \caption{Non-uniform mesh for the isentropic vortex problem.
    \label{fig:ivmesh}}
\end{figure}

\begin{figure}[!htbp]
  \centering
    \begin{subfigure}[b]{\columnwidth}
    \includegraphics[width=0.61\columnwidth]{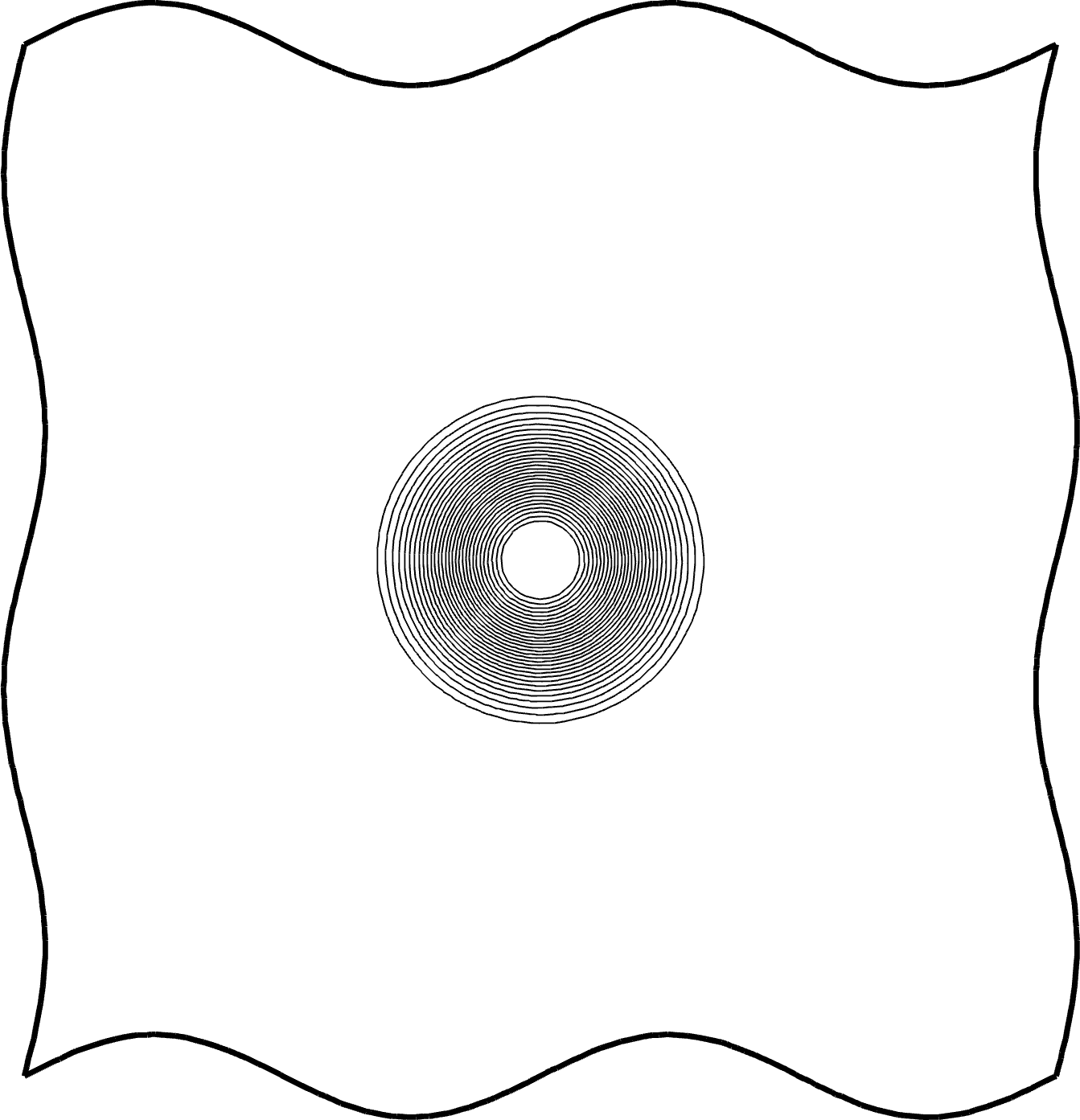}
    \caption{Third-order hybrid CLS-CWENO scheme.}
    \end{subfigure}
    \begin{subfigure}[b]{\columnwidth}
    \includegraphics[width=0.61\columnwidth]{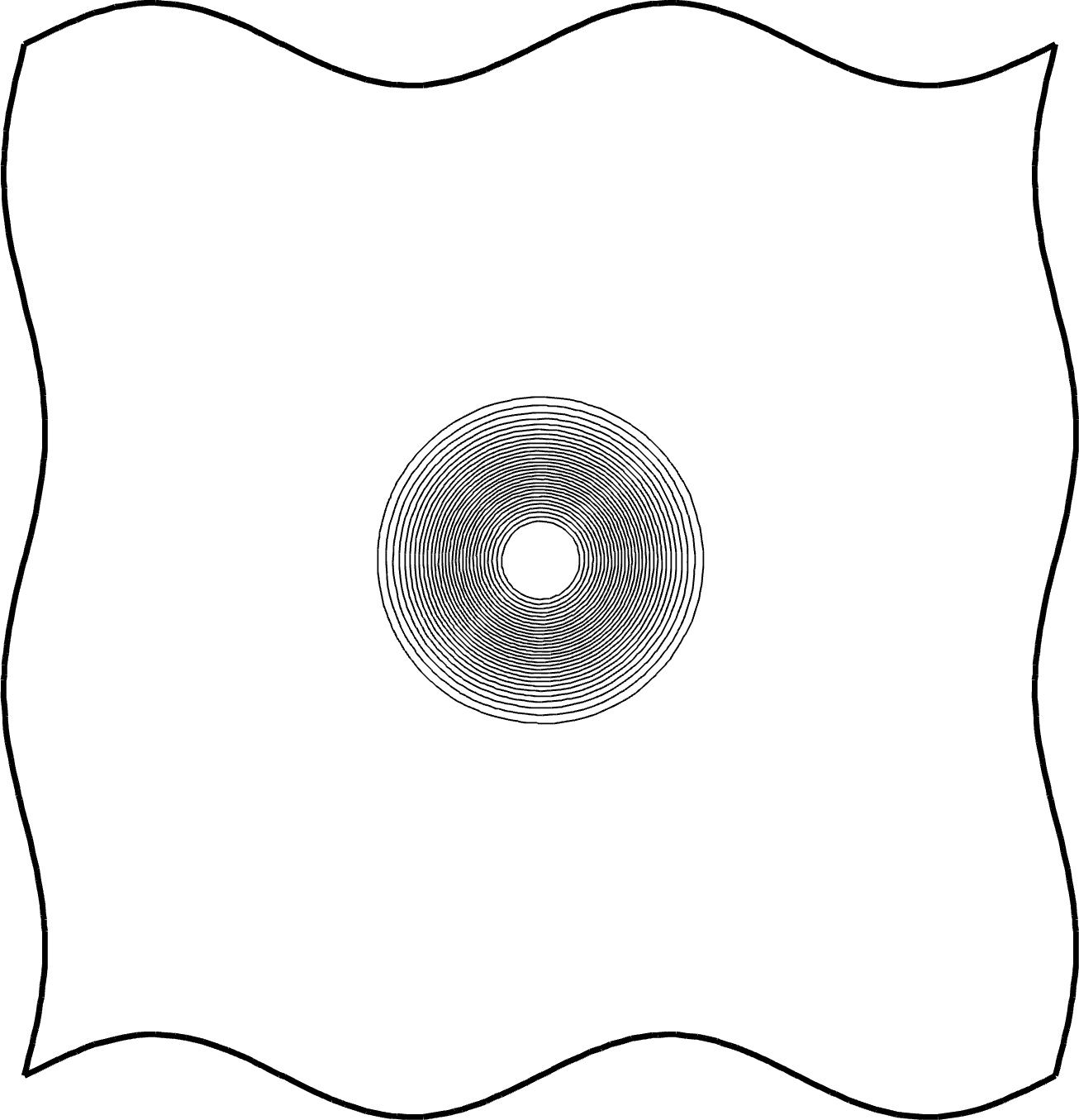}
    \caption{Fifth-order hybrid CLS-CWENO scheme.}
    \end{subfigure}
    \caption{Contours for density of isentropic vortex on non-uniform curvilinear grids with 30 lines from 0.55 to 0.95.
    \label{fig:ivcontoru}}
\end{figure}

Curvilinear grids defined as in Sec. 4.3.1 in the work of Wang \cite{wang2015accurate} are utilized to check the performance of the proposed hybrid schemes on non-uniform grids as shown in Fig. \ref{fig:ivmesh}. The number of control volumes is $128\times 128$.
The density contours at time $t = 50$ is shown as in Fig. \ref{fig:ivcontoru}. As illustrated, the proposed schemes can be applied on structured non-uniform curvilinear grids with high resolution.

\subsection{Double shear layer problem\label{sec:dsl}}
\begin{figure}[!htbp]
  \centering
    \begin{subfigure}[b]{\columnwidth}
    \includegraphics[width=0.61\columnwidth]{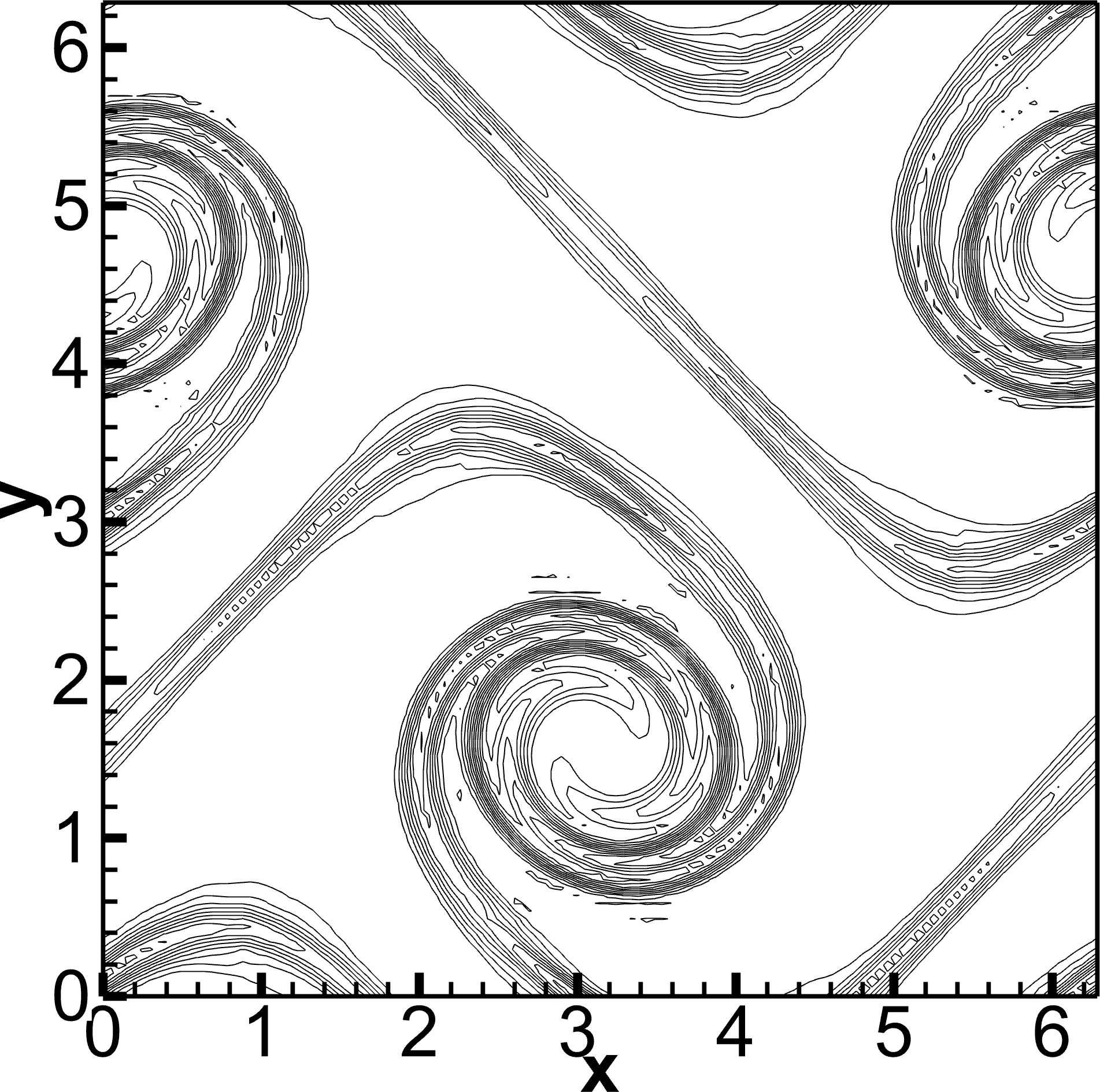}
    \caption{Third-order hybrid CLS-CWENO scheme.}
    \end{subfigure}
    \begin{subfigure}[b]{\columnwidth}
    \includegraphics[width=0.61\columnwidth]{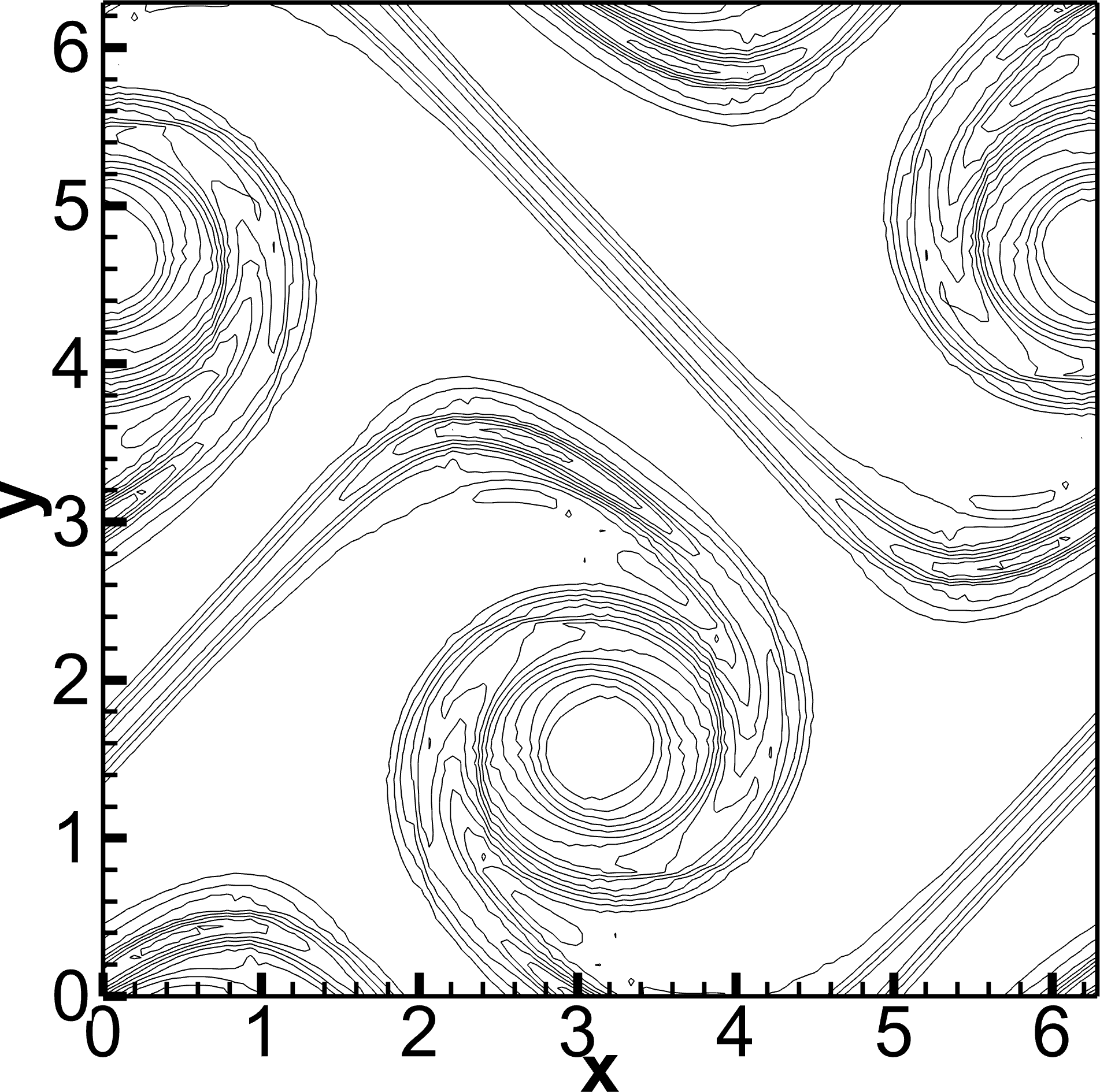}
    \caption{Third-order CWENO scheme.\label{fig:dsl_cweno_3rd}}
    \end{subfigure}
    \begin{subfigure}[b]{\columnwidth}
    \includegraphics[width=0.61\columnwidth]{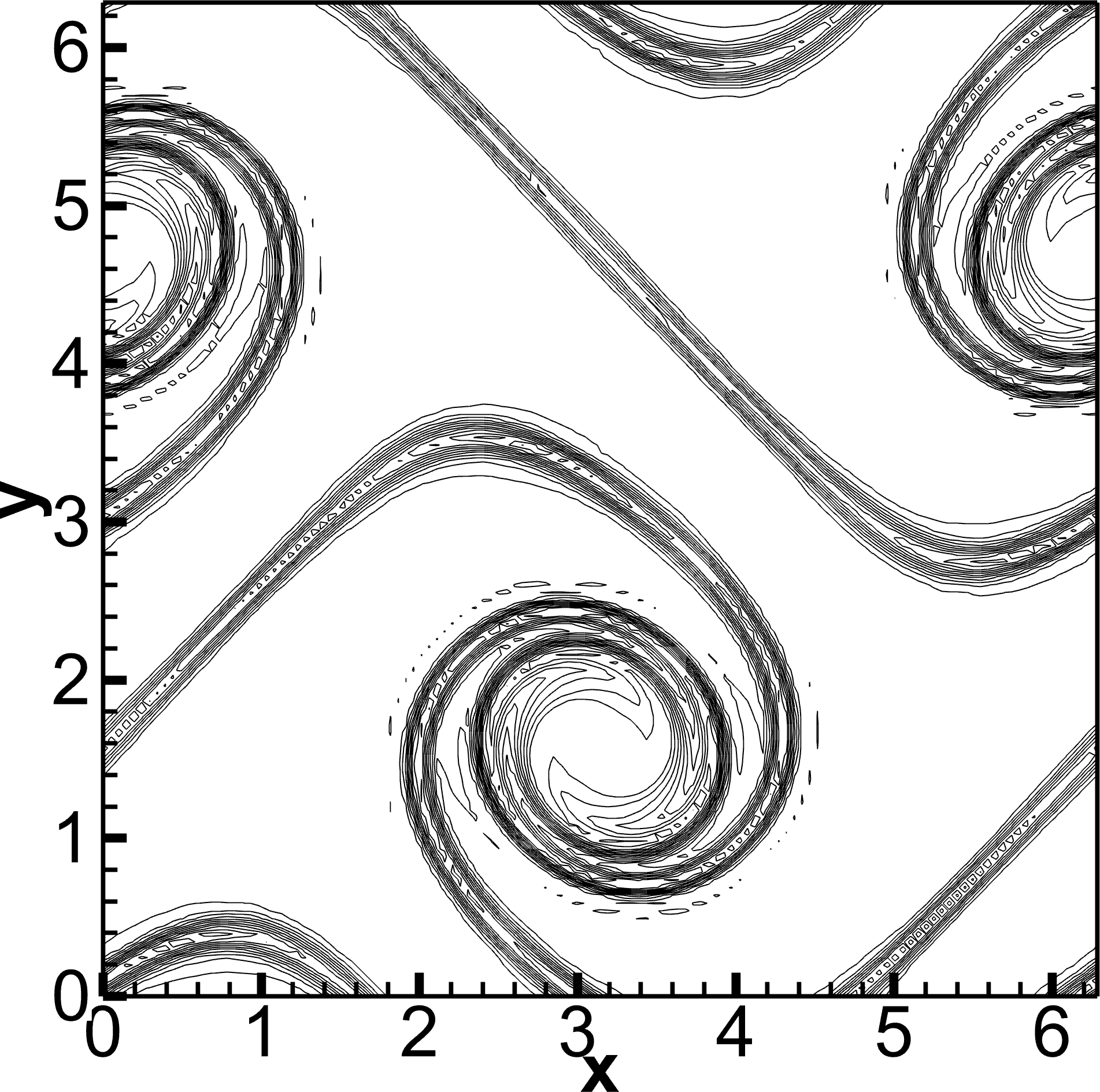}
    \caption{Fifth-order hybrid CLS-CWENO scheme.}
    \end{subfigure}
    \begin{subfigure}[b]{\columnwidth}
    \includegraphics[width=0.61\columnwidth]{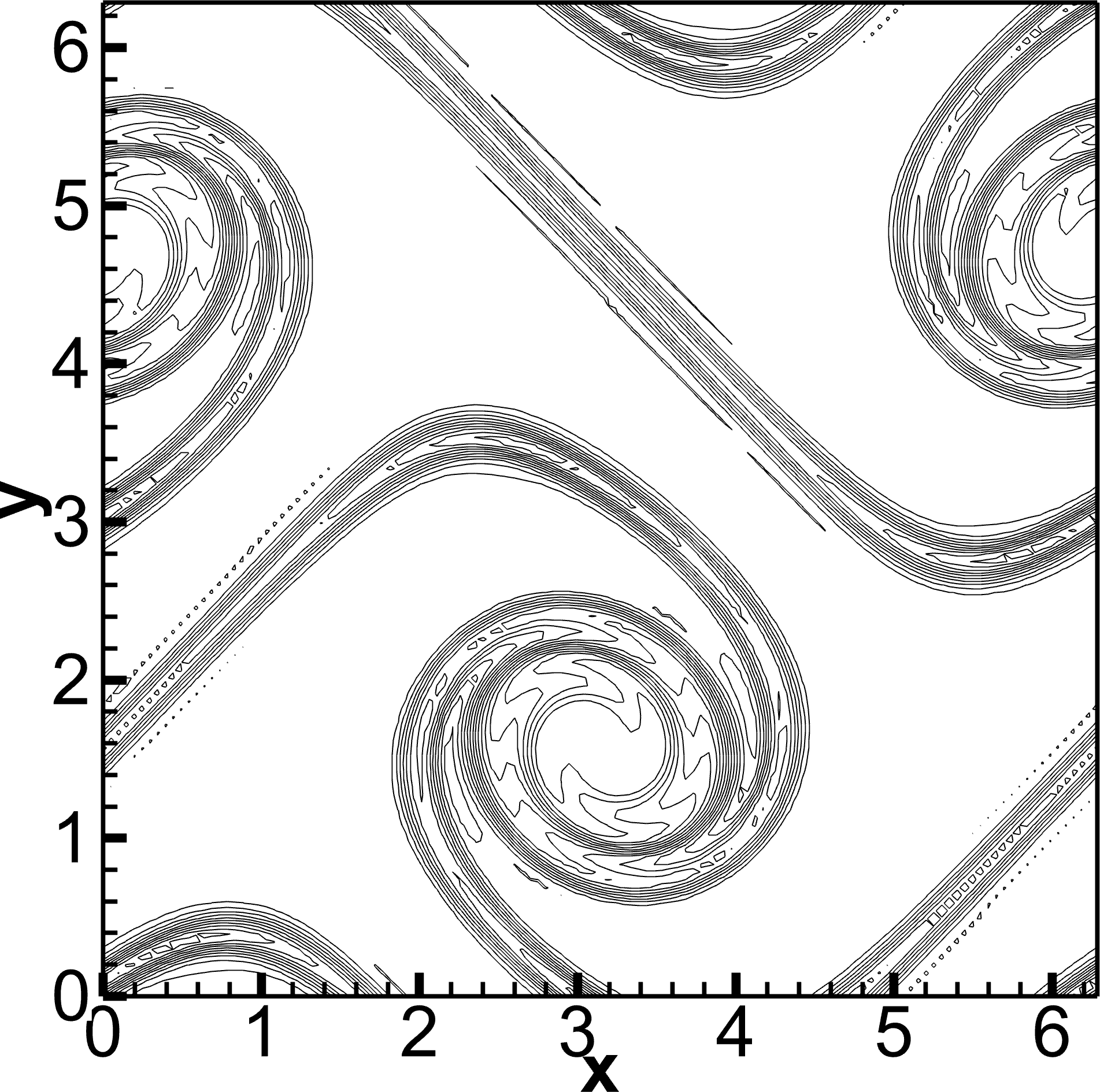}
    \caption{Fifth-order CWENO scheme.\label{fig:dsl_cweno_5th}}
    \end{subfigure}
    \caption{Z-vorticity contours for the double shear layer problem with 30 lines ranging from -4.3 to 4.3.
    \label{fig:dsl}}
\end{figure}

2D Euler equations with two shear layers are solved in this case to validate the performance of the proposed hybrid schemes.
The initial condition is
\begin{align}
  u(x,y) & = \left\{
  \begin{array}{ll}
    \tanh\left(\frac{15(y-0.5\pi)}{\pi}\right), & y\leq \pi, \\
    \tanh\left(\frac{15(1.5\pi-y)}{\pi}\right), & y >  \pi,
  \end{array}
  \right. \\
  v(x,y) & = 0.05 \sin(x),\\
  \rho(x,y) &= 1.0.
\end{align}

The initial pressure $p$ is uniform and assigned such that the Mach number  based on the mean velocity $\overline{\sqrt{u^2+v^2}}$ is 0.1, where $\overline{\left(\cdot \right)}$ is an average operator over the whole computational domain. The computational domain is $\Omega = [0,2\pi]\times[0,2\pi]$ discretized by $128\times 128$ uniform control volumes. The simulation end time is $t_{end}=8.0$ and the Courant number is 0.8.

Figures \ref{fig:dsl} shows the contours of z-vorticity for the hybrid CLS-CWENO schemes. As a comparison, the results of the third- and fifth-order CWENO scheme are also presented as in Figs. \ref{fig:dsl_cweno_3rd} and \ref{fig:dsl_cweno_5th}, respectively.  Thirty contour lines ranging from -4.3 to 4.3 are plotted. As illustrated, the hybrid CLS-CWENO schemes can resolve finer shear layers than the CWENO schemes of the same order.

\subsection{Double Mach reflection\label{sec:dmr}}

\begin{figure}[!htbp]
  \centering
    \begin{subfigure}[b]{\columnwidth}
    \includegraphics[width=1.0\columnwidth]{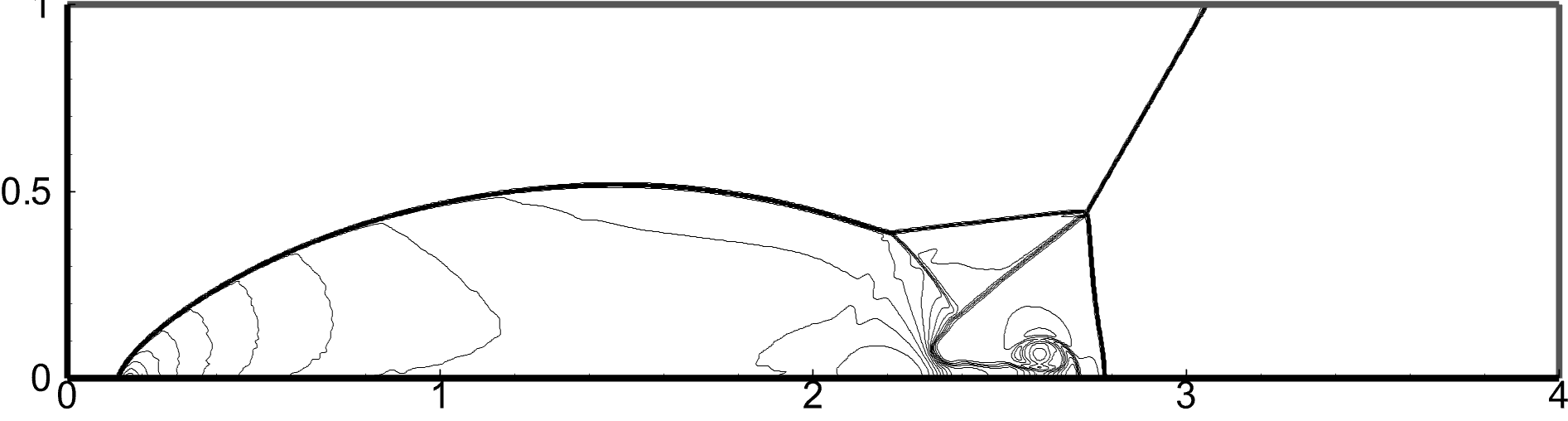}
    \caption{Third-order hybrid CLS-CWENO scheme.\label{fig:dmr_clscweno_3rd}}
    \end{subfigure}
    \begin{subfigure}[b]{\columnwidth}
    \includegraphics[width=1.0\columnwidth]{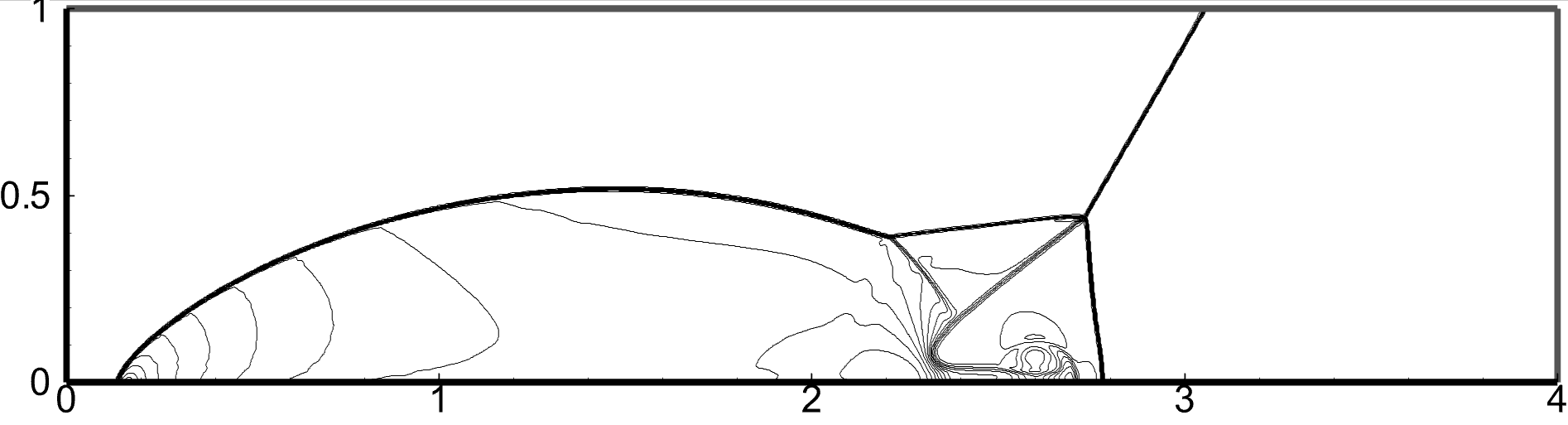}
    \caption{Third-order CWENO scheme.\label{fig:dmr_cweno_3rd}}
    \end{subfigure}
    \begin{subfigure}[b]{\columnwidth}
    \includegraphics[width=1.0\columnwidth]{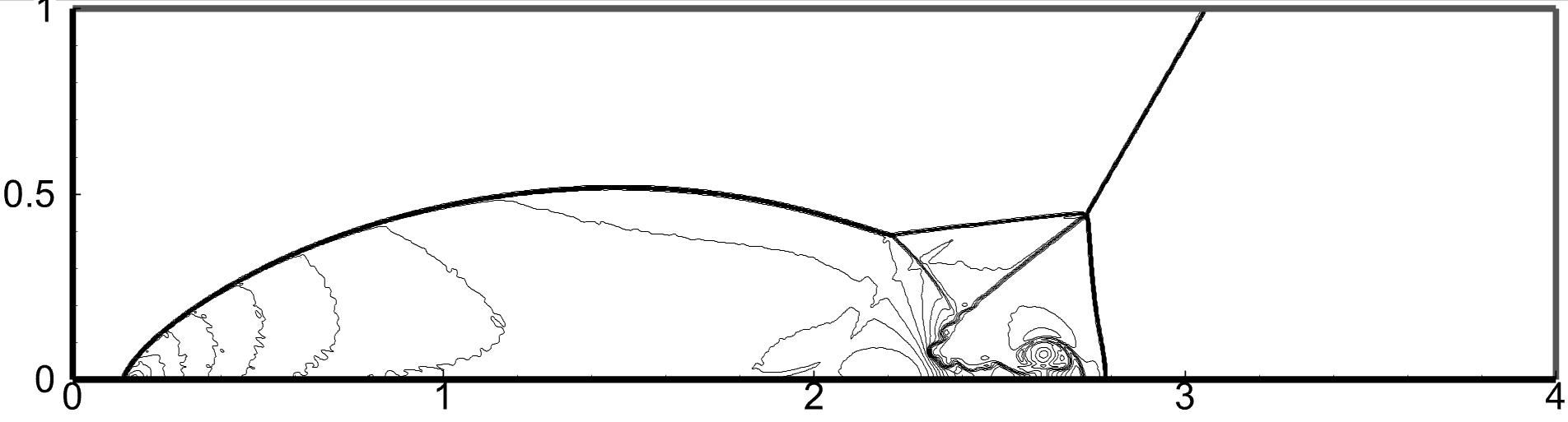}
    \caption{Fifth-order hybrid CLS-CWENO scheme.\label{fig:dmr_clscweno_5th}}
    \end{subfigure}
    \begin{subfigure}[b]{\columnwidth}
    \includegraphics[width=1.0\columnwidth]{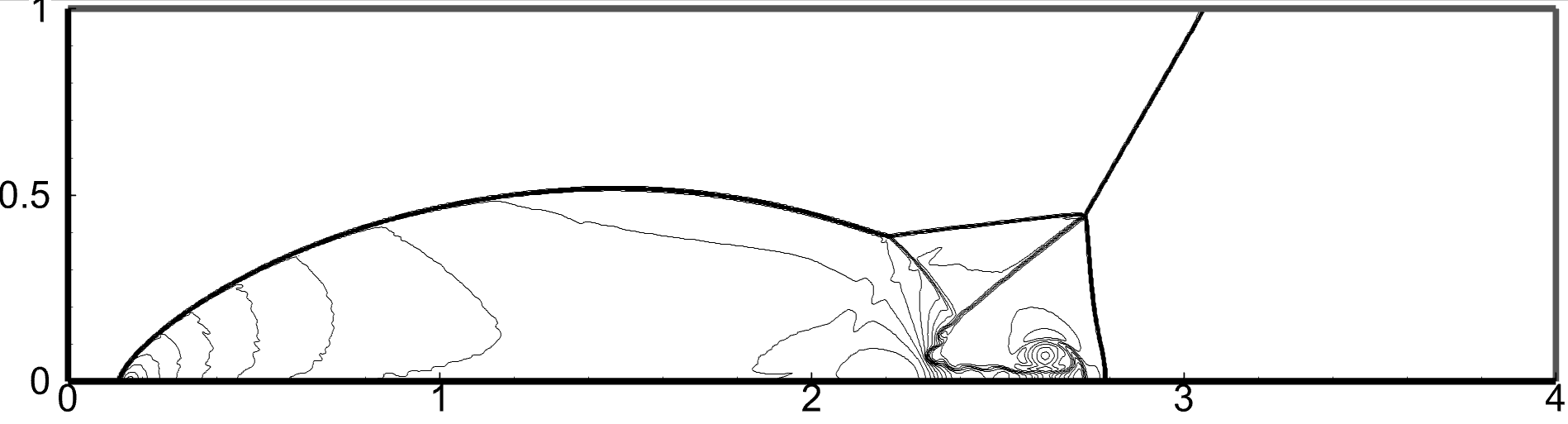}
    \caption{Fifth-order CWENO scheme.\label{fig:dmr_cweno_5th}}
    \end{subfigure}
    \caption{Density contours for the double Mach reflection problem with 30 lines ranging from 3.1 to 22. Cell numer is $960\times 240$.
    \label{fig:dmr}}
\end{figure}

\begin{figure}[!htbp]
  \centering
    \begin{subfigure}[b]{\columnwidth}
    \includegraphics[width=1.0\columnwidth]{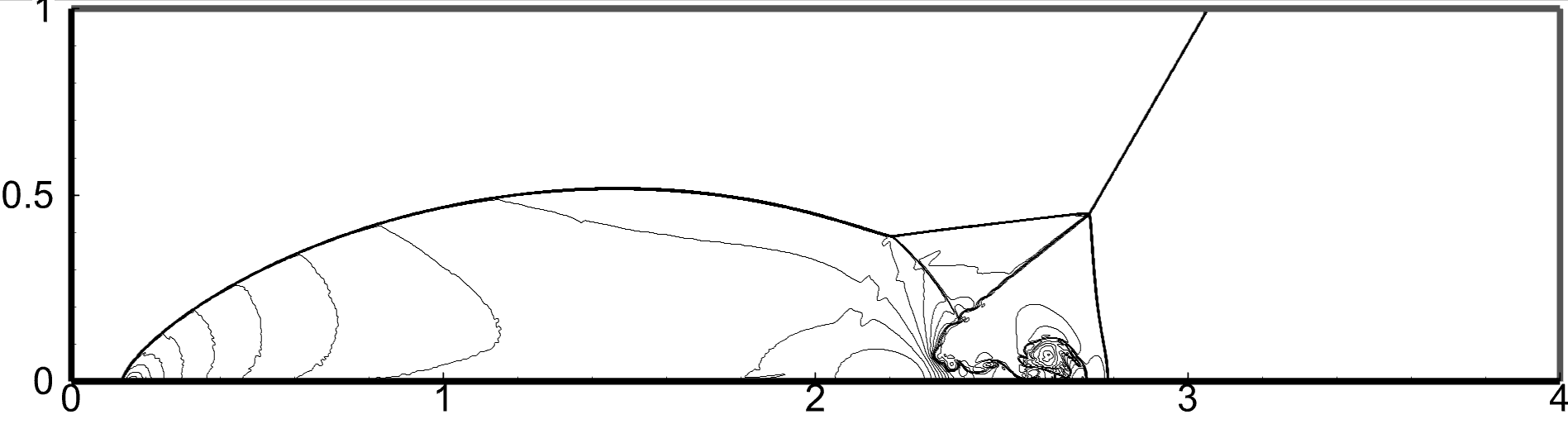}
    \caption{Third-order hybrid CLS-CWENO scheme.\label{fig:dmr_clscweno_3rd_480}}
    \end{subfigure}
    \begin{subfigure}[b]{\columnwidth}
    \includegraphics[width=1.0\columnwidth]{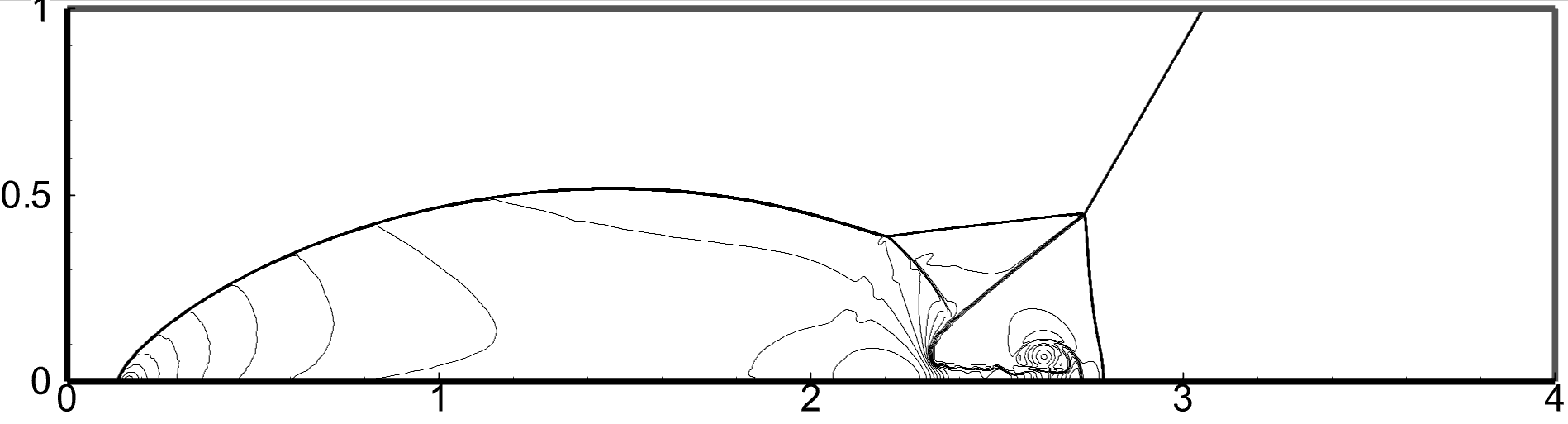}
    \caption{Third-order CWENO scheme.\label{fig:dmr_cweno_3rd_480}}
    \end{subfigure}
    \begin{subfigure}[b]{\columnwidth}
    \includegraphics[width=1.0\columnwidth]{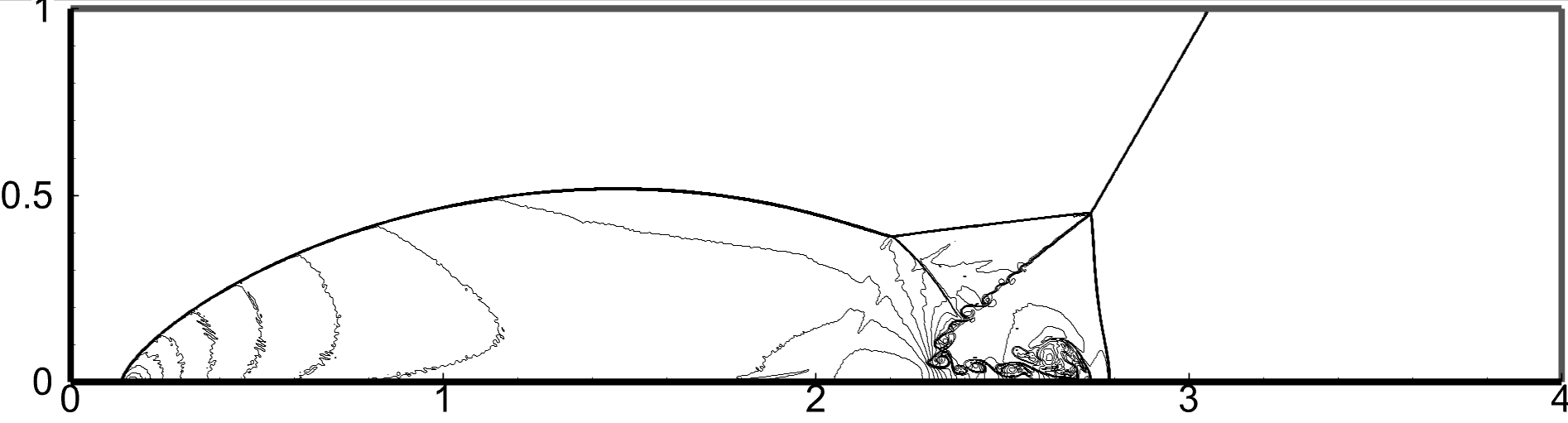}
    \caption{Fifth-order hybrid CLS-CWENO scheme.\label{fig:dmr_clscweno_5th_480}}
    \end{subfigure}
    \begin{subfigure}[b]{\columnwidth}
    \includegraphics[width=1.0\columnwidth]{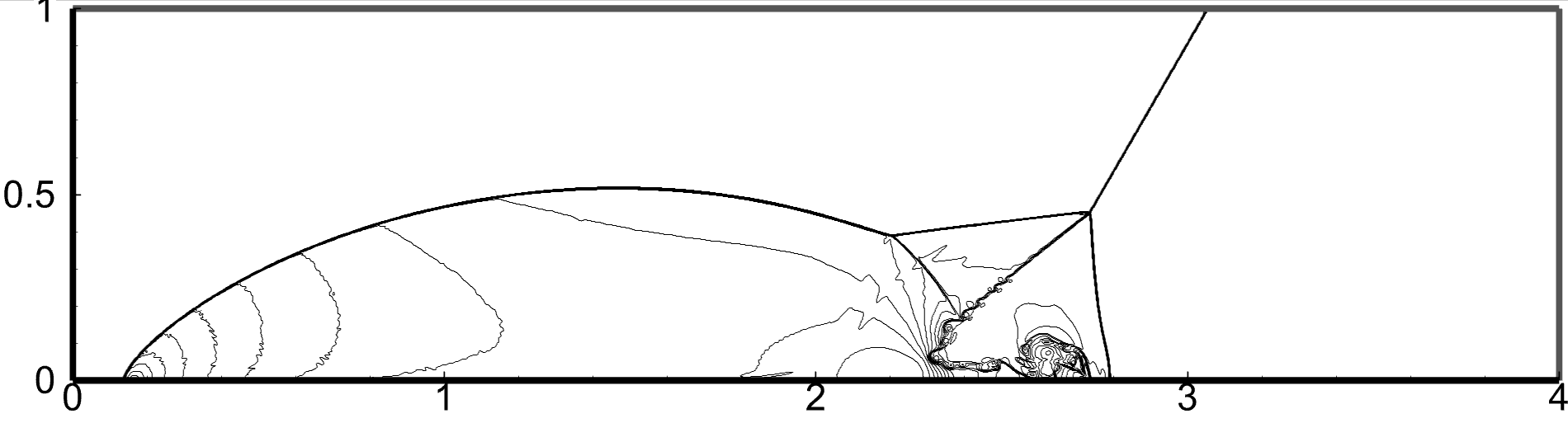}
    \caption{Fifth-order CWENO scheme.\label{fig:dmr_cweno_5th_480}}
    \end{subfigure}
    \caption{Density contours for the double Mach reflection problem with 30 lines ranging from 3.1 to 22. Cell numer is $1920\times 480$.
    \label{fig:dmr_480}}
\end{figure}

\begin{figure}[!htbp]
  \centering
    \begin{subfigure}[b]{\columnwidth}
    \includegraphics[width=0.61\columnwidth]{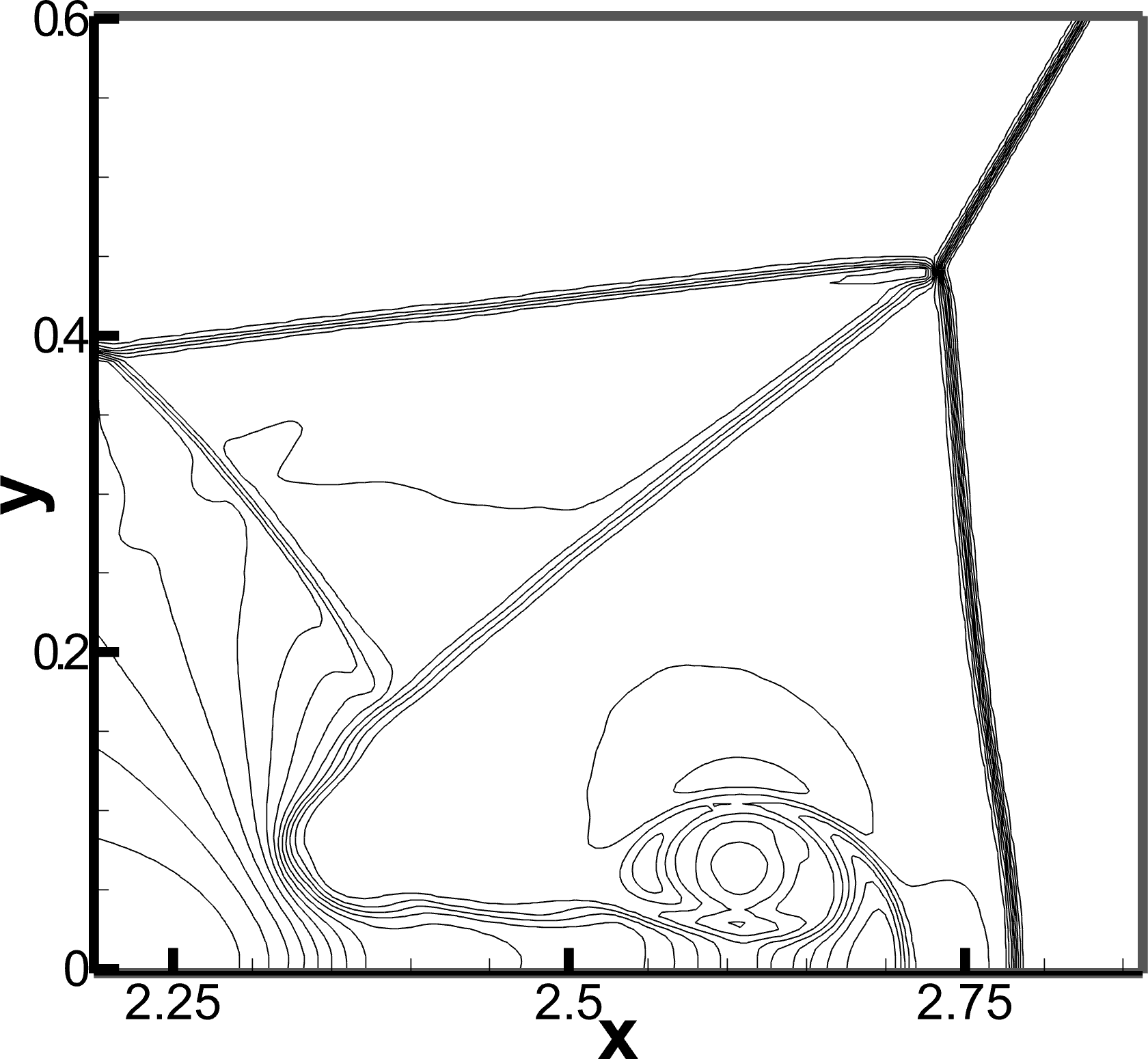}
    \caption{Third-order hybrid CLS-CWENO scheme.\label{fig:dmr_clscweno_3rd_cv}}
    \end{subfigure}
    \begin{subfigure}[b]{\columnwidth}
    \includegraphics[width=0.61\columnwidth]{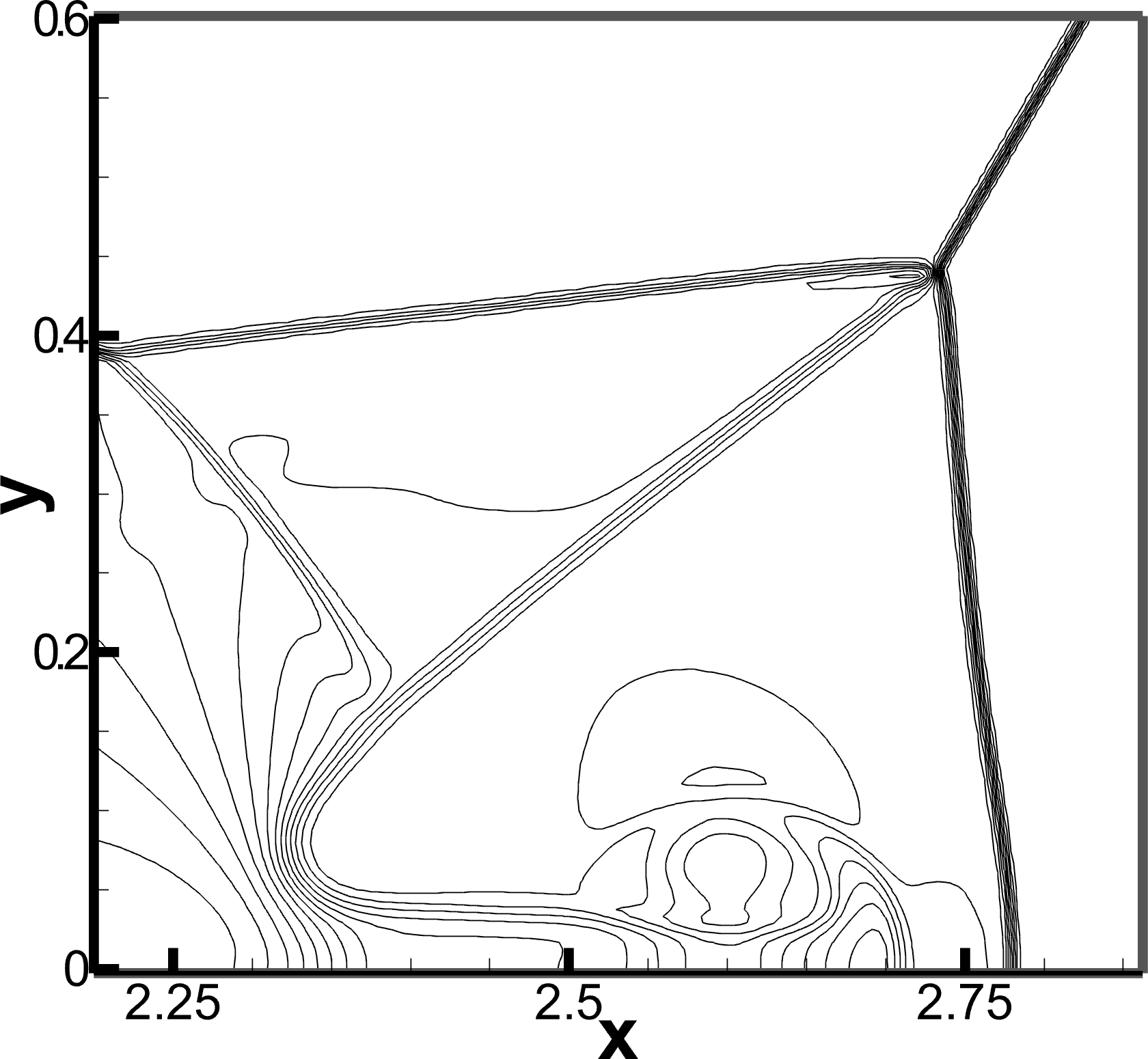}
    \caption{Third-order CWENO scheme.\label{fig:dmr_cweno_3rd_cv}}
    \end{subfigure}
    \begin{subfigure}[b]{\columnwidth}
    \includegraphics[width=0.61\columnwidth]{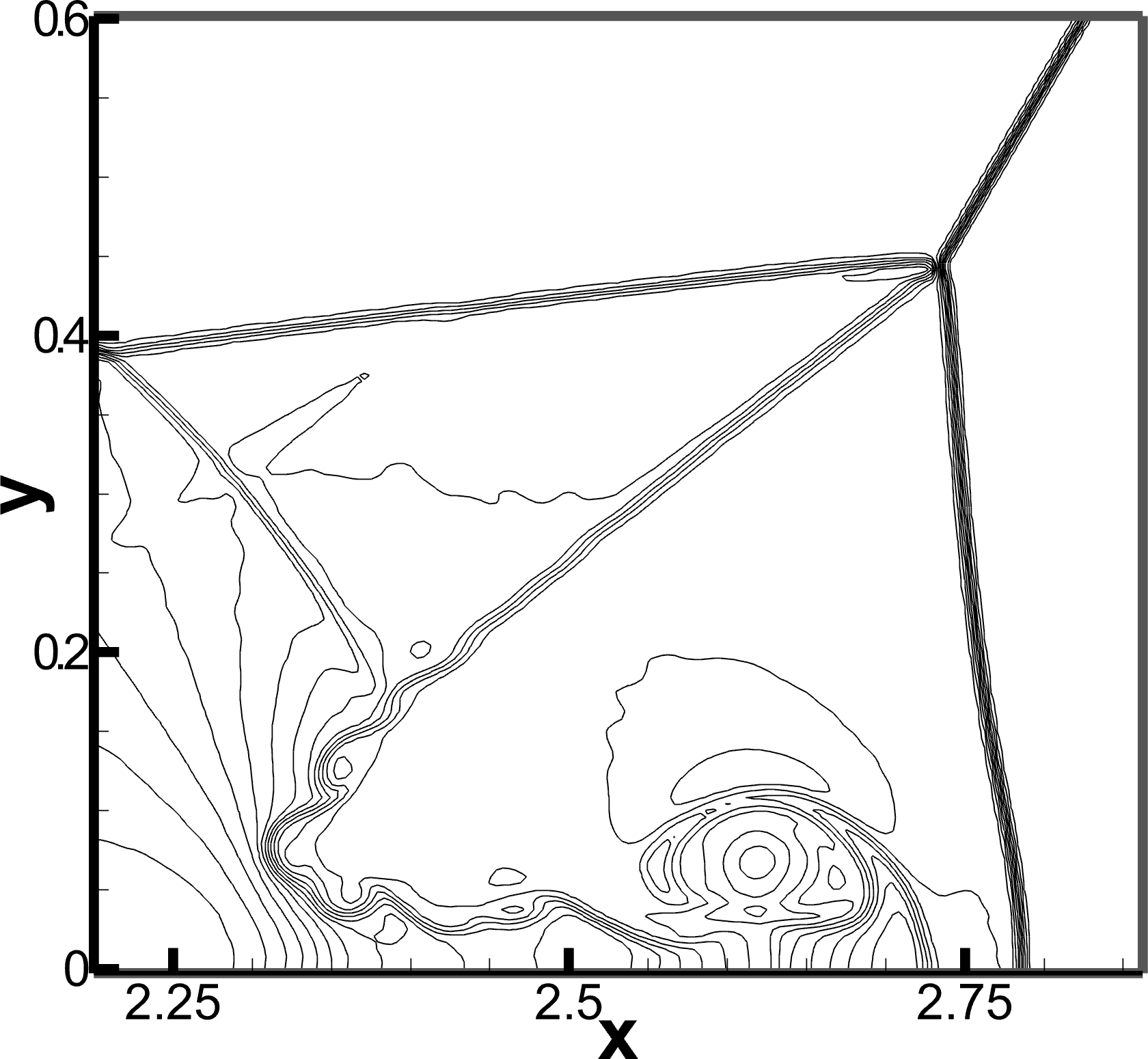}
    \caption{Fifth-order hybrid CLS-CWENO scheme.\label{fig:dmr_clscweno_5th_cv}}
    \end{subfigure}
    \begin{subfigure}[b]{\columnwidth}
    \includegraphics[width=0.61\columnwidth]{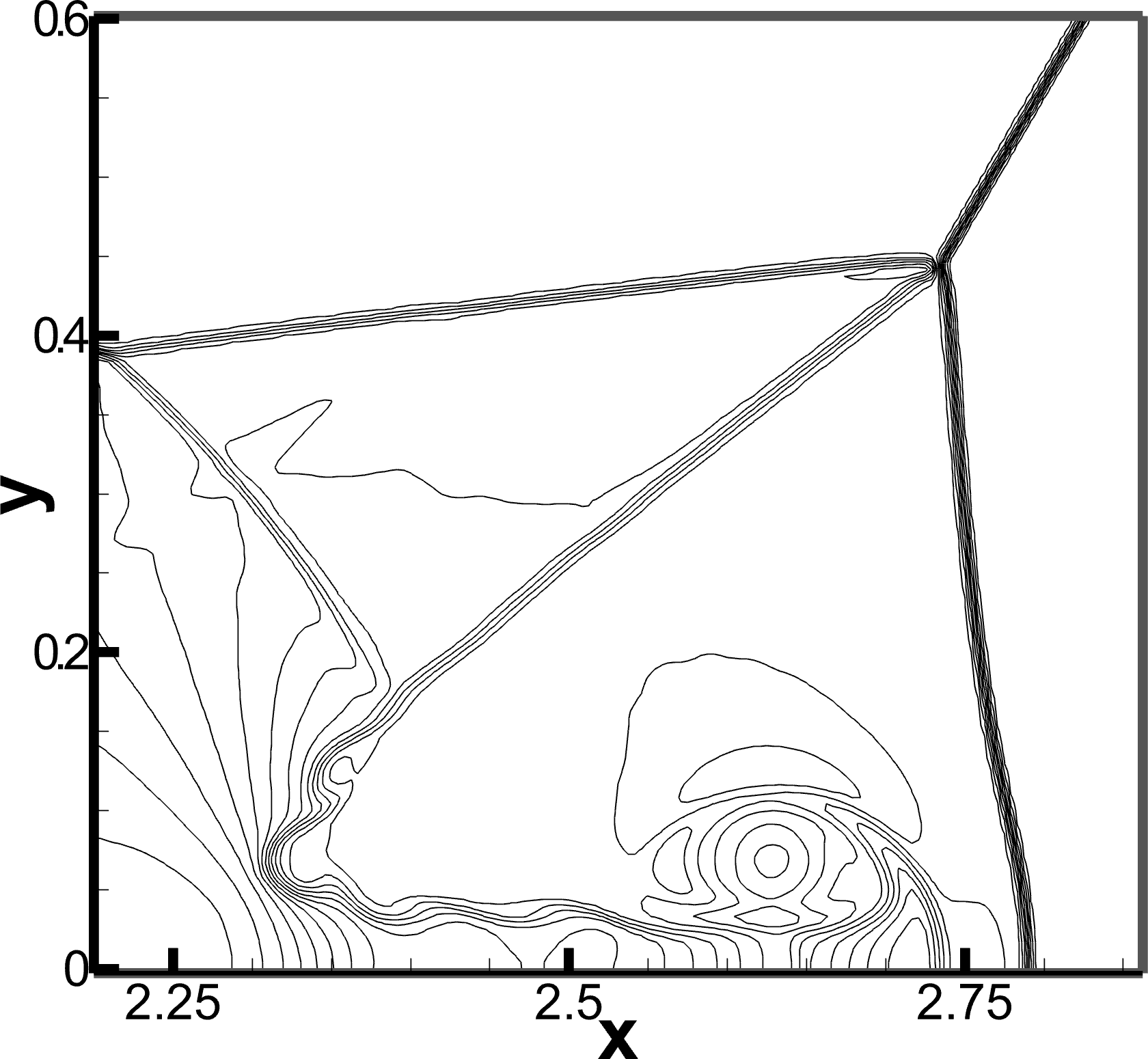}
    \caption{Fifth-order CWENO scheme.\label{fig:dmr_cweno_5th_cv}}
    \end{subfigure}
    \caption{Close view for the density contours of the double Mach reflection problem. Cell numer is $960\times 240$.
    \label{fig:dmr_close_view}}
\end{figure}

\begin{figure}[!htbp]
  \centering
    \begin{subfigure}[b]{\columnwidth}
    \includegraphics[width=0.61\columnwidth]{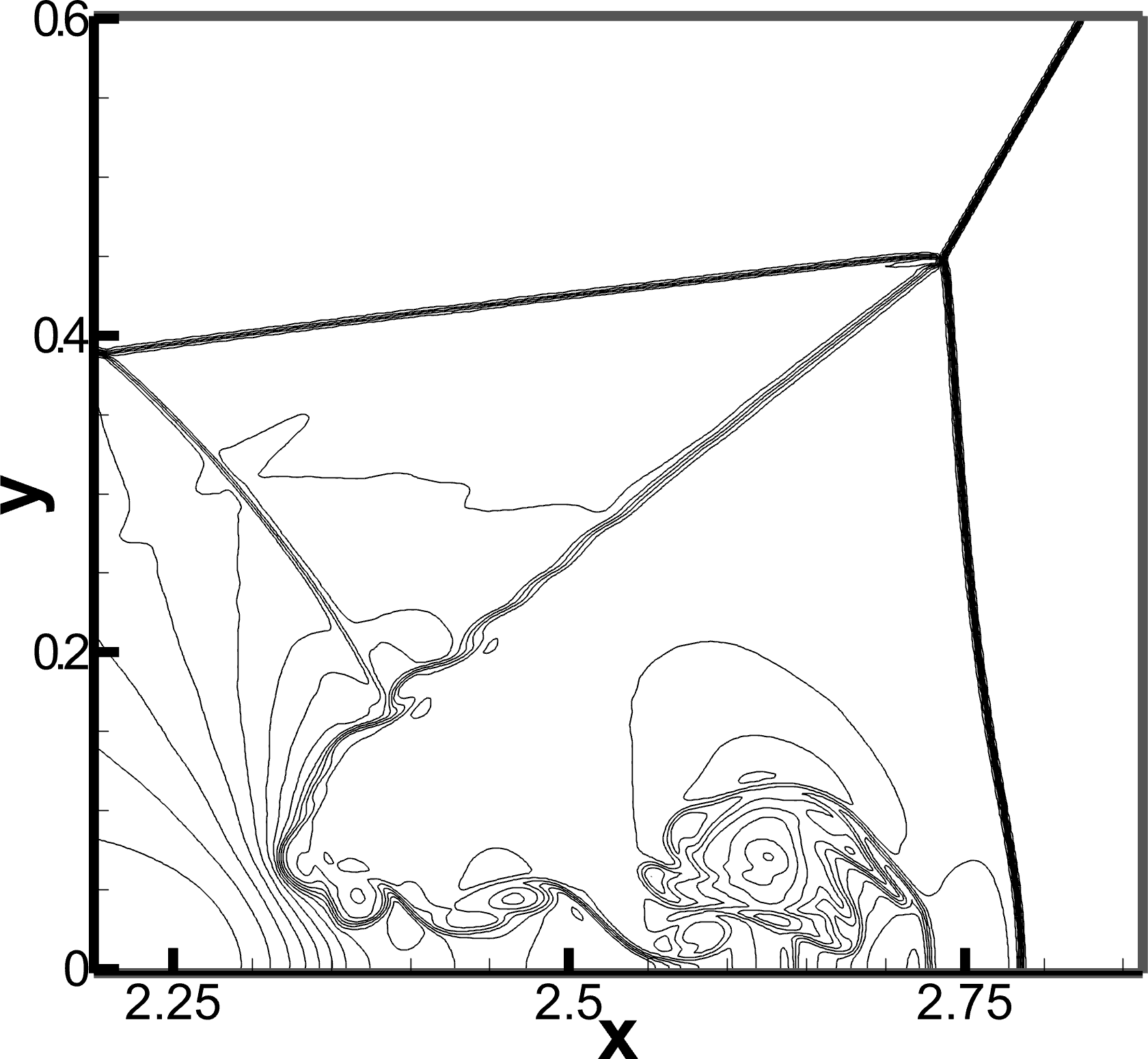}
    \caption{Third-order hybrid CLS-CWENO scheme.\label{fig:dmr_clscweno_3rd_480_cv}}
    \end{subfigure}
    \begin{subfigure}[b]{\columnwidth}
    \includegraphics[width=0.61\columnwidth]{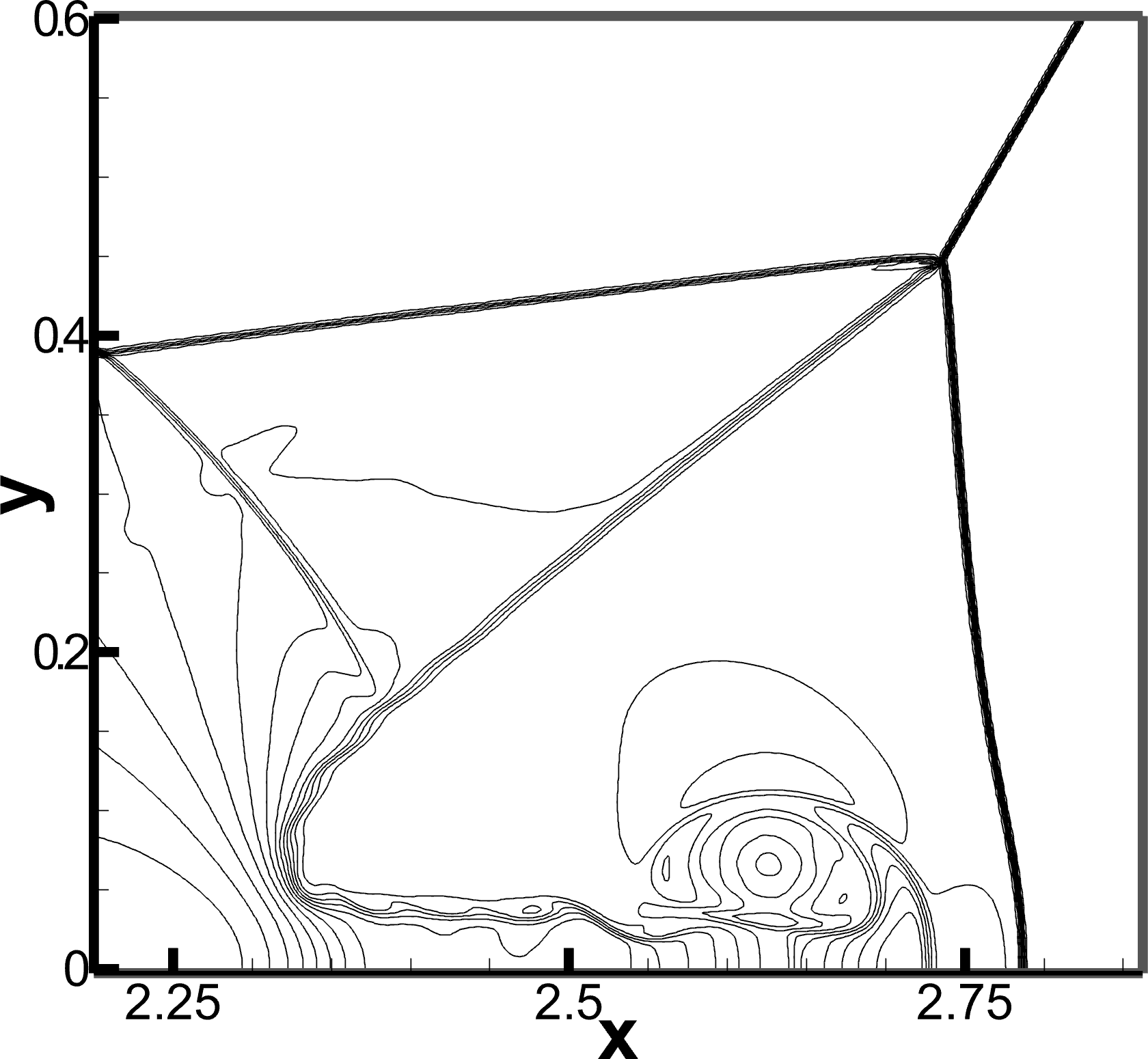}
    \caption{Third-order CWENO scheme.\label{fig:dmr_cweno_3rd_480_cv}}
    \end{subfigure}
    \begin{subfigure}[b]{\columnwidth}
    \includegraphics[width=0.61\columnwidth]{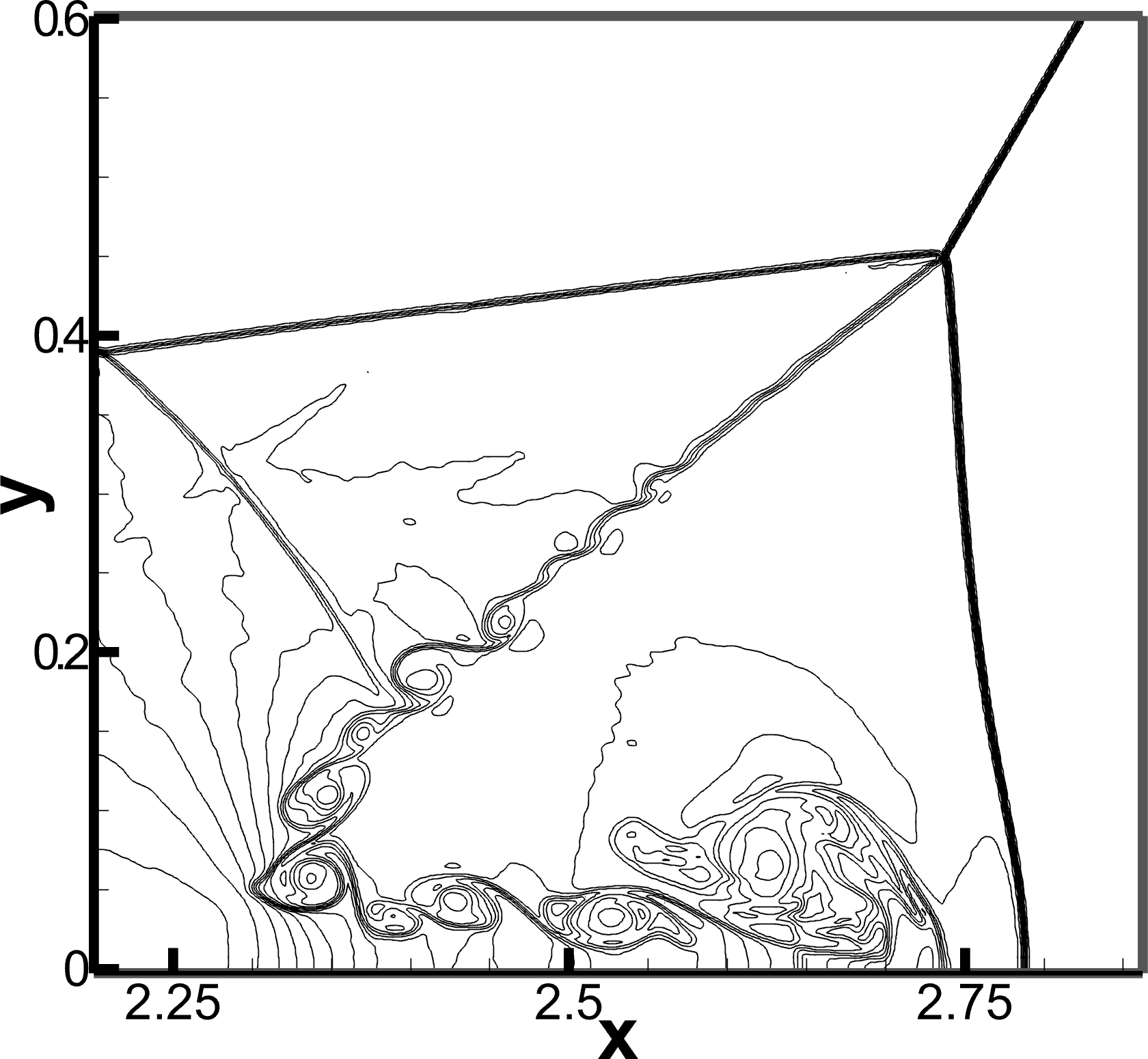}
    \caption{Fifth-order hybrid CLS-CWENO scheme.\label{fig:dmr_clscweno_5th_480_cv}}
    \end{subfigure}
    \begin{subfigure}[b]{\columnwidth}
    \includegraphics[width=0.61\columnwidth]{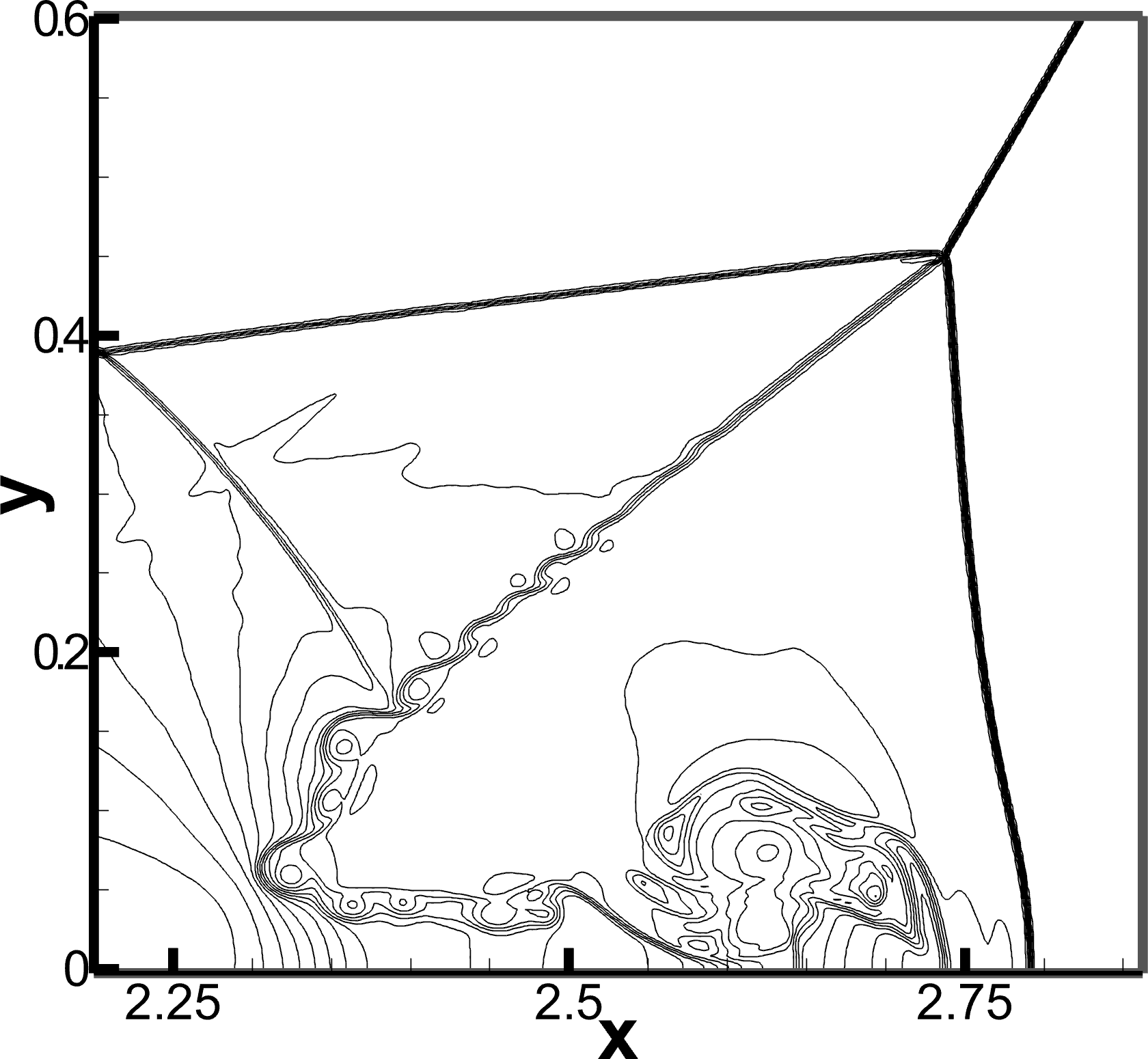}
    \caption{Fifth-order CWENO scheme.\label{fig:dmr_cweno_5th_480_cv}}
    \end{subfigure}
    \caption{Close view for the density contours of the double Mach reflection problem. Cell numer is $1920\times 480$.
    \label{fig:dmr_close_view_480}}
\end{figure}

\begin{figure}[!htbp]
  \centering
    \begin{subfigure}[b]{\columnwidth}
    \includegraphics[width=1.0\columnwidth]{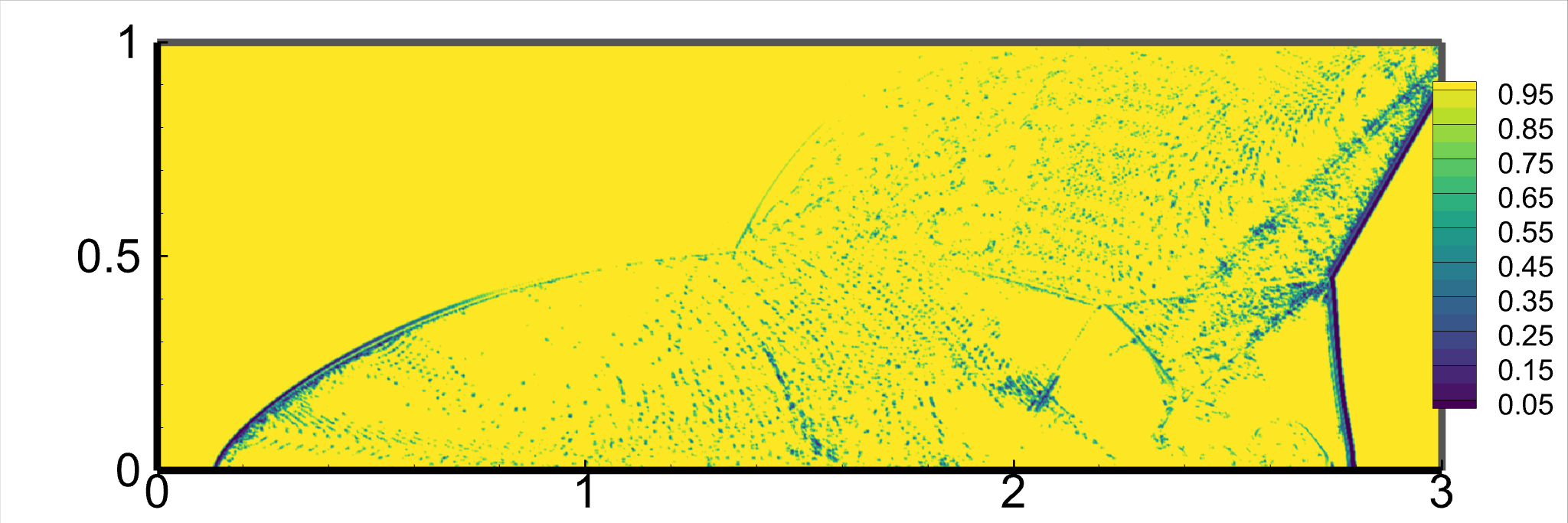}
    \caption{$\sigma^{\mathrm{Li}}$ in x direction. Third-order hybrid scheme.}
    \end{subfigure}
    \begin{subfigure}[b]{\columnwidth}
    \includegraphics[width=1.0\columnwidth]{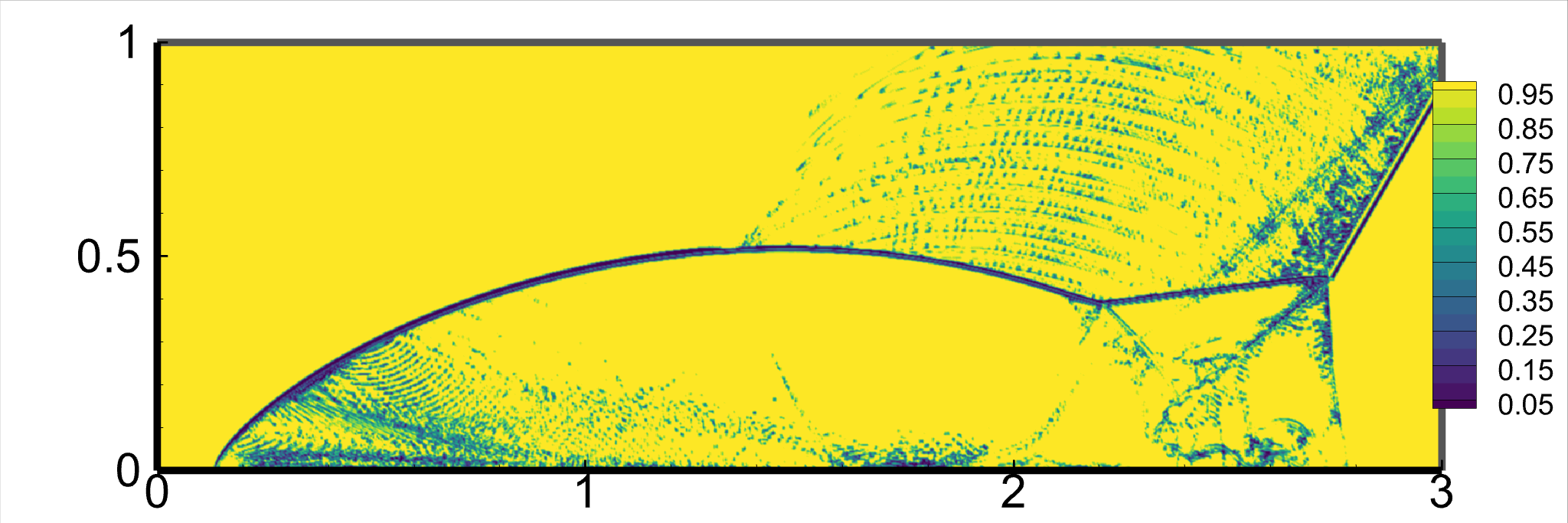}
    \caption{$\sigma^{\mathrm{Li}}$ in y direction. Third-order hybrid scheme.}
    \end{subfigure}
    \begin{subfigure}[b]{\columnwidth}
    \includegraphics[width=1.0\columnwidth]{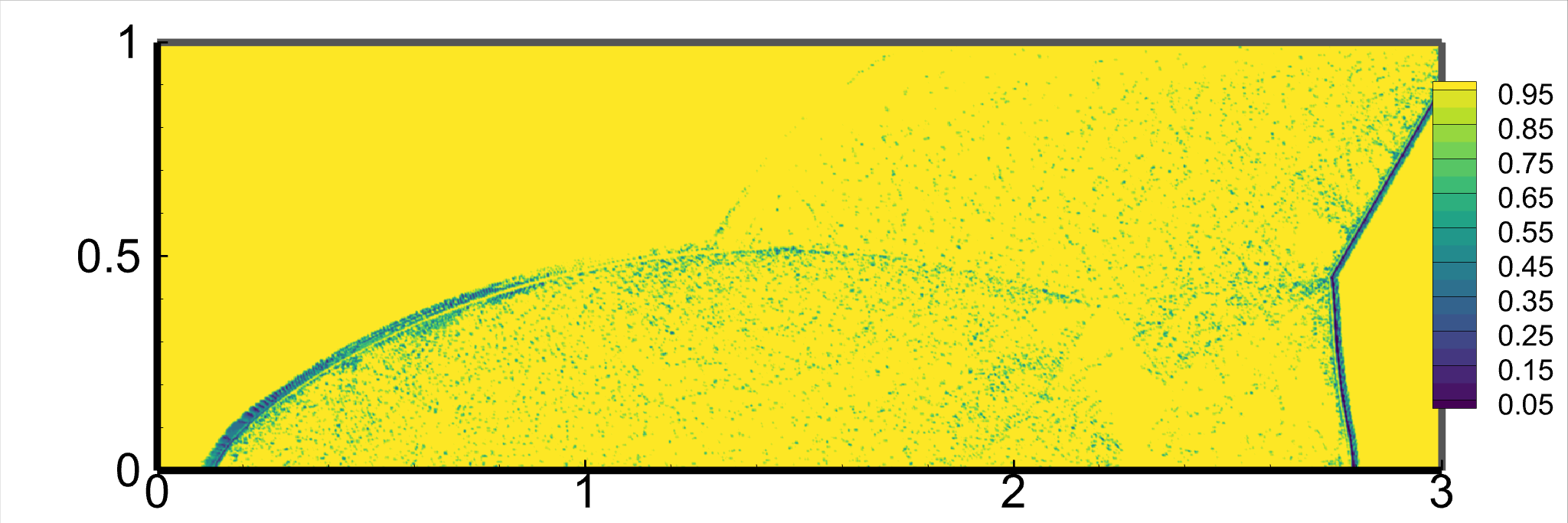}
    \caption{$\sigma^{\mathrm{Li}}$ in x direction. Fifth-order hybrid scheme.}
    \end{subfigure}
    \begin{subfigure}[b]{\columnwidth}
    \includegraphics[width=1.0\columnwidth]{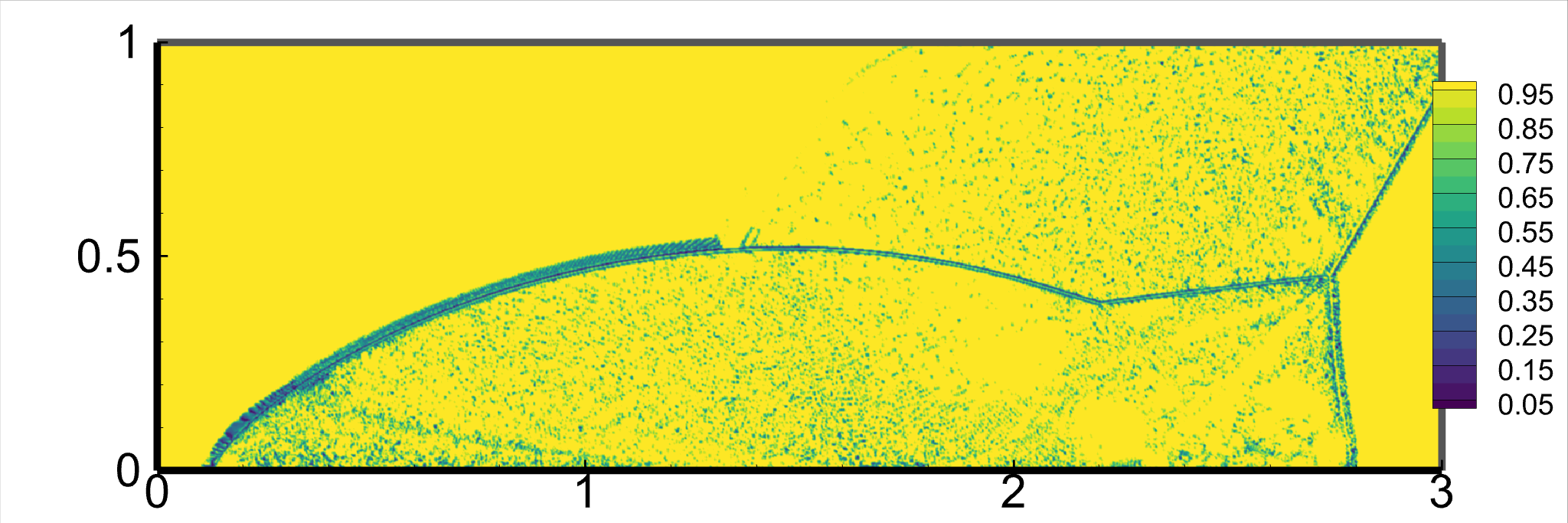}
    \caption{$\sigma^{\mathrm{Li}}$ in y direction. Fifth-order hybrid scheme.}
    \end{subfigure}
    \caption{Distribution of the shock detector in the hybrid scheme. Cell numer is $1920\times 480$.
    \label{fig:dmr_sd}}
\end{figure}

Double Mach reflection is a classic benchmark test for the compressible high-resolution schemes. A shock of Mach 10 moves right-downward with an inclined angle of 30 degrees with x-axis. The simulation domain is $[0,4]\times [0,1]$ and an inviscid wall locates at the bottom boundary along $x \geq 1/6$. The initial condition is
\begin{equation}
  \left[\rho, u, v, p\right] = \left\{
  \begin{array}{ll}
    8.0, 7.1447, -4.125, 116.5, & x \leq 1/6 + \sqrt{3}y/3,\\
    1.4, 0, 0, 1, & x > 1/6 + \sqrt{3}y/3,
  \end{array}
  \right.
\end{equation}

Figures \ref{fig:dmr} and \ref{fig:dmr_480} show the density contours at time $t_{end}=0.2$ with different schemes on grids with $\Delta x = 1/240$ and $\Delta x = 1/480$, respectively. Figures \ref{fig:dmr_close_view} and \ref{fig:dmr_close_view_480} show the close view around the Mach stem. With the increase of grid resolution, finer vortex structures are triggered along the slip line. For both the third- and fifth-order schemes, the hybrid one performs better than the CWENO schemes with the same order of accuracy.

Figure \ref{fig:dmr_sd} presents the distribution of the proposed hybrid schemes in both $x$ and $y$ directions. The discontinuities are clearly identified. However, it should also be noted that, some smooth cells with high-frequency variables are also treated as troubled cell where CWENO schemes are applied.

\subsection{2D Riemann problem\label{sec:2drp}\cite{huang2018high}}
\begin{figure}[!htbp]
  \centering
    \begin{subfigure}[b]{\columnwidth}
    \includegraphics[width=0.61\columnwidth]{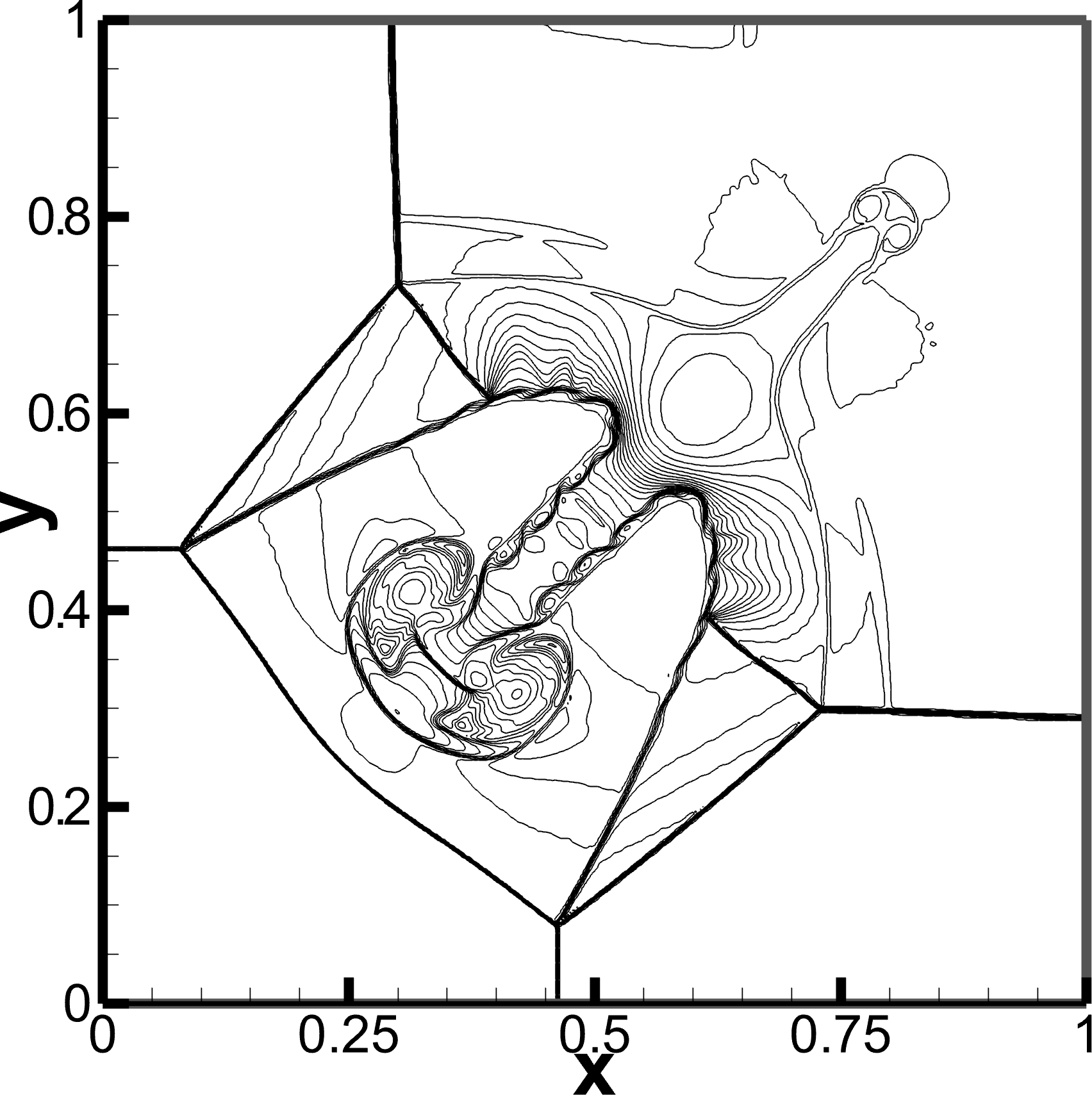}
    \caption{Third-order hybrid CLS-CWENO scheme.\label{fig:2d_clscweno_3rd}}
    \end{subfigure}
    \begin{subfigure}[b]{\columnwidth}
    \includegraphics[width=0.61\columnwidth]{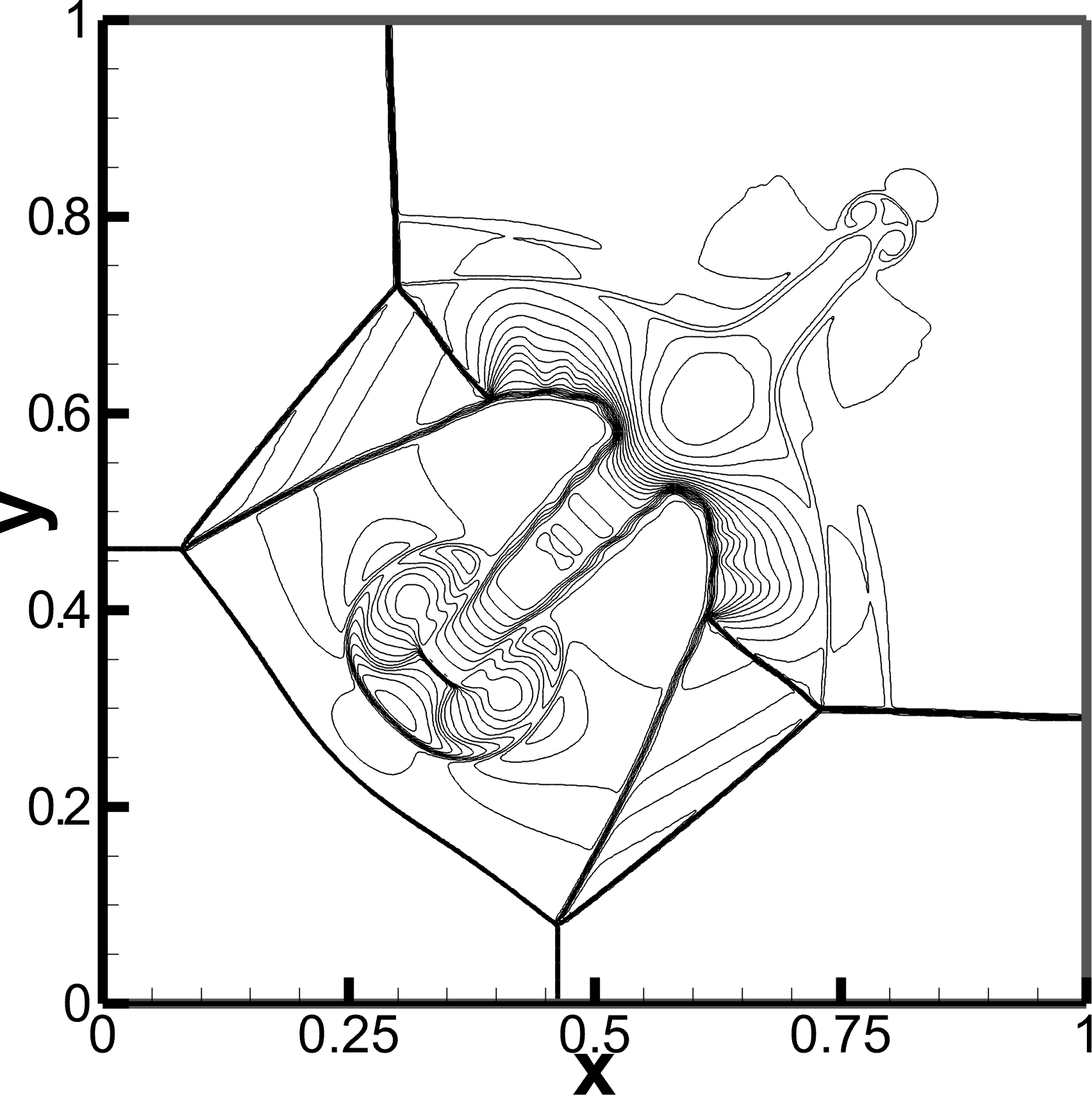}
    \caption{Third-order CWENO scheme.\label{fig:2d_cweno_3rd}}
    \end{subfigure}
    \begin{subfigure}[b]{\columnwidth}
    \includegraphics[width=0.61\columnwidth]{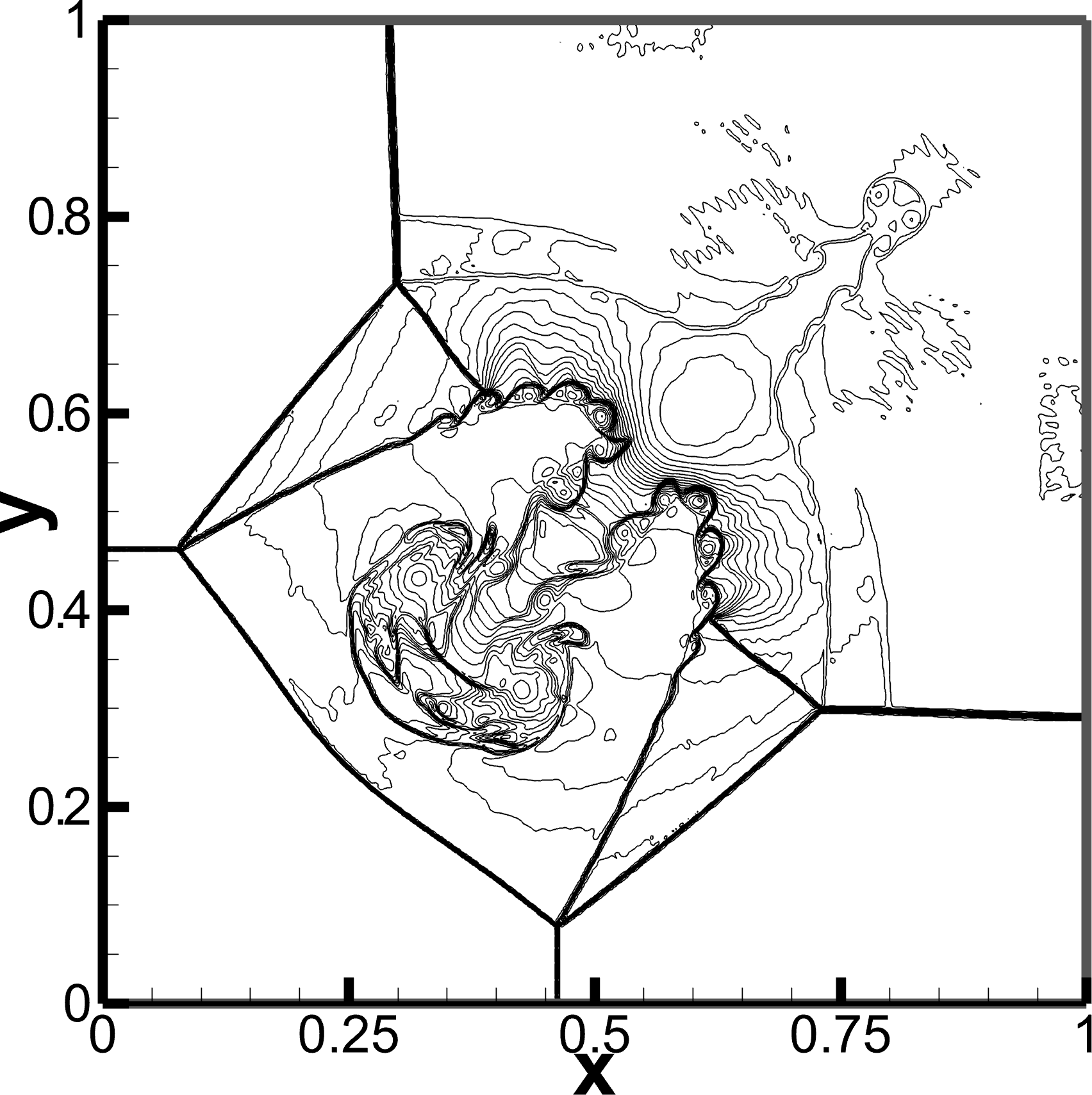}
    \caption{Fifth-order hybrid CLS-CWENO scheme.\label{fig:2d_clscweno_5th}}
    \end{subfigure}
    \begin{subfigure}[b]{\columnwidth}
    \includegraphics[width=0.61\columnwidth]{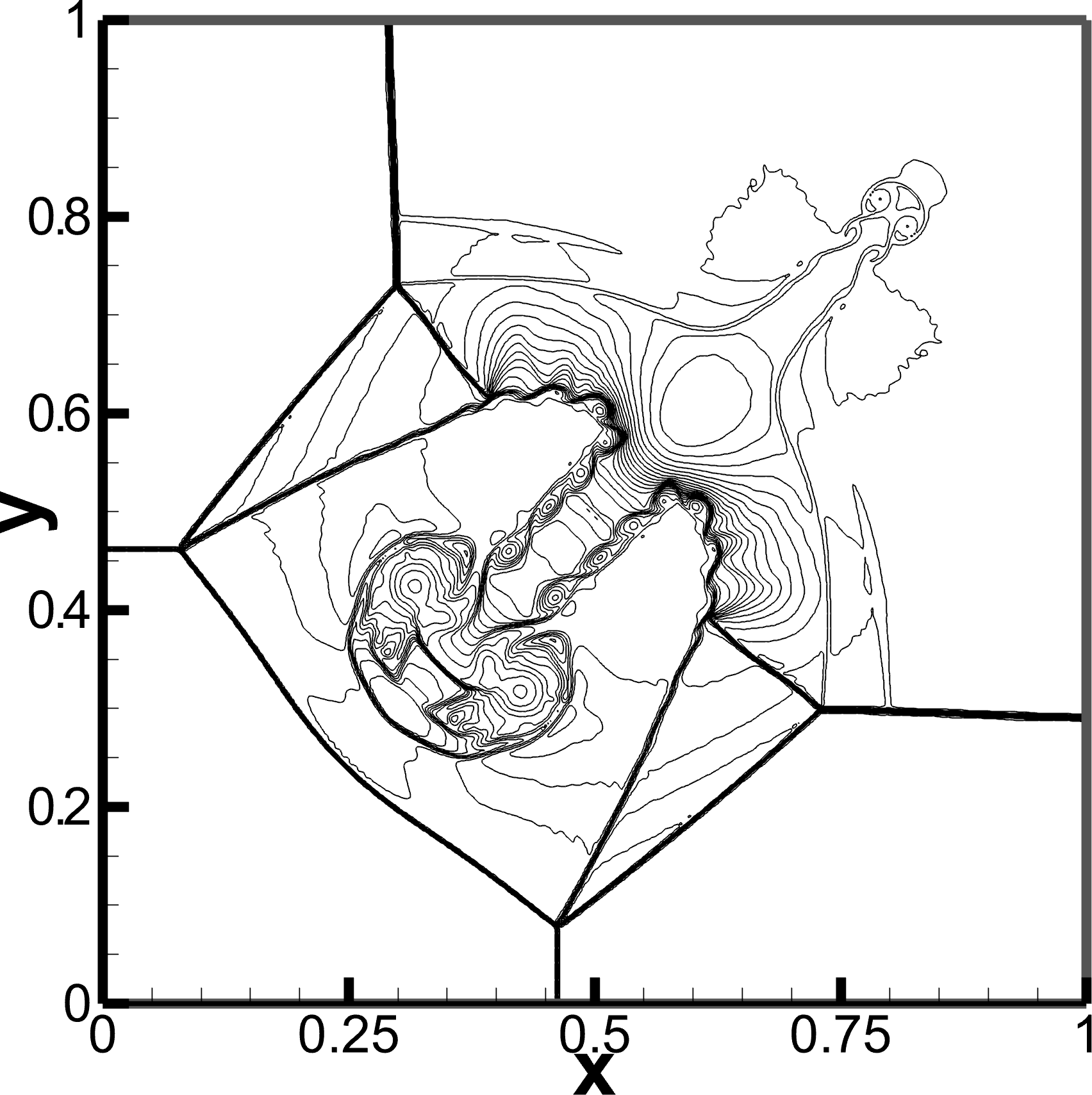}
    \caption{Fifth-order CWENO scheme.\label{fig:2d_cweno_5th}}
    \end{subfigure}
    \caption{Density contours for the 2D Riemann problem with 30 lines from 0.18 to 1.7.
    \label{fig:2driemann}}
\end{figure}

The initial condition for the 2D Riemann problem is
\begin{equation}
  \begin{aligned}
  &\left[\rho, u, v, p\right]  = \\
  &\,\, \left\{
  \begin{array}{ll}
    1.5, 0, 0, 1.5, & 0.8 \leq x \leq 1, 0.8 \leq y \leq 1,\\
    0.5323, 1.206, 0, 0.3, &0 \leq x < 0.8, 0.8 \leq y \leq 1,\\
    0.138, 1.206, 1.206, 0.029, &0 \leq x < 0.8, 0 \leq y < 0.8,\\
    0.5323, 0, 1.206, 0.3, & 0.8 \leq x \leq 1, 0 \leq y < 0.8.\\
  \end{array}
  \right.
\end{aligned}
\end{equation}

Four shock waves propagate and interact with each other, forming complex shock diffraction patterns, leading to the emergence of Mach stem structures and Kelvin-Helmholtz roll-ups along vertical slip lines. 

The computational domain consists of $400\times 400$ uniform cells.  The simulation advances to $t_{end} = 0.8$ with Courant number as $0.6$.

Figure \ref{fig:2driemann} presents the density contours of different schemes.  As shown in Fig. \ref{fig:2driemann}, the proposed third- and fifth-order hybrid CLS-CWENO schemes capture the discontinuities robustly and resolve the fine structures with high resolution.
Compared with the CWENO schemes, the hybrid CLS-CWENO schemes show superiority in describing the multiscale flow structures.

\begin{figure}[!htbp]
  \centering
    \begin{subfigure}[b]{\columnwidth}
    \includegraphics[width=0.61\columnwidth]{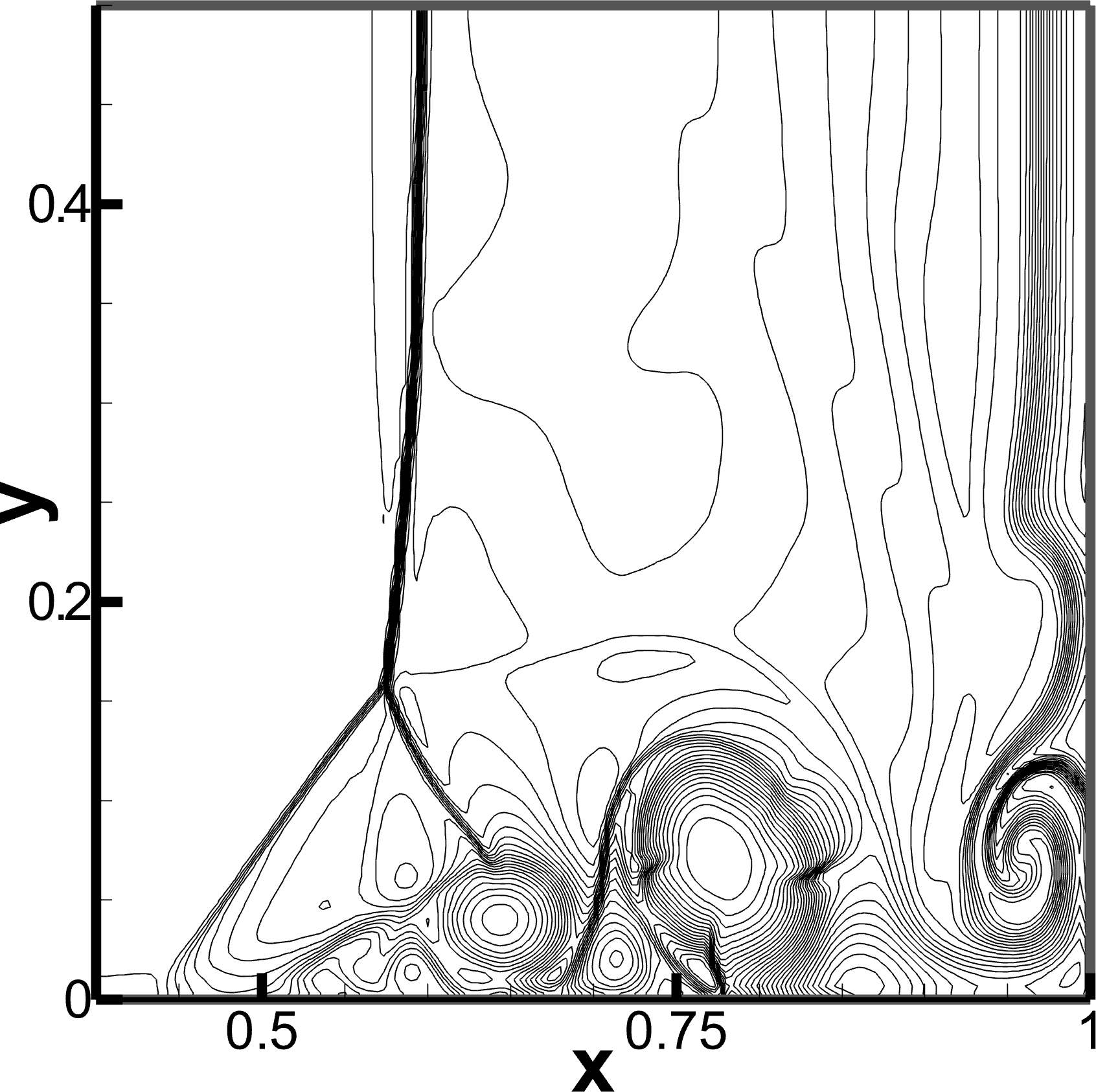}
    \caption{Third-order hybrid CLS-CWENO scheme.\label{fig:vst_clscweno_3rd}}
    \end{subfigure}
    \begin{subfigure}[b]{\columnwidth}
    \includegraphics[width=0.61\columnwidth]{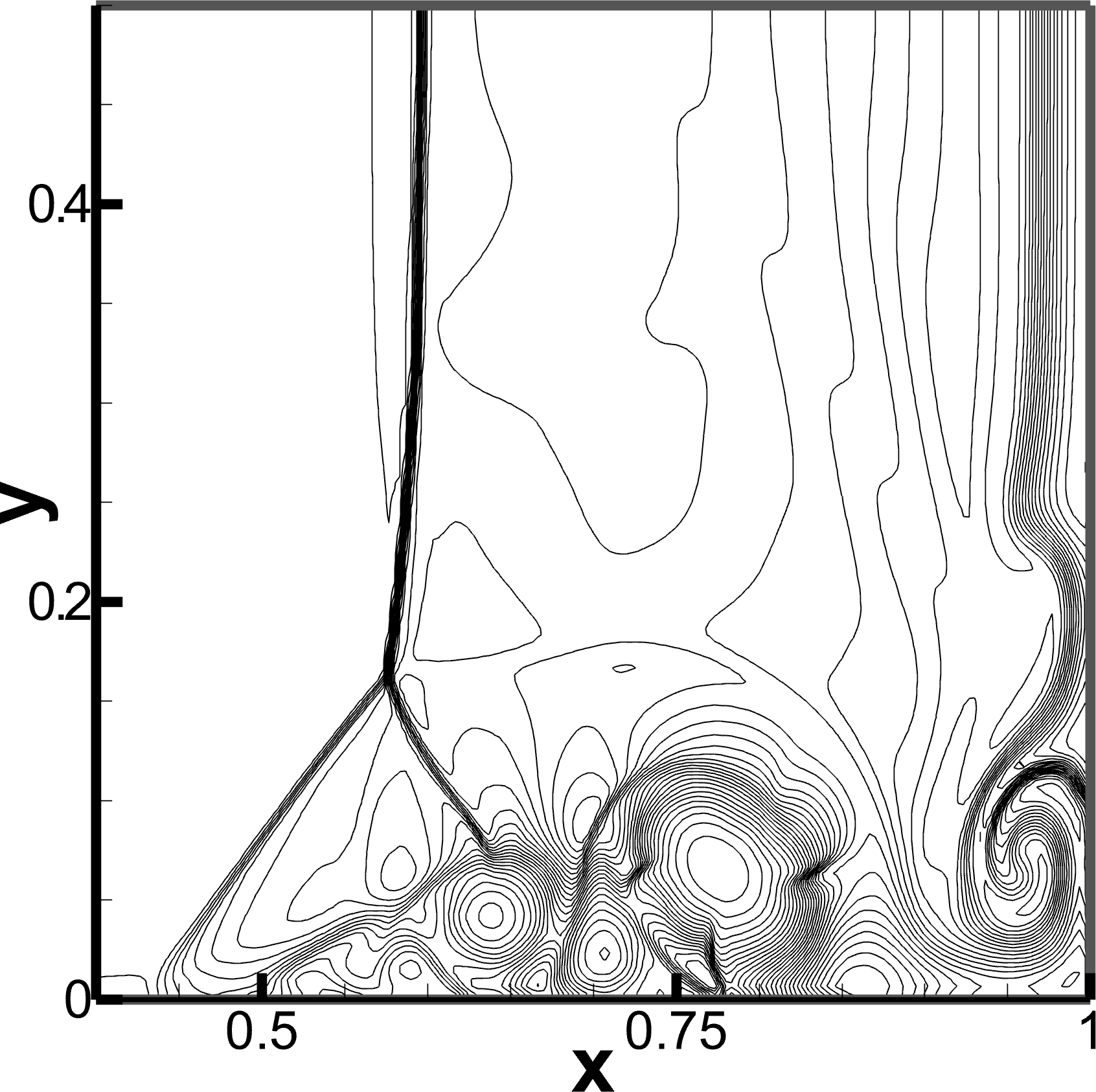}
    \caption{Third-order CWENO scheme.\label{fig:vst_cweno_3rd}}
    \end{subfigure}
    \begin{subfigure}[b]{\columnwidth}
    \includegraphics[width=0.61\columnwidth]{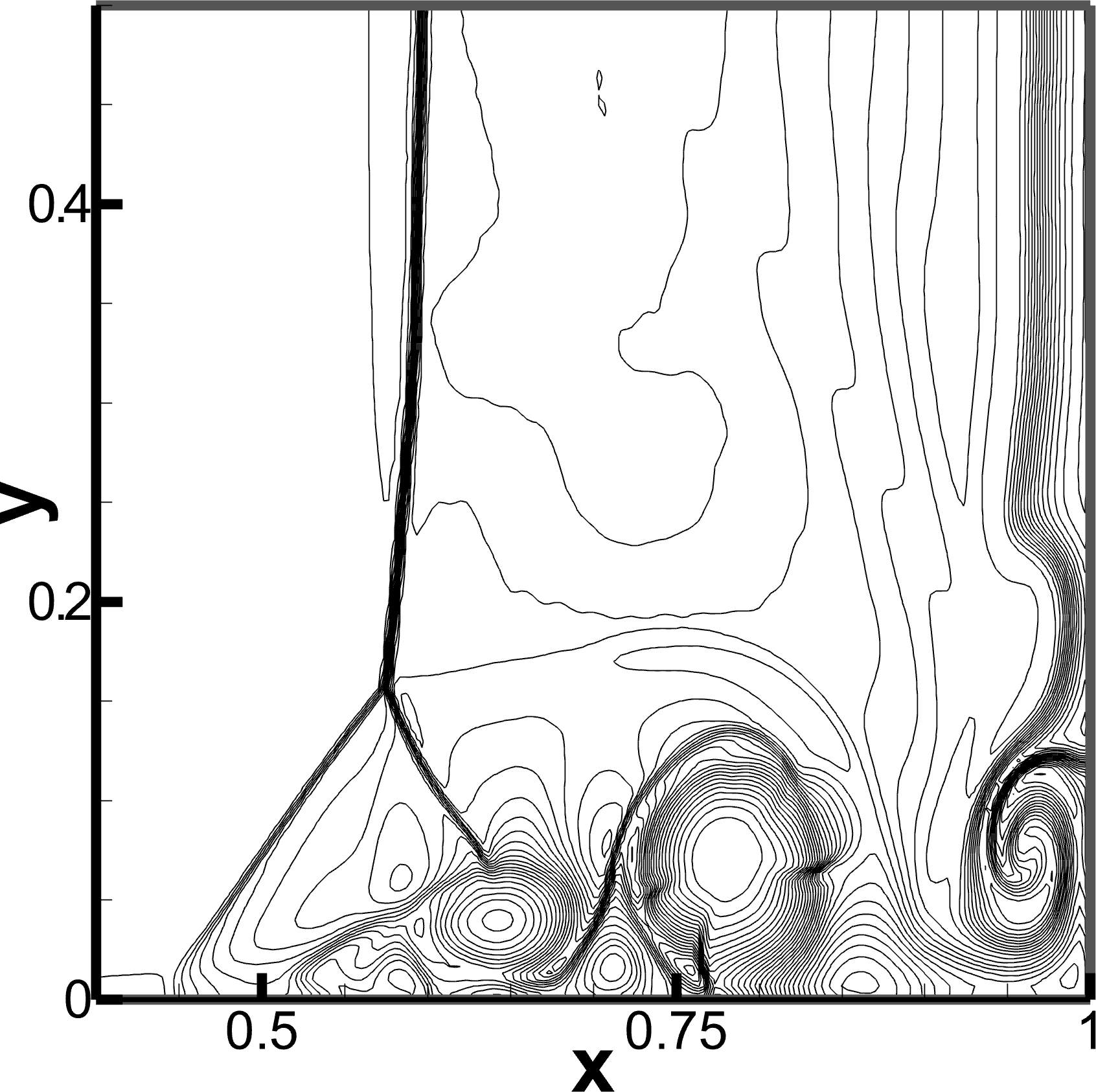}
    \caption{Fifth-order hybrid CLS-CWENO scheme.\label{fig:vst_clscweno_5th}}
    \end{subfigure}
    \begin{subfigure}[b]{\columnwidth}
    \includegraphics[width=0.61\columnwidth]{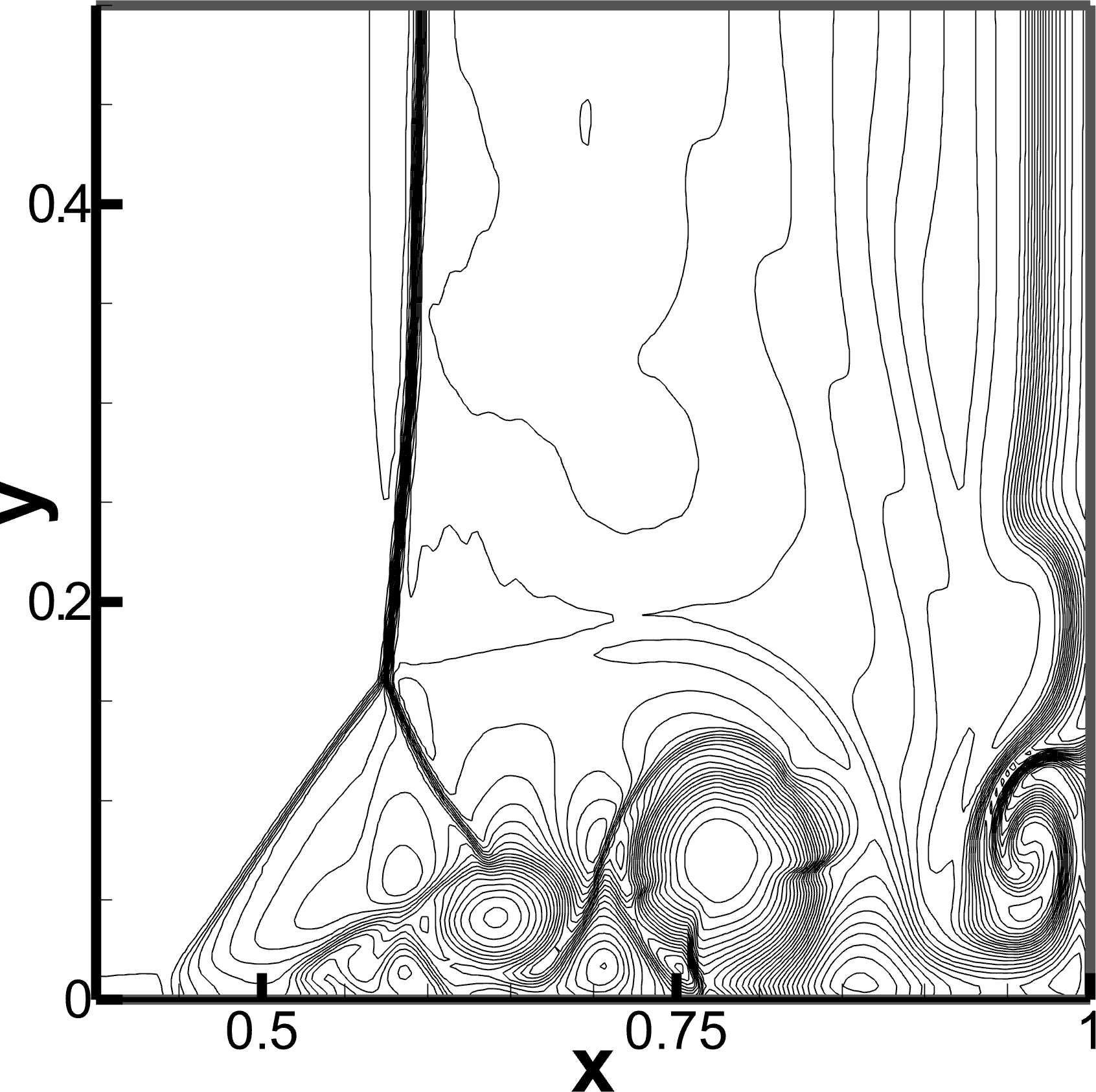}
    \caption{Fifth-order CWENO scheme.\label{fig:vst_cweno_5th}}
    \end{subfigure}
    \caption{Density contours for the viscous shock tube with $N_x \times N_y = 300 \times 300$ with 30 lines from 21 to 131.
    \label{fig:vst}}
\end{figure}

\begin{figure}[!htbp]
  \centering
    \begin{subfigure}[b]{\columnwidth}
    \includegraphics[width=0.61\columnwidth]{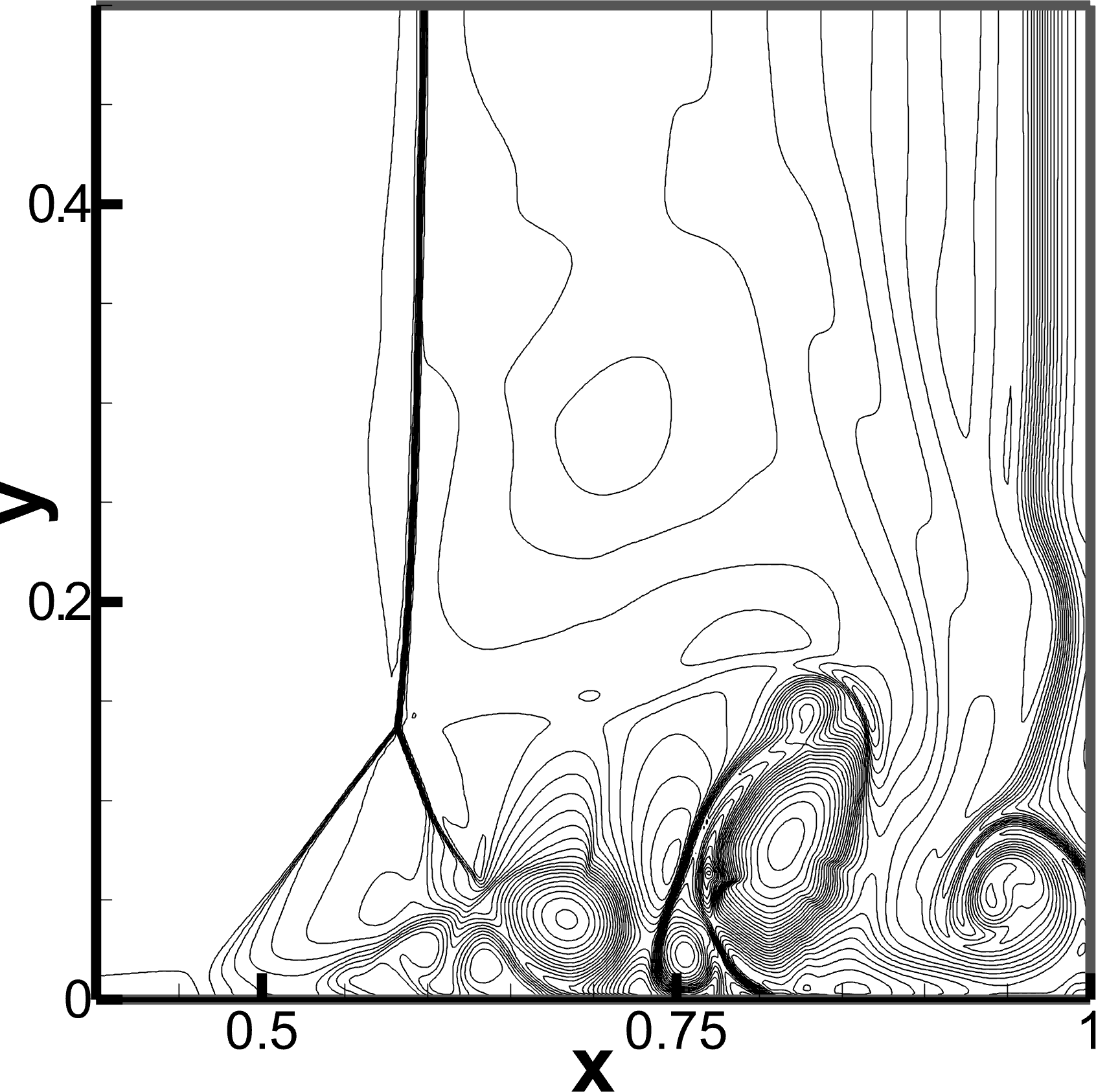}
    \caption{Third-order hybrid CLS-CWENO scheme.\label{fig:vst_clscweno_3rd_600}}
    \end{subfigure}
    \begin{subfigure}[b]{\columnwidth}
    \includegraphics[width=0.61\columnwidth]{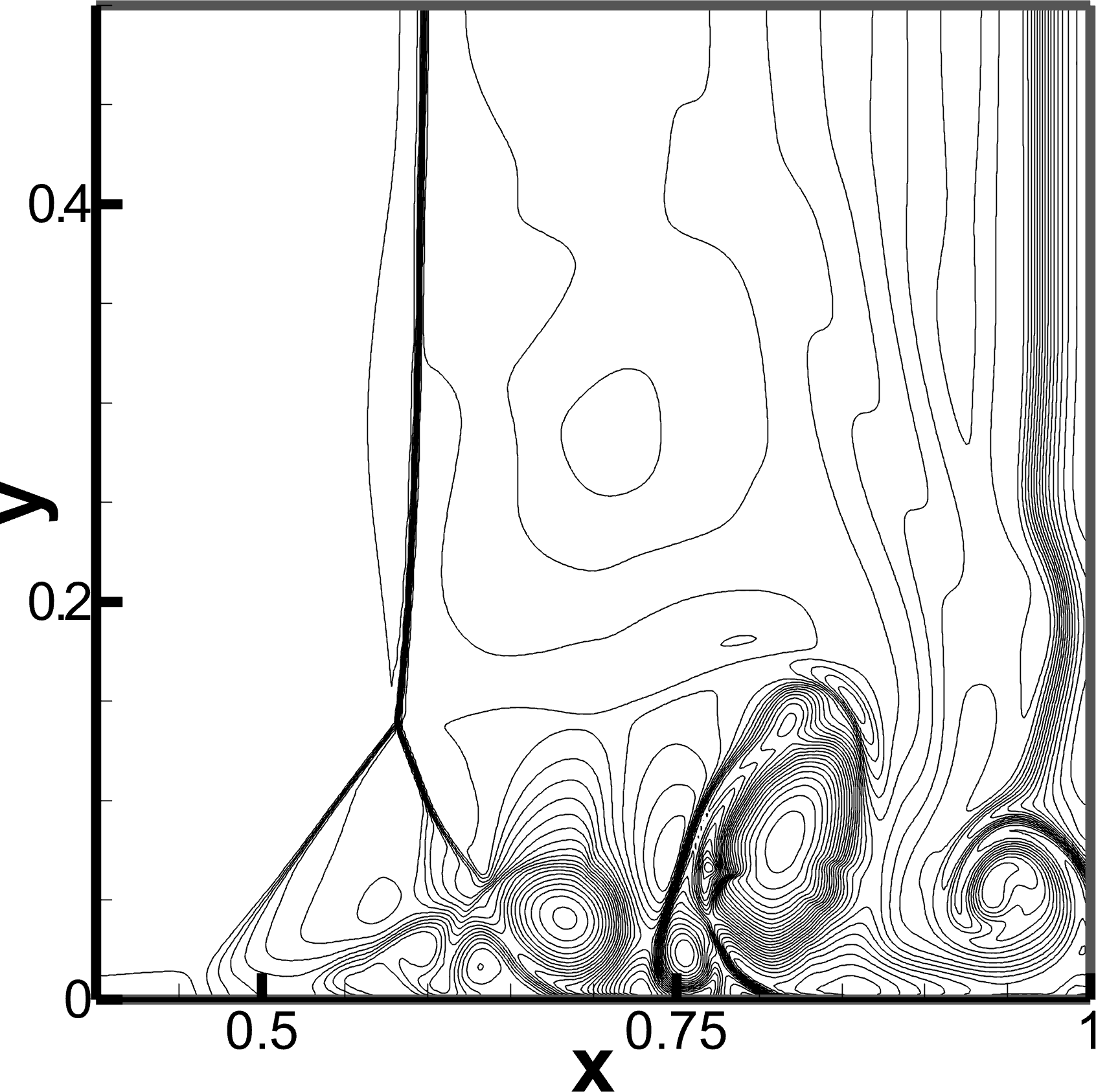}
    \caption{Third-order CWENO scheme.\label{fig:vst_cweno_3rd_600}}
    \end{subfigure}
    \begin{subfigure}[b]{\columnwidth}
    \includegraphics[width=0.61\columnwidth]{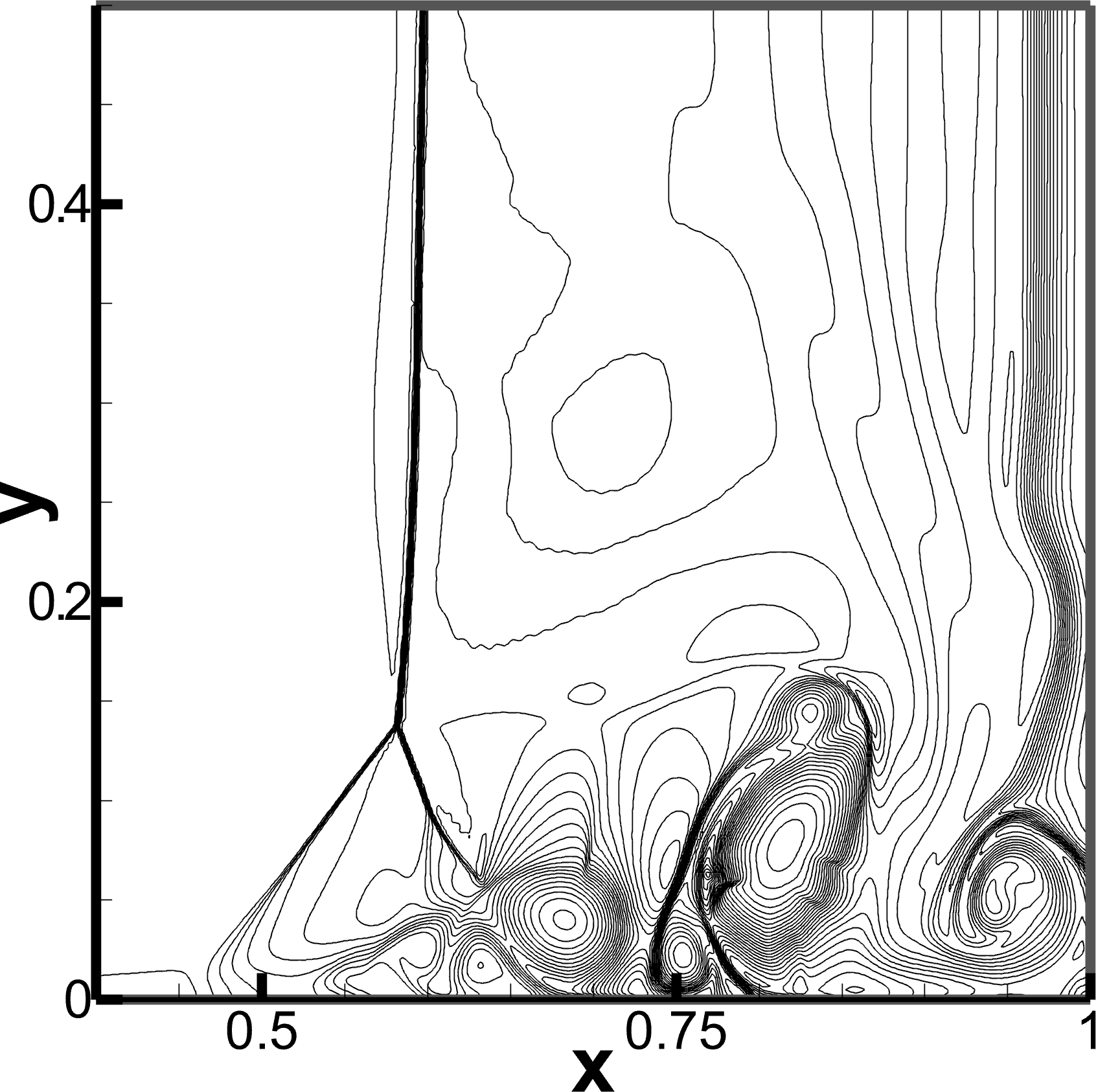}
    \caption{Fifth-order hybrid CLS-CWENO scheme.\label{fig:vst_clscweno_5th_600}}
    \end{subfigure}
    \begin{subfigure}[b]{\columnwidth}
    \includegraphics[width=0.61\columnwidth]{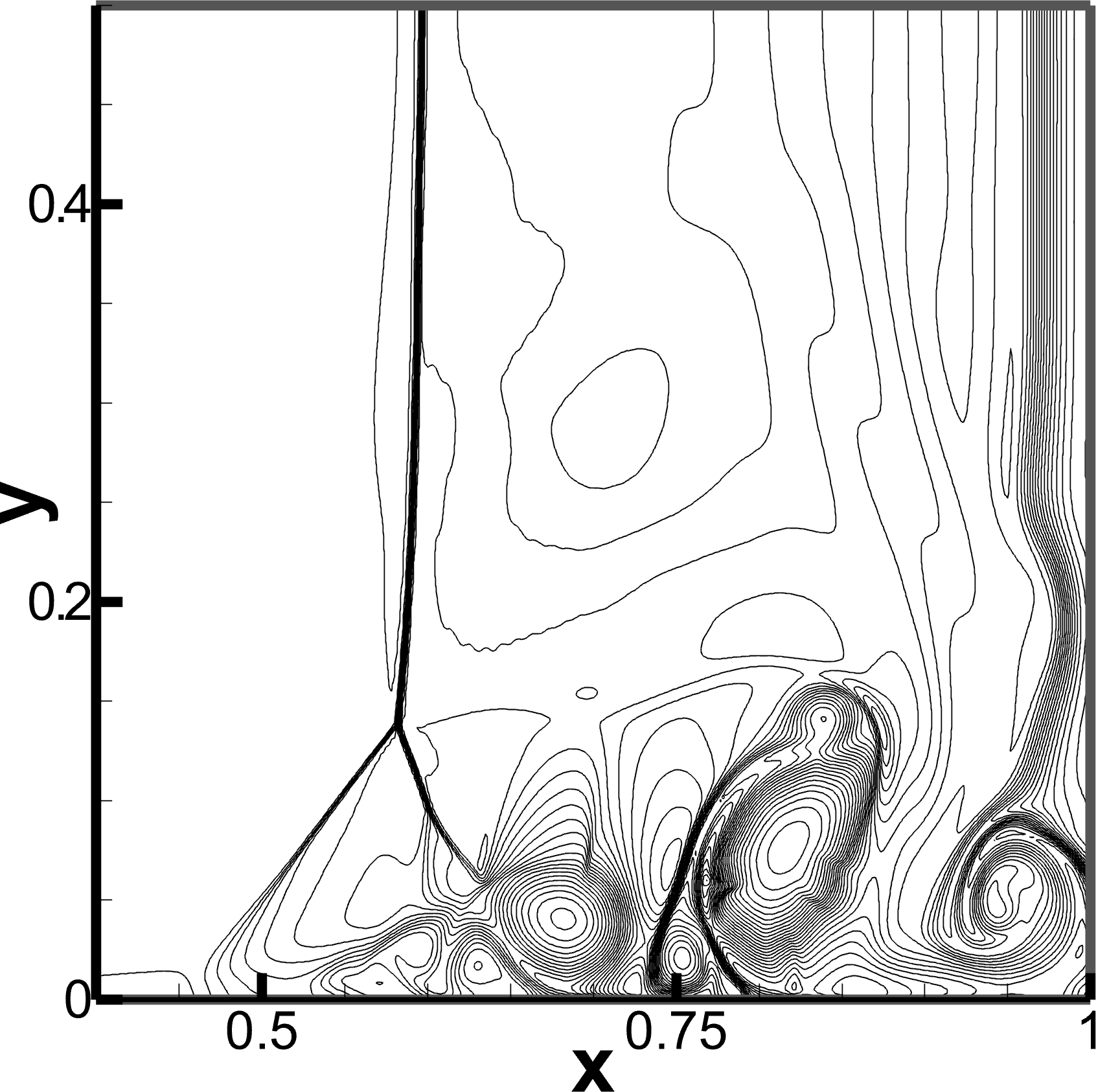}
    \caption{Fifth-order CWENO scheme.\label{fig:vst_cweno_5th_600}}
    \end{subfigure}
    \caption{Density contours for the viscous shock tube with $N_x \times N_y = 600 \times 600$ with 30 lines from 21 to 131.
    \label{fig:vst_600}}
\end{figure}

\subsection{Viscous shock tube\label{sec:vst}}

\begin{table}
\caption{Height of main vortex in viscous shock tube.\label{tab:viscousbubbleheight}}
\begin{ruledtabular}
\begin{tabular}{ccccc}
Grid &CLS-CWENO3 & CWENO3 & CLS-CWENO5 & CWENO5\\
\hline
$300\times 300$ & 0.134 & 0.121 & 0.137 & 0.136  \\
$600\times 600$ & 0.164 & 0.162 & 0.164 & 0.159  \\
\end{tabular}
\end{ruledtabular}
\end{table}

Eventually, a viscous shock tube problem is taken to test the performance of the proposed hybrid CLS-CWENO schemes in viscous problem.
The viscous shock tube problem is characterized by the interaction of strong shock, boundary layer, and vortex. The computational domain is a square defined by $[0,1]\times[0,1]$ with adiabatic walls. The initial condition is given by
\begin{equation}
  \left[\rho, u, v, p\right]  = \left\{
    \begin{array}{ll}
      120,0,0,\frac{120}{\gamma},& x < 0.5,\\
      1.2,0,0,\frac{1.2}{\gamma}, & x\geq 0.5,
    \end{array}
  \right.
\end{equation}
where $\gamma=1.4$.
Initially, the perfect gas is separated into left and right parts with different pressure.  Upon start, a shock wave propagates rightward, followed by a contact discontinuity. The moving shock and contact discontinuity interact with the horizontal viscous walls, forming a thin boundary layer. When the shock reflects from the right wall, it interacts with both the contact discontinuity and the boundary layer, generating a $\lambda$-shaped shock structure. This structure further induces boundary layer separation bubbles, unstable slip lines, and vortex structures downstream of the bubbles.

The Reynolds number tested in this section is 200 and the Prandtl number is 0.73. The computational domain is discretized by $300\times 300$ and $600\times 600$ uniform control volumes. The simulation end time is $t_{end} = 1.0$ with Courant number as 0.6. 
When the number of grids is $300 \times 300$, the boundary layer is not fully resolved and the main vortex does not converge to the converging result shown as in the work of Daru and Tenaud\cite{daru_numerical_2009}. When the grid number is increased to $600 \times 600$, there is a secondary vortex generated near the main vortex and the height of the main vortex becomes closer to the reference result\cite{daru_numerical_2009}, which is about 0.165.

Table \ref{tab:viscousbubbleheight} shows the height of the main bubble in Figs. \ref{fig:vst} and \ref{fig:vst_600}, which confirms the better performance of the proposed hybrid schemes.

In the end, the simulation time of all the 2D cases is shown in Tab. \ref{tab:simtime}. The time excludes the subroutines of pre- and post-processes. Compared to the accuracy improvements of the hybrid schemes, the slight overhead in simulation time is thought to be acceptable in authors' view.
\begin{table}
\caption{Simulation time of 2D cases. Unit: s.\label{tab:simtime}}
\begin{ruledtabular}
\begin{tabular}{lcccc}
Case &CLS-CWENO3 & CWENO3 & CLS-CWENO5 & CWENO5\\
\hline
Sec. \ref{sec:iv} & 193 &	153 &	247 &	211 \\
Sec. \ref{sec:dsl} & 207	& 178& 	280& 	253.2\\
Sec. \ref{sec:dmr} & 2346	& 1574	& 3412& 	2181\\
Sec. \ref{sec:2drp} & 1545	& 1293	& 2502	& 1776\\
Sec. \ref{sec:vst} &1402 &1184 	& 	2007& 	1530 \\
\end{tabular}
\end{ruledtabular}
\end{table}

\section{Conclusion}
In this paper, a series of third- and fifth-order compact least-squares schemes hybridized with CWENO schemes are proposed. The hybrid schemes not only capture the discontinuities robustly but also resolve the flow structures in a broad range of bandwidths.
Improvements are made in several aspects. Firstly, the free parameters of the compact least-squares schemes are optimized; secondly, a more accurate shock detector proposed in our previous work is analyzed and validated; thirdly, the compact least-squares schemes are hybridized with the CWENO schemes. At last, the proposed hybrid schemes are tested through a series of carefully and thoroughly testing including linear and nonlinear, smooth and non-smooth, inviscid and viscous equations for both 1D and 2D problems. The scheme is promising in the simulation of direct numerical simulation of compressible flows.

\section*{Availability of data}
The data that support the findings of this study are available from the corresponding author upon reasonable request.

\section*{Acknowledgements}
This work was supported by the National Natural Science Foundation of China (No. 12102211), the Science and Technology Innovation 2025 Major Project of Ningbo, China (No. 2022Z213), the Major Program of the National Natural Science Foundation of China (Nos. 12292980,12292982) and Project in Northeast Normal University (No. GFPY202505).
\section*{References}
\bibliography{CLSWENO}

\begin{thebibliography}{47}%
\makeatletter
\providecommand \@ifxundefined [1]{%
 \@ifx{#1\undefined}
}%
\providecommand \@ifnum [1]{%
 \ifnum #1\expandafter \@firstoftwo
 \else \expandafter \@secondoftwo
 \fi
}%
\providecommand \@ifx [1]{%
 \ifx #1\expandafter \@firstoftwo
 \else \expandafter \@secondoftwo
 \fi
}%
\providecommand \natexlab [1]{#1}%
\providecommand \enquote  [1]{``#1''}%
\providecommand \bibnamefont  [1]{#1}%
\providecommand \bibfnamefont [1]{#1}%
\providecommand \citenamefont [1]{#1}%
\providecommand \href@noop [0]{\@secondoftwo}%
\providecommand \href [0]{\begingroup \@sanitize@url \@href}%
\providecommand \@href[1]{\@@startlink{#1}\@@href}%
\providecommand \@@href[1]{\endgroup#1\@@endlink}%
\providecommand \@sanitize@url [0]{\catcode `\\12\catcode `\$12\catcode `\&12\catcode `\#12\catcode `\^12\catcode `\_12\catcode `\%12\relax}%
\providecommand \@@startlink[1]{}%
\providecommand \@@endlink[0]{}%
\providecommand \url  [0]{\begingroup\@sanitize@url \@url }%
\providecommand \@url [1]{\endgroup\@href {#1}{\urlprefix }}%
\providecommand \urlprefix  [0]{URL }%
\providecommand \Eprint [0]{\href }%
\providecommand \doibase [0]{http://dx.doi.org/}%
\providecommand \selectlanguage [0]{\@gobble}%
\providecommand \bibinfo  [0]{\@secondoftwo}%
\providecommand \bibfield  [0]{\@secondoftwo}%
\providecommand \translation [1]{[#1]}%
\providecommand \BibitemOpen [0]{}%
\providecommand \bibitemStop [0]{}%
\providecommand \bibitemNoStop [0]{.\EOS\space}%
\providecommand \EOS [0]{\spacefactor3000\relax}%
\providecommand \BibitemShut  [1]{\csname bibitem#1\endcsname}%
\let\auto@bib@innerbib\@empty
\bibitem [{\citenamefont {Tam}\ and\ \citenamefont {Webb}(1993)}]{tam1993dispersion}%
  \BibitemOpen
  \bibfield  {author} {\bibinfo {author} {\bibfnamefont {C.~K.}\ \bibnamefont {Tam}}\ and\ \bibinfo {author} {\bibfnamefont {J.~C.}\ \bibnamefont {Webb}},\ }\bibfield  {title} {\enquote {\bibinfo {title} {Dispersion-relation-preserving finite difference schemes for computational acoustics},}\ }\href@noop {} {\bibfield  {journal} {\bibinfo  {journal} {Journal of Computational Physics}\ }\textbf {\bibinfo {volume} {107}},\ \bibinfo {pages} {262--281} (\bibinfo {year} {1993})}\BibitemShut {NoStop}%
\bibitem [{\citenamefont {Cheong}\ and\ \citenamefont {Lee}(2001)}]{cheong2001grid}%
  \BibitemOpen
  \bibfield  {author} {\bibinfo {author} {\bibfnamefont {C.}~\bibnamefont {Cheong}}\ and\ \bibinfo {author} {\bibfnamefont {S.}~\bibnamefont {Lee}},\ }\bibfield  {title} {\enquote {\bibinfo {title} {Grid-optimized dispersion-relation-preserving schemes on general geometries for computational aeroacoustics},}\ }\href@noop {} {\bibfield  {journal} {\bibinfo  {journal} {Journal of Computational Physics}\ }\textbf {\bibinfo {volume} {174}},\ \bibinfo {pages} {248--276} (\bibinfo {year} {2001})}\BibitemShut {NoStop}%
\bibitem [{\citenamefont {Sun}\ \emph {et~al.}(2011)\citenamefont {Sun}, \citenamefont {Ren}, \citenamefont {Larricq}, \citenamefont {Zhang},\ and\ \citenamefont {Yang}}]{sun2011class}%
  \BibitemOpen
  \bibfield  {author} {\bibinfo {author} {\bibfnamefont {Z.-S.}\ \bibnamefont {Sun}}, \bibinfo {author} {\bibfnamefont {Y.-X.}\ \bibnamefont {Ren}}, \bibinfo {author} {\bibfnamefont {C.}~\bibnamefont {Larricq}}, \bibinfo {author} {\bibfnamefont {S.-Y.}\ \bibnamefont {Zhang}}, \ and\ \bibinfo {author} {\bibfnamefont {Y.-C.}\ \bibnamefont {Yang}},\ }\bibfield  {title} {\enquote {\bibinfo {title} {A class of finite difference schemes with low dispersion and controllable dissipation for {DNS} of compressible turbulence},}\ }\href@noop {} {\bibfield  {journal} {\bibinfo  {journal} {Journal of Computational Physics}\ }\textbf {\bibinfo {volume} {230}},\ \bibinfo {pages} {4616--4635} (\bibinfo {year} {2011})}\BibitemShut {NoStop}%
\bibitem [{\citenamefont {Sun}\ \emph {et~al.}(2014)\citenamefont {Sun}, \citenamefont {Luo}, \citenamefont {Ren},\ and\ \citenamefont {Zhang}}]{sun2014sixth}%
  \BibitemOpen
  \bibfield  {author} {\bibinfo {author} {\bibfnamefont {Z.-S.}\ \bibnamefont {Sun}}, \bibinfo {author} {\bibfnamefont {L.}~\bibnamefont {Luo}}, \bibinfo {author} {\bibfnamefont {Y.-X.}\ \bibnamefont {Ren}}, \ and\ \bibinfo {author} {\bibfnamefont {S.-Y.}\ \bibnamefont {Zhang}},\ }\bibfield  {title} {\enquote {\bibinfo {title} {A sixth order hybrid finite difference scheme based on the minimized dispersion and controllable dissipation technique},}\ }\href@noop {} {\bibfield  {journal} {\bibinfo  {journal} {Journal of Computational Physics}\ }\textbf {\bibinfo {volume} {270}},\ \bibinfo {pages} {238--254} (\bibinfo {year} {2014})}\BibitemShut {NoStop}%
\bibitem [{\citenamefont {Zeng}\ \emph {et~al.}(2025)\citenamefont {Zeng}, \citenamefont {Liu}, \citenamefont {Zeng}, \citenamefont {Pan}, \citenamefont {Yin},\ and\ \citenamefont {Ren}}]{zeng2025high}%
  \BibitemOpen
  \bibfield  {author} {\bibinfo {author} {\bibfnamefont {W.-G.}\ \bibnamefont {Zeng}}, \bibinfo {author} {\bibfnamefont {L.}~\bibnamefont {Liu}}, \bibinfo {author} {\bibfnamefont {L.-J.}\ \bibnamefont {Zeng}}, \bibinfo {author} {\bibfnamefont {J.-H.}\ \bibnamefont {Pan}}, \bibinfo {author} {\bibfnamefont {J.-P.}\ \bibnamefont {Yin}}, \ and\ \bibinfo {author} {\bibfnamefont {Y.-X.}\ \bibnamefont {Ren}},\ }\bibfield  {title} {\enquote {\bibinfo {title} {High-order alternative formulation of weighted essentially non-oscillatory scheme with minimized dispersion and controllable dissipation for compressible flows},}\ }\href@noop {} {\bibfield  {journal} {\bibinfo  {journal} {International Journal for Numerical Methods in Fluids}\ } (\bibinfo {year} {2025})}\BibitemShut {NoStop}%
\bibitem [{\citenamefont {Li}\ and\ \citenamefont {Ren}(2023)}]{li2023scale}%
  \BibitemOpen
  \bibfield  {author} {\bibinfo {author} {\bibfnamefont {Y.}~\bibnamefont {Li}}\ and\ \bibinfo {author} {\bibfnamefont {Y.-X.}\ \bibnamefont {Ren}},\ }\bibfield  {title} {\enquote {\bibinfo {title} {A scale-aware dispersion-relation-preserving finite difference scheme for computational aeroacoustics},}\ }\href@noop {} {\bibfield  {journal} {\bibinfo  {journal} {Physics of Fluids}\ }\textbf {\bibinfo {volume} {35}} (\bibinfo {year} {2023})}\BibitemShut {NoStop}%
\bibitem [{\citenamefont {Sengupta}, \citenamefont {Lakshmanan},\ and\ \citenamefont {Vijay}(2009)}]{sengupta2009new}%
  \BibitemOpen
  \bibfield  {author} {\bibinfo {author} {\bibfnamefont {T.~K.}\ \bibnamefont {Sengupta}}, \bibinfo {author} {\bibfnamefont {V.}~\bibnamefont {Lakshmanan}}, \ and\ \bibinfo {author} {\bibfnamefont {V.}~\bibnamefont {Vijay}},\ }\bibfield  {title} {\enquote {\bibinfo {title} {A new combined stable and dispersion relation preserving compact scheme for non-periodic problems},}\ }\href@noop {} {\bibfield  {journal} {\bibinfo  {journal} {Journal of Computational Physics}\ }\textbf {\bibinfo {volume} {228}},\ \bibinfo {pages} {3048--3071} (\bibinfo {year} {2009})}\BibitemShut {NoStop}%
\bibitem [{\citenamefont {Popescu}, \citenamefont {Shyy},\ and\ \citenamefont {Garbey}(2005)}]{popescu2005finite}%
  \BibitemOpen
  \bibfield  {author} {\bibinfo {author} {\bibfnamefont {M.}~\bibnamefont {Popescu}}, \bibinfo {author} {\bibfnamefont {W.}~\bibnamefont {Shyy}}, \ and\ \bibinfo {author} {\bibfnamefont {M.}~\bibnamefont {Garbey}},\ }\bibfield  {title} {\enquote {\bibinfo {title} {Finite volume treatment of dispersion-relation-preserving and optimized prefactored compact schemes for wave propagation},}\ }\href@noop {} {\bibfield  {journal} {\bibinfo  {journal} {Journal of Computational Physics}\ }\textbf {\bibinfo {volume} {210}},\ \bibinfo {pages} {705--729} (\bibinfo {year} {2005})}\BibitemShut {NoStop}%
\bibitem [{\citenamefont {Wang}\ \emph {et~al.}(2013)\citenamefont {Wang}, \citenamefont {Ren}, \citenamefont {Sun},\ and\ \citenamefont {Sun}}]{wang2013low}%
  \BibitemOpen
  \bibfield  {author} {\bibinfo {author} {\bibfnamefont {Q.}~\bibnamefont {Wang}}, \bibinfo {author} {\bibfnamefont {Y.}~\bibnamefont {Ren}}, \bibinfo {author} {\bibfnamefont {Z.}~\bibnamefont {Sun}}, \ and\ \bibinfo {author} {\bibfnamefont {Y.}~\bibnamefont {Sun}},\ }\bibfield  {title} {\enquote {\bibinfo {title} {Low dispersion finite volume scheme based on reconstruction with minimized dispersion and controllable dissipation},}\ }\href@noop {} {\bibfield  {journal} {\bibinfo  {journal} {Science China Physics, Mechanics and Astronomy}\ }\textbf {\bibinfo {volume} {56}},\ \bibinfo {pages} {423--431} (\bibinfo {year} {2013})}\BibitemShut {NoStop}%
\bibitem [{\citenamefont {Lele}(1992)}]{lele1992compact}%
  \BibitemOpen
  \bibfield  {author} {\bibinfo {author} {\bibfnamefont {S.~K.}\ \bibnamefont {Lele}},\ }\bibfield  {title} {\enquote {\bibinfo {title} {Compact finite difference schemes with spectral-like resolution},}\ }\href@noop {} {\bibfield  {journal} {\bibinfo  {journal} {Journal of Computational Physics}\ }\textbf {\bibinfo {volume} {103}},\ \bibinfo {pages} {16--42} (\bibinfo {year} {1992})}\BibitemShut {NoStop}%
\bibitem [{\citenamefont {Zhong}\ and\ \citenamefont {Tatineni}(2003)}]{zhong2003high}%
  \BibitemOpen
  \bibfield  {author} {\bibinfo {author} {\bibfnamefont {X.}~\bibnamefont {Zhong}}\ and\ \bibinfo {author} {\bibfnamefont {M.}~\bibnamefont {Tatineni}},\ }\bibfield  {title} {\enquote {\bibinfo {title} {High-order non-uniform grid schemes for numerical simulation of hypersonic boundary-layer stability and transition},}\ }\href@noop {} {\bibfield  {journal} {\bibinfo  {journal} {Journal of Computational Physics}\ }\textbf {\bibinfo {volume} {190}},\ \bibinfo {pages} {419--458} (\bibinfo {year} {2003})}\BibitemShut {NoStop}%
\bibitem [{\citenamefont {Spotz}(1998)}]{spotz1998formulation}%
  \BibitemOpen
  \bibfield  {author} {\bibinfo {author} {\bibfnamefont {W.}~\bibnamefont {Spotz}},\ }\bibfield  {title} {\enquote {\bibinfo {title} {Formulation and experiments with high-order compact schemes for nonuniform grids},}\ }\href@noop {} {\bibfield  {journal} {\bibinfo  {journal} {International Journal of Numerical Methods for Heat \& Fluid Flow}\ }\textbf {\bibinfo {volume} {8}},\ \bibinfo {pages} {288--303} (\bibinfo {year} {1998})}\BibitemShut {NoStop}%
\bibitem [{\citenamefont {Ge}\ and\ \citenamefont {Zhang}(2001)}]{ge2001high}%
  \BibitemOpen
  \bibfield  {author} {\bibinfo {author} {\bibfnamefont {L.}~\bibnamefont {Ge}}\ and\ \bibinfo {author} {\bibfnamefont {J.}~\bibnamefont {Zhang}},\ }\bibfield  {title} {\enquote {\bibinfo {title} {High accuracy iterative solution of convection diffusion equation with boundary layers on nonuniform grids},}\ }\href@noop {} {\bibfield  {journal} {\bibinfo  {journal} {Journal of Computational Physics}\ }\textbf {\bibinfo {volume} {171}},\ \bibinfo {pages} {560--578} (\bibinfo {year} {2001})}\BibitemShut {NoStop}%
\bibitem [{\citenamefont {Wang}\ \emph {et~al.}(2023)\citenamefont {Wang}, \citenamefont {Zhu}, \citenamefont {Wang},\ and\ \citenamefont {Zhao}}]{wang2023efficient}%
  \BibitemOpen
  \bibfield  {author} {\bibinfo {author} {\bibfnamefont {Z.}~\bibnamefont {Wang}}, \bibinfo {author} {\bibfnamefont {J.}~\bibnamefont {Zhu}}, \bibinfo {author} {\bibfnamefont {C.}~\bibnamefont {Wang}}, \ and\ \bibinfo {author} {\bibfnamefont {N.}~\bibnamefont {Zhao}},\ }\bibfield  {title} {\enquote {\bibinfo {title} {An efficient hybrid multi-resolution {WCNS} scheme for solving compressible flows},}\ }\href@noop {} {\bibfield  {journal} {\bibinfo  {journal} {Journal of Computational Physics}\ }\textbf {\bibinfo {volume} {477}},\ \bibinfo {pages} {111877} (\bibinfo {year} {2023})}\BibitemShut {NoStop}%
\bibitem [{\citenamefont {Gamet}\ \emph {et~al.}(1999)\citenamefont {Gamet}, \citenamefont {Ducros}, \citenamefont {Nicoud},\ and\ \citenamefont {Poinsot}}]{gamet1999compact}%
  \BibitemOpen
  \bibfield  {author} {\bibinfo {author} {\bibfnamefont {L.}~\bibnamefont {Gamet}}, \bibinfo {author} {\bibfnamefont {F.}~\bibnamefont {Ducros}}, \bibinfo {author} {\bibfnamefont {F.}~\bibnamefont {Nicoud}}, \ and\ \bibinfo {author} {\bibfnamefont {T.}~\bibnamefont {Poinsot}},\ }\bibfield  {title} {\enquote {\bibinfo {title} {Compact finite difference schemes on non-uniform meshes. {A}pplication to direct numerical simulations of compressible flows},}\ }\href@noop {} {\bibfield  {journal} {\bibinfo  {journal} {International Journal for Numerical Methods in Fluids}\ }\textbf {\bibinfo {volume} {29}},\ \bibinfo {pages} {159--191} (\bibinfo {year} {1999})}\BibitemShut {NoStop}%
\bibitem [{\citenamefont {Shukla}, \citenamefont {Tatineni},\ and\ \citenamefont {Zhong}(2007)}]{shukla2007very}%
  \BibitemOpen
  \bibfield  {author} {\bibinfo {author} {\bibfnamefont {R.~K.}\ \bibnamefont {Shukla}}, \bibinfo {author} {\bibfnamefont {M.}~\bibnamefont {Tatineni}}, \ and\ \bibinfo {author} {\bibfnamefont {X.}~\bibnamefont {Zhong}},\ }\bibfield  {title} {\enquote {\bibinfo {title} {Very high-order compact finite difference schemes on non-uniform grids for incompressible {N}avier-{S}tokes equations},}\ }\href@noop {} {\bibfield  {journal} {\bibinfo  {journal} {Journal of Computational Physics}\ }\textbf {\bibinfo {volume} {224}},\ \bibinfo {pages} {1064--1094} (\bibinfo {year} {2007})}\BibitemShut {NoStop}%
\bibitem [{\citenamefont {Shukla}\ and\ \citenamefont {Zhong}(2005)}]{shukla2005derivation}%
  \BibitemOpen
  \bibfield  {author} {\bibinfo {author} {\bibfnamefont {R.~K.}\ \bibnamefont {Shukla}}\ and\ \bibinfo {author} {\bibfnamefont {X.}~\bibnamefont {Zhong}},\ }\bibfield  {title} {\enquote {\bibinfo {title} {Derivation of high-order compact finite difference schemes for non-uniform grid using polynomial interpolation},}\ }\href@noop {} {\bibfield  {journal} {\bibinfo  {journal} {Journal of Computational Physics}\ }\textbf {\bibinfo {volume} {204}},\ \bibinfo {pages} {404--429} (\bibinfo {year} {2005})}\BibitemShut {NoStop}%
\bibitem [{\citenamefont {Gaitonde}\ and\ \citenamefont {Shang}(1997)}]{gaitonde1997optimized}%
  \BibitemOpen
  \bibfield  {author} {\bibinfo {author} {\bibfnamefont {D.}~\bibnamefont {Gaitonde}}\ and\ \bibinfo {author} {\bibfnamefont {J.}~\bibnamefont {Shang}},\ }\bibfield  {title} {\enquote {\bibinfo {title} {Optimized compact-difference-based finite-volume schemes for linear wave phenomena},}\ }\href@noop {} {\bibfield  {journal} {\bibinfo  {journal} {Journal of Computational Physics}\ }\textbf {\bibinfo {volume} {138}},\ \bibinfo {pages} {617--643} (\bibinfo {year} {1997})}\BibitemShut {NoStop}%
\bibitem [{\citenamefont {Kobayashi}(1999)}]{kobayashi1999class}%
  \BibitemOpen
  \bibfield  {author} {\bibinfo {author} {\bibfnamefont {M.~H.}\ \bibnamefont {Kobayashi}},\ }\bibfield  {title} {\enquote {\bibinfo {title} {On a class of {P}ad{\'e} finite volume methods},}\ }\href@noop {} {\bibfield  {journal} {\bibinfo  {journal} {Journal of Computational Physics}\ }\textbf {\bibinfo {volume} {156}},\ \bibinfo {pages} {137--180} (\bibinfo {year} {1999})}\BibitemShut {NoStop}%
\bibitem [{\citenamefont {Pereira}, \citenamefont {Kobayashi},\ and\ \citenamefont {Pereira}(2001)}]{pereira2001fourth}%
  \BibitemOpen
  \bibfield  {author} {\bibinfo {author} {\bibfnamefont {J.~M.}\ \bibnamefont {Pereira}}, \bibinfo {author} {\bibfnamefont {M.}~\bibnamefont {Kobayashi}}, \ and\ \bibinfo {author} {\bibfnamefont {J.~C.}\ \bibnamefont {Pereira}},\ }\bibfield  {title} {\enquote {\bibinfo {title} {A fourth-order-accurate finite volume compact method for the incompressible {N}avier-{S}tokes solutions},}\ }\href@noop {} {\bibfield  {journal} {\bibinfo  {journal} {Journal of Computational Physics}\ }\textbf {\bibinfo {volume} {167}},\ \bibinfo {pages} {217--243} (\bibinfo {year} {2001})}\BibitemShut {NoStop}%
\bibitem [{\citenamefont {Piller}\ and\ \citenamefont {Stalio}(2008)}]{piller2008compact}%
  \BibitemOpen
  \bibfield  {author} {\bibinfo {author} {\bibfnamefont {M.}~\bibnamefont {Piller}}\ and\ \bibinfo {author} {\bibfnamefont {E.}~\bibnamefont {Stalio}},\ }\bibfield  {title} {\enquote {\bibinfo {title} {Compact finite volume schemes on boundary-fitted grids},}\ }\href@noop {} {\bibfield  {journal} {\bibinfo  {journal} {Journal of Computational physics}\ }\textbf {\bibinfo {volume} {227}},\ \bibinfo {pages} {4736--4762} (\bibinfo {year} {2008})}\BibitemShut {NoStop}%
\bibitem [{\citenamefont {Fosso}\ \emph {et~al.}(2010)\citenamefont {Fosso}, \citenamefont {Deniau}, \citenamefont {Sicot},\ and\ \citenamefont {Sagaut}}]{fosso2010curvilinear}%
  \BibitemOpen
  \bibfield  {author} {\bibinfo {author} {\bibfnamefont {A.}~\bibnamefont {Fosso}}, \bibinfo {author} {\bibfnamefont {H.}~\bibnamefont {Deniau}}, \bibinfo {author} {\bibfnamefont {F.}~\bibnamefont {Sicot}}, \ and\ \bibinfo {author} {\bibfnamefont {P.}~\bibnamefont {Sagaut}},\ }\bibfield  {title} {\enquote {\bibinfo {title} {Curvilinear finite-volume schemes using high-order compact interpolation},}\ }\href@noop {} {\bibfield  {journal} {\bibinfo  {journal} {Journal of Computational Physics}\ }\textbf {\bibinfo {volume} {229}},\ \bibinfo {pages} {5090--5122} (\bibinfo {year} {2010})}\BibitemShut {NoStop}%
\bibitem [{\citenamefont {Lacor}, \citenamefont {Smirnov},\ and\ \citenamefont {Baelmans}(2004)}]{lacor2004finite}%
  \BibitemOpen
  \bibfield  {author} {\bibinfo {author} {\bibfnamefont {C.}~\bibnamefont {Lacor}}, \bibinfo {author} {\bibfnamefont {S.}~\bibnamefont {Smirnov}}, \ and\ \bibinfo {author} {\bibfnamefont {M.}~\bibnamefont {Baelmans}},\ }\bibfield  {title} {\enquote {\bibinfo {title} {A finite volume formulation of compact central schemes on arbitrary structured grids},}\ }\href@noop {} {\bibfield  {journal} {\bibinfo  {journal} {Journal of Computational Physics}\ }\textbf {\bibinfo {volume} {198}},\ \bibinfo {pages} {535--566} (\bibinfo {year} {2004})}\BibitemShut {NoStop}%
\bibitem [{\citenamefont {Wang}\ and\ \citenamefont {Ren}(2015)}]{wang2015accurate}%
  \BibitemOpen
  \bibfield  {author} {\bibinfo {author} {\bibfnamefont {Q.}~\bibnamefont {Wang}}\ and\ \bibinfo {author} {\bibfnamefont {Y.-X.}\ \bibnamefont {Ren}},\ }\bibfield  {title} {\enquote {\bibinfo {title} {An accurate and robust finite volume scheme based on the spline interpolation for solving the {E}uler and {N}avier-{S}tokes equations on non-uniform curvilinear grids},}\ }\href@noop {} {\bibfield  {journal} {\bibinfo  {journal} {Journal of Computational Physics}\ }\textbf {\bibinfo {volume} {284}},\ \bibinfo {pages} {648--667} (\bibinfo {year} {2015})}\BibitemShut {NoStop}%
\bibitem [{\citenamefont {Huang}\ \emph {et~al.}(2018)\citenamefont {Huang}, \citenamefont {Ren}, \citenamefont {Wang},\ and\ \citenamefont {Jiang}}]{huang2018high}%
  \BibitemOpen
  \bibfield  {author} {\bibinfo {author} {\bibfnamefont {W.-F.}\ \bibnamefont {Huang}}, \bibinfo {author} {\bibfnamefont {Y.-X.}\ \bibnamefont {Ren}}, \bibinfo {author} {\bibfnamefont {Q.}~\bibnamefont {Wang}}, \ and\ \bibinfo {author} {\bibfnamefont {X.}~\bibnamefont {Jiang}},\ }\bibfield  {title} {\enquote {\bibinfo {title} {High resolution finite volume scheme based on the quintic spline reconstruction on non-uniform grids},}\ }\href@noop {} {\bibfield  {journal} {\bibinfo  {journal} {Journal of Scientific Computing}\ }\textbf {\bibinfo {volume} {74}},\ \bibinfo {pages} {1816--1852} (\bibinfo {year} {2018})}\BibitemShut {NoStop}%
\bibitem [{\citenamefont {Huang}\ \emph {et~al.}(2022{\natexlab{a}})\citenamefont {Huang}, \citenamefont {Ren}, \citenamefont {Tu}, \citenamefont {Xianxu},\ and\ \citenamefont {Jianqiang}}]{huang2022adaptive}%
  \BibitemOpen
  \bibfield  {author} {\bibinfo {author} {\bibfnamefont {W.}~\bibnamefont {Huang}}, \bibinfo {author} {\bibfnamefont {Y.}~\bibnamefont {Ren}}, \bibinfo {author} {\bibfnamefont {G.}~\bibnamefont {Tu}}, \bibinfo {author} {\bibfnamefont {Y.}~\bibnamefont {Xianxu}}, \ and\ \bibinfo {author} {\bibfnamefont {C.}~\bibnamefont {Jianqiang}},\ }\bibfield  {title} {\enquote {\bibinfo {title} {An adaptive artificial viscosity method for quintic spline reconstruction scheme},}\ }\href@noop {} {\bibfield  {journal} {\bibinfo  {journal} {Computers \& Fluids}\ }\textbf {\bibinfo {volume} {240}},\ \bibinfo {pages} {105435} (\bibinfo {year} {2022}{\natexlab{a}})}\BibitemShut {NoStop}%
\bibitem [{\citenamefont {Wang}, \citenamefont {Ren},\ and\ \citenamefont {Li}(2016{\natexlab{a}})}]{wang2016compact1}%
  \BibitemOpen
  \bibfield  {author} {\bibinfo {author} {\bibfnamefont {Q.}~\bibnamefont {Wang}}, \bibinfo {author} {\bibfnamefont {Y.-X.}\ \bibnamefont {Ren}}, \ and\ \bibinfo {author} {\bibfnamefont {W.}~\bibnamefont {Li}},\ }\bibfield  {title} {\enquote {\bibinfo {title} {Compact high order finite volume method on unstructured grids {I}: {B}asic formulations and one-dimensional schemes},}\ }\href@noop {} {\bibfield  {journal} {\bibinfo  {journal} {Journal of Computational Physics}\ }\textbf {\bibinfo {volume} {314}},\ \bibinfo {pages} {863--882} (\bibinfo {year} {2016}{\natexlab{a}})}\BibitemShut {NoStop}%
\bibitem [{\citenamefont {Wang}, \citenamefont {Ren},\ and\ \citenamefont {Li}(2016{\natexlab{b}})}]{wang2016compact2}%
  \BibitemOpen
  \bibfield  {author} {\bibinfo {author} {\bibfnamefont {Q.}~\bibnamefont {Wang}}, \bibinfo {author} {\bibfnamefont {Y.-X.}\ \bibnamefont {Ren}}, \ and\ \bibinfo {author} {\bibfnamefont {W.}~\bibnamefont {Li}},\ }\bibfield  {title} {\enquote {\bibinfo {title} {Compact high order finite volume method on unstructured grids {II}: {E}xtension to two-dimensional {E}uler equations},}\ }\href@noop {} {\bibfield  {journal} {\bibinfo  {journal} {Journal of Computational Physics}\ }\textbf {\bibinfo {volume} {314}},\ \bibinfo {pages} {883--908} (\bibinfo {year} {2016}{\natexlab{b}})}\BibitemShut {NoStop}%
\bibitem [{\citenamefont {Cockburn}\ and\ \citenamefont {Shu}(1994)}]{cockburn1994nonlinearly}%
  \BibitemOpen
  \bibfield  {author} {\bibinfo {author} {\bibfnamefont {B.}~\bibnamefont {Cockburn}}\ and\ \bibinfo {author} {\bibfnamefont {C.-W.}\ \bibnamefont {Shu}},\ }\bibfield  {title} {\enquote {\bibinfo {title} {Nonlinearly stable compact schemes for shock calculations},}\ }\href@noop {} {\bibfield  {journal} {\bibinfo  {journal} {SIAM Journal on Numerical Analysis}\ }\textbf {\bibinfo {volume} {31}},\ \bibinfo {pages} {607--627} (\bibinfo {year} {1994})}\BibitemShut {NoStop}%
\bibitem [{\citenamefont {Adams}\ and\ \citenamefont {Shariff}(1996)}]{adams1996high}%
  \BibitemOpen
  \bibfield  {author} {\bibinfo {author} {\bibfnamefont {N.~A.}\ \bibnamefont {Adams}}\ and\ \bibinfo {author} {\bibfnamefont {K.}~\bibnamefont {Shariff}},\ }\bibfield  {title} {\enquote {\bibinfo {title} {A high-resolution hybrid compact-{ENO} scheme for shock-turbulence interaction problems},}\ }\href@noop {} {\bibfield  {journal} {\bibinfo  {journal} {Journal of Computational Physics}\ }\textbf {\bibinfo {volume} {127}},\ \bibinfo {pages} {27--51} (\bibinfo {year} {1996})}\BibitemShut {NoStop}%
\bibitem [{\citenamefont {Pirozzoli}(2002)}]{pirozzoli2002conservative}%
  \BibitemOpen
  \bibfield  {author} {\bibinfo {author} {\bibfnamefont {S.}~\bibnamefont {Pirozzoli}},\ }\bibfield  {title} {\enquote {\bibinfo {title} {Conservative hybrid compact-{WENO} schemes for shock-turbulence interaction},}\ }\href@noop {} {\bibfield  {journal} {\bibinfo  {journal} {Journal of Computational Physics}\ }\textbf {\bibinfo {volume} {178}},\ \bibinfo {pages} {81--117} (\bibinfo {year} {2002})}\BibitemShut {NoStop}%
\bibitem [{\citenamefont {Deng}\ and\ \citenamefont {Maekawa}(1997)}]{deng1997compact}%
  \BibitemOpen
  \bibfield  {author} {\bibinfo {author} {\bibfnamefont {X.}~\bibnamefont {Deng}}\ and\ \bibinfo {author} {\bibfnamefont {H.}~\bibnamefont {Maekawa}},\ }\bibfield  {title} {\enquote {\bibinfo {title} {Compact high-order accurate nonlinear schemes},}\ }\href@noop {} {\bibfield  {journal} {\bibinfo  {journal} {Journal of Computational Physics}\ }\textbf {\bibinfo {volume} {130}},\ \bibinfo {pages} {77--91} (\bibinfo {year} {1997})}\BibitemShut {NoStop}%
\bibitem [{\citenamefont {Guo}, \citenamefont {Shi},\ and\ \citenamefont {Li}(2016)}]{guo2016fifth}%
  \BibitemOpen
  \bibfield  {author} {\bibinfo {author} {\bibfnamefont {Y.}~\bibnamefont {Guo}}, \bibinfo {author} {\bibfnamefont {Y.-f.}\ \bibnamefont {Shi}}, \ and\ \bibinfo {author} {\bibfnamefont {Y.-m.}\ \bibnamefont {Li}},\ }\bibfield  {title} {\enquote {\bibinfo {title} {A fifth-order finite volume weighted compact scheme for solving one-dimensional {B}urgers' equation},}\ }\href@noop {} {\bibfield  {journal} {\bibinfo  {journal} {Applied Mathematics and Computation}\ }\textbf {\bibinfo {volume} {281}},\ \bibinfo {pages} {172--185} (\bibinfo {year} {2016})}\BibitemShut {NoStop}%
\bibitem [{\citenamefont {Li}, \citenamefont {PAN},\ and\ \citenamefont {Zeng}()}]{PANWCLS}%
  \BibitemOpen
  \bibfield  {author} {\bibinfo {author} {\bibfnamefont {L.}~\bibnamefont {Li}}, \bibinfo {author} {\bibfnamefont {J.}~\bibnamefont {PAN}}, \ and\ \bibinfo {author} {\bibfnamefont {W.-G.}\ \bibnamefont {Zeng}},\ }\bibfield  {title} {\enquote {\bibinfo {title} {A third-order weighted non-oscillatory compact least-squares scheme for hyperbolic conservation laws on non-uniform grids},}\ }\href@noop {} {\bibinfo  {journal} {Submission in progress}\ }\BibitemShut {NoStop}%
\bibitem [{\citenamefont {Wang}\ \emph {et~al.}(2017)\citenamefont {Wang}, \citenamefont {Ren}, \citenamefont {Pan},\ and\ \citenamefont {Li}}]{wang_compact_2017}%
  \BibitemOpen
\bibfield  {journal} {  }\bibfield  {author} {\bibinfo {author} {\bibfnamefont {Q.}~\bibnamefont {Wang}}, \bibinfo {author} {\bibfnamefont {Y.-X.}\ \bibnamefont {Ren}}, \bibinfo {author} {\bibfnamefont {J.}~\bibnamefont {Pan}}, \ and\ \bibinfo {author} {\bibfnamefont {W.}~\bibnamefont {Li}},\ }\bibfield  {title} {\enquote {\bibinfo {title} {Compact high order finite volume method on unstructured grids {III}: {Variational} reconstruction},}\ }\href@noop {} {\bibfield  {journal} {\bibinfo  {journal} {Journal of Computational Physics}\ }\textbf {\bibinfo {volume} {337}},\ \bibinfo {pages} {1--26} (\bibinfo {year} {2017})}\BibitemShut {NoStop}%
\bibitem [{\citenamefont {Gottlieb}, \citenamefont {Ketcheson},\ and\ \citenamefont {Shu}(2011)}]{gottlieb2011strong}%
  \BibitemOpen
  \bibfield  {author} {\bibinfo {author} {\bibfnamefont {S.}~\bibnamefont {Gottlieb}}, \bibinfo {author} {\bibfnamefont {D.}~\bibnamefont {Ketcheson}}, \ and\ \bibinfo {author} {\bibfnamefont {C.-W.}\ \bibnamefont {Shu}},\ }\href@noop {} {\emph {\bibinfo {title} {Strong stability preserving {R}unge-{K}utta and multistep time discretizations}}}\ (\bibinfo  {publisher} {World Scientific},\ \bibinfo {year} {2011})\BibitemShut {NoStop}%
\bibitem [{\citenamefont {PAN}\ and\ \citenamefont {Li}(2023)}]{PAN202324}%
  \BibitemOpen
  \bibfield  {author} {\bibinfo {author} {\bibfnamefont {J.}~\bibnamefont {PAN}}\ and\ \bibinfo {author} {\bibfnamefont {L.}~\bibnamefont {Li}},\ }\bibfield  {title} {\enquote {\bibinfo {title} {Third-order unconditional positivity-preserving schemes for reactive flows keeping both mass and mole balance},}\ }\href@noop {} {\bibfield  {journal} {\bibinfo  {journal} {Chinese Journal of Aeronautics}\ }\textbf {\bibinfo {volume} {36}},\ \bibinfo {pages} {24--41} (\bibinfo {year} {2023})}\BibitemShut {NoStop}%
\bibitem [{\citenamefont {Huang}\ \emph {et~al.}(2022{\natexlab{b}})\citenamefont {Huang}, \citenamefont {Ren}, \citenamefont {Wang},\ and\ \citenamefont {Pan}}]{huang_high-order_2022}%
  \BibitemOpen
  \bibfield  {author} {\bibinfo {author} {\bibfnamefont {Q.-M.}\ \bibnamefont {Huang}}, \bibinfo {author} {\bibfnamefont {Y.-X.}\ \bibnamefont {Ren}}, \bibinfo {author} {\bibfnamefont {Q.}~\bibnamefont {Wang}}, \ and\ \bibinfo {author} {\bibfnamefont {J.-H.}\ \bibnamefont {Pan}},\ }\bibfield  {title} {\enquote {\bibinfo {title} {High-order compact finite volume schemes for solving the {Reynolds} averaged {Navier}-{Stokes} equations on the unstructured mixed grids with a large aspect ratio},}\ }\href@noop {} {\bibfield  {journal} {\bibinfo  {journal} {Journal of Computational Physics}\ }\textbf {\bibinfo {volume} {467}},\ \bibinfo {pages} {111458} (\bibinfo {year} {2022}{\natexlab{b}})}\BibitemShut {NoStop}%
\bibitem [{\citenamefont {Hu}\ \emph {et~al.}(2012)\citenamefont {Hu}, \citenamefont {Tritschler}, \citenamefont {Pirozzoli},\ and\ \citenamefont {Adams}}]{hu2012dispersion}%
  \BibitemOpen
  \bibfield  {author} {\bibinfo {author} {\bibfnamefont {X.}~\bibnamefont {Hu}}, \bibinfo {author} {\bibfnamefont {V.}~\bibnamefont {Tritschler}}, \bibinfo {author} {\bibfnamefont {S.}~\bibnamefont {Pirozzoli}}, \ and\ \bibinfo {author} {\bibfnamefont {N.}~\bibnamefont {Adams}},\ }\bibfield  {title} {\enquote {\bibinfo {title} {Dispersion-dissipation condition for finite difference schemes},}\ }\href@noop {} {\bibfield  {journal} {\bibinfo  {journal} {arXiv preprint arXiv:1204.5088}\ } (\bibinfo {year} {2012})}\BibitemShut {NoStop}%
\bibitem [{\citenamefont {Baeza}\ \emph {et~al.}(2019)\citenamefont {Baeza}, \citenamefont {Bürger}, \citenamefont {Mulet},\ and\ \citenamefont {Zorío}}]{baeza_central_2019}%
  \BibitemOpen
  \bibfield  {author} {\bibinfo {author} {\bibfnamefont {A.}~\bibnamefont {Baeza}}, \bibinfo {author} {\bibfnamefont {R.}~\bibnamefont {Bürger}}, \bibinfo {author} {\bibfnamefont {P.}~\bibnamefont {Mulet}}, \ and\ \bibinfo {author} {\bibfnamefont {D.}~\bibnamefont {Zorío}},\ }\bibfield  {title} {\enquote {\bibinfo {title} {Central {WENO} schemes through a global average weight},}\ }\href@noop {} {\bibfield  {journal} {\bibinfo  {journal} {Journal of Scientific Computing}\ }\textbf {\bibinfo {volume} {78}},\ \bibinfo {pages} {499--530} (\bibinfo {year} {2019})}\BibitemShut {NoStop}%
\bibitem [{\citenamefont {Ren}, \citenamefont {Liu},\ and\ \citenamefont {Zhang}(2003)}]{ren_characteristic-wise_2003}%
  \BibitemOpen
  \bibfield  {author} {\bibinfo {author} {\bibfnamefont {Y.-X.}\ \bibnamefont {Ren}}, \bibinfo {author} {\bibfnamefont {M.}~\bibnamefont {Liu}}, \ and\ \bibinfo {author} {\bibfnamefont {H.}~\bibnamefont {Zhang}},\ }\bibfield  {title} {\enquote {\bibinfo {title} {A characteristic-wise hybrid compact-{WENO} scheme for solving hyperbolic conservation laws},}\ }\href@noop {} {\bibfield  {journal} {\bibinfo  {journal} {Journal of Computational Physics}\ }\textbf {\bibinfo {volume} {192}},\ \bibinfo {pages} {365--386} (\bibinfo {year} {2003})}\BibitemShut {NoStop}%
\bibitem [{\citenamefont {Baeza}\ \emph {et~al.}(2020)\citenamefont {Baeza}, \citenamefont {Bürger}, \citenamefont {Mulet},\ and\ \citenamefont {Zorío}}]{baeza2020efficient}%
  \BibitemOpen
  \bibfield  {author} {\bibinfo {author} {\bibfnamefont {A.}~\bibnamefont {Baeza}}, \bibinfo {author} {\bibfnamefont {R.}~\bibnamefont {Bürger}}, \bibinfo {author} {\bibfnamefont {P.}~\bibnamefont {Mulet}}, \ and\ \bibinfo {author} {\bibfnamefont {D.}~\bibnamefont {Zorío}},\ }\bibfield  {title} {\enquote {\bibinfo {title} {An efficient third-order {WENO} scheme with unconditionally optimal accuracy},}\ }\href@noop {} {\bibfield  {journal} {\bibinfo  {journal} {SIAM Journal on Scientific Computing}\ }\textbf {\bibinfo {volume} {42}},\ \bibinfo {pages} {A1028--A1051} (\bibinfo {year} {2020})}\BibitemShut {NoStop}%
\bibitem [{\citenamefont {Sanders}, \citenamefont {Morano},\ and\ \citenamefont {Druguet}(1998)}]{sanders1998multidimensional}%
  \BibitemOpen
  \bibfield  {author} {\bibinfo {author} {\bibfnamefont {R.}~\bibnamefont {Sanders}}, \bibinfo {author} {\bibfnamefont {E.}~\bibnamefont {Morano}}, \ and\ \bibinfo {author} {\bibfnamefont {M.-C.}\ \bibnamefont {Druguet}},\ }\bibfield  {title} {\enquote {\bibinfo {title} {Multidimensional dissipation for upwind schemes: stability and applications to gas dynamics},}\ }\href@noop {} {\bibfield  {journal} {\bibinfo  {journal} {Journal of Computational Physics}\ }\textbf {\bibinfo {volume} {145}},\ \bibinfo {pages} {511--537} (\bibinfo {year} {1998})}\BibitemShut {NoStop}%
\bibitem [{\citenamefont {Lax}(1954)}]{lax}%
  \BibitemOpen
  \bibfield  {author} {\bibinfo {author} {\bibfnamefont {P.~D.}\ \bibnamefont {Lax}},\ }\bibfield  {title} {\enquote {\bibinfo {title} {Weak solutions of nonlinear hyperbolic equations and their numerical computation},}\ }\href@noop {} {\bibfield  {journal} {\bibinfo  {journal} {Communications on Pure and Applied Mathematics}\ }\textbf {\bibinfo {volume} {7}},\ \bibinfo {pages} {159--193} (\bibinfo {year} {1954})}\BibitemShut {NoStop}%
\bibitem [{\citenamefont {Shu}\ and\ \citenamefont {Osher}(1989)}]{shu1989efficient}%
  \BibitemOpen
  \bibfield  {author} {\bibinfo {author} {\bibfnamefont {C.-W.}\ \bibnamefont {Shu}}\ and\ \bibinfo {author} {\bibfnamefont {S.}~\bibnamefont {Osher}},\ }\bibfield  {title} {\enquote {\bibinfo {title} {Efficient implementation of essentially non-oscillatory shock-capturing schemes, {II}},}\ }\href@noop {} {\bibfield  {journal} {\bibinfo  {journal} {Journal of computational physics}\ }\textbf {\bibinfo {volume} {83}},\ \bibinfo {pages} {32--78} (\bibinfo {year} {1989})}\BibitemShut {NoStop}%
\bibitem [{\citenamefont {Woodward}\ and\ \citenamefont {Colella}(1984)}]{woodward_numerical_1984}%
  \BibitemOpen
  \bibfield  {author} {\bibinfo {author} {\bibfnamefont {P.}~\bibnamefont {Woodward}}\ and\ \bibinfo {author} {\bibfnamefont {P.}~\bibnamefont {Colella}},\ }\bibfield  {title} {\enquote {\bibinfo {title} {The numerical simulation of two-dimensional fluid flow with strong shocks},}\ }\href@noop {} {\bibfield  {journal} {\bibinfo  {journal} {Journal of Computational Physics}\ }\textbf {\bibinfo {volume} {54}},\ \bibinfo {pages} {115--173} (\bibinfo {year} {1984})}\BibitemShut {NoStop}%
\bibitem [{\citenamefont {Daru}\ and\ \citenamefont {Tenaud}(2009)}]{daru_numerical_2009}%
  \BibitemOpen
  \bibfield  {author} {\bibinfo {author} {\bibfnamefont {V.}~\bibnamefont {Daru}}\ and\ \bibinfo {author} {\bibfnamefont {C.}~\bibnamefont {Tenaud}},\ }\bibfield  {title} {\enquote {\bibinfo {title} {Numerical simulation of the viscous shock tube problem by using a high resolution monotonicity-preserving scheme},}\ }\href@noop {} {\bibfield  {journal} {\bibinfo  {journal} {Computers \& Fluids}\ }\textbf {\bibinfo {volume} {38}},\ \bibinfo {pages} {664--676} (\bibinfo {year} {2009})}\BibitemShut {NoStop}%
\end{thebibliography}%

\appendix
\section{Coefficients in the modified non-dimensional wavenumber of fifth-order CLS-CD scheme}
The detailed expressions of $n_{15}, n_{16}, \cdots, n_{25}$ and $d_{12}, d_{13}, \cdots, d_{16}$ in Eq. (\ref{eq:dispdisswang5}) are 
\begin{align}
n_{15} & = \left(720 W_2^2+\left(12672 W_3+18144 W_4+\frac{1925}{4}\right) W_2 \right. \nonumber\\
& \left. +\frac{37587 W_3}{10}+3960 W_4+5184 W_3 \left(5 W_3+7 W_4\right) \right.\nonumber \\
& \left. +\frac{37}{6}\right) W_1^2 +\left(\left(18144 W_3+36288 W_4+\frac{39513}{10}\right)\right. \nonumber \\
& \left.\cdot W_2^2  +\left(36288 W_3^2+\left(54432 W_4+\frac{758337}{20}\right) W_3\right.\right.\nonumber\\
& \left. \left. +44640 W_4+\frac{3281}{15}\right) W_2+37368 W_3^2+948 W_4\right.\nonumber\\
&\left. +W_3 \left(49356 W_4+\frac{9579}{10}\right)\right) W_1+432 W_3 \left(17 W_3 \right. \nonumber \\ 
& \left. +22 W_4\right) +3 W_2^2 \left(7344 W_3+14688 W_4+455\right)\nonumber\\
&+\frac{3}{5} W_2 \left(W_3 \left(73440 W_3+14387\right)+360 \left(294 W_3 \right.\right.\nonumber \\
& \left.\left.+41\right) W_4\right),
\end{align}
\begin{align}
n_{16} &=\frac{1}{60} \left(\left(-34560 W_2^2-9 \left(72960 W_3+120960 W_4\right.\right. \right. \nonumber\\
&\left.\left.\left. -263\right) W_2 -270 W_3 \left(4608 W_3+61\right)-4320 \left(504 W_3 \right.\right.\right. \nonumber \\
&\left.\left.\left. +31\right) W_4-8\right) W_1^2+2 \left(-19440 \left(56 W_2+37\right) W_3^2\right.\right. \nonumber \\
&\left.\left. -9 \left(224 W_2 \left(270 W_2+191\right)+227\right) W_3\right.\right.\nonumber\\
&\left.\left.-11 W_2 \left(1143 W_2+11\right)-360 \left(3708 W_3 \right.\right.\right.\nonumber\\
&\left.\left.\left. +12 W_2 \left(252 W_2+432 W_3+223\right)+43\right) W_4\right) W_1\right.\nonumber\\
&\left.+72 \left(-145 W_2^2-4 \left(4050 W_2+463\right) W_3 W_2\right.\right.\nonumber\\
&\left.\left.-3600 \left(9 W_2+1\right) W_3^2\right.\right.\nonumber\\
&\left.\left.-180 \left(38 W_3+3 W_2 \left(60 W_2+104 W_3+9\right)\right) W_4\right)\right),
\end{align}
\begin{align}
n_{17} &=\left(144 W_2^2+\left(3456 W_3+7776 W_4+\frac{109}{20}\right) W_2 \right.\nonumber \\
& \left.+\frac{195 W_3}{2}+216 W_4+5184 W_3 \left(W_3+3 W_4\right)+\frac{1}{30}\right) W_1^2\nonumber\\
& +\frac{3}{40} \left(4 \left(25920 W_3+51840 W_4+383\right) W_2^2\nonumber \right.\\
&\left. +3 \left(69120 W_3^2+13943 W_3+17280 \left(9 W_3+2\right) W_4\right.\right.\nonumber\\
&\left.\left. +8\right) W_2+4 W_3 \left(12240 W_3+53\right)+240 \left(873 W_3+2\right) \right.\nonumber \\
&\left. \cdot W_4\right) W_1+432 W_3 \left(W_3+6 W_4\right)\nonumber \\
&+3 W_2^2 \left(2160 W_3+4320 W_4+7\right)\nonumber\\
&+\frac{9}{5} W_2 \left(W_3 \left(7200 W_3+197\right)+120 \left(153 W_3+5\right) W_4\right),
\end{align}
\begin{align}
n_{18}&=\frac{1}{120} \left(3 \left(-462 W_3-1440 \left(72 W_3+1\right) W_4\right.\right.\nonumber\\
&\left.\left.-5 W_2 \left(2304 W_3+10368 W_4+1\right)\right) W_1^2\right.\nonumber\\
&\left.+2 \left(-6480 \left(24 W_2+1\right) W_3^2-27 \left(64 W_2 \left(45 W_2+8\right)\right.\right.\right.\nonumber \\
& \left.\left.\left. +3\right) W_3-W_2 \left(27 W_2+1\right)-360 \left(324 W_3+12 W_2 \right.\right.\right.\nonumber \\
&\left.\left.\left. \cdot \left(36 W_2 +144 W_3+13\right)+1\right) W_4\right) W_1 \right.\nonumber\\
&\left. +288 \left(-W_2 W_3 \left(270 W_2 +540 W_3+7\right)\right.\right.\nonumber \\
& \left.\left. -45 \left(2 W_3+W_2 \left(12 W_2+72 W_3+1\right)\right) W_4\right)\right),
\end{align}
\begin{align}
n_{19}&=\frac{3}{40} W_3 \left(720 \left(W_1+12 \left(2 W_1+1\right) W_2\right) W_4-W_1 W_2\right),
\end{align}
\begin{align}
n_{20}&=\frac{1}{24} \left(-34560 W_2^2-9 \left(72960 W_3+137\right) W_2 \right.\nonumber\\
& \left. -270 W_3 \left(4608 W_3+61\right)-30240 \left(36 W_2+72 W_3 \right.\right.\nonumber\\
&\left.\left. +1\right) W_4-8\right) W_1^2+\frac{1}{60} \left(-680400 \left(8 W_2+3\right) W_3^2\right.\nonumber\\
&\left. -9 \left(W_2 \left(302400 W_2+132317\right)+835\right) W_3\right.\nonumber\\
&\left.-5 W_2 \left(13653 W_2+211\right)-2520 \left(1422 W_3\right.\right.\nonumber\\
&\left.\left.+12 W_2 \left(180 W_2+324 W_3+65\right)+5\right) W_4\right) W_1\nonumber\\
&-6 \left(35 \left(108 W_3+216 W_4+1\right) W_2^2\right.\nonumber\\
&\left. +2 \left(4 W_3 \left(945 W_3+52\right)+63 \left(108 W_3+5\right) W_4\right) W_2\right.\nonumber\\
&\left. +180 W_3 \left(4 W_3+7 W_4\right)\right),
\end{align}
\begin{align}
n_{21} &= \frac{1}{10} \left(21600 W_2^2+420480 W_3 W_2+775 W_2\right.\nonumber \\
&\left.+777600 W_3^2+10698 W_3+20160 \left(36 W_2+72 W_3+1\right) \right.\nonumber \\
&\left. \cdot W_4+5\right) W_1^2+\frac{1}{60} \left(108 \left(40320 W_3+80640 W_4+949\right) \right.\nonumber \\
&\left. \cdot W_2^2+\left(8709120 W_3^2+1834695 W_3+120960 \left(135 W_3\right.\right.\right.\nonumber\\
&\left.\left.\left. +26\right) W_4+1586\right) W_2+108 W_3 \left(28560 W_3+107\right)\right.\nonumber\\
&\left. +5040 \left(1143 W_3+4\right) W_4\right) W_1\nonumber\\
&+\frac{6}{5} W_2 W_3 \left(60480 W_3+3169\right)\nonumber\\
& +3024 W_2 \left(45 W_3+2\right) W_4+432 W_3 \left(15 W_3+28 W_4\right)\nonumber\\
& +63 W_2^2 \left(576 W_3+1152 W_4+5\right),
\end{align}
\begin{align}
n_{22} &= \frac{1}{30} \left(-3 \left(8640 W_2^2+\left(184320 W_3+317\right) W_2\right.\right.\nonumber\\
& \left.\left.+6 W_3 \left(51840 W_3+821\right)+2\right) W_1^2\right.\nonumber\\
&\left.-30240 \left(36 W_2+72 W_3+1\right) W_4 W_1^2\right.\nonumber\\
&\left. +2 \left(-45360 \left(24 W_2+7\right) W_3^2-9 \left(W_2 \left(60480 W_2\right.\right.\right.\right.\nonumber\\
&\left.\left.\left.\left. +22537\right)+141\right) W_3-W_2 \left(10287 W_2+160\right)\right.\right.\nonumber\\
&\left.\left.-360 \left(2034 W_3+12 W_2 \left(252 W_2+540 W_3+91\right)\right.\right.\right.\nonumber\\
&\left.\left.\left. +7\right) W_4\right) W_1+36 \left(-105 W_2^2-2 \left(7560 W_2+673\right)\right.\right.\nonumber\\
&\left.\left.\cdot W_3 W_2-2160 \left(14 W_2+1\right) W_3^2-360 \left(14 W_3\right.\right.\right.\nonumber\\
&\left.\left.\left. +W_2 \left(84 W_2+180 W_3+7\right)\right) W_4\right)\right),
\end{align}
\begin{align}
n_{23} &= \frac{1}{30} \left(3618 W_3+8640 W_4+9 \left(480 W_2^2+\left(13440 W_3 \right.\right.\right.\nonumber\\
&\left.\left.\left. +34560 W_4+19\right) W_2+17280 W_3 \left(W_3+4 W_4\right)\right)+1\right) W_1^2\nonumber\\
&+\left(\left(10368 W_3+20736 W_4+\frac{579}{5}\right) W_2^2+\left(20736 W_3^2 \right.\right.\nonumber\\
&\left.\left.+\left(58320 W_4+\frac{120873}{40}\right) W_3+7488 W_4+\frac{11}{6}\right) W_2 \right.\nonumber\\
&\left. +\frac{93 W_3}{5}+48 W_4+54 W_3 \left(72 W_3+269 W_4\right)\right) W_1 \nonumber\\
& +432 W_3 \left(W_3+4 W_4\right)+3 W_2^2 \left(1728 W_3+3456 W_4+7\right) \nonumber\\
& +\frac{6}{5} W_2 \left(W_3 \left(8640 W_3+271\right)\right.\nonumber\\
& \left.+180 \left(135 W_3+4\right) W_4\right),
\end{align}
\begin{align}
n_{24} & = \frac{1}{40} \left(-11520 W_3 W_2-5 W_2-462 W_3 \right.\nonumber\\
&\left. -1440 \left(36 W_2+72 W_3+1\right) W_4\right) W_1^2 \nonumber\\
& +\frac{1}{60} \left(-6480 \left(24 W_2+1\right) W_3^2 \right. \nonumber\\
&\left.-9 \left(5 W_2 \left(1728 W_2+307\right)+9\right) W_3 \right.\nonumber\\
&\left.-W_2 \left(27 W_2+1\right)\right) W_1\nonumber\\
&-6 \left(342 W_3+12 W_2 \left(36 W_2+180 W_3+13\right)+1\right) W_4 W_1 \nonumber\\
&-\frac{12}{5} W_2 W_3 \left(270 W_2+540 W_3+7\right)\nonumber\\
&-108 \left(12 W_2^2+60 W_3 W_2+W_2+2 W_3\right) W_4 ,
\end{align}
\begin{align}
n_{25} &= 54 \left(24 W_2 W_1+W_1+12 W_2\right) W_3 W_4-\frac{3}{40} W_1 W_2 W_3,
\end{align}
\begin{align}
d_{12}&=\left(864 W_2^2+9 \left(1792 W_3+2880 W_4+37\right) W_2 \right.\nonumber\\
& \left. +1728 W_4+18 W_3 \left(1728 W_3+2880 W_4+131\right)+4\right) W_1^2 \nonumber\\
& +18 \left(\left(1440 W_3+2880 W_4+157\right) W_2^2+\left(1616 W_4 \right.\right.\nonumber\\
& \left.\left.+W_3 \left(2880 W_3+5040 W_4+1493\right)+8\right) W_2+24 W_4 \right.\nonumber\\
& \left. +2 W_3 \left(792 W_3+1224 W_4+17\right)\right) W_1+18 \left(2 \left(432 W_3\right.\right.\nonumber\\
&\left.\left. +864 W_4+25\right) W_2^2+3 \left(96 W_4+W_3 \left(576 W_3+960 W_4 \right.\right.\right.\nonumber\\
&\left.\left.\left. +107\right)\right) W_2+144 W_3 \left(2 W_3+3 W_4\right)\right),
\end{align}
\begin{align}
d_{13}&=-2 \left(576 W_2^2+2 \left(5616 W_3+9720 W_4-55\right) W_2 \right.\nonumber\\
& \left. +576 W_4+108 W_3 \left(192 W_3+360 W_4-5\right)-1\right) W_1^2\nonumber\\
& -\left(432 \left(90 W_3+180 W_4-1\right) W_2^2+\left(32832 W_4 \right.\right.\nonumber\\
&\left.\left. +9 W_3 \left(8640 W_3+16128 W_4+985\right)-70\right) W_2 \right.\nonumber\\
&\left. +36 \left(8 W_4+W_3 \left(864 W_3+1692 W_4-7\right)\right)\right) W_1\nonumber\\
& -36 \left(\left(576 W_3+1152 W_4-5\right) W_2^2+4 \left(36 W_4 \right.\right.\nonumber\\
& \left.\left.+W_3 \left(288 W_3+540 W_4+5\right)\right) W_2\right.\nonumber\\
& \left. +144 W_3 \left(W_3+2 W_4\right)\right),
\end{align}
\begin{align}
d_{14}&=\left(288 W_2^2+6912 W_3 W_2+23 W_2+10368 W_3^2+18 W_3\right.\nonumber\\
& \left. +576 \left(27 W_2+54 W_3-1\right) W_4\right) W_1^2+2 \left(9 \left(864 W_3\right.\right.\nonumber\\
&\left.\left. +1728 W_4+11\right) W_2^2+2160 W_4 W_2+9 W_3 \left(1728 W_3\right.\right.\nonumber\\
&\left.\left. +4032 W_4+155\right) W_2+W_2-72 W_4\right.\nonumber\\
&\left.+1296 W_3 \left(W_3+7 W_4\right)\right) W_1\nonumber\\
& +18 W_2 W_3 \left(288 W_2+576 W_3+7\right)\nonumber\\
& +2592 \left(W_3+4 W_2 \left(W_2+3 W_3\right)\right) W_4,
\end{align}
\begin{align}
d_{15}&=-9 W_1 W_2 W_3 \left(64 W_1+288 W_2+576 W_3+7\right)\nonumber\\
& -144 \left(2 W_1 W_2 \left(9 W_1+18 W_2+2\right)\right.\nonumber\\
& \left.+9 \left(4 W_1+1\right) \left(W_1+4 W_2\right) W_3\right) W_4,
\end{align}
\begin{align}
d_{16}&=2592 W_1 W_2 W_3 W_4.
\end{align}

\end{document}